\documentclass[11pt]{article}
\usepackage{amsfonts}
\usepackage{graphicx}
\usepackage{amsmath}
\usepackage{amssymb}
\usepackage{makecell}
\usepackage{subfigure}
\usepackage{subcaption} 
\usepackage{enumitem}
\usepackage{epsfig}
\usepackage{graphicx}
\usepackage{color}
\usepackage{indentfirst}
\usepackage{setspace}
\usepackage{braket}
\usepackage{slashed}
\usepackage{cite}
\usepackage{url}
\usepackage{multirow}
\usepackage{cancel}
\usepackage{caption}
\usepackage{array}
\usepackage{appendix}
\usepackage{titlesec}
\usepackage{etoolbox}
\usepackage{hyperref}
\usepackage{newtxtext}
\usepackage{txfonts}
\usepackage[normalem]{ulem}
\usepackage{comment}
\usepackage{makecell}
\usepackage{mathrsfs}
\allowdisplaybreaks
\titleformat{\section}{\normalfont\large\bfseries}{\thesection}{1em}{} 
\titleformat{\subsection}[runin]{\normalfont\bfseries\itshape}{\thesubsection}{1em}{} 
\titleformat{\subsubsection}[runin]{\normalfont\itshape}{\thesubsubsection}{1em}{} 
\hypersetup{colorlinks=true, citecolor=blue, urlcolor=blue, linkcolor=red, linktocpage=true}
\begin{document}
\title{Baryons and baryoniums in the perspective of QCD sum rules}
\author{Sheng-Qi Zhang$ ^{1, 2} $, and Cong-Feng Qiao$^{1}\footnote{Contact author: qiaocf@ucas.ac.cn.}$\\\\
\normalsize{\textit{$^{1}$School of Physical Sciences, University of Chinese Academy of Sciences, Beijing 100049, China}}
\\
\normalsize{\textit{$^{2}$Center for High Energy Physics, Peking University, Beijing 100871, China}}\\\\}
\date{}
\maketitle

\begin{abstract}
Following the experimental confirmation of tetraquark and pentaquark states, the search for hexaquark states has emerged as a new frontier in hadron physics. The recent observation of $X(1840)$ and $X(1880)$ by BESIII collaboration has provided evidence for the predicted $p\bar{p}$ bound states. Such baryon-antibaryon configurations, encompassing both bound states and resonances, are commonly referred to as baryoniums and are regarded as promising hexaquark candidates. In this article, we provide a comprehensive review of the investigations into baryonium states and their constituents, i.e., baryons, within the framework of QCD sum rules. We delineate the fundamental calculation procedures of this method to facilitate its practical application and benchmark the theoretical predictions against alternative models as well as the latest experimental data. 

\end{abstract}
%\tableofcontents
\vspace{1cm}

\section{Introduction}\label{sec:introduction}

Hadrons, constituting the cornerstone of visible matter, occupy a focal point in particle physics research. According to the quark model, these composite entities arise from quarks and gluons governed by the strong interaction~\cite{Gell-Mann:1964ewy,Zweig:1964ruk,Zweig:1964jf}. Structurally, hadrons are categorized into the conventional and exotic sectors. The conventional sector includes mesons and baryons, characterized by valence quark-antiquark ($q\bar{q}$) and three-quark ($qqq$) configurations, respectively. Conversely, the exotic sector includes multiquark systems, hybrids, and glueballs composed entirely of gluonic matter.

Due to the color confinement property of QCD, free quarks and gluons cannot exist in isolation, and thus must be confined within color-singlet hadrons. As a result, hadrons are the only strong interaction particles that can be observed in experiments. Furthermore, hadrons are formed in the low-energy regime, where the momentum transfer in their decay processes is typically not large enough for perturbative methods to be applicable. Instead, nonperturbative effects become significant. Hence, it is of critical importance to develop nonperturbative models, which should satisfy the fundamental principles of QCD and remain consistent with experimental data. At present, several well-established nonperturbative theoretical models are widely employed, including various quark models~\cite{Gell-Mann:1964ewy, Zweig:1964jf, Zweig:1964ruk, Godfrey:1985xj, Capstick:1986ter}, the MIT bag model~\cite{Chodos:1974je}, the flux tube model~\cite{Isgur:1984bm}, light-cone sum rules~\cite{Balitsky:1989ry}, lattice QCD~\cite{Wilson:1974sk}, and QCD sum rules~\cite{Shifman:1978bx, Shifman:1978by}, among others. However, all theoretical calculations in QCD are performed at the quark and gluon level, which differs from the hadronic observables measured in experiments. To bridge this gap between theoretical predictions and experimental observations, Poggio, Quinn, and Weinberg proposed the smearing method in 1975~\cite{Poggio:1975af}, which suggests that certain inclusive hadronic cross sections at high energies, when appropriately averaged over an energy range, can approximately coincide with those calculated using quark-gluon perturbation theory. Since its proposal, this idea has been widely applied across various areas of QCD, including QCD sum rules, which are mainly discussed in this review.

QCD sum rules, which were established over four decades ago by Shifman, Vainshtein, and Zakharov~\cite{Shifman:1978bx,Shifman:1978by}, have become one of the most widely used methods in the study of hadron physics. To date, the two seminal papers have been cited over 9,000 times in total, highlighting their profound and lasting impact on the field. Rather than relying on constituent quark or gluon degrees of freedom, QCD sum rules employ interpolating currents built from well-defined quark and gluon fields to describe hadrons. An approximate phenomenological treatment of nonperturbative effects is achieved through the operator product expansion, which allows the correlation functions of the interpolating currents to be calculated at the quark-gluon level. The long-distance interaction can be parametrized in terms of the vacuum expectation values of quark and gluon composite operators, commonly referred to as vacuum condensates. These parameters, which are independent of the specific hadron under consideration, are universal and depend only on the energy scale. At the hadronic level, the same correlation function is expressed in terms of physical observables such as masses and coupling constants. By invoking dispersion relations and quark-hadron duality, the two descriptions can be matched, enabling the extraction of hadronic parameters. QCD sum rules thereby provide a powerful theoretical tool for predicting hadronic properties. The fundamental principles are presented in chapter~\ref{sec:method}. The readers are also referred to the following reviews~\cite{Novikov:1977dq,Reinders:1984sr,Narison:1989aq,Shuryak:1993kg,deRafael:1997ea,Colangelo:2000dp,Shifman:2000jv,Shifman:2001qm} for detailed discussions of this method.

QCD sum rules were originally formulated for the study of charmoniums~\cite{Shifman:1978bx,Shifman:1978by}, which were subsequently extended to baryons and electromagnetic form factors by Ioffe, Chung, Nesterenko, and their collaborators~\cite{Ioffe:1981kw,Chung:1981wm,Ioffe:1982ia,Nesterenko:1982gc,Ioffe:1982qb}. Since the observation of the $X(3872)$ in 2003, QCD sum rules have also played a remarkably important role in the study of exotic hadron spectrum. Here we highlight two representative examples: 1) Qiao et al. constructed interpolating currents for the unconventional $0^{--}$ trigluon glueballs and predicted their masses, production mechanisms, and possible decay modes~\cite{Qiao:2014vva}, which has significantly advanced the study of glueball physics. 2) Chen et al. provided a theoretical interpretation of the resonances $P_c(4380)$ and $P_c(4450)$ observed by LHCb, identifying them as hidden-charm pentaquark candidates~\cite{Chen:2015moa}. Their analysis suggests that $P_c(4380)$ may be interpreted as a $\bar{D}^* \Sigma_c$ molecular state with $J^P = \frac{3}{2}^{-}$, while $P_c(4450)$ could be a mixture of $\bar{D}^* \Lambda_c$ and $\bar{D} \Sigma_c^*$ with $J^P = \frac{5}{2}^{+}$. The applications of QCD sum rules to other exotic states can be found in recent papers~\cite{Tang:2015twt,Agaev:2016mjb,Azizi:2016dhy,Agaev:2016dev,Wang:2017jtz,Tang:2019nwv,Chen:2019bip,Wan:2020oxt,Chen:2020aos,Yang:2020wkh,Wan:2020fsk,Chen:2020uif,Wang:2020dgr,Wang:2020xyc,Albuquerque:2020hio,Albuquerque:2020ugi,Albuquerque:2021tqd,Xin:2021wcr,Chen:2021bck,Agaev:2021vur,Tang:2021zti,Wan:2022xkx,Su:2022fqr,Zhang:2022obn,Lian:2023cgs,Wan:2024ykm,Tang:2024zvf,Wan:2024pet,Duan:2024uuf,Tang:2024kmh,Wan:2024dmi,Zhang:2024jvv,Wan:2024fam,Xu:2025oqn} and reviews~\cite{Nielsen:2009uh,Albuquerque:2018jkn,Wang:2025sic}. Additionally, a review for recent status of the QCD sum rule approach in vacuum and in hot or dense matter can be found in Ref.~\cite{Gubler:2018ctz}. Due to the generally encouraging results and their consistency with experimental data, Colangelo and Khodjamirian stated in their review that QCD sum rule predictions are among the most reliable ones when estimating unknown hadronic parameters~\cite{Colangelo:2000dp}. 

In the past two decades, the existence of tetraquark and pentaquark states has been firmly established by experiments. Although their internal structures remain controversial, such as whether they are compact multiquark states or loosely bound molecular configurations, a series of experimental observations unambiguously confirms the existence of exotic hadrons beyond the traditional quark model. However, there is still no conclusive evidence for the existence of hexaquark states apart from the well-known deuteron, which highlights the theoretical and experimental importance of exploring bound states with larger numbers of quarks. Encouragingly, recent experimental observations from the BESIII collaboration have confirmed the existence of $X(1840)$ and $X(1880)$ as new particles~\cite{BESIII:2023vvr}, which are promising candidates for protonium~\cite{Ma:2024gsw}. Bound states composed of a baryon and an antibaryon, such as protonium, are generally referred to as baryoniums. Here, we aim to clarify the terminology of ``baryonium". The term is inspired by ``quarkonium", which traditionally refers to bound states formed by a quark and its corresponding antiquark. In this review, we adopt the term ``baryonium" to denote bound states formed by a baryon-antibaryon pair. In addition, the constituent baryons in baryoniums are in principle not restricted to color-singlet configurations, which is analogous to quarkonium where the constituent quarks themselves are not color singlets. This is slightly different from the situation when discussing hadronic molecular states, whose component hadrons are strictly restricted to color singlets. Such a picture suggests that baryonium may involve more compact and intrinsically multiquark configurations beyond simple hadronic molecules.

Since the beginning of the new millennium, in addition to the candidates for protonium, several newly discovered exotic hadrons such as the $Y(4260)$, $Y(4360)$, and $Y(4630)$ which contain a $c\bar{c}$ component, have been interpreted as hidden-charm baryonium states~\cite{Qiao:2005av,Qiao:2007ce,Lee:2011rka,Chen:2011cta,Meguro:2011nr,Li:2012bt,Chen:2013sba}. However, the possibility that these states are hidden-charm tetraquark configurations cannot be ruled out. In fact, the discussion of baryonium states always faces various theoretical approaches and experimental observations, which are often difficult to reconcile with one another, or even internally within a given framework. In this review, we will provide a brief overview of the historical studies on baryonium, with particular emphasis on the structural analysis and mass predictions from the perspective of QCD sum rules.

As the constituent particles of baryoniums, baryons play a crucial role in determining their properties. The proton and the neutron, being the two relatively stable nucleons, are not only the most thoroughly studied baryons but also the best understood hadrons overall. Since the pioneering calculation performed by Ioffe et al. within QCD sum rules for the masses and decay constants of the proton and the $\Delta$ baryon, this approach has been increasingly applied in the study of baryons. Besides, the form factors, which are the most important physical quantities in semileptonic decays, can be fully determined using QCD sum rules. Once the form factors are obtained, one can construct experimentally relevant observables such as branching ratios and asymmetry parameters. Certain observables are particularly useful for testing the standard model or probing its possible extensions. Therefore, the QCD sum rules approach also finds significant applications in the study of flavor physics. Most existing reviews on QCD sum rules focus on mesons or exotic hadrons, while systematic summaries on baryon spectroscopy and their semileptonic decays remain relatively scarce. In this review, we aim to fill this gap by providing a dedicated discussion of these topics, which are presented in chapters~\ref{sec:spectra} and~\ref{sec:decay}, respectively.

\section{Fundamental principles}\label{sec:method}
QCD sum rules (QCDSR) were established by M.A. Shifman, A.I. Vainshtein, and V.I. Zakharov in 1979~\cite{Shifman:1978bx, Shifman:1978by}. It is an analytic method based on QCD first principle. Over the past few decades, this approach has achieved significant success in describing hadron structure and decay dynamics. In this chapter, we outline the general procedure for applying QCD sum rules to extract valuable information about hadrons, such as masses, coupling constants, form factors, and more.

\subsection{Correlation functions}\label{sec2.1}\

QCD sum rules begin with constructing the correlation functions, with a general two-point correlation function expressed as:
\begin{align}
\Pi(q^2) = i\int d^4 x\, e^{iq\cdot x}\bra{0}T\{j(x),\,j^\dagger(0) \}\ket{0}\,.
\label{2ptcf}
\end{align}
Here, $ q $ is the total four-momentum carried by the hadrons and $T$ represents the time-ordered product of the operators. The operators $ j(x) $, composed of quark and gluon fields, are designed to capture the internal structure and quantum numbers of the hadrons being studied. They are often referred to as interpolating currents. $ \ket{0} $ denotes the QCD vacuum.

The two-point correlation functions~(\ref{2ptcf}) are used to extract the mass and decay constant of the hadrons, while the three-point correlation functions are needed to study hadron decays. The three-point correlation function is defined as:
\begin{align}
\Pi_\mu(q_1^2, q_2^2, q^2)=i^2 \int d^4 x\; d^4 y\; e^{i(-q_1 x+q_2 y)} \langle0|T\{j_{B}(y)j_\mu(0)j^{\dagger}_{A}(x) \}|0\rangle
\label{3ptcf-SL}
\end{align}
for a semileptonic decay $ A\to B+\ell+\nu_\ell $, and 
\begin{align}
\Pi(q_1^2, q_2^2, q^2) = i^2\int d^4x d^4y e^{iq_1 \cdot x} e^{iq_2 \cdot y}
\bra{0} T\{j_{_B}(x)j_{_C}(y)j_{_A}^\dagger(0)\}\ket{0}
\label{3ptcf-strong}
\end{align}
for a strong decay $ A\to B+C $. $ j_\mu(0) = \bar{q}_f \gamma_\mu\left(1-\gamma_5\right) q_i $ denotes the weak transition current, where $ q_{i} $ and $ q_{f} $ represent the quarks in the initial and final state of transformation.

Theoretical predictions in QCD are formulated in terms of quark and gluon degrees of freedom, whereas experimental observations are restricted to hadronic states due to color confinement. The method of QCD sum rules exploits the correlation functions in Eq.~(\ref{2ptcf}) to link these two regimes. This is achieved by evaluating the correlation function in two distinct domains: the QCD representation and the phenomenological representation.

The QCD representation is described by the operator product expansion (OPE), where the nonperturbative effects of QCD are characterized in terms of the vacuum expectation values of composite operators so-called condensates. The phenomenological representation is obtained using the creation and annihilation operators of the hadron, where the experimental physical observables such as hadron masses are included. The concept of quark-hadron duality is then introduced to assert the consistency between the OPE-calculated correlation function and its hadronic counterpart. Matching these representations allows for the extraction of specific hadron parameters, as detailed in the sections that follow.

\subsection{Quark level: QCD representation}\label{sec2.2}\

In QCD, only the perturbative regime is well-defined, allowing for reliable derivation of the perturbative contributions in Eq.~(\ref{2ptcf}). However, the effects of soft gluons and quark fields present in the QCD vacuum cannot be ignored, indicating that the expectation values of the operators associated with these fields are non-zero. One approach to addressing the complexities of the QCD vacuum is to use the operator product expansion (OPE)\cite{Wilson:1969zs}, which effectively separates the short-distance and long-distance scales.

\begin{figure}[ht]
\centering
\includegraphics[width=1\linewidth]{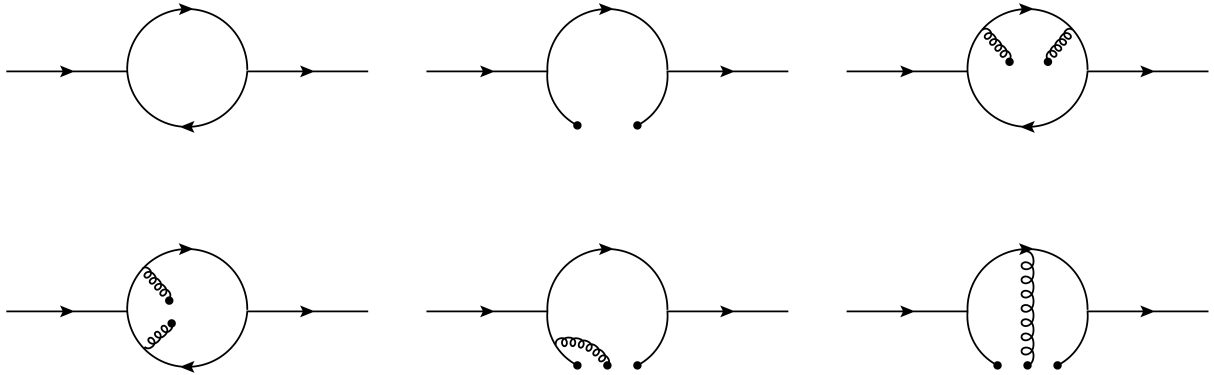}
\caption{Typical Feynman diagrams of the operator product expansion up to dimension 5.}
\label{fig:ope}
\end{figure}

Using the OPE, correlation functions can be expressed as a sum of Wilson coefficients multiplied by the expectation values of composite operators. The perturbative contributions are encoded in these coefficients, which are derived from perturbative QCD. The complex structure of the QCD vacuum resides in the condensates. Specially, Eq.~(\ref{2ptcf}) can be expanded in a series of local operators:
\begin{align}
\Pi(q^2) = \sum_{d}C_d(q^2)\bra{0}\hat{O}_d(0)\ket{0}\,,
\end{align}
where $ C_d(q^2) $ is so-called the Wilson coefficients, characterizing the short-distance contributions. The vacuum expectation value of the composite local operators $ \bra{0}\hat{O}_d(0)\ket{0} $ parameterize the nonperturbative effects in the QCD vacuum, which represent the long-distance contributions. The index $ d $ denotes the dimension of the local operators $ \hat{O}_d $. The lowest dimension $ \hat{O}_0=1 $ and $ C_0(q^2) $ describes the perturbative contributions. The operators with $ d\leq 6 $ are listed as follows:
\begin{align}
\nonumber
\hat{O}_{3}=& :\bar{q}(0) q(0): \equiv\bar{q}q \,,\\\nonumber
\hat{O}_{4}=& : g_s^2 G^{a}_{\mu\nu}(0) G^{a}_{\mu\nu}(0):\equiv g_s^2 G^2 \,,\\\nonumber
\hat{O}_{5}=&:\bar{q}(0) g_s \frac{\lambda^a}{2}\sigma^{\mu\nu} G^{a}_{\mu\nu}(0) q(0): \equiv \bar{q}Gq \,, \\\nonumber
\hat{O}^q_{6} =& :\bar{q}(0)q(0) \bar{q}(0) q(0):\equiv \bar{q}q\bar{q}q \,,\\
\hat{O}^G_{6} =& :f_{abc} g_s^3 G^{a}_{\mu\nu}(0) G^{b}_{\nu\rho}(0) G^{c}_{\rho\mu}(0):\equiv g_s^3 G^3\,,
\end{align}
where the symbol “::” is the normal ordering of the operators. $ q(0) $ denotes the quark field, $ G^{a}_{\mu\nu}(0) $ is the gluon field strength tensor, $ f_{abc} $ represents the structure constant, $ \lambda_a $ is the Gell-Mann matrix, and $ \sigma^{\mu\nu} = \frac{i}{2}[\gamma_\mu,\gamma_\nu] $. The vacuum expectation value of the local operators $ \bra{0}\hat{O}_d(0)\ket{0} $ gives the quark condensate $ \langle \bar{q}q\rangle$, gluon condensate $ \langle g_s^2 G^2\rangle$, quark-gluon condensate $ \langle \bar{q}Gq \rangle$, four-quark condensate $ \langle \bar{q}q\bar{q}q \rangle$, and triple gluon condensate $ \langle g_s^3 G^3\rangle$. So the generic expression of the correlation function in the QCD representation can be written as follows:
\begin{align}
\Pi^{\text{QCD}}(q^2)=C_0(q^2)+C_3(q^2)\langle \bar{q}q\rangle+C_4(q^2)\langle g_s^2 G^2\rangle+C_5(q^2)\langle \bar{q}Gq \rangle+...\, ,
\label{ope-expansion}
\end{align}
where the symbol “...” denotes the contributions of higher-dimensional operators. The Wilson coefficients can be derived by calculating the typical diagrams in Fig~\ref{fig:ope} and the values of the condensate can be extracted from Lattice calculations or instanton model~\cite{Gubler:2018ctz, Davies:2018hmw, Forkel:2000fd}. Additionally, the quark condensate $ \langle \bar{q}q \rangle $ plays a crucial role in QCD as it accounts for the spontaneous breaking of chiral symmetry. Its value was established well before it found application in QCD sum rules~\cite{Reinders:1984sr, Colangelo:2000dp}:
\begin{align}
\langle\bar{q} q\rangle=-\frac{f_\pi^2 m_\pi^2}{\left(m_u+m_d\right)}\,,\,\,\,\,\,q=u,d\,,
\end{align}
where $ f_\pi $ and $ m_\pi $ are the decay constant and mass of the pion. $ m_{u(d)} $ is the mass of up (down) quark. At renormalization scale $ \mu=1 \,\mathrm{GeV} $, the value of quark condensate is estimated to be $\langle\bar{q} q\rangle=-(0.24 \pm 0.01)^3 \,\mathrm{GeV}^3$~\cite{Shifman:1978bx, Narison:1989aq, Colangelo:2000dp}

Rewrite the $ \Pi^{\text{QCD}}(q^2) $ in Eq.~(\ref{ope-expansion}) with dispersion relation, we can obtain:
\begin{align}
\Pi^{\text{QCD}}(q^2) = \int^{+\infty}_{s_{min}} \frac{\rho^{\text{QCD}}(s)}{s - q^2} \, ds\,,
\label{2pt-QCD}
\end{align}
where $\rho^{\text{QCD}}(s)=\frac{1}{\pi}\text{Im}[\Pi^{\text{QCD}}(s)]$ is the spectral density and $ s_{min} $ is a kinematic limit related to the square of the sum of the quark masses in the hadron current. Using the dispersion relation allows for a clearer comparison between the QCD and phenomenological representations, enabling the separation of the ground-state hadron and extraction of hadron parameters.

\subsection{Hadron level: phenomenological representation}\label{sec2.3}\

On the other hand, we can apply the hadron parameters, such as mass and decay constants, to relate the correlation function~(\ref{2ptcf}) to the hadron state. The interpolating currents can be interpreted as the creation and annihilation operators of a hadron. After inserting a complete set of intermediate hadronic states and employing dispersion relation again, one can obtain the general form:
\begin{align}
\Pi^{\text{phe}}(q^2)=\int_{0}^{+\infty}\frac{\rho(s)}{s-q^2+i\epsilon}+\cdot\cdot\cdot\,,
\end{align}
where “$ \cdot\cdot\cdot $” represents the subtraction terms and the hadron spectral density is defined as:
\begin{align}
\rho(s)\equiv\sum_{H}|\bra{0}j(0)\ket{H}|^2\delta(s-E_H^2)\,.
\label{hadron spectral density}
\end{align}
In the hadron spectral density, the ground-state hadron is typically narrow, whereas the higher excited and continuum states are broader. This allows the ground-state hadron to be isolated as a distinct sharp pole. The spectral density~(\ref{hadron spectral density}) can then be expressed as the sum of contributions from the ground-state, the continuum spectrum, and the higher excited states:
\begin{align}
\rho(s)&=|\bra{0}j(0)\ket{H_0}|^2\delta(s-M_{H}^2)+\sum_{H^\prime}|\bra{0}j(0)\ket{H^\prime}|^2\delta(s-E_{H^\prime}^2)\nonumber\\
&\equiv\lambda_H^2\delta(s-M_{H}^2)+\theta(s-s_0)\rho^{\text{cont}}(s)\,,
\label{hadron spectral density-2}
\end{align}
where $ \lambda_H \equiv \bra{0}j(0)\ket{H_0} $ denotes the coupling constant of the ground-state hadron to the interpolating current, commonly referred to as the decay constant. $ M_H $ represents the mass of the ground-state hadron, while $ s_0 $ is the threshold parameter for the continuum spectrum. The term $ \rho^{\text{cont}}(s) $ accounts for the contributions from high excited states and the continuum spectrum. Because of quark-hadron duality, $ \rho^{\text{cont}}(s) $ is assumed to be equal to the results obtained by OPE, i.e., $ \rho^{\text{cont}}(s) = \rho^{\text{QCD}}(s)$. With these definitions, the phenomenological representation of the correlation functions can be expressed as:
\begin{align}
\Pi^{\text{phe}}(q^2)=\frac{\lambda_H^2}{M_H^2-q^2}+\int_{s_0}^{+\infty}\frac{\rho^{\text{QCD}}(s)}{s-q^2}ds+\cdot\cdot\cdot\,.
\label{2pt-phe}
\end{align}
As for the investigation of decay process, the three-point correlation functions are used to derive the coupling constants and form factors, which will be discussed in Sec~\ref{sec2.6}.

\subsection{Quark-hadron duality and Borel transformation}\label{sec2.4}\

Now, two representations of the correlation functions have been derived. The first is the QCD representation~(\ref{2pt-QCD}), obtained by directly calculating Feynman diagrams. The second is the phenomenological representation~(\ref{2pt-phe}), which encapsulates the hadron information of interest. It is natural to posit that if these two representations are equivalent, the physical quantities of interest could be extracted from them. Unfortunately, no rigorous mathematical proof currently exists to confirm this equivalence. Therefore, we must rely on an assumption known as the quark-hadron duality.

The quark-hadron duality allows the correlation functions at the quark level to be equivalent to those at the hadron level within a certain energy scale. This idea was originally proposed by Poggio, Quinn and Weinberg~\cite{Poggio:1975af}, which suggested that theoretical calculations and experimental measurements can be averaged over an appropriate energy range to achieve consistency, also called the smearing method. Quark-hadron duality is widely utilized in various aspects of QCD, including applications such as $ e^+e^- $ annihilation, deep inelastic scattering, jet physics, $ \tau $ decays and so on. The details of the quark-hadron duality can be found in the original article~\cite{Poggio:1975af}, as well as in the lectures by Shifman~\cite{Shifman:2000jv, Shifman:2001qm}. The dynamical realization of quark-hadron duality at high energies was further investigated by Blok, Shifman, and Zhang~\cite{Blok:1998zx} using the 't Hooft model, providing an illustrative example of how smearing mechanisms can give rise to local duality.

Based on the quark-hadron duality, we can obtain:
\begin{align}
\Pi^{\mathrm{QCD}}(q^2) \simeq \Pi^{\mathrm{phe}}(q^2)\,.
\label{quark-hadron-duality}
\end{align}
The calculation of $ \Pi^{\mathrm{QCD}}(q^2) $ is performed in the deep spacelike region by using OPE, and the result is approximate. Since it is impractical to include condensates up to infinite dimensions, they must be truncated at some finite dimensions. However, the accuracy of $ \Pi^{\mathrm{QCD}}(q^2) $ significantly impacts the validity of Eq.~(\ref{quark-hadron-duality})~\cite{Shifman:2001qm}. Therefore, a specialized operator is required to minimize the contribution of condensates, bringing the calculation results closer to the “real” values. On the other hand, the phenomenological representation requires minimal contributions from the higher excited and continuum states. Moreover, the subtraction term in Eq.~(\ref{2pt-phe}), being an unknown polynomial in $q^2$, must be excluded from the calculations. To achieve the above objectives, the authors of Refs~\cite{Shifman:1978bx, Shifman:1978by} proposed to employ the Borel transformation:
\begin{align}
\mathcal{B}[\Pi(q^2)] \equiv \Pi(\tau^2)=\lim _{\substack{-q^2, n \to \infty \\[3pt] -q^2 / n=\tau^2}} \frac{(-q^2)^{n+1}}{n!}\left(\frac{\partial}{\partial q^2}\right)^n \Pi(q^2)\,,
\label{Borel-transformation}
\end{align}
where $ \tau^2 $ is Borel parameter introduced after Borel transformation. Some commonly used formulas are listed:
\begin{align}
\label{Borel-1}
\mathcal{B}[(q^2)^k] & =0\,, \\[5pt]
\label{Borel-2}
\mathcal{B}[\frac{1}{(s-q^2)^k}] & =\frac{1}{(k-1)!}\left(\frac{1}{\tau^2}\right)^{k-1} e^{-s / \tau^2}\,.
\end{align}

Taking the expressions of $ \Pi^{\mathrm{QCD}}(q^2) $ and $ \Pi^{\mathrm{phe}}(q^2) $ in Eqs.~(\ref{2pt-QCD}) and (\ref{2pt-phe}), and performing the Borel transformation, we can get the basic equations of QCD sum rules:
\begin{align}
\int^{+\infty}_{s_{min}} \! \: \frac{\rho^{\text{QCD}}(s)}{s - q^2} ds\simeq\frac{\lambda_H^2}{M_H^2-q^2}+\int_{s_0}^{+\infty}\frac{\rho^{\text{QCD}}(s)}{s-q^2}ds+\cdot\cdot\cdot\,,
\label{quark-hadron-1}
\end{align}
\begin{align}
\int_{s_{m i n}}^{+\infty} \rho^{\mathrm{QCD}}(s)\, e^{-s/\tau^2} \,ds =\lambda_H^2 \,e^{-M_H^2/\tau^2}+\int_{s_{0}}^{+\infty} \rho^{\mathrm{QCD}}(s)\, e^{-s/\tau^2} \,ds\,.
\label{quark-hadron-Borel}
\end{align}
It is evident that both sides of Eq.~(\ref{quark-hadron-Borel}) are suppressed factorially and exponentially, which implies that contributions from high-dimension condensates in the operator product expansion, as well as those from higher excited states and the continuum in the hadronic spectrum, are significantly suppressed. Furthermore, the Borel transformation eliminates the subtraction terms in Eq.~(\ref{quark-hadron-1}) by applying Eq.~(\ref{Borel-1}). In conclusion, the application of the Borel transformation facilitates a coincidence between the “quark” and “hadron”, improving the consistency of both sides in Eq.~(\ref{quark-hadron-duality}).

\subsection{Two-point sum rules}\label{sec2.5}\

Rewrite Eq.~(\ref{quark-hadron-Borel}) into a more concise form:
\begin{align}
\lambda_H^2 \,e^{-M_H^2/\tau^2}=\int_{s_{min}}^{s_0} \rho^{\mathrm{QCD}}(s)\, e^{-s/\tau^2} \,ds\,.
\label{qcdsr-equation}
\end{align}
In practice, there are cases where the correlation function associated with a condensate of a certain dimension lacks an imaginary part but yields a nontrivial contribution after applying the Borel transformation. Therefore, an additional term may appear in the R.H.S. of Eq.~(\ref{qcdsr-equation}) to characterize such contributions. Taking the derivative of both sides of Eq.~(\ref{qcdsr-equation}) with respect to $1/\tau^2$ (where $d/d(1/\tau^2) = -\tau^4 \, d/d\tau^2$), one obtains:
\begin{align}
\lambda_H^2 \,M_H^2\,e^{-M_H^2/\tau^2}=\int_{s_{min}}^{s_0} s\,\rho^{\mathrm{QCD}}(s)\, e^{-s/\tau^2} \,ds\,.
\label{qcdsr-equation-2}
\end{align}
Then dividing Eq.~(\ref{qcdsr-equation-2}) by Eq.~(\ref{qcdsr-equation}), we can obtain the expression for the hadron mass:
\begin{align}
M_H^2=\frac{\int_{s_{min}}^{s_0} s\, \rho^{\text{QCD}}(s) \,e^{-s / \tau^2}\,ds}{\int_{s_{min}}^{s_0} \rho^{\text{QCD}}(s) \,e^{-s / \tau^2}\,ds}\,.
\label{mass-function}
\end{align}
One drawback of this equation is the dependence on the Borel parameter. The hadron mass, being an intrinsic physical observable, should not rely on any auxiliary parameters. Therefore, to obtain a reliable hadron mass, it is essential to impose constraints on the Borel parameter. 

First, as discussed in the previous section, the condensates must be truncated at finite dimensions. Thus, to ensure the convergence and validity of the OPE, the contribution of the highest-dimension condensate must be sufficiently small. Typically, the contribution of the highest-order condensate should not exceed 10\% to 30\%. This criterion can be formulated as follows:
\begin{align}
R^{\mathrm{OPE}}=\frac{\int_{s_{min}}^{s_0} \rho^{\text{QCD,\,N}}(s) \,e^{-s / \tau^2}\,ds}{\int_{s_{min}}^{s_0} \rho^{\text{QCD}}(s) \,e^{-s / \tau^2}\,ds}\,,
\label{ope}
\end{align}
where $ \rho^{\text{QCD,\,N}}(s) $ represents the spectral density of highest-dimension condensate.

Second, to evaluate the contribution of ground-state hadrons, it is crucial that the pole contribution dominates the spectrum. Therefore, a selection of the pole contribution larger than $40\%\sim 60\%$ is often made, formulated as follows:
\begin{align}
R^{\mathrm{PC}}=\frac{\int_{s_{min}}^{s_0} \rho^{\text{QCD}}(s) \,e^{-s / \tau^2}\,ds}{\int_{s_{min}}^{+\infty} \rho^{\text{QCD}}(s) \,e^{-s / \tau^2}\,ds}\,.
\label{pole}
\end{align}

From Eq.~(\ref{Borel-transformation}), one can see that the Borel parameter is inherently connected to the energy scale of the hadron system. A small $\tau^2$ corresponds to a low-energy regime, where higher-dimensional condensates become more significant and may spoil the convergence of the OPE. The OPE convergence criterion in Eq.~(\ref{ope}) therefore imposes a lower bound $\tau^2_{\min}$. Conversely, a large $\tau^2$ corresponds to a high-energy regime, where the suppression of continuum and excited states is weaker, allowing their contributions to dominate over the ground-state pole. Hence the pole dominance criterion in Eq.~(\ref{pole}) imposes an upper bound $\tau^2_{\max}$. 

Finally, we need to identify a range for $ \tau^2 $ within the two constraints mentioned above, where the hadron mass and other physical observables exhibit minimal dependence on $ \tau^2 $. However, sometimes it is not possible to find a stable range for $ \tau^2 $, which implies that reliable conclusions cannot be drawn from QCD sum rules in such situations. In practice, incorporating higher-order condensates in OPE may help enhance the stability of $ \tau^2 $~\cite{Albuquerque:2018jkn}.

In Eq.~(\ref{mass-function}), another critical parameter is $ s_0 $, which represents the threshold where the continuum and higher excited states begin to contribute. Ideally, $ s_0 $ should be determined from the hadron spectrum. However, as the hadron spectrum for most hadrons remains unclear, a reasonable estimate of $ s_0 $ is necessary. In many cases, a good approximation for $ s_0 $ is around the square of the first excited state's mass. Given that the energy level spacing for most hadrons is typically less than $ 1\,\mathrm{GeV} $, a practical constraint is $ s_0 < \left(M_H + 1\,\mathrm{GeV}\right)^2 $. Within this restriction, the optimal choice of $ s_0 $ should ensure maximum stability for $ \tau^2 $. The selection of $ s_0 $ is often a range, which inherently introduces systematic uncertainty. In Refs.~\cite{Wu:2021tzo, Wu:2022qwd, Wu:2023ntn,Wang:2017qvg}, the authors propose a dependence of the threshold parameter $s_0$ on $\tau^2$, thereby determining suitable values for both $\tau^2$ and $s_0$.

In addition, by using the obtained mass and Eq.~(\ref{qcdsr-equation}), the decay constant $ \lambda_H $ can also be determined:
\begin{align}
\lambda_H^2 =e^{M_H^2/\tau^2}\int_{s_{min}}^{s_0} \rho^{\mathrm{QCD}}(s)\, e^{-s/\tau^2} \,ds\,.
\label{decay-constant}
\end{align}

\subsection{Three-point sum rules}\label{sec2.6}\

In this section, we extend the previous framework to explore the application of three-point correlation functions. These functions connect up to three hadrons and are closely tied to hadron decays. They enable the extraction of decay-related quantities, such as coupling constants for strong decays and form factors for semileptonic decays, which can be used to compute transition matrix elements and, ultimately, decay widths. The final states in strong decays and semileptonic decays involve different numbers of hadrons, leading to slight differences in the construction of QCD sum rules. This section focuses on the calculation framework for semileptonic decays, which are the main subject of this review. For QCD sum rule studies of strong decays, we refer the readers to the review~\cite{Albuquerque:2018jkn} and related works~\cite{Lian:2023cgs, Wang:2016wkj, Dias:2013qga, Wan:2020oxt, Wang:2023sii}. 

\subsubsection{Transition matrix elements and form factors}\

Semileptonic decays are crucial for probing strong and weak interactions, with the former involving parton hadronization and the latter linked to flavor-changing processes. Their relative theoretical simplicity makes them valuable for measuring fundamental standard model (SM) parameters and studying decay dynamics in detail~\cite{Richman:1995wm}. A key parameter is the Cabibbo-Kobayashi-Maskawa (CKM) matrix element, which
can be used to rigorously test the SM and its potential extensions.

\begin{figure}[ht]
\centering
\includegraphics[width=15cm]{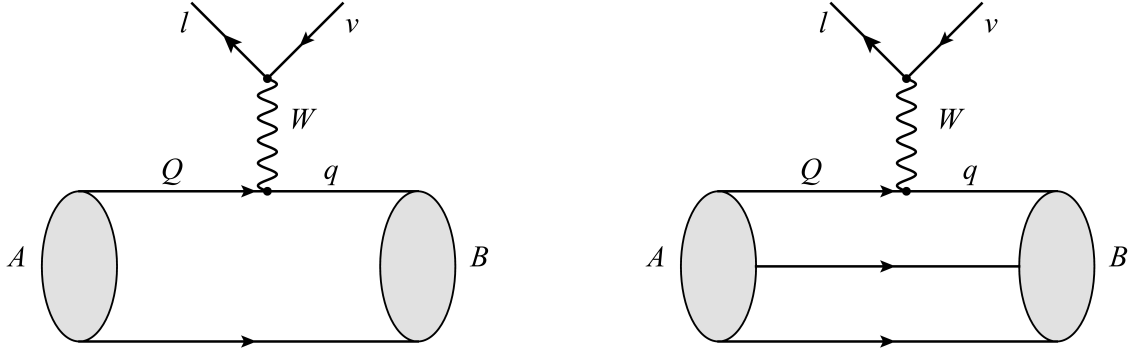}
\caption{Feynman diagrams of the mesonic (left) and baryonic (right) semileptonic decays.}
\label{fig:SL-decay}
\end{figure}

Semileptonic decay $ A\to B+\ell+\nu_\ell $ is a typical multi-scale process, with its Feynman diagram shown in Fig~\ref{fig:SL-decay}. For such decays, the factorization theorem allows independent calculations in the separation of different energy scales. Consequently, the semileptonic decay $ A\to B+\ell+\nu_\ell $ can be viewed as $ A $ decaying into $ B $ with the emission of a $ W $ boson, which subsequently decays into a lepton-neutrino pair. Thus, the transition matrix element for this process can be factorized as leptonic part and hadronic part:
\begin{align}
\mathcal{M}_{A \to B \ell \nu_{\ell}}=\frac{G_F}{\sqrt{2}} V_{Q q}[\bar{\ell} \gamma^\mu(1-\gamma_5) \nu_{\ell}]\bra{B(q_2)} \bar{q} \gamma_\mu(1-\gamma_5) Q\ket{A(q_1)}\,,
\label{SL-matrix-element}
\end{align}
where $ Q $ and $ q $ represent the flavor of the quarks involved in the initial and final states of the decay, respectively. $ V_{Qq} $ denotes the CKM matrix element. The four-momenta of the initial and final hadrons are given by $ q_1 $ and $ q_2 $, respectively. $ G_F $ is the Fermi constant. The leptonic part of the transition matrix element corresponds to a high-energy process and can be calculated using electroweak perturbation theory. However, the hadronic part involves the low-energy aspects of QCD and can not be treated perturbatively. Instead, non-perturbative effects must be included. Generally, the hadronic transition matrix element is parameterized in terms of a set of form factors, which encapsulate the non-perturbative QCD information. QCD sum rules are essentially used to calculate these form factors. 

\subsubsection{QCD representation}\

\begin{figure}[ht]
\centering
\includegraphics[width=15cm]{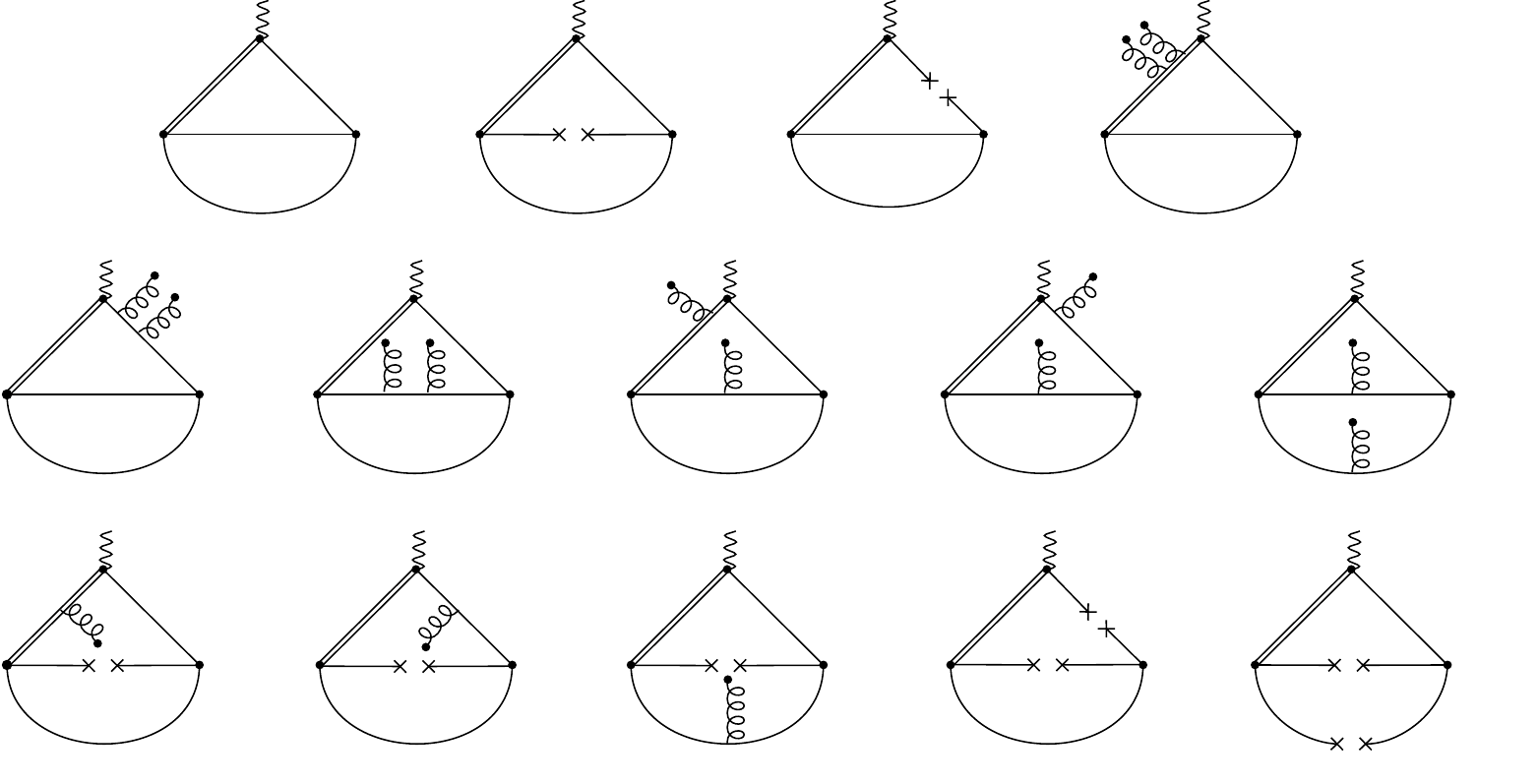}
\caption{Typical Feynman diagrams for baryonic semileptonic decays.}
\label{fig:ope-decay}
\end{figure}

In the QCD representation, the three-point correlation functions for semileptonic decays~(\ref{3ptcf-SL}) can be expressed by double dispersion relations~\cite{Ball:1991bs, MarquesdeCarvalho:1999bqs, Yang:2005bv, Du:2003ja, Zhao:2020mod, Shi:2019hbf, Xing:2021enr}:
\begin{align}
\Pi_\mu^{\mathrm{QCD}}(q_1^2, q_2^2, q^2)=\int_{s_1^{min }}^{\infty} d s_1 \int_{s_2^{min }}^{\infty} d s_2 \frac{\rho_\mu^{\mathrm{QCD}}(s_1, s_2, q^2)}{(s_1-q_1^2)(s_2-q_2^2)}\,,
\label{3pt-SL-QCD}
\end{align}
where the condensates in the spectral density are generally calculated up to dimension-6. The typical Feynman diagrams for baryonic semileptonic decays are shown in Fig~\ref{fig:ope-decay}, while the meson-type can be derived by reducing one quark line and corresponding condensates. Besides, other tensorial structures, metric tensors, and the four momenta which carry Lorentz indices may appear in $ \rho_\mu^{\mathrm{QCD}}(s_1, s_2, q^2) $ when meson currents are taken into account. Due to the use of the dispersion relation, we are primarily concerned with the imaginary part of the correlation function. Thus, it is natural to consider applying Cutkosky cutting rules to extract the imaginary part of the spectral density~\cite{Cutkosky:1960sp}. According to Cutkosky cutting rules, the quark denominators can be substituted to $ \delta $ functions:
\begin{align}
\frac{1}{p^2-m^2+i \varepsilon} \to-2 \pi i \delta(p^2-m^2)\,.
\end{align}
Then the loop integral can be transformed into a phase-space-like integral. The details for the application of Cutkosky cutting rules to the QCD sum rules can be found in meson-type~\cite{Du:2003ja, Ball:1991bs, Wang:2012hu, Wu:2024gcq} and baryon-type~\cite{Shi:2019hbf}.

\subsubsection{Phenomenological representation}\

In the phenomenological representation, by inserting the completeness relations for the initial and final hadron states into Eq.~(\ref{3ptcf-SL}) and applying the double dispersion relation, the phenomenological representation of the three-point correlation function can be derived as:
\begin{align}
\Pi_\mu^{\text {phe}}(q_1^2, q_2^2, q^2)=\sum_{\text {spins }} \frac{ \bra{0} j_B\ket{B(q_2)}\bra{ B(q_2)} j_\mu\ket{A(q_1)}\bra{A(q_1)} j_A^\dagger\ket{0}}{\big(q_1^2-M_A^2\big)\big(q_2^2-M_B^2\big)}+\cdot\cdot\cdot\,.
\label{3pt-SL-phe-1}
\end{align}
The transition amplitudes from vacuum to hadron can be parameterized by decay constants:
\begin{align}
\bra{0}j_A\ket{A(q_1)}\sim\lambda_A\,,
\label{SL-decay-constant}
\end{align}
and the hadronic transition matrix element $ \bra{ B(q_2)} j_\mu\ket{A(q_1)} $ is described by the form factors. 

For baryonic decays, these two transition matrix elements are:
\begin{align}
\bra{0}j_{A}\ket{A(q_1)}&=\lambda_A u_A(q_1)\,,\\[5pt]
\bra{B(q_2)} \bar{q} \gamma_\mu(1-\gamma_5) Q\ket{A(q_1)}&= \bar{u}_B(q_2)\big[f_1(q^2) \gamma_\mu+f_2(q^2) i\sigma_{\mu \nu}q^\nu/M_A+f_3(q^2) q_\mu/M_A\big] u_A(q_1) \nonumber\\[5pt]
& -\bar{u}_B(q_2)\big[g_1(q^2) \gamma_\mu+g_2(q^2) i\sigma_{\mu \nu}q^\nu/M_A+g_3(q^2) q_\mu/M_A\big] \gamma_5 u_A(q_1),
\label{baryon-form-factros}
\end{align}
where the $ f_i(g_i) $ represent the form factors. Due to time reversal invariance, they are real functions of the momentum transfer squared, $ q^2 \equiv (q_1 - q_2)^2 $~\cite{Gaillard:1984ny}. $ u_{A(B)} $ are the Dirac spinors for the initial and final states. The redundant Lorentz structures of the form factors in baryonic transition matrix element can be addressed by introducing negative-parity baryons~\cite{Jido:1996zw, Jido:1996ia, Zhao:2020mod, Shi:2019hbf}, as well as the projection method provided in Ref.~\cite{MarquesdeCarvalho:1999bqs}. The form factors $ f_3(q^2) $ and $ g_3(q^2) $ are usually neglected in the electron and muon mode since their contributions to the decay width are suppressed significantly by the factor $ (m_{\ell}/M_A)^2 $, where $ m_\ell(\ell=e,\mu) $ denotes the lepton mass.

For mesonic decays, the vacuum-to-meson transition amplitudes are:
\begin{align}
\bra{0}j_{A}\ket{A(q_1)}_{pse}=\frac{\lambda_{A}\;M_A^2}{m_1+m_2}\,,\;\; \bra{0}j_{A}\ket{A(q_1)}_{vec}=\lambda_{A}M_A\epsilon_\nu\,,\; \bra{0}j_{A}\ket{A(q_1)}_{axi}=i\lambda_{A}q_{1_\nu}\,,
\end{align}
for pseudoscalar, vector, and axial-vector meson currents, respectively. Here, $ m_{1(2)} $ represents the masses of the two quarks that make up $ j_A $. The specific form of meson transition matrix element $ \bra{ B(q_2)} j_\mu\ket{A(q_1)} $ is determined by the spin of final state mesons, where 
\begin{align}
\bra{ B(q_2)} \bar{q} \gamma_\mu Q\ket{A(q_1)} = f_{+}(q^2)(q_{1_\mu}+q_{2_\mu}-\frac{M_A^2-M_B^2}{q^2} q_\mu)+f_0(q^2) \frac{M_A^2-M_B^2}{q^2} q_\mu
\end{align}
for $ B $ a pseudoscalar meson~\cite{Leljak:2019fqa, Wu:2024gcq, Wang:2012hu} and 
\begin{align}
\bra{ B(q_2)} \bar{q} \gamma_\mu(1-\gamma_5) Q\ket{A(q_1)} &= \frac{2V(q^2)}{M_A+M_B}\varepsilon_{\mu\nu\alpha\beta}\epsilon^{*\nu}q_1^\alpha q_2^\beta-(M_A+M_B)(\epsilon_\mu^*-\frac{\epsilon^*\cdot q}{q^2}q_\mu) A_1(q^2)\nonumber\\[5pt]
&+i \frac{(\epsilon^* \cdot q)}{M_A+M_B}\big((q_1+q_2)_\mu-\frac{M_A^2-M_B^2}{q^2} q_\mu\big) A_2(q^2)\nonumber\\[5pt]
&-i\frac{2 M_B (\epsilon^*\cdot q) q_\mu}{q^2}A_0(q^2)
\end{align}
for $ B $ a vector meson~\cite{Leljak:2019fqa, Wu:2024gcq, Wang:2012hu, Du:2003ja}. Here, $ f_+ $, $ f_0 $, $ V $, and $ A_i $ are form factors. The tensorial structures, metric tensors, and four-momenta in the mesonic transition matrix element can be matched in the QCD representation~(\ref{3pt-SL-QCD}). The quark-hadron duality principle underlying three-point sum rules was discussed in Ref.~\cite{Blok:1998zz}. By aligning the two representations and applying a Borel transformation, the form factors can be analytically derived. The general expression for the form factors is given as:
\begin{align}
f^{A \to B}(q^2)=\frac{e^{M_A^2 / \tau_1^2} e^{M_B^2 / \tau_2^2}}{\lambda_A \lambda_B} \int_{s_1^{min }}^{s_1^0} d s_1 \int_{s_2^{min }}^{s_2^0} d s_2\, \rho(s_1, s_2, q^2) e^{-s_1 / \tau_1^2} e^{-s_2 / \tau_2^2}\,.
\label{general-form-fator}
\end{align}
Here, the lower integration limit $s_1^{min}$ is typically set by the constraint that all internal quarks are on their mass shell~\cite{Landau:1959fi, Ball:1991bs, Du:2003ja}. Since the decay process involves an initial and a final hadron, the Borel transformation introduces two separate Borel parameters $ \tau_1^2 $ and $ \tau_2^2 $. The values of the form factors depend on these two Borel parameters. In principle, they should be determined by selecting an optimal region in the free $ \tau_1^2 $ and $ \tau_2^2 $ plane based on the limitations stated in Sec~\ref{sec2.5}. Another way is to simplify these two Borel parameters by following ratio\cite{Ball:1991bs, Shi:2019hbf, MarquesdeCarvalho:1999bqs, Leljak:2019fqa}:
\begin{align}
\frac{\tau_1^2}{\tau_2^2}=\frac{M_{A}^2-m_Q^2}{M_{B}^2-m_q^2}\;,
\label{simplify-two-borel}
\end{align}
which has been found a good approximation with free Borel parameters scheme in Ref.~\cite{Shi:2019hbf}. 

Once the Borel parameters, the threshold parameters, and other parameters are fixed, the value of the form factors at specific $ q^2 $ points can be obtained using Eq.~(\ref{general-form-fator}). Since the usage of OPE, QCD sum rules only works in the large recoil region, where $ q^2\sim 0 $. However, the decay process requires the distribution of form factors across the entire physical range. Therefore, after calculating the form factors at several specific $ q^2 $ points, an appropriate analytic function must be used to extrapolate the form factors to the full kinematic region. A dipole~\cite{Jaus:1996np, Cheng:2003sm} and $ z $-series~\cite{Bourrely:2008za} parameterization scheme are always taken into account, which read:
\begin{align}
\label{dipole}
f(q^2)=\frac{f(0)}{(1-q^2 / M_A^2)\big[1-a_1(q^2 / M_A^2)+a_2(q^2 / M_A^2)^2\big]}
\end{align}
for dipole parameterization and
\begin{align}
\label{BCL}
f(q^2)&=\frac{f(0)}{1-q^2/(m_{pole})^2}\Bigl\{1+a_1(z(q^2,t_0)-z(0,t_0))\Bigr\}\,,\nonumber\\[5pt]
z(q^2,t_0)&=\frac{\sqrt{t_+-q^2}-\sqrt{t_+-t_0}}{\sqrt{t_+-q^2}+\sqrt{t_+-t_0}}\,.
\end{align}
for $ z $-series parameterization. Here, $ f(0) $ and $ a_{1(2)} $ are fitting parameters. $ m_{pole} $ represents the mass of the meson related to quark flavor transitions. The variables $ t_{\pm}\equiv (M_{A}\pm M_B)^2$, and $ t_0=t_+-\sqrt{t_+-t_-} \sqrt{t_+-t_{min}}$, where $ t_{min} $ is the choice of minimal value of $ q^2 $ for the calculation of form factors using QCD sum rules. The nonlinear least squares ($\chi^2$) method is always employed in the fitting analysis, where the uncertainties of the fitting parameters will bring additional errors to the decays.

\subsection{Input parameters}\

In the numerical calculations, the selection of parameters can refer to Refs.~\cite{Colangelo:2000dp, Narison:1989aq, ParticleDataGroup:2024cfk, Shifman:1978bx, Shifman:1978by, Nielsen:2009uh, Albuquerque:2018jkn}, as listed in Table~\ref{input parameters}. The uncertainties in these parameters also contribute to the errors in the numerical results.

\begin{table}[ht]
\centering
\caption{Input parameters of QCD sum rules.}
\renewcommand{\arraystretch}{1.5}
\begin{tabular}{lclc}
\hline\hline
Parameter & Value & Parameter & Value \\ \hline
$ m_u $ & $ (2.16\pm 0.07) \,\mathrm{MeV}$&$\langle\bar{q} q\rangle$ & $ -(0.24 \pm 0.01)^3 \,\mathrm{GeV}^3 $ \\
$ m_d $ & $ (4.70\pm 0.07) \,\mathrm{MeV}$ & $ \left\langle g_s \bar{q} \sigma \cdot G q\right\rangle $ & $ m_0^2\langle\bar{q} q\rangle $ \\
$ m_s $ & $ (93.5\pm 0.8) \,\mathrm{MeV} $ & $ \langle\bar{s} s\rangle $ & $ (0.8\pm 0.1)\langle\bar{q} q\rangle $ \\
$ m_c $ & $ (1.27\pm 0.02) \,\mathrm{GeV}$ & $ \left\langle g_s \bar{s} \sigma \cdot G s\right\rangle $ & $ m_0^2\langle\bar{s} s\rangle $ \\
$ m_b $ & $ (4.18\pm 0.01) \,\mathrm{GeV}$ & $ \langle g_s^2 G^2\rangle $ & $ (0.88\pm 0.25)\,\mathrm{GeV}^4 $ \\ 
$ m_0^2 $ & $(0.8\pm 0.2) \,\mathrm{GeV}^2 $ & $ \langle g_s^2 G^3\rangle $ & $ (0.045\pm 0.0213)\,\mathrm{GeV}^4 $ \\ 
\hline\hline
\end{tabular}
\label{input parameters}
\end{table}

These parameters are scale-dependent. The quark masses are provided in the $ \overline{\text{MS}} $ renormalization scheme, while the other condensate parameters are specified at $ \mu=1\,\mathrm{GeV} $. Therefore, the actual values used in specific physical problems may differ from those listed in Table~\ref{input parameters} due to different energy scales. Additionally, studies of decay processes involve hadronic parameters and other decay kinematics parameters.

Once all the parameters are fixed, QCD sum rules can be used to derive various physical quantities, such as hadron masses, form factors, thereby enabling the investigation of relevant phenomenological issues. In the following chapters, we will discuss the application of QCD sum rules in exploring the structure and decay properties of hadrons.

\subsection{Discussion on the sensitivity of Borel window selection}\

In the QCD sum rule approach, the selection of the Borel window is crucial for obtaining reliable results. The Borel window is defined by the lower and upper bounds of the Borel parameter $ \tau^2 $, which are determined by the OPE convergence and pole dominance criteria, respectively.

\subsubsection{OPE convergence}\

Theoretically, the Wilson coefficients preceding higher-dimensional operators $ \hat{O}_n $ are relatively small since both sides of Eq.~(\ref{ope-expansion}) must maintain consistent dimensions. In high-energy processes, contributions from higher-dimensional operators are further suppressed by their Wilson coefficients, so calculating only the first few terms captures the main expansion contributions, aligning with perturbation theory principles. However, to ensure the convergence of the OPE in practice, one should let the contribution from the highest-order term remain minimal. 

Condensates with $ d \leq 6 $ were provided in the original work for the charmoniums and good convergence was exhibited~\cite{Shifman:1978bx, Shifman:1978by}. The OPE for a series of mesons was detailed in Refs.~\cite{Reinders:1984sr,Colangelo:2005hv}. In the vast majority of these cases, the expansion was truncated at dimension-6, which was sufficient to demonstrate good convergence and establish a stable Borel window. Furthermore, Nikolaev and Radyushkin extended the charmonium analysis to incorporate the contributions of gluonic operators with dimensions $d=6$ and $d=8$~\cite{Nikolaev:1981ff,Nikolaev:1982rq,Nikolaev:1982ra}.

However, for exotic states involving high dimension currents, terms of higher dimensions up to dimension-10 may be included since such contributions cannot be safely ignored~\cite{Matheus:2006xi,Wang:2013vex}. As indicated by Eq.~\eqref{ope-expansion}, standard meson correlation functions are characterized by a lower overall dimension, which leads to a significant suppression of higher-dimensional condensates by factors of $1/q^2$. Conversely, for high-dimension systems like tetraquarks or pentaquarks, the correlation functions exhibit a higher overall dimension, rendering this suppression mechanism less effective.

When dealing with higher-dimensional condensates, the vacuum saturation approximation (VSA) is a commonly used method~\cite{Shifman:1978by}. In essence, this approximation factorizes complex higher-dimensional condensates into products of simpler, lower-dimensional ones, such as $\langle \bar{q}q\bar{q}q\rangle \simeq \langle\bar{q}q\rangle^2$ and $\langle \bar{q}q \bar{q}Gq\rangle \simeq \langle \bar{q}q \rangle \langle \bar{q}Gq\rangle$. A more rigorous approach involves introducing a parameter $\kappa$ to quantify the deviation from the vacuum saturation hypothesis~\cite{Narison:1989aq}:
\begin{align}
\langle \bar{q}q\bar{q}q\rangle &\simeq \kappa \langle\bar{q}q\rangle^2 \,,\\
\langle \bar{q}q \bar{q}Gq\rangle &\simeq \kappa\langle \bar{q}q \rangle \langle \bar{q}Gq\rangle \,.
\end{align}
It is important to emphasize that the VSA is not an arbitrary assumption but is grounded in a profound physical understanding of the QCD vacuum structure. Within the framework of the $1/N_c$ expansion, the VSA becomes strictly valid in the large-$N_c$ limit, with corrections scaling as $O(1/N_c)$. While the VSA typically yields reasonable estimates for lower-dimensional condensates, non-factorizable contributions can become significant when dealing with higher-dimensional condensates (e.g., $d=10$ or higher). Consequently, the direct application of the VSA in high-dimensional truncations may introduce substantial systematic errors. Thus, its applicability must be carefully evaluated by introducing phenomenological parameters, such as the factor $\kappa$.

\subsubsection{Pole contribution}\

The pole contribution measures the ground-state fraction of the total spectral density. While a value exceeding 50\% is ideal to ensure ground-state dominance over continuum and excited states, strict adherence to this limit is often challenging in multiquark studies due to the proximity of the continuum threshold $s_0$ to the ground-state mass. Such a stringent requirement can severely restrict or eliminate the valid Borel window. As a result, a more flexible threshold of 30\% or 40\% is frequently adopted in the literature. This compromise allows for the determination of a reliable Borel window without compromising the significance of the ground-state contribution.

In the case of hexaquark systems, a lower pole contribution, such as 15\%, is often deemed acceptable. This concession is not a relaxation of standards but an intrinsic necessity imposed by the specific mathematical structure of these states. A hexaquark current possesses a mass dimension of 9, leading to a perturbative spectral density that scales with a high power of $s$. Given that the continuum contribution is expressed as $\int_{s_0}^{\infty} \rho(s) e^{-s / \tau^2} ds$, the rapid growth of $\rho(s)$ results in a dominant high-energy contribution. This can only be suppressed by a sufficiently small Borel parameter $\tau^2$. However, since the OPE for hexaquark states involves high-dimensional condensates, an excessive reduction in $\tau^2$ undermines the convergence of the expansion. Consequently, a pole contribution in the range of 15\%–20\% emerges as the optimal balance between OPE convergence and pole dominance. Demanding a threshold as high as 40\% would inevitably lead to the disappearance of the Borel window.

\subsection{Extension of QCD sum rules}\

The preceding section described the canonical approach to constructing QCD sum rules: the Borel transform method, commonly known as the SVZ sum rules or Laplace sum rules. The name ``Laplace sum rule" arises because the Borel transform \eqref{Borel-transformation} in fact represents the algebraic form of an inverse Laplace operator. Narison also contributed the distinctive term ``QCD spectral sum rules'' (QSSR), a witty allusion to the USSR that honors the Soviet roots of the method's three founding fathers~\cite{Narison:2014wqa}. In addition, the continuous development of QCD sum rules has led to the establishment of several distinct variants and extended forms. A concise introduction to these alternative approaches is presented in the concluding part of this section.

\subsubsection{Finite energy sum rules}\

A central challenge in QCD is bridging the gap between asymptotic freedom at short distances, described by quarks and gluons, and color confinement at long distances, manifested as hadronic resonances. Finite Energy Sum Rules (FESR)~\cite{Shankar:1977ap} offer a robust framework to address this issue, grounded in complex analysis and Cauchy's theorem. The cornerstone of this approach is quark-hadron duality. By performing a contour integral of the vacuum polarization function $\Pi(s)$ in the complex plane, FESR relates the integral of the hadronic spectral function (derived from experimental data) to the QCD theoretical prediction, which includes both perturbative and non-perturbative corrections.

\begin{figure}[ht]
\centering
\includegraphics[width=0.4\linewidth]{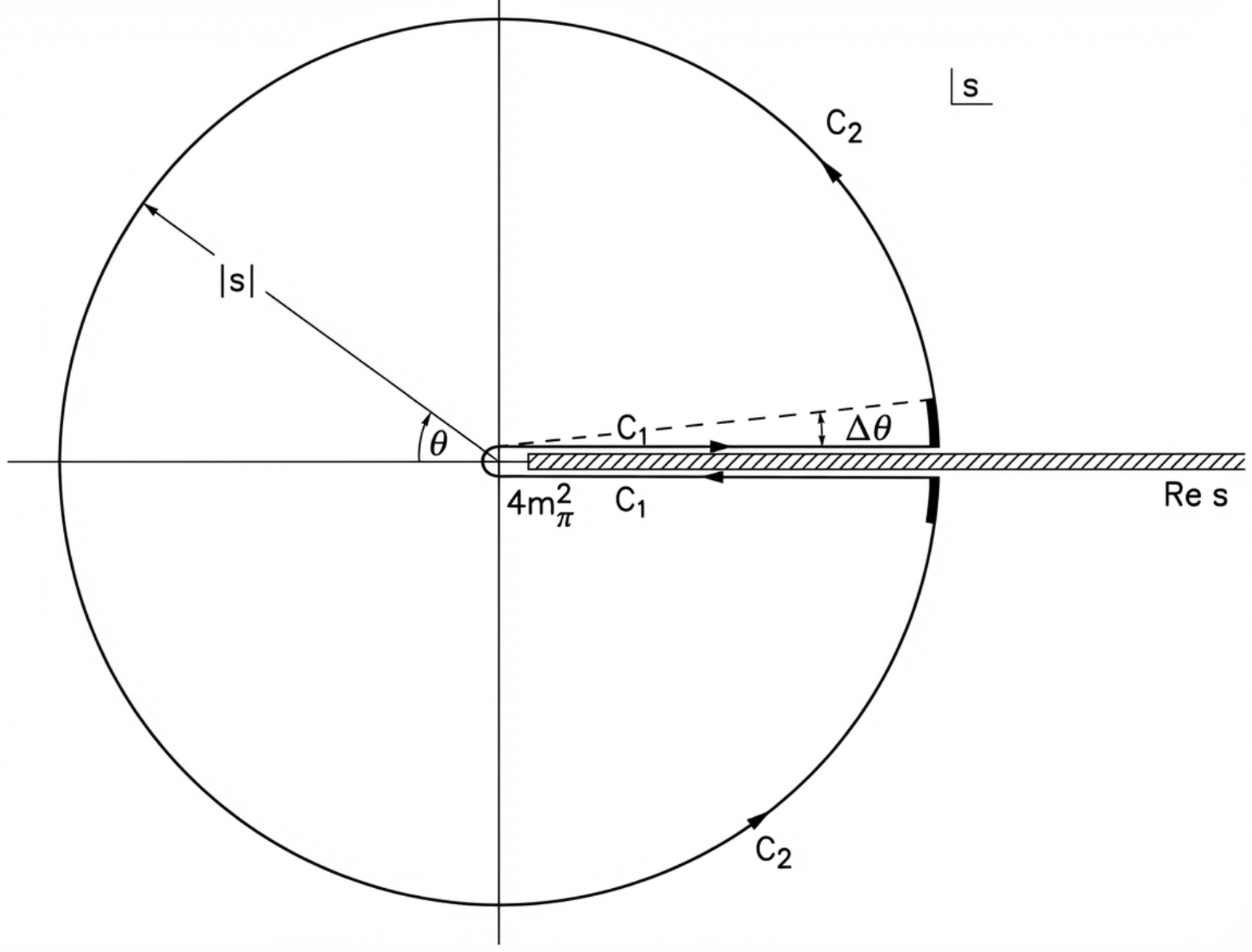}
\caption{The contour integration in the complex $s$-plane for finite energy sum rules~\cite{Shankar:1977ap}.}
\label{fig:Integral}
\end{figure}

The fundamental basis of FESR lies in the analytic properties of the vacuum polarization tensor $\Pi(s)$. Invoking Cauchy's theorem, the integral of an analytic function over a closed contour devoid of singularities vanishes, as illustrated in Fig~\ref{fig:Integral}. Consequently, the integration of $\Pi(s)$ along the closed contour formed by $C_1 + C_2$ is zero:
\begin{align}
\oint_{C_1+C_2} \Pi(s) ds = 0 \quad \Rightarrow \quad \int_{C_1} \Pi(s) ds = - \int_{C_2} \Pi(s) ds \,. 
\end{align}
Along the physical contour $C_1$, the imaginary part of $\Pi(s)$ relates directly to the physical observable $R$ (e.g., the cross-section ratio for $e^+e^- \to \text{hadrons}$), implying that the contour integral represents an integration over the experimental data $R^{\text{exp}}(s)$. On the theoretical contour $C_2$, however, the effective coupling $\alpha_s(s)$ becomes negligible provided $|s|$ is sufficiently large and the contour avoids the resonance singularities on the real axis. This regime allows $\Pi(s)$ to be described by the renormalization-group-improved perturbative QCD series:
\begin{align}
\Pi^{\text {theor }}(s)=\Pi^0(s)+\frac{4}{3} \alpha(s,\Lambda) \Pi^{(1)}(s)+\cdots\,,
\end{align}
where $\Lambda$ is the QCD scale parameter. Then the central equation of FESR can be established~\cite{Shankar:1977ap}:
\begin{align}
\int_{4 m_\pi^2}^s R^{\exp } d s =\int_0^s R^0 d s+\frac{4}{3} \alpha(s,\Lambda) \int_0^s R^{(1)} d s\,.
\label{eq:FESR}
\end{align}
Based on the core FESR equation \eqref{eq:FESR}, Shankar extracted the fundamental QCD scale parameter $\Lambda \approx 700$ MeV, which exhibits good self-consistency when extending the test to higher energy regions. This result demonstrates the effectiveness of perturbative QCD on complex planes.

Established prior to the advent of QCD sum rules, FESR serves as both a precursor and a foundational form of the method. However, the original FESR approach relied primarily on perturbative QCD and lacked a systematic treatment of non-perturbative effects, which limited its application to low-energy strong interaction phenomena. Following the development of QCD sum rules, non-perturbative effects were also incorporated into the FESR framework~\cite{Novikov:1976tn,Bertlmann:1987ty}. Bertlmann et al. explicitly formulated the $n$-th order FESR as follows~\cite{Bertlmann:1987ty}:
\begin{align}
(-)^{n-1} C_{2 n}\left\langle 0_{2 n}\right\rangle= 8 \pi^2 \int_0^{s_0} d s s^{n-1} \frac{1}{\pi} \operatorname{Im} \pi(s) -\frac{s_0^n}{n} F_{2 n}\left(s_0\right)\,.
\label{eq:FESR2}
\end{align}
Eq.~\eqref{eq:FESR2} demonstrates that the isolation of vacuum condensates of specific dimensions can be achieved through the modulation of the weight factor $s^{n-1}$. In contrast to standard QCD sum rules, the FESR defined in Eq.~\eqref{eq:FESR2} exhibits a pronounced sensitivity to the continuum threshold $s_0$. The significance of stability criteria was underscored by Bertlmann et al., who posited that physical observables should remain robust against marginal variations in $s_0$. The extraction of optimal physical parameters was therefore achieved by locating a stability plateau with respect to $s_0$. Applying Eq.~\eqref{eq:FESR2} to the vector meson channel, Bertlmann et al. achieved a successful determination of the four-quark and gluon condensates~\cite{Bertlmann:1987ty}, which provided complementary insights to QCD sum rule analysis~\cite{Shifman:1978bx, Shifman:1978by}.

According to a recent analysis~\cite{Matheus:2006xi}, the FESR formalism can be recovered from Eq.~\eqref{qcdsr-equation-2} by imposing the limit $1/\tau^2 \to 0$ and equating the coefficients of identical powers of $1/\tau^2$ on both sides of the equation.
\begin{align}
\lambda_H^2 \,M_H^2\, \sum_{n=0}^{+\infty} \frac{(-1)^n }{n!\,\tau^{2 n}} M_H^{2 n}=\sum_{n=0}^{+\infty} \frac{(-1)^n }{n!\,\tau^{2 n}} \int_{s_{min}}^{s_0} d s \,s^n\, \rho^{\mathrm{QCD}}(s)\,.
\label{FESR3}
\end{align}
Equating the coefficients of the polynomial in $1 / \tau^2$ in both sides of Eq. \eqref{FESR3} gives $n$ equations:
\begin{align}
\lambda_H^2 \,M_H^{2 n+2}= \int_{s_{min}}^{s_0} d s \,s^n\, \rho^{\mathrm{QCD}}(s)\,.
\end{align}
Dividing the equation for $n+1$ by that for $n$ allows the hadron mass to be expressed as:
\begin{align}
M_H^2=\frac{\int_{s_{min}}^{s_0} d s \,s^{n+1}\, \rho^{\mathrm{QCD}}(s)}{\int_{s_{min}}^{s_0} d s \,s^n\, \rho^{\mathrm{QCD}}(s)}\,.
\label{FESR-mass}
\end{align}
As indicated by Eq.~\eqref{FESR-mass}, the hadron mass $M_H$ exhibits no dependence on the Borel parameter $\tau^2$, a feature that eliminates the necessity of identifying a Borel window. Conversely, the determination of the continuum threshold $s_0$ is of paramount importance, given its direct impact on the numerical results. Rigorous stability criteria regarding $s_0$ must therefore be imposed to guarantee robust predictions, despite the inherent difficulty in establishing such stability within the FESR framework~\cite{Matheus:2006xi}.

\subsubsection{Gaussian sum rules}\

The standard QCD sum rule excels at isolating ground-state hadron properties; however, its exponential kernel $e^{-s/\tau^2}$ suppresses higher-mass contributions, hindering the study of excited resonances. To overcome this monotonic suppression and enable spectroscopic analysis, Bertlmann et al. proposed the Gaussian sum rules (GSR)~\cite{Bertlmann:1984ih}.

Questioning whether QCD sum rules represent the definitive approach in QCD, the authors proposed the Gauss-Weierstrass transform as the optimal spectral function transformation. Specifically,
\begin{align}
G(\hat{s}, \sigma)=\frac{1}{\sqrt{4 \pi \sigma}} \int_0^{\infty} \mathrm{d} s \exp \left(-\frac{(s-\hat{s})^2}{4 \sigma}\right) \frac{1}{\pi} \operatorname{Im} \Pi(s)\,,
\label{GSR}
\end{align}
where $\hat{s}$ represents the center of the Gauss function and $\sigma$ is the width parameter. Eq~\eqref{GSR} can be regarded as the convolution of the spectral function with a Gaussian centered at an arbitrary value $\hat{s}$ with a finite width $\sqrt{2\sigma}$. The primary advantage of this transform over the conventional Laplace transform stems from two elements of enhanced control over the analysis kinematics:
\begin{itemize}
\item Whereas the kernel of the Laplace transform, $e^{-s/\tau^2}$, preferentially weights the low-energy threshold at $s=0$, the Gaussian kernel can be centered at an arbitrary energy scale $\hat{s}$. This freedom liberates the analysis from a sole focus on the ground state, permitting a comprehensive scan of the hadronic spectrum, including resonances and excited states.
\item The Gaussian width, $\sigma$, provides a continuously adjustable parameter to control the resolution of the spectral probe. A fine resolution (small $\sigma$) allows for a detailed examination of the spectral structure around $\hat{s}$ but may be susceptible to instabilities from local OPE fluctuations. In contrast, a coarse resolution (large $\sigma$) ensures theoretical stability by integrating over a wider spectral region. Thus, $\sigma$ functions as the central analysis scale, analogous to the Borel mass $\tau^2$ or the momentum scale $Q^2$ in traditional sum rule approaches.
\end{itemize}
Additionally, the Gaussian-Weierstrass transform provides a formal mechanism to realize the concept of local duality in a quantifiable manner. Local duality implies a point-wise identity between the theoretical (QCD) and the physical spectral functions for all $s$. Such an equivalence is, however, unachievable in practice due to the fundamentally approximate nature of the operator product expansion. The Gaussian sum rule resolves this impasse by constructing a theoretical observable, $G(\hat{s}, \sigma)$, which corresponds to the physical spectrum probed at a finite resolution $\sigma$ around the energy scale $\hat{s}$. Improvements in the theoretical QCD input, such as the incorporation of higher-order corrections, systematically allow for a reduction of $\sigma$, thereby enabling the prediction $G(\hat{s}, \sigma)$ to progressively approach the exact physical spectrum.

A direct application of GSR involves the investigation of the complex structure of the scalar meson spectrum~\cite{Orlandini:2000nv}. Initially, the method was applied to the vector $\rho$ meson channel, yielding a mass of $m_\rho=(0.75 \pm 0.07) \,\mathrm{GeV}$, which is in agreement with experimental data. Subsequently, the approach was extended to non-strange scalar meson channels by introducing a two-resonance model. For the $I=0$ channel, the ground and first excited state masses were determined to be $0.97 \,\mathrm{GeV}$ and $1.43\,\mathrm{GeV}$, respectively. The ground state mass aligns with the $f_0(980)$, supporting a significant non-strange quark component, while the excited state corresponds to the $f_0(1370)$ or $f_0(1500)$ region. In the $I=1$ channel, masses of $1.44 \,\mathrm{GeV}$ and $1.81 \,\mathrm{GeV}$ were obtained. These results support the identification of the $a_0(1450)$, rather than the $a_0(980)$, as the lightest non-strange quark scalar state, and suggest interpreting the $X(1775)$ as a radially excited scalar meson. This work not only demonstrates the unique capability of GSR in resolving excited states but also provides important theoretical constraints on the quark structure of scalar mesons.

GSR is also particularly well-suited for the investigation of glueball physics~\cite{Harnett:2000fy,Wen:2010qoe,Harnett:2008cw,Xian:2014jpa,Ho:2019org,Li:2025hsp}. The intricate nature of glueball spectroscopy, often characterized by minimal mass splittings between ground and excited states, presents significant challenges for conventional QCD sum rule analyses. However, the adjustable centroid and width parameters inherent to GSR facilitate a precise spectral scan across varying energy regions, thereby effectively resolving these near-degenerate states. In the context of glueball-meson mixing, the simultaneous analysis of diagonal and non-diagonal correlation functions enables the unambiguous separation of distinct mass eigenstates~\cite{Harnett:2008cw} and the determination of their coupling strengths. Furthermore, while glueballs are generally broad resonances, traditional analyses often rely on the ``narrow resonance approximation'', which frequently leads to inconsistencies for scalar glueballs~\cite{Harnett:2000fy,Wen:2010qoe}. In contrast, GSR provides a robust framework that accommodates finite-width phenomenological models, allowing for precise constraints on the width and line shape of these resonances.

\subsubsection{Double ratio of sum rules}\

In 1988, Narison pointed out that for heavy-light quark systems (such as $B$ mesons), standard QCD sum rules suffer from poor series convergence in certain channels, where the continuum contribution is difficult to suppress, leading to significant uncertainties~\cite{Narison:1988ep}. Furthermore, the dominant heavy quark mass term ($m_Q$) tends to mask the subtle effects of QCD interactions. Consequently, calculating mass splittings by directly subtracting the results of two independent sum rules yields substantial errors. To address these challenges, Narison proposed the double ratio of sum rules (DRSR). Here the modern version of DRSR is presented~\cite{Albuquerque:2009pr,Narison:2010py, Narison:2010pd, Dias:2011mi,Albuquerque:2012zy}:
\begin{align}
R_{HH^{\prime}}=\frac{\mathcal{R}_H}{\mathcal{R}_{H^{\prime}}} \simeq \frac{M_H^2}{M_{H^{\prime}}^2}\,,
\end{align}
where $\mathcal{R}_H$ and $\mathcal{R}_{H^{\prime}}$ are the mass sum rules~\eqref{mass-function} for hadrons $H$ and $H^{\prime}$, respectively. The essence of this formalism lies in the cancellation mechanism inherent in its expansion. Taking heavy quarkonium as an example, the double ratio can be expanded as follows~\cite{Narison:1995tw}:
\begin{align}
\mathcal{R}_{H H^{\prime}}(x) \equiv \frac{\mathcal{R}_H}{\mathcal{R}_{H^{\prime}}} \simeq \frac{M_H^2}{M_{H^{\prime}}^2}=\Delta_0^{H H^{\prime}}\Big[1+\alpha_s \Delta_{\alpha_s}^{H H^{\prime}}+\frac{4 \pi}{9}\big\langle\alpha_s G^2\big\rangle \sigma^2 x^2 \Delta_G^{H H^{\prime}}\Big]\,,
\end{align}
where $\Delta_0^{H H^{\prime}}$ is the leading-order term ratio, $\Delta_{\alpha_s}^{H H^{\prime}}$ represents the ratio of perturbative $\alpha_s$ correction, and $\Delta_G^{H H^{\prime}}$ encapsulates the ratio from non-perturbative gluon condensate contribution. 

By defining the double ratio, the leading heavy quark mass terms in the numerator and denominator are effectively cancelled, rendering the results largely independent of the heavy quark mass scheme. Similarly, the continuum contributions, which manifest comparably in both parts of the ratio, are suppressed, resulting in high stability across a broad threshold range. This cancellation mechanism isolates minute physical effects, including $SU(3)$ symmetry breaking and hyperfine splittings. Moreover, DRSR facilitates the direct comparison of mass ratios for distinct configurations, such as molecular and compact currents. A ratio near unity implies degeneracy within sum rule precision, whereas deviations reveal the underlying mass hierarchy. Note that the efficacy of the cancellation mechanism in DRSR is strictly predicated on the structural and physical resemblance between the two hadronic states (or correlation functions) under comparison. The arbitrary pairing of unrelated hadrons invalidates the underlying assumption of correlated uncertainties, thereby nullifying the advantages of the method and potentially yielding spurious results.

In practice, the DRSR method is primarily applied in the field of heavy flavor hadron physics. In the original paper~\cite{Narison:1988ep}, the authors explored the mass splitting between $B_c$ and $B$ mesons, interpreting it as an effect of $SU(4)$ symmetry breaking. The mass ratios of spin-$1/2$ and spin-$3/2$ doubly charmed baryons, along with flavor $SU(3)$ breaking effects, were evaluated in Ref.~\cite{Narison:2010py}, where the splitting was found to follow a $1/m_Q$ behavior. Furthermore, DRSR has been employed to investigate the internal structure and mass ratios of exotic states, including the $X(3872)$, $T_{cc}$, and a series of vector $Y$ states~\cite{Narison:2010pd,Dias:2011mi,Albuquerque:2012zy}.

To summarize, since its establishment in 1979, QCD sum rules have evolved significantly, becoming a robust tool for investigating hadronic properties. The continuous extension and adaptation of this formalism allow it to address diverse physical scenarios. As previously discussed, the Borel transform effectively suppresses continuum contributions to isolate ground states; the Gaussian kernel facilitates the spectral scanning of excited states and finite widths; and the double ratio method mitigates heavy quark mass uncertainties to probe mass splittings. This flexibility—tailoring the sum rule kernel to the specific physical objective—underpins the enduring relevance of the theory. Another significant extension of the theory that we did not mention is the light-cone sum rules (LCSR). In contrast to the standard SVZ formalism, which parameterizes non-perturbative dynamics via local vacuum condensates using a short-distance operator product expansion, LCSR operates through an expansion near the light cone. Consequently, the non-perturbative inputs in this framework are described by hadronic light-cone distribution amplitudes (LCDAs) rather than constant vacuum expectation values. As a detailed treatment of LCSR is beyond the scope of this review, readers are referred to Refs.~\cite{Balitsky:1989ry,Braun:1988qv,Chernyak:1990ag,Khodjamirian:1997lay,Ball:1998kk,Colangelo:2000dp,Khodjamirian:2023wol} for comprehensive discussions. 

In the following sections, we will present a systematic review of QCD sum rule applications to baryons, baryon semileptonic decays, and baryonium states. Relevant experimental progress and alternative theoretical approaches are also discussed to provide a broader context.

\section{Baryon spectrum}\label{sec:spectra}
In the past decade, hadron physics has witnessed a renaissance in baryon spectroscopy, driven by a wealth of high-precision data from facilities such as LHCb, Belle II, and BESIII. The discovery of numerous excited charmed and bottom baryons, has challenged our traditional understanding of the quark model. These experimental breakthroughs demand rigorous theoretical interpretations regarding the internal structure and quantum numbers of these new states. Unlike mesons, which are simplified by their two-body nature, baryons offer a unique laboratory for studying the three-body dynamics of QCD~\cite{Isgur:2000ad}. A central question in this field is the organization of the three valence quarks: do they behave as independent entities within a mean field, or do they form highly correlated diquark clusters? This internal geometry significantly influences the excitation spectrum and decay properties. The method of QCD sum rules is particularly well-suited to address this issue, as the choice of the interpolating current allows one to selectively probe different internal correlations, thereby distinguishing between competing structural models. In this chapter, we concentrate on the study of baryons within the framework of QCD sum rules. Comprehensive reviews covering other theoretical and experimental aspects of baryon physics can be found in Refs.~\cite{Copley:1979wj,Korner:1994nh,Capstick:2000qj,Karliner:2008sv,Klempt:2009pi, Crede:2013kia, Chen:2016spr,Chen:2022asf, Cheng:2021qpd}.

Baryons are fermions composed of three valence quarks, so the wave function must be antisymmetric under interchange of any two equal-mass quark, which can be written as 
\begin{align}
\ket{qqq}_A=\ket{ \text{color} }_A \times \ket{ \text{space, spin, flavor} }_S\,,
\end{align}
where the subscripts `$S$' and `$A$' denote symmetry and antisymmetry under interchange of any two equal-mass quarks, respectively. Since baryons are color singlets, the spin-statistics relation requires that their color component of the wave function must be completely antisymmetric. 

For the light-flavor baryons made up of $u$, $d$, and $s$ quarks, light baryons can be included in different $ SU(3) $ multiplets after considering the flavor symmetry:
\begin{align}
3 \otimes 3 \otimes 3=10_S \oplus 8_M \oplus 8_M \oplus 1_A\,,
\end{align}
where the index `$M$' represents mixed symmetry. The nucleons we are familiar with are members of baryon octet. Other spin-$\frac{3}{2}$ baryons such as $ \Delta $ and $\Omega$ belong to baryon decuplet. Furthermore, many excited baryon states can also form additional octet and decuplet multiplets~\cite{ParticleDataGroup:2024cfk}.

For the heavy baryons containing $c$, and $b$ quarks, the addition of $c$ or $b$ quark to the light quarks extends the flavor symmetry to $SU(4)$ or $SU(5)$. Notably, due to the significantly larger masses of the 
$ c $ and $ b $ quarks compared to light quarks, these symmetries are strongly broken. Nevertheless, they remain effective for classifying baryons. For example, the flavor $ SU(4) $ multiplets can be expressed as:
\begin{align}
4 \otimes 4 \otimes 4=20_S \oplus 20_M \oplus 20_M \oplus \bar{4}_A\,.
\end{align}
For heavy hadrons with a single heavy quark, an effective QCD theory based on heavy quark symmetry in the heavy quark limit $ m_{Q}\to \infty $~\cite{Isgur:1989vq}, known as heavy quark effective theory (HQET), has been proposed~\cite{Georgi:1990um,Falk:1990yz}. In HQET, the light quarks and heavy quarks are treated separately. The picture is similar to the atomic system, where the properties of heavy baryons are primarily determined by the light degrees of freedom~\cite{Chen:2022asf}. We recommend that readers refer to Refs.~\cite{Neubert:1993mb, Voloshin:1986dir, Politzer:1988wp, Grinstein:1990mj, Eichten:1989zv, Isgur:1991wq} for more detailed discussions about HQET.

In 1981, Ioffe and Chung et al. published two groundbreaking papers that extended the QCD sum rules from the quark-antiquark system to the three-quark system~\cite{Ioffe:1981kw,Chung:1981wm}, providing crucial evidence of the universality of this method. For spin-$ \frac{1}{2} $ baryon, the two-point correlation function~(\ref{2ptcf}) has two invariant functions:
\begin{align}
\Pi(q^2) = i\int d^4 x e^{iq\cdot x}\bra{0}T\{j(x),\,j^\dagger(0) \}\ket{0}=\slashed{q}\Pi_1(q^2)+\Pi_2(q^2)\,,
\label{2ptcf-1/2-baryon}
\end{align}
where the interpolating currents of baryons is composed of three quark fields. Both invariant functions can be obtained by QCD sum rules. The baryon mass can be determined by the ratio
\begin{align}
R_i=\frac{\int_{s_{min}}^{s_0} s\, \rho_i^{\text{QCD}}(s) \,e^{-s / \tau^2}\,ds}{\int_{s_{min}}^{s_0} \rho_i^{\text{QCD}}(s) \,e^{-s / \tau^2}\,ds}\,,\quad i=1,2\,.
\label{mass-baryon}
\end{align}
Here, $ \rho_i^{\text{QCD}}(s) $ denotes the spectral density corresponding to the invariant function $ \Pi_i(q^2) $. At present, which structure $ R_i $ is most appropriate for determining the baryon mass remains unresolved. In Ref.\cite{Wang:2009ozr}, the authors found that the OPE converges better in the structure $ \Pi_1(q^2) $ than in $ \Pi_2(q^2) $, and the masses of heavy baryons $ \Omega_{c,b} $ obtained from $ \Pi_1(q^2) $ agree more closely with experimental data. Structure $ \Pi_1(q^2) $ is also adopted in Refs.\cite{Aliev:2012ru,Wang:2017qvg,Agaev:2017lip,Azizi:2020azq,Azizi:2022dpn,ShekariTousi:2024mso} for heavy baryons. While in Refs.~\cite{ Zhang:2008rt, Zhang:2009re,Zhang:2008pm, Zhang:2008iz}, Zhang and Huang take the average of the results obtained from both two structures to reduce systematic errors. Furthermore, some earlier studies provide a rough estimate that, the baryon mass satisfies $ M_{\text{baryon}} \simeq \sqrt{R_i} \simeq R_{2}/R_{1}$ at the $\tau$-stability points~\cite{Bagan:1991sc, Bagan:1992tp, Bagan:1992za} with double ratio of sum rules~\cite{Narison:1988ep, Albuquerque:2009pr, Narison:2010py}. 

The case of spin-$ \frac{3}{2} $ baryons is more complex, where more Lorentz structures are involved in the correlation functions compared to Eq.~(\ref{2ptcf-1/2-baryon}). In practice, since the invariant functions multiplied by $ g_{\mu\nu} $ and $ g_{\mu\nu}\slashed{q} $ receive contributions only from spin-$ \frac{3}{2} $ component, the structure containing the metric tensor $ g_{\mu\nu} $ are chosen for the analysis, which can be generally expressed as
\begin{align}
\Pi(q^2) = i\int d^4 x e^{iq\cdot x}\bra{0}T\{j(x),\,j^\dagger(0) \}\ket{0}=g_{\mu\nu}\big[\slashed{q}\Pi_1(q^2)+\Pi_2(q^2)\big]+\cdot\cdot\cdot\,.
\label{2ptcf-3/2-baryon}
\end{align}

In this chapter, we begin with light baryon sector, focusing on the $SU(3)$ flavor octet and decuplet states, along with their corresponding excitations. We then turn to heavy baryons, systematically discussing the spectral properties of singly, doubly, and triply heavy systems. Finally, we examine a distinct class of baryonic configurations-hybrid baryons, which incorporate explicit gluonic degrees of freedom.

\subsection{Light baryons}\label{light baryon}\

Over 100 light baryons are now listed in Particle Data Group~\cite{ParticleDataGroup:2024cfk}, including the ground and excited states of the octet and decuplet. The mass of the ground baryon octet and decuplet is listed in Table~\ref{table:mass-baryon-octet-decuplet}. Historically, there has been significant debate over light baryons in QCD sum rules, particularly regarding the choice of the most suitable currents~\cite{Chung:1981cc, Chung:1982rd, Ioffe:1982ce, Dosch:2010zz, Belyaev:1982sa}. 

\begin{table}[ht]
\centering
\renewcommand{\arraystretch}{1.5}
\caption{Mass of baryon octet and decuplet~\cite{ParticleDataGroup:2024cfk}.}
\begin{tabular}{ccccccccccc}
\hline\hline
& $ p $ &$ n $& $ \Lambda $ & $ \Sigma^+ $& $ \Sigma^0 $ & $ \Sigma^- $& $ \Xi^0 $& $ \Xi^- $ \\ \hline
$ M(\text{GeV})$&0.938&0.940&1.116&1.189&1.193&1.197&1.315 & 1.322 \\\hline
& $ \Delta^{++} $ & $ \Delta^+ $ & $ \Delta^0 $& $ \Delta^- $ & $ \Sigma^{*+} $& $ \Sigma^{*0} $& $ \Sigma^{*-} $& $ \Xi^{*0} $& $ \Xi^{*-} $& $ \Omega^- $ \\ \hline
$ M(\text{GeV})$&1.232&1.232&1.233 & 1.232 & 1.383&1.384&1.387&1.532&1.535&1.672\\
\hline\hline
\end{tabular}
\label{table:mass-baryon-octet-decuplet}
\end{table}

Ioffe proposed two forms of the proton current~\cite{Ioffe:1981kw, Ioffe:1982ce}
\begin{align}
\label{proton-1}
j_p(x)&=\varepsilon_{a b c}\big(u^{a T}(x) \mathcal{C} \gamma^\mu u^b(x)\big) \gamma_5 \gamma_\mu d^c(x)\,, \\[5pt]
\label{proton-2}
j_{p}(x)&=\varepsilon_{a b c}\big(u^{a T}(x) \mathcal{C} \sigma^{\mu \nu} u^b(x)\big) \gamma_5 \sigma_{\mu \nu} d^c(x)\,,
\end{align}
with following criteria. First, the contributions from excited states should be minimized, which requires using a current with the fewest possible derivatives. Second, within the chosen range of the Borel parameter, the neglected power corrections in the OPE must remain sufficiently small. Finally, the current should couple strongly to the targeted baryon state. Since the current~(\ref{proton-2}) does not receive contributions from the lowest-dimensional chiral symmetry breaking operators, the resonance is expected to couple weakly to this current~\cite{Reinders:1984sr}. Therefore, Ioffe argued that the most suitable current for the determination of the proton mass is the current~(\ref{proton-1}). This conclusion is supported by subsequent analysis~\cite{Espriu:1983hu}, which considers linear combinations of interpolating currents with suitably chosen coefficients to maximize the overlap.

Using the current~(\ref{proton-1}) and neglecting the power correction, a simple form of the proton mass and decay constant can be derived:
\begin{align}
M_p &=\big[-2(2 \pi)^2\langle\bar{q} q\rangle\big]^{1/3}\cong 1\,\text{GeV}\,,\\[5pt]
\lambda_{p}^2&=M_{p}^6 e / 2(2 \pi)^4 \cong 1.2\times 10^{-3}\,\text{GeV}\,,
\end{align}
where the uncertainties in the mass and decay constant are approximately 20\% and 50\%, respectively. Replacing the $ d $ quark in current~(\ref{proton-1}) with an $ s $ quark allows for the treatment of the $ \Sigma $ hyperon, leading to the mass splitting with $M_{\Sigma}-M_{p}=240\, \mathrm{MeV}$. Both results show significant accuracy with the experimental data.

The Ioffe-type currents for other light baryon octet and decuplet states can also be constructed:
\begin{align} 
& j_{\Sigma}(x)=\varepsilon_{a b c}\big(u^{aT}(x) C \gamma^\mu u^b(x)\big) \gamma_5 \gamma_\mu s^c(x)\,, \\[3pt]
& j_{\Lambda}(x)=\sqrt{\frac{2}{3}} \,\varepsilon_{a b c}\Big[\big(u^{aT}(x) C \gamma^\mu s^b(x)\big) \gamma_5 \gamma_\mu d^c(x)-\big(d^{aT}(x) C \gamma^\mu s^b(x)\big) \gamma_5 \gamma_\mu u^c(x)\Big]\,, \\[3pt]
& j_{\Xi}(x)=-\varepsilon_{a b c}\big(s^{aT}(x) C \gamma^\mu s^b(x)\big) \gamma_5 \gamma_\mu u^c(x)\,,\\[3pt]
&j_{\Delta}^\mu(x)=\varepsilon_{a b c}\big(u^{aT}(x) C \gamma^\mu u^b(x)\big) u^c(x)\,,\\[3pt]
&j_{\Sigma^*}^\mu(x)=\sqrt{\frac{1}{3}}\, \varepsilon_{a b c}\Big[2\big(u^{aT}(x) C \gamma^\mu s^b(x)\big) u^c(x)+\big(u^{aT}(x) C \gamma^\mu u^b(x)\big) s^c(x)\Big]\,,\\[3pt]
&j_{\Xi^*}^\mu(x)=\sqrt{\frac{1}{3}}\, \varepsilon_{a b c}\Big[2\big(s^{aT}(x) C \gamma^\mu u^b(x)\big) s^c(x)+\big(s^{aT}(x) C \gamma^\mu s^b(x)\big) u^c(x)\Big]\,,\\[3pt]
&j_{\Omega}^\mu(x)=\varepsilon_{a b c}\big(s^{aT}(x) C \gamma^\mu s^b(x)\big) s^c(x)\,.
\end{align}
In Ref.~\cite{Ioffe:1981kw}, Ioffe also derived the mass and decay constant of $ \Delta $ baryon, as well as the mass splitting within the decuplet:
\begin{align}
M_{\Delta}=1.4 \,\mathrm{GeV} \pm 15 \%\,,\quad \lambda_{\Delta}^2=2.5 \times 10^{-3} \,\mathrm{GeV}^6\,,\quad M_{\Sigma^*}-M_{\Delta}=125\, \mathrm{MeV}\,.
\end{align}
However, no reliable results were obtained for $ \Lambda $ and $ \Xi $ hyperon, this failure was attributed to the effects of instantons in the nonperturbative regime~\cite{Ioffe:1981kw}.

Another form of effective interpolating current was introduced by Chung et al. based on the positivity conditions of the spectral function~\cite{Chung:1981cc,Chung:1981wm,Chung:1984gr,Dosch:1988vv}, which is constructed as a linear combination of different currents:
\begin{align}
j_p(x)&=\sqrt{\frac{1}{2}}\varepsilon_{a b c}\big[\big(u^{aT}(x) C \gamma^5 u^b(x)\big) d^c(x)+\beta\big(u^{aT}(x) C u^b(x)\big)\gamma^5 d^c(x)\big]\,,\\
j_\Delta^\mu(x)&=\sqrt{\frac{1}{2}}\varepsilon_{a b c}\,u^{aT}(x) C \gamma^5 \gamma_\rho d^b(x)\big(g^{\mu \rho}-\frac{1}{4} \gamma^\mu \gamma^\rho\big) d^c(x)\,.
\end{align}
The mixing parameter is chosen as $ \beta = -1/5 $ with the maximum contribution from the ground hadron~\cite{Chung:1981cc,Chung:1981wm}. With this choice, the authors in Ref.~\cite{Dosch:1988vv} obtained the complete mass spectrum of the baryon octet and decuplet:
\begin{align}
&M_N=1.05\,\text{GeV}\,,\quad M_{\Lambda}=1.24\,\text{GeV}\,,\quad M_{\Sigma}=1.16\,\text{GeV}\,,\quad M_{\Xi}=1.33\,\text{GeV}\,,\nonumber\\
&M_{\Delta}=1.21\,\text{GeV}\,,\quad M_{\Sigma^*}=1.35\,\text{GeV}\,,\quad M_{\Xi^*}=1.48\,\text{GeV}\,,\quad M_{\Omega}=1.61\,\text{GeV}\,.
\end{align}

It is found that these two types of currents lead to slight differences in the results. In principle, if the OPE were calculated to infinite order and the spectral function included the complete set of excited states, any valid interpolating current would yield the identical physical hadron mass. However, in practical applications, the OPE series must be truncated at a finite order. This truncation inevitably leads to a dependence of the extracted results on the specific choice of the interpolating current. Additionally, the Borel parameter and threshold parameter have a non-negligible impact on the outcome. Therefore, reasonable constraints must be applied in actual calculations to obtain reliable results using QCD sum rules.

Other studies of ground baryon octet and decuplet can be found in Refs.~\cite{Nesterenko:1983ef,Yang:1993bp,Hwang:1994vp,Jido:1996ia,Jido:1996zw,Lee:2002jb,Lee:2006bu,Chen:2008qv,Chen:2009sf,Chen:2012ex}, with different choices of interpolating currents. In Refs.~\cite{Jido:1996ia,Jido:1996zw}, Jido et al. presented studies of both positive and negative parity baryons, which will be discussed later. In Refs.~\cite{Yang:1993bp,Hwang:1994vp}, the authors investigated the mass differences between the neutron-proton, $ \Delta $-$ N $, and $ \Sigma $-$ \Lambda $, finding that the mass splittings of $ \Delta $-$ N $ and $ \Sigma $-$ \Lambda $ are primarily driven by the quark-gluon condensate. In Refs.~\cite{Lee:2002jb,Lee:2006bu}, Lee et al. calculated the mass spectrum of the baryon octet and decuplet using both the conventional method and a parity-projection approach. They also assessed the predictive accuracy of the sum rules through a Monte Carlo-based analysis, which suggested that a chiral-odd sum rules are more effective when dealing with baryon correlation functions. In Refs.~\cite{Chen:2008qv,Chen:2009sf,Chen:2012ex}, Chen et al. classified baryon interpolating fields as local products of three quarks based on their flavor and chiral structures, from which they derived the masses of ground baryons in the octet and decuplet for different chiral representations. The mass and decay constant of ground baryon octet and decuplet calculated in papers are summarized in Table~\ref{table:QCDSR-baryon-octet-decuplet}. 

\begin{table}[ht]
\centering
\caption{The masses (GeV) and decay constants of ground light baryons in the octet and decuplet obtained from QCD sum rules, along with experimental values~\cite{ParticleDataGroup:2024cfk}. Multiple entries in a cell correspond to different choices of interpolating currents.}
\renewcommand{\arraystretch}{1.5}
\resizebox{0.9\textwidth}{!}{
\tiny
\begin{tabular}{cccc}
\hline\hline
State & Exp & Mass &Decay constant ($ \text{GeV}^3 $)\\
\hline
\multirow{2}{*}{$\ket{N,\frac{1}{2}^+} $} &\multirow{2}{*}{0.94}&0.99~\cite{Ioffe:1981kw}, 1.05~\cite{Dosch:1988vv}, 0.94~\cite{Jido:1996zw},&0.035~\cite{Ioffe:1981kw}, 0.0255~\cite{Nesterenko:1983ef}, \\
&&1.06$\pm$0.11~\cite{Lee:2002jb}, 0.95$\pm$0.06~\cite{Chen:2012ex}& 0.02~\cite{Chung:1984gr}, 0.024$\pm$0.004~\cite{Chen:2012ex}\\\hline

\multirow{2}{*}{$\ket{\Lambda,\frac{1}{2}^+} $} &\multirow{2}{*}{1.12}&1.24~\cite{Dosch:1988vv}, 1.118~\cite{Hwang:1994vp}, 1.12~\cite{Jido:1996zw},&0.0324~\cite{Hwang:1994vp},\\
&&1.23$\pm$0.12~\cite{Lee:2002jb}, 1.11$\pm$0.06~\cite{Chen:2012ex}&0.027$\pm$0.004~\cite{Chen:2012ex}\\\hline

\multirow{2}{*}{$\ket{\Sigma,\frac{1}{2}^+} $} &\multirow{2}{*}{1.19}&1.16~\cite{Dosch:1988vv}, 1.184~\cite{Hwang:1994vp}, 1.21~\cite{Jido:1996zw},&0.0347~\cite{Hwang:1994vp},\\
&&1.16$\pm$0.12~\cite{Lee:2002jb}, 1.23$\pm$0.06~\cite{Chen:2012ex},&0.037$\pm$0.004~\cite{Chen:2012ex}\\\hline

\multirow{2}{*}{$\ket{\Xi,\frac{1}{2}^+} $} &\multirow{2}{*}{1.32}&1.33~\cite{Dosch:1988vv}, 1.32~\cite{Jido:1996zw},&\multirow{2}{*}{0.032$\pm$0.004~\cite{Chen:2012ex}}\\
&&1.32$\pm$0.14~\cite{Lee:2002jb}, 1.31$\pm$0.07~\cite{Chen:2012ex}&\\\hline

\multirow{2}{*}{$\ket{\Delta,\frac{3}{2}^+} $} &\multirow{2}{*}{1.23}&1.4~\cite{Ioffe:1981kw}, 1.21~\cite{Dosch:1988vv}, 1.38~\cite{Hwang:1994vp},&0.05~\cite{Ioffe:1981kw}, 0.04~\cite{Hwang:1994vp},\\
&&1.21$\pm$0.06~\cite{Lee:2006bu}, 1.21$\pm$0.06~\cite{Chen:2012ex}&0.040$\pm$0.016~\cite{Lee:2006bu}, 0.070$\pm$0.005~\cite{Chen:2012ex}\\\hline

\multirow{2}{*}{$\ket{\Sigma^*,\frac{3}{2}^+} $} &\multirow{2}{*}{1.38}&1.35~\cite{Dosch:1988vv}, 1.38$\pm$0.06~\cite{Lee:2006bu}, &0.047$\pm$0.017~\cite{Lee:2006bu},\\
&&1.38$\pm$0.07~\cite{Chen:2012ex}&0.086$\pm$0.007~\cite{Chen:2012ex}\\\hline

\multirow{2}{*}{$\ket{\Xi^*,\frac{3}{2}^+} $} &\multirow{2}{*}{1.53}&1.48~\cite{Dosch:1988vv}, 1.47$\pm$0.05~\cite{Lee:2006bu},&0.052$\pm$0.017~\cite{Lee:2006bu},\\
&&1.52$\pm$0.08~\cite{Chen:2012ex}&0.10$\pm$0.01~\cite{Chen:2012ex}\\\hline

\multirow{2}{*}{$\ket{\Omega,\frac{3}{2}^+} $} &\multirow{2}{*}{1.67}&1.61~\cite{Dosch:1988vv}, 1.68$\pm$0.05~\cite{Lee:2006bu},&0.068$\pm$0.020~\cite{Lee:2006bu},\\
&&1.66$\pm$0.09~\cite{Chen:2012ex}&0.13$\pm$0.01~\cite{Chen:2012ex}\\\hline
\hline
\end{tabular}
}
\label{table:QCDSR-baryon-octet-decuplet}
\end{table}

With advancements in experimental techniques, several highly excited baryons have been discovered, such as $ N(1535) $, $ \Lambda(1405) $, $ \Lambda(1520) $, $ \Sigma(1620) $, $ \Xi(1690) $, and $ \Omega(2012) $. These discoveries have expanded the known baryon spectrum, highlighting discrepancies with quark model predictions and drawing increased attention to the study of highly excited states. In QCD sum rules, such excited baryons are typically treated as part of the `continuum', with their contributions incorporated accordingly. Moreover, when considering excited states, the role of negative-parity baryons must also be taken into account. Therefore, let us take some time to discuss the issue of negative-parity baryons.

The spinor structure inherent to the interpolating operator generally does not possess a well-defined parity. This implies that the baryon current exhibits a non-vanishing overlap with both the positive-parity ground state, characterized by the spinor $u(p)$, and the negative-parity excitations represented by $\gamma_5 u(p)$. Within the phenomenological representation, this mixing manifests as a superposition of positive and negative parity states, which differ only by the sign of the mass terms in the Dirac decomposition. Such a distinction is physically significant, as it underscores the mass splitting of parity doublets induced by the spontaneous breaking of chiral symmetry in QCD. Hence, negative-parity baryons can be described by introducing an additional $\gamma_5$ matrix in the QCD sum rules framework~\cite{Chung:1981cc}. In Ref.~\cite{Bagan:1993ii}, Bagan et al. unambiguously separated the positive and negative parity contributions to baryonic correlation functions, from which they examined the reliability of baryon sum rules in the heavy quark effective theory. In 1996, a novel approach was proposed by Jido et al. to separate the contribution of the negative-parity light baryons from the positive-parity ones~\cite{Jido:1996ia, Jido:1996zw}. According to Lorentz covariance, the correlation functions containing the negative-parity baryon component can be written as:
\begin{align}
\Pi_{\pm}(q^2) =i\int d^4 x e^{iq\cdot x}\bra{0}T\{j_{\pm}(x),\,j^\dagger_{\pm}(0) \}\ket{0}=\slashed{q}\Pi_1(q^2)\pm\Pi_2(q^2)\,,
\label{negative-baryon}
\end{align}
where $ j_{-}=i\gamma_5 j_+ $ and the subscripts `$ \pm $' denote the positive and negative parity, respectively. The invariant functions $ \Pi_1(q^2) $ and $ \Pi_2(q^2) $ are same in the positive and negative parity sectors, with the only difference being the sign in front of $ \Pi_2(q^2) $. The current $ j_+ $ couples to both positive and negative parity baryons
\begin{align}
\bra{0} j_{+}\ket{B^{\pm}}\bra{B^{\pm}} j^\dagger_{+}\ket{0}=-\gamma_5\bra{0} j_{-}\ket{B^{\pm}}\bra{B^{\pm}} j^\dagger_{-}\ket{0}\gamma_5\,,
\end{align}
where $ \ket{B^{\pm}} $ denotes a single baryon state with positive and negative parity, respectively. Then the phenomenological representation can be obtained:
\begin{align}
\Pi_{+}(q^2)=\lambda_{+}^2 \frac{\slashed{q}+M_{+}}{M_{+}^2-q^2}+\lambda_{-}^2 \frac{\slashed{q}-M_{-}}{M_{-}^2-q^2}+\cdots,
\end{align}
where $ M_{\pm} $ represents the mass of the positive and negative parity baryons, respectively. The decay constant is defined as $ \bra{0} j_{\pm}\ket{B^{\pm}}=\lambda_{\pm}u_{\pm} $, where $ u_{\pm} $ are the Dirac spinor. When focusing on the lowest-lying positive-parity baryon, excited states are treated as part of the ``continuum" contribution. As a result, terms from negative-parity baryons do not appear explicitly. Therefore, separating the contribution of negative-parity baryons is essential. In this review, we do not present the details of this separation, and we refer the reader to Refs.~\cite{Jido:1996ia, Jido:1996zw} for further details. The masses of negative-parity light baryons obtained from QCD sum rules and possible corresponding resonances (in brackets) are~\cite{Jido:1996ia, Jido:1996zw}:
\begin{align}
&M_{N(\frac{1}{2}^-)}=1.54\,\text{GeV}\big[N(1535)\big]\,,\quad M_{\Lambda(\frac{1}{2}^-)}=1.55\,\text{GeV}\big[\Lambda(1670)\big]\,,\quad M_{\Lambda_{S(\frac{1}{2}^-)}}=1.31\,\text{GeV}\big[\Lambda(1405)\big]\,,\nonumber\\[5pt]
&M_{\Sigma(\frac{1}{2}^-)}=1.63\,\text{GeV}\big[\Sigma(1620)\big]\,,\quad M_{\Xi(\frac{1}{2}^-)}=1.63\,\text{GeV}\big[\Xi(1690)\big]\,.
\end{align}
Here, $ \Lambda_{S(\frac{1}{2}^-)} $ denotes a flavor-singlet baryon with negative parity. The spin-parity of $ \Xi(1690) $ remains uncertain, but some evidence suggests that it has $ J^P = \frac{1}{2}^- $~\cite{BaBar:2008myc}. 

Later in Refs.~\cite{Lee:2002jb,Lee:2006bu}, the authors provided information of $ J^{PC}=\frac{3}{2}^- $ light baryons, where the baryon currents they used are
\begin{alignat}{2}
j^{3 / 2, \mu}_N &= \varepsilon_{a b c} 
\big[\big(u^{a T} C \sigma_{\rho \lambda} d^b\big) \sigma^{\rho \lambda} \gamma^\mu u^c
-\big(u^{a T} C \sigma_{\rho \lambda} u^b\big) \sigma^{\rho \lambda} \gamma^\mu d^c\big]\,, \\
j^{3 / 2, \mu}_{\Sigma} &= \varepsilon_{a b c} 
\big[\big(u^{a T} C \sigma_{\rho \lambda} s^b\big) \sigma^{\rho \lambda} \gamma^\mu u^c
-\big(u^{a T} C \sigma_{\rho \lambda} u^b\big) \sigma^{\rho \lambda} \gamma^\mu s^c\big]\,, \\
j^{3 / 2, \mu}_{\Xi} &= \varepsilon_{a b c} 
\big[\big(s^{a T} C \sigma_{\rho \lambda} u^b\big) \sigma^{\rho \lambda} \gamma^\mu s^c
-\big(s^{a T} C \sigma_{\rho \lambda} s^b\big) \sigma^{\rho \lambda} \gamma^\mu u^c\big]\,, \\
j^{3 / 2, \mu}_{\Lambda_O} &= \sqrt{\frac{1}{6}} \varepsilon_{a b c} 
\big[2\big(u^{a T} C \sigma_{\rho \lambda} d^b\big) \sigma^{\rho \lambda} \gamma^\mu s^c
+\big(u^{a T} C \sigma_{\rho \lambda} s^b\big) \sigma^{\rho \lambda} \gamma^\mu d^c -\big(d^{a T} C \sigma_{\rho \lambda} s^b\big) \sigma^{\rho \lambda} \gamma_\mu u^c\big]\,, \\
j^{3 / 2, \mu}_{\Lambda_S} &= \sqrt{\frac{1}{3}} \varepsilon_{a b c} 
\big[\big(u^{a T} C \sigma_{\rho \lambda} d^b\big) \sigma^{\rho \lambda} \gamma^\mu s^c
-\big(u^{a T} C \sigma_{\rho \lambda} s^b\big) \sigma^{\rho \lambda} \gamma^\mu d^c +\big(d^{a T} C \sigma_{\rho \lambda} s^b\big) \sigma^{\rho \lambda} \gamma_\mu u^c\big]\,, \\
j^{3 / 2, \mu}_{\Lambda_C} &= \sqrt{\frac{1}{2}} \varepsilon_{a b c} 
\big[\big(u^{a T} C \sigma_{\rho \lambda} s^b\big) \sigma^{\rho \lambda} \gamma^\mu d^c
-\big(d^{a T} C \sigma_{\rho \lambda} s^b\big) \sigma^{\rho \lambda} \gamma^\mu u^c\big]\,,\\
j^\mu_{\Delta}(x)&=\varepsilon_{a b c}\big(u^{a T}(x) C \sigma_\mu u^b(x)\big) u^c(x)\,,\\
j^\mu_{\Sigma^*}&=\sqrt{\frac{1}{3}} \varepsilon_{a b c}\big[2\big(u^{a T} C \sigma^\mu s^b\big) u^c+\big(u^{a T} C \sigma^\mu u^b\big) s^c\big]\,,\\
j^\mu_{\Xi^*}&=\sqrt{\frac{1}{3}} \varepsilon_{a b c}\big[2\big(s^{a T} C \sigma^\mu u^b\big) s^c+\big(s^{a T} C \sigma^\mu s^b\big) u^c\big]\,,\\
j^\mu_{\Omega^{-}}&=\varepsilon_{a b c}\big(s^{a T} C \sigma^\mu s^b\big) s^c\,.
\end{alignat}
Here, the subscripts`$O$', `$S$', and `$C$' denote octet, flavor-singlet, and common $ \Lambda $ states, respectively. The parities are projected by considering the `forward-propagating' version of the correlation function. Their results provide theoretical insights into the structure of excited light baryons. For instance, for the two baryons $ \Xi(1950) $ and $ \Omega(2380) $, whose $ J^{PC} $ quantum numbers remain undetermined, they offered interpretations based on different threshold parameter choices and quantum numbers. In Ref.\cite{Aliev:2018hre}, the properties of the ground $ \Xi $ baryon and its first orbitally and radially excited states were analyzed. The obtained mass of $ \Xi(1690) $ agrees well with previous studies\cite{Jido:1996zw,Lee:2002jb,Lee:2006bu} and experimental data, whereas notable discrepancies were found in the decay constant.

In Refs.~\cite{Azizi:2019dfh,Azizi:2024mmb}, Azizi et al. conducted a comprehensive study for the excited states of the nucleon and $ \Lambda $ baryon with $ J^P=\frac{3}{2}^{\pm} $. Their theoretical results align with the mass ranges listed in the Particle Data Group\cite{ParticleDataGroup:2024cfk} for the resonances $ N(1520) $, $ N(1700) $, $ N(1720) $, $ \Lambda(1520) $, $ \Lambda(1690) $, and $ \Lambda(1890) $. They also derived the corresponding decay constants.

For the recently observed resonance $ \Omega(2012) $~\cite{Belle:2018mqs,Belle:2021gtf,Belle:2022mrg}, the authors in Refs.~\cite{Aliev:2018syi,Su:2024lzy} interpreted it as an excited state of the $ \Omega $ baryon with three strange quarks and provided theoretical analysis of its mass and decay constant. In Ref.~\cite{Su:2024lzy}, the authors chose the currents containing a derivative for $ \Omega(2012) $
\begin{align}
\begin{aligned}
J & =-2 \varepsilon_{a b c}\big[\big(D^\mu s_a^T\big) C \gamma_5 s_b\big] \gamma_\mu s_c\,, \\
J_\mu & =-2 \varepsilon_{a b c}\big[\big(D^\nu s_a^T\big) C \gamma_5 s_b\big]\big(g_{\mu \nu}-\frac{1}{4} \gamma_\mu \gamma_\nu\big) s_c\,.
\end{aligned}
\end{align}
With this derivative, the diquark acquires a well-defined $ P $-wave nature, providing clearer sum rule signals~\cite{Su:2024lzy}. Their results suggested that the $ \Omega(2012) $ could possess negative parity. Additionally, other possible excited $ \Omega $ states were discussed in Ref.~\cite{Su:2024lzy}.

The QCD sum rule results for excited light baryons are summarized in Table~\ref{table:QCDSR-lightbaryon-excited}. It is evident that the current precision is insufficient to clearly distinguish light resonances, especially considering their small mass splittings. Future studies may need to include next-to-leading order contributions. In addition to the conventional interpretation as nucleon excitations, these states are also hypothesized as possible candidates for pentaquark molecules~\cite{Lebed:2015dca, He:2017aps, Huang:2018ehi,Lin:2018kcc, Ben:2024qeg, Wu:2023ywu, Ben:2023uev, Suo:2025rty, Tian:2025bkx,Ben:2025wqn}, which are beyond the scope of this review.

\begin{table}[ht]
\centering
\caption{Same caption with Table~\ref{table:QCDSR-baryon-octet-decuplet}, but for excited light baryons.}
\renewcommand{\arraystretch}{1.4}
\resizebox{1\textwidth}{!}{
\tiny
\begin{tabular}{cccc}
\hline\hline
State & Exp & Mass &Decay constant (GeV$^3$) \\\hline

$\ket{N(1520),\frac{3}{2}^-} $&1.52&1.44$\pm$0.13~\cite{Lee:2002jb}, 1.505$\pm$0.025~\cite{Azizi:2019dfh}&0.13$\pm$0.09~\cite{Lee:2002jb}, 0.127$\pm$0.004~\cite{Azizi:2019dfh} \\\hline

$\ket{N(1535),\frac{1}{2}^-} $&1.54&1.54~\cite{Jido:1996zw}& \\\hline

$\ket{N(1700),\frac{3}{2}^-} $&1.70&1.70~\cite{Lee:2006bu}, 1.701$\pm$0.062~\cite{Azizi:2019dfh}&0.032$\pm$0.023~\cite{Lee:2006bu}, 0.129$\pm$0.011~\cite{Azizi:2019dfh} \\\hline

$\ket{N(1720),\frac{3}{2}^+} $&1.72&1.709$\pm$0.042~\cite{Azizi:2019dfh}&0.111$\pm$0.041~\cite{Azizi:2019dfh} \\\hline

$\ket{\Delta(1700),\frac{3}{2}^-} $&1.70&1.70~\cite{Lee:2006bu}&0.032$\pm$0.023~\cite{Lee:2006bu} \\\hline

$\ket{\Delta(1910),\frac{1}{2}^+} $&1.91&2.10$\pm$0.54~\cite{Lee:2006bu}&$-$ \\\hline

$\ket{\Lambda(1405),\frac{1}{2}^-} $&1.67&1.31~\cite{Jido:1996zw}, 1.43$\pm$0.13~\cite{Lee:2002jb}&0.16$\pm$0.10~\cite{Lee:2002jb} \\\hline

\multirow{2}{*}{$\ket{\Lambda(1520),\frac{3}{2}^-} $}&\multirow{2}{*}{1.52}&1.71$\pm$0.15~\cite{Lee:2002jb}, &0.11$\pm$0.08~\cite{Lee:2002jb}, 0.034$\pm$0.002~\cite{Azizi:2024mmb} \\
&&1.514(1.470)~\cite{Azizi:2024mmb}&0.049$\pm$0.002~\cite{Azizi:2024mmb}\\\hline

$\ket{\Lambda(1670),\frac{1}{2}^-} $&1.67&1.55~\cite{Jido:1996zw}, 1.38$\pm$0.32~\cite{Lee:2002jb}&0.20$\pm$0.17~\cite{Lee:2002jb} \\\hline

\multirow{2}{*}{$\ket{\Lambda(1690),\frac{3}{2}^-} $}&\multirow{2}{*}{1.69}&1.80$\pm$0.16~\cite{Lee:2002jb},&0.10$\pm$0.07~\cite{Lee:2002jb}, 0.021$\pm$0.003~\cite{Azizi:2024mmb}, \\
&&1.688(1.733)~\cite{Azizi:2024mmb}&0.029$\pm$0.004~\cite{Azizi:2024mmb}\\\hline

\multirow{2}{*}{$\ket{\Lambda(1890),\frac{3}{2}^+} $}&\multirow{2}{*}{1.89}&\multirow{2}{*}{1.882(1.843)~\cite{Azizi:2024mmb}}&0.021$\pm$0.003~\cite{Azizi:2024mmb}\\
&&&0.030$\pm$0.004~\cite{Azizi:2024mmb}\\\hline

$\ket{\Sigma(1580),\frac{3}{2}^-}$&1.58&1.69$\pm$0.14~\cite{Lee:2002jb}, 1.58~\cite{Lee:2006bu}&0.17$\pm$0.11~\cite{Lee:2002jb}, 0.027$\pm$0.020~\cite{Lee:2006bu} \\\hline

$\ket{\Sigma(1620),\frac{1}{2}^-}$&1.62&1.63~\cite{Jido:1996zw}, 1.54$\pm$0.34~\cite{Lee:2002jb}&0.24$\pm$0.18~\cite{Lee:2002jb} \\\hline

$\ket{\Sigma(1670),\frac{3}{2}^-}$&1.67&1.67~\cite{Lee:2006bu}&0.029$\pm$0.021~\cite{Lee:2006bu} \\\hline

$\ket{\Sigma(1880),\frac{1}{2}^+} $&1.88&1.97$\pm$0.62~\cite{Lee:2006bu}&$-$ \\\hline

\multirow{2}{*}{$\ket{\Xi(1690),?} $}&\multirow{2}{*}{1.69}&1.63~\cite{Jido:1996zw}, 1.55$\pm$0.38~\cite{Lee:2002jb},&0.27$\pm$0.23~\cite{Lee:2002jb}, 0.019$\pm$0.004~\cite{Aliev:2018hre}, \\
&&1.69~\cite{Lee:2006bu}, 1.685$\pm$0.069~\cite{Aliev:2018hre}&0.055$\pm$0.010~\cite{Aliev:2018hre}, 0.027$\pm$0.020~\cite{Lee:2006bu}\\\hline

$\ket{\Xi(1820),\frac{3}{2}^-} $&1.82&1.84$\pm$0.16~\cite{Lee:2002jb}, 1.82~\cite{Lee:2006bu}&0.23$\pm$0.17~\cite{Lee:2002jb}, 0.030$\pm$0.023~\cite{Lee:2006bu}\\\hline

$\ket{\Xi(1950),?} $&1.95&2.11$\pm$0.54~\cite{Lee:2006bu}, 1.95~\cite{Lee:2006bu}&0.034$\pm$0.026~\cite{Lee:2006bu} \\\hline

$ \ket{\Omega(2012),?} $&2.012&$ 2.02^{+0.02}_{-0.03} $~\cite{Aliev:2018syi}, $ 2.05^{+0.09}_{-0.10} $~\cite{Su:2024lzy}&$0.094_{-0.004}^{+0.003}$~\cite{Aliev:2018syi}, $0.037_{-0.007}^{+0.007}$~\cite{Su:2024lzy}\\\hline

$\ket{\Omega(2380),?} $&2.38&2.37$\pm$0.41~\cite{Lee:2006bu}, 2.38~\cite{Lee:2006bu}&0.047$\pm$0.038~\cite{Lee:2006bu} \\\hline

$\ket{\Omega(2470),?} $&2.47&2.47~\cite{Lee:2006bu}&0.053$\pm$0.042~\cite{Lee:2006bu} \\\hline

\end{tabular}
}
\label{table:QCDSR-lightbaryon-excited}
\end{table}

\subsection{Heavy baryons}\label{heavy baryon}\

After the discovery of the charmonium state $ J/\psi $, the existence of charmed baryons was also anticipated. In 1976, Fermilab observed the first charmed baryon $ \Lambda_c $~\cite{Knapp:1976qw}, followed by the discovery of the bottom baryon $ \Lambda_b $~\cite{Basile:1981wr,Bari:1991ty,Bari:1991in,UA1:1991vse}. These discoveries marked the beginning of a new era in baryon research. Up to now, the 2024 edition of the Particle Data Group lists 36 charmed baryons and 27 bottom baryons~\cite{ParticleDataGroup:2024cfk}, a notable increase from the 2014 edition, which included only 24 charmed baryons and 9 bottom baryons~\cite{ParticleDataGroup:2014cgo}. Furthermore, a significant breakthrough occurred in 2017 with the definitive observation of the doubly heavy baryon $\Xi_{cc}^{++}$~\cite{LHCb:2017iph}, which contains two charm quarks. These rapid experimental progress has rekindled the interest of physicists in heavy baryons.

Heavy baryons are generally expected to have narrow widths and minimal overlap with each other, making their detection and isolation easier compared to light baryons. However, a higher center-of-mass energy is also required for their production. Among the heavy baryons listed in the Particle Data Group, only $ \Lambda_c $, $ \Sigma_c(2455) $, and $ \Xi_c^0 $ have a 4-star assignment. The remaining heavy baryons have 3-star assignments or less, indicating that their properties such as spin and parity are not yet well understood. 

In heavy baryons, the heavy quark masses are much larger than the typical scale of the strong interaction. This allows HQET to naturally incorporate the symmetries emerging in the heavy quark limit and systematically describe deviations from this idealized scenario. Generally, HQET is more naturally applied to singly heavy baryons. For doubly heavy baryons, the heavy quark pair $QQ$ can bind into a point-like color antitriplet in the limit that the mass of both heavy quarks becomes infinite. Then the doubly heavy baryons can be regarded as mesons with the light quark $q$ acting as an accompanying particle~\cite{Savage:1990di,White:1991hz}. Moreover, since the heavy quark mass is actually finite, HQET is more effective for bottom baryons than for charmed baryons. 

By combining HQET with QCD sum rules, one can systematically study the properties of heavy baryons, such as their masses, decay constants, and form factors. The details of QCD sum rules with HQET will not be discussed here, and readers may refer to Refs.~\cite{Chen:2016spr,Shuryak:1981fza,Huang:2000tn,Colangelo:1995qp,Grozin:1992td,Bagan:1993ii,Dai:1995bc,Wang:2002ts,Liu:2007fg,Wang:2003zp,Chen:2015kpa,Mao:2015gya,Chen:2016phw,Mao:2017wbz,Cui:2019dzj,Mao:2020jln,Yang:2020zjl,Yang:2020zrh,Yang:2021lce,Yang:2023fsc,Tan:2023opd,Groote:1996em,Groote:1997yr,Nishikawa:2024lnh} for more information. Additionally, QCD sum rules have also been applied within full QCD, and comparing these results with HQET-based calculations provides insight into the applicability of heavy quark symmetry.

In this section, we will first discuss singly heavy baryons, followed by doubly heavy baryons, and finally triply heavy baryons.

\subsubsection{Singly heavy baryons}\

The most well-known singly heavy baryon is the charmed baryon $ \Lambda_c $, which was the first heavy baryon discovered and remains the lightest. As the final decay product of many bottom baryons and heavier charmed baryons, it becomes the cornerstone of the heavy baryon spectrum. Its mass and lifetime listed in Particle Data Group~\cite{ParticleDataGroup:2024cfk} are
\begin{align}
M_{\Lambda_c}=2286.46\pm0.14\,\text{MeV}\,,\quad \tau_{\Lambda_c}=(2.026 \pm 0.010) \times 10^{-13} s
\end{align}
The hierarchy of charmed hadron lifetimes was analyzed from the QCD perspective by Blok and Shifman~\cite{Blok:1993zz}, who argued that the observed pattern reflects the intimate features of hadronic structure. The decay constant of $ \Lambda_c $ is also of interest, as it plays a role in its weak decays. However, no experimental measurement is available yet, and its value depends on phenomenological models such as QCD sum rules.

The first QCD sum rule study of heavy baryons focused on $ \Lambda_c $ and $ \Sigma_c $, which was conducted by Shuryak in 1982~\cite{Shuryak:1981fza}. Shortly thereafter, the QCD sum rule analysis of charmed baryons was further developed~\cite{Belyaev:1986zz} and extended to strange flavors~\cite{Blok:1986zz}. The HQET framework was employed to expand the mass of the singly heavy baryons as $ M_H=m_Q+\bar{\Lambda}+\mathcal{O}\left(1 / m_Q\right) $, where $ \bar{\Lambda} $ denotes the heavy baryon binding energy in the leading-order. Shuryak constructed a current for $ \Lambda_Q $ related to heavy quark limit, which was later widely adopted:
\begin{align}
j_{\Lambda_Q} = \varepsilon_{abc}\,\big(q_1^{aT}C\gamma_5 q_2^b\big)\,Q^c\,,
\label{current-Lambda_Q}
\end{align}
where $ q_{1,2} $ denotes the light quark. Along with current \eqref{current-Lambda_Q}, other commonly used currents for singly heavy baryons in the literature are:
\begin{align}
\label{current-Sigma_Q}
&j_{\Sigma_Q} = \varepsilon_{abc}\,\big(q_1^{aT}C\gamma_\mu q_2^b\big)\gamma^\mu \gamma_5 Q^c\,,\\
\label{current-Xi_Q}
&j_{\Xi_Q} = \varepsilon_{abc}\,\big(q_1^{aT}C\gamma_5 s^b\big)\,Q^c\,,\\
\label{current-Omega_Q}
&j_{\Omega_Q} = \varepsilon_{abc}\,\big(s^{aT}C\gamma_\mu s^b\big)\gamma^\mu \gamma_5 Q^c\,.
\end{align}
Following Ref.~\cite{Shuryak:1981fza}, the mass sum rules in the leading-order were further analyzed in Refs.~\cite{Huang:2000tn,Colangelo:1995qp,Grozin:1992td,Bagan:1993ii}. Higher-order corrections in the heavy quark expansion can also be systematically incorporated, with next-to-leading order $1/m_Q$ corrections included in Refs.\cite{Dai:1995bc,Wang:2002ts,Liu:2007fg,Wang:2003zp,Chen:2015kpa,Mao:2015gya,Chen:2016phw,Mao:2017wbz,Cui:2019dzj,Mao:2020jln,Yang:2020zjl,Yang:2020zrh,Yang:2021lce,Yang:2023fsc,Tan:2023opd} and $\alpha_s$ corrections in Refs.\cite{Groote:1996em,Groote:1997yr,Nishikawa:2024lnh}. Another form to study singly heavy baryons using QCD sum rules is in full QCD, which can be found in Refs.~\cite{Bagan:1991sc,Bagan:1992tp,Wang:2007sqa,Duraes:2007te,Zhang:2008iz,Zhang:2008pm,Wang:2009ozr,Wang:2009cr,Albuquerque:2009pr,Wang:2010fq,Azizi:2020ljx,Wang:2020mxk,Wang:2010vn,Wang:2010it,Wang:2015kua,Azizi:2015ksa,Wang:2017zjw,Agaev:2017jyt,Aliev:2018vye,Agaev:2017lip,Aliev:2018lcs,Wang:2020pri,Wang:2023wii,Xin:2023usd,Xu:2020cht,Wang:2017xam,Aliev:2017led,Wang:2017vtv,Azizi:2020tgh,Yu:2021zvl,Azizi:2022dpn}. The results from these studies have been extensively used to interpret a series of ground and excited singly heavy baryons observed in experiments.

The mass sum rules of ground singly heavy baryons were investigated in Refs.~\cite{Grozin:1992td,Bagan:1993ii,Colangelo:1995qp,Dai:1995bc,Wang:2002ts,Groote:1996em,Groote:1997yr,Liu:2007fg,Nishikawa:2024lnh,Bagan:1991sc,Bagan:1992tp,Wang:2007sqa,Duraes:2007te,Zhang:2008iz,Zhang:2008pm,Wang:2009ozr,Albuquerque:2009pr,Wang:2009cr,Wang:2010fq,Wang:2010vn,Agaev:2017lip,Azizi:2020ljx,Wang:2020mxk}. For comparison, the mass results from both HQET and full QCD are presented in Table~\ref{table:QCDSR-singly-heavybaryon}. Since the charm quark is not sufficiently heavy, the convergence of the $1/m_Q$ expansion is not guaranteed, making HQET, in principle, ineffective in the charm sector. However, the results in Table~\ref{table:QCDSR-singly-heavybaryon} show that the masses of charmed baryons obtained from HQET and full QCD remain highly consistent and agree with experimental data.

\begin{table}[ht]
\centering
\caption{The masses (GeV) and decay constants of ground singly heavy baryons obtained from QCD sum rules in HQET and full QCD, along with experimental values~\cite{ParticleDataGroup:2024cfk}. The HQET column shows results with the heavy quark expansion, while ``full QCD'' refers to calculations without this expansion.}
\renewcommand{\arraystretch}{2.0}
\resizebox{1\textwidth}{!}{
\tiny
\begin{tabular}{ccccc}
\hline\hline
State & Exp & HQET~\cite{Liu:2007fg} & Full QCD & Decay constant (GeV$^3$) \\ \hline

$\ket{\Lambda_c,\frac{1}{2}^+}$ & 2.286 & 2.271$^{+0.067}_{-0.049}$ &2.31$\pm$0.19~\cite{Zhang:2008iz}, 2.26$\pm$0.27~\cite{Wang:2010fq} &0.028$\pm$0.002~\cite{Groote:1996em}, 0.022$\pm$0.008~\cite{Wang:2010fq} \\ \hline

$\ket{\Sigma_c,\frac{1}{2}^+}$ & 2.454 & 2.411$^{+0.093}_{-0.081}$ &2.40$\pm$0.31~\cite{Zhang:2008iz}, 2.40$\pm$0.26~\cite{Wang:2009cr} & 0.039$\pm$0.003~\cite{Groote:1996em}, 0.045$\pm$0.015~\cite{Wang:2009cr} \\ \hline

$\ket{\Sigma_c,\frac{3}{2}^+}$ & 2.518 &2.534$^{+0.096}_{-0.081}$ &2.56$\pm$0.24~\cite{Zhang:2008iz}, 2.48$\pm$0.25~\cite{Wang:2010vn} &0.033$\sim$0.035~\cite{Bagan:1992tp}, 0.027$\pm$0.008~\cite{Wang:2010vn} \\ \hline

$\ket{\Xi_c,\frac{1}{2}^+}$ & 2.468 & 2.432$^{+0.079}_{-0.068}$ &2.48$\pm$0.21~\cite{Zhang:2008pm}, 2.50$\pm$0.20~\cite{Duraes:2007te} & 0.027$\pm$0.008~\cite{Wang:2010fq} \\ \hline

$\ket{\Xi_c^\prime,\frac{1}{2}^+}$ & 2.578 & 2.508$^{+0.097}_{-0.091}$ &2.50$\pm$0.29~\cite{Zhang:2008pm}, 2.56$\pm$0.22~\cite{Wang:2009cr} &0.055$\pm$0.016~\cite{Wang:2009cr} \\ \hline

$\ket{\Xi_c,\frac{3}{2}^+}$ & 2.645 & 2.634$^{+0.102}_{-0.094}$ &2.64$\pm$0.22~\cite{Zhang:2008pm}, 2.65$\pm$0.20~\cite{Wang:2010vn} &0.033$\pm$0.008~\cite{Wang:2010vn} \\ \hline

$\ket{\Omega_c,\frac{1}{2}^+}$ & 2.695 & 2.657$^{+0.102}_{-0.099}$ &2.62$\pm$0.29~\cite{Zhang:2008pm}, 2.65$\pm$0.25~\cite{Duraes:2007te} &0.093$\pm$0.023~\cite{Wang:2009cr}, 0.062$\pm$0.018~\cite{Agaev:2017lip} \\ \hline

$\ket{\Omega_c,\frac{3}{2}^+}$ & 2.770 & 2.790$^{+0.109}_{-0.105}$ &2.74$\pm$0.23~\cite{Zhang:2008pm}, 2.79$\pm$0.19~\cite{Wang:2010vn} & 0.056$\pm$0.012~\cite{Wang:2010vn}, 0.071$\pm$0.010~\cite{Agaev:2017lip} \\ \hline\hline

%%%

$\ket{\Lambda_b,\frac{1}{2}^+}$ & 5.620 & 5.637$^{+0.068}_{-0.056}$ &5.69$\pm$0.13~\cite{Zhang:2008iz}, 5.65$\pm$0.20~\cite{Wang:2010fq}&0.030$\pm$0.009~\cite{Wang:2010fq}, \\ \hline

$\ket{\Sigma_b,\frac{1}{2}^+}$ & 5.816 & 5.809$^{+0.082}_{-0.076}$ &5.73$\pm$0.21~\cite{Zhang:2008iz}, 5.80$\pm$0.19~\cite{Wang:2009cr} &0.032$\sim$0.067~\cite{Bagan:1991sc}, 0.062$\pm$0.018~\cite{Wang:2009cr} \\ \hline

$\ket{\Sigma_b,\frac{3}{2}^+}$ & 5.835 & 5.835$^{+0.082}_{-0.077}$ &5.81$\pm$0.19~\cite{Zhang:2008iz}, 5.85$\pm$0.20~\cite{Wang:2010vn}&0.045$\sim$0.073~\cite{Bagan:1992tp}, 0.038$\pm$0.011~\cite{Wang:2010vn} \\ \hline

$\ket{\Xi_b,\frac{1}{2}^+}$ & 5.797 & 5.780$^{+0.073}_{-0.068}$ &5.75$\pm$0.13~\cite{Zhang:2008pm}, 5.75$\pm$0.25~\cite{Duraes:2007te} &0.032$\pm$0.009~\cite{Wang:2010fq}, \\ \hline

$\ket{\Xi_b^\prime,\frac{1}{2}^+}$ & 5.935 & 5.903$^{+0.081}_{-0.079}$ &5.87$\pm$0.20~\cite{Zhang:2008pm}, 5.96$\pm$0.17~\cite{Wang:2009cr} &0.079$\pm$0.020~\cite{Wang:2009cr} \\ \hline

$\ket{\Xi_b,\frac{3}{2}^+}$ & 5.956 & 5.929$^{+0.083}_{-0.079}$ &5.94$\pm$0.17~\cite{Zhang:2008pm}, 6.02$\pm$0.17~\cite{Wang:2010vn} &0.049$\pm$0.012~\cite{Wang:2010vn} \\ \hline

$\ket{\Omega_b,\frac{1}{2}^+}$ & 6.046 & 6.036$^{+0.081}_{-0.081}$ &5.89$\pm$0.18~\cite{Zhang:2008pm}, 5.82$\pm$0.23~\cite{Duraes:2007te} &0.134$\pm$0.030~\cite{Wang:2009cr} \\ \hline

$\ket{\Omega_b,\frac{3}{2}^+}$ & & 6.063$^{+0.083}_{-0.082}$ &6.00$\pm$0.16~\cite{Zhang:2008pm}, 6.17$\pm$0.15~\cite{Wang:2010vn} &0.083$\pm$0.018~\cite{Wang:2010vn} \\ \hline

\end{tabular}
}
\label{table:QCDSR-singly-heavybaryon}
\end{table}

The case of excited singly heavy baryons is much more complex, with those rated 3 stars or higher in the Particle Data Group~\cite{ParticleDataGroup:2024cfk} listed as follows:
\begin{itemize}
\item $\Lambda_c(2595)$, $\Lambda_c(2625)$, $\Lambda_c(2860)$, $\Lambda_c(2880)$, $\Lambda_c(2940)$, \\[5pt]
$ \Lambda_b(5912) $, $ \Lambda_b(5920) $, $ \Lambda_b(6070) $, $ \Lambda_b(6146) $, $ \Lambda_b(6152) $. 
\item $\Sigma_c(2800)$, $\Sigma_b(6097)$.
\item $\Xi_c(2790)$, $\Xi_c(2815)$, $\Xi_c(2970)$, $\Xi_c(3055)$, $\Xi_c(3080)$, \\[5pt]
$\Xi_b(6087)$, $\Xi_b(6095)$, $\Xi_b(6100)$, $\Xi_b(6227)$, $\Xi_b(6327)$, $\Xi_b(6333)$.
\item $ \Omega_c(3000) $, $ \Omega_c(3050) $, $ \Omega_c(3065) $, $ \Omega_c(3090) $, $ \Omega_c(3120) $, $ \Omega_c(3185) $, $ \Omega_c(3327) $,\\[5pt]
$ \Omega_b(6316) $, $ \Omega_b(6330) $, $ \Omega_b(6340) $, $ \Omega_b(6350) $.
\end{itemize}
The nature of excited heavy baryon states was studied with mass sum rules in Refs.~\cite{Azizi:2020azq,Huang:2000tn,Wang:2003zp,Chen:2015kpa,Mao:2015gya,Chen:2016phw,Mao:2017wbz,Cui:2019dzj,Mao:2020jln,Yang:2020zjl,Yang:2020zrh,Yang:2021lce,Yang:2023fsc,Tan:2023opd,Nishikawa:2024lnh,Wang:2010fq,Wang:2020mxk,Wang:2010vn,Wang:2010it,Wang:2015kua,Azizi:2015ksa,Wang:2017zjw,Agaev:2017jyt,Aliev:2018vye,Agaev:2017lip,Aliev:2018lcs,Wang:2020pri,Wang:2023wii,Xin:2023usd,Xu:2020cht,Azizi:2020ljx,Wang:2017xam,Aliev:2017led,Wang:2017vtv,Azizi:2020tgh,Yu:2021zvl,Azizi:2022dpn}, where part of the results are summarized in Table~\ref{table:QCDSR-singly-charmbaryon-excited} and \ref{table:QCDSR-singly-bottombaryon-excited}. Additionally, some of these resonances lie slightly above the threshold of two hadrons and may be interpreted as hadronic molecular states. This topic goes beyond the scope of this review, and more details can be found in the review of hadronic molecules~\cite{Guo:2017jvc}. Here, we shall provide a brief review of QCD sum rule studies on the resonances with undetermined $J^P$ listed in the Particle Data Group~\cite{ParticleDataGroup:2024cfk}. Further details for these resonances can be found in the reviews~\cite{Chen:2016spr,Chen:2022asf,Cheng:2021qpd,Klempt:2009pi,Crede:2013kia,Karliner:2008sv}. 
\begin{table}[ht]
\centering
\caption{Same caption with Table~\ref{table:QCDSR-singly-heavybaryon}, but for excited singly charmed baryons.}
\renewcommand{\arraystretch}{1.3}
\resizebox{1\textwidth}{!}{
\tiny
\begin{tabular}{ccccc}
\hline\hline
State & Exp & HQET & Full QCD & Decay constant \\ \hline

\multirow{2}{*}{$\ket{\Lambda_c(2595),\frac{1}{2}^-} $} &\multirow{2}{*}{2.59}&\multirow{2}{*}{2.60$\pm$0.14~\cite{Chen:2015kpa}} &\multirow{2}{*}{2.61$\pm$0.21~\cite{Wang:2010fq} } &0.07$\pm$0.03 GeV$^4$~\cite{Chen:2015kpa}, \\
&&&&0.035$\pm$0.009 GeV$^3$~\cite{Wang:2010fq}, \\\hline

\multirow{2}{*}{$\ket{\Lambda_c(2625),\frac{3}{2}^-} $} &\multirow{2}{*}{2.63}&\multirow{2}{*}{2.65$\pm$0.14~\cite{Chen:2015kpa}} &\multirow{2}{*}{2.62$\pm$0.18~\cite{Wang:2015kua}} &0.07$\pm$0.03 GeV$^4$~\cite{Chen:2015kpa}, \\
&&&&0.041$\pm$0.014 GeV$^4$~\cite{Wang:2015kua}\\\hline

\multirow{2}{*}{$\ket{\Lambda_c(2860),\frac{3}{2}^+} $} &\multirow{2}{*}{2.86}&\multirow{2}{*}{2.81$^{+0.33}_{-0.18}$~\cite{Chen:2016phw}} &\multirow{2}{*}{2.83$^{+0.15}_{-0.24}$~\cite{Wang:2017vtv}} &0.012 GeV$^4$~\cite{Chen:2016phw}, \\
&&&&0.084$^{+0.032}_{-0.033}$ GeV$^5$~\cite{Wang:2017vtv}\\\hline

\multirow{2}{*}{$\ket{\Lambda_c(2880),\frac{5}{2}^+} $} &\multirow{2}{*}{2.88}&\multirow{2}{*}{2.84$^{+0.37}_{-0.20}$~\cite{Chen:2016phw}} &\multirow{2}{*}{2.88$^{+0.18}_{-0.29}$~\cite{Wang:2017vtv}} &0.012 GeV$^4$~\cite{Chen:2016phw}, \\
&&&&0.025$^{+0.009}_{-0.009}$ GeV$^5$~\cite{Wang:2017vtv}\\\hline

\multirow{2}{*}{$\ket{\Lambda_c(2940),\frac{3}{2}^-} $} &\multirow{2}{*}{2.94}&\multirow{2}{*}{3.07$\pm$0.11~\cite{Yang:2023fsc}} &\multirow{2}{*}{} &\multirow{2}{*}{0.027$\pm$0.007 GeV$^4$~\cite{Yang:2023fsc}} \\
&&&&\\\hline

\multirow{2}{*}{$\ket{\Sigma_c(2800),?} $} &\multirow{2}{*}{2.80}&\multirow{2}{*}{2.75$\pm$0.18~\cite{Chen:2015kpa}} &\multirow{2}{*}{2.74$\pm$0.20~\cite{Wang:2010it}} &0.06$\pm$0.03 GeV$^4$~\cite{Chen:2015kpa}, \\
&&&&0.037$\pm$0.009 GeV$^3$~\cite{Wang:2010it}\\\hline

\multirow{2}{*}{$\ket{\Xi_c(2790),\frac{1}{2}^-} $} &\multirow{2}{*}{2.79}&\multirow{2}{*}{2.79$\pm$0.15~\cite{Chen:2015kpa}} &\multirow{2}{*}{2.76$\pm$0.18~\cite{Wang:2010fq} } &0.11$\pm$0.04 GeV$^4$~\cite{Chen:2015kpa}, \\
&&&&0.042$\pm$0.009 GeV$^3$~\cite{Wang:2010fq}, \\\hline

\multirow{2}{*}{$\ket{\Xi_c(2815),\frac{3}{2}^-} $} &\multirow{2}{*}{2.82}&\multirow{2}{*}{2.83$\pm$0.15~\cite{Chen:2015kpa}} &\multirow{2}{*}{2.83$\pm$0.17~\cite{Wang:2015kua}} &0.11$\pm$0.04 GeV$^4$~\cite{Chen:2015kpa}, \\
&&&&0.065$\pm$0.022 GeV$^4$~\cite{Wang:2015kua}\\\hline

\multirow{2}{*}{$\ket{\Xi_c(2970),\frac{1}{2}^+} $} &\multirow{2}{*}{2.97}&\multirow{2}{*}{2.98$\pm$0.15~\cite{Chen:2015kpa}} &\multirow{2}{*}{} &\multirow{2}{*}{0.10$\pm$0.03 GeV$^4$~\cite{Chen:2015kpa}} \\
&&&&\\\hline

\multirow{2}{*}{$\ket{\Xi_c(3055),?} $} &\multirow{2}{*}{3.06}&\multirow{2}{*}{3.04$^{+0.15}_{-0.15}$~\cite{Chen:2016phw}} &\multirow{2}{*}{3.06$^{+0.11}_{-0.13}$~\cite{Wang:2017vtv}} &0.025 GeV$^4$~\cite{Chen:2016phw}, \\
&&&&0.147$^{+0.037}_{-0.035}$ GeV$^5$~\cite{Wang:2017vtv}\\\hline

\multirow{2}{*}{$\ket{\Xi_c(3080),?} $} &\multirow{2}{*}{3.08}&\multirow{2}{*}{3.05$^{+0.15}_{-0.16}$~\cite{Chen:2016phw}} &\multirow{2}{*}{3.09$^{+0.13}_{-0.15}$~\cite{Wang:2017vtv}} & 0.025 GeV$^4$~\cite{Chen:2016phw}, \\
&&&&0.037$^{+0.009}_{-0.009}$ GeV$^5$~\cite{Wang:2017vtv}\\\hline

\multirow{2}{*}{$\ket{\Omega_c(3000),?} $} &\multirow{2}{*}{3.00}&\multirow{2}{*}{2.99$^{+0.15}_{-0.15}$~\cite{Yang:2021lce}} &2.99$\pm$0.13~\cite{Agaev:2017lip}, & 0.105$\pm$0.023 GeV$^4$~\cite{Yang:2021lce}, \\
&&&2.98$\pm$0.16~\cite{Wang:2010it}&0.119$\pm$0.028 GeV$^3$~\cite{Agaev:2017lip}, 0.136$\pm$0.027 GeV$^3$~\cite{Wang:2010it}\\\hline

\multirow{2}{*}{$\ket{\Omega_c(3050),?} $} &\multirow{2}{*}{3.05}&\multirow{2}{*}{3.04$^{+0.11}_{-0.09}$~\cite{Yang:2021lce}} &3.06$\pm$0.10~\cite{Agaev:2017lip}, &0.069$\pm$0.011 GeV$^4$~\cite{Yang:2021lce}, \\
&&&3.05$\pm$0.11~\cite{Wang:2017zjw}&0.161$\pm$0.018 GeV$^3$~\cite{Agaev:2017lip}, 0.234$\pm$0.050 GeV$^4$~\cite{Wang:2017zjw}\\\hline

\multirow{2}{*}{$\ket{\Omega_c(3065),?} $} &\multirow{2}{*}{3.07}&\multirow{2}{*}{3.07$^{+0.10}_{-0.09}$~\cite{Yang:2021lce}} &3.08$\pm$0.14~\cite{Agaev:2017lip},&0.032$\pm$0.005 GeV$^4$~\cite{Yang:2021lce}, \\
&&&3.06$\pm$0.11~\cite{Wang:2017zjw}&0.171$\pm$0.034 GeV$^3$~\cite{Agaev:2017lip}, 0.103$\pm$0.023 GeV$^4$~\cite{Wang:2017zjw}\\\hline

\multirow{2}{*}{$\ket{\Omega_c(3090),?} $} &\multirow{2}{*}{3.09}&\multirow{2}{*}{3.08$^{+0.22}_{-0.17}$~\cite{Yang:2021lce}} &3.08$\pm$0.14~\cite{Agaev:2017lip},&0.103$\pm$0.026 GeV$^4$~\cite{Yang:2021lce}, \\
&&&3.06$\pm$0.10~\cite{Wang:2017zjw}&0.171$\pm$0.034 GeV$^3$~\cite{Agaev:2017lip}, 0.247$\pm$0.047 GeV$^4$~\cite{Wang:2017zjw}\\\hline

\multirow{2}{*}{$\ket{\Omega_c(3120),?} $} &\multirow{2}{*}{3.12}&\multirow{2}{*}{3.14$^{+0.21}_{-0.15}$~\cite{Yang:2021lce}} &3.12$\pm$0.11~\cite{Agaev:2017lip},&0.062$\pm$0.016 GeV$^4$~\cite{Yang:2021lce}, \\
&&&3.11$\pm$0.10~\cite{Wang:2017zjw}&0.250$\pm$0.031 GeV$^3$~\cite{Agaev:2017lip}, 0.107$\pm$0.017 GeV$^4$~\cite{Wang:2017zjw}\\\hline

\end{tabular}
}
\label{table:QCDSR-singly-charmbaryon-excited}
\end{table}

\begin{itemize}
\item $\Sigma_c(2800)$. The $\Sigma_c(2800)$ was observed in the $ \Lambda_c^+\pi $ invariant mass distribution by the Belle collaboration in 2005~\cite{Belle:2004zjl}, which is currently the only excited $\Sigma_c$ state observed experimentally. In Ref.~\cite{Wang:2010it}, the masses of two excited $\Sigma_c$ states with $J^P=\frac{1}{2}^-$ and $\frac{3}{2}^-$ were found to be close to experimental values, but their predicted decay constants differed significantly. In Ref.~\cite{Chen:2015kpa}, the authors identified two $\Sigma_c(2800)$ states with $J^P=\frac{1}{2}^-$ and $J^P=\frac{3}{2}^-$, classifying them within the $SU(3)\, \mathbf{6}_F$ multiplets. They also provided a mass difference of $14 \pm 7\,\text{MeV}$ as a distinguishing feature. Besides the conventional baryon picture $ cqq $, the $\Sigma_c(2800)$ was also interpreted as a $DN$ molecular state in Ref.~\cite{Zhang:2012jk} using QCD sum rules.

\item $\Xi_c(3055)$, $\Xi_c(3080)$. The $\Xi_c(3080)$ was first observed by the Belle Collaboration in the $\Lambda_c^{+} K^{-} \pi^{+}$ final state~\cite{Belle:2006edu}, and was later confirmed by the BaBar Collaboration~\cite{BaBar:2007zjt}, which also reported evidence for the $\Xi_c(3055)$. In Refs.~\cite{Wang:2017vtv,Chen:2016phw}, the $\Xi_c(3055)$ and $\Xi_c(3080)$ were assigned as $D$-wave charmed baryons with $J^P = \frac{3}{2}^{+}$ and $J^P = \frac{5}{2}^{+}$, respectively, based on QCD sum rules in the frameworks of HQET and full QCD. A recent LHCb measurement determined the spin-parity of the $\Xi_c(3055)$ to be $ \frac{3}{2}^{+} $ with a significance exceeding $6.5\sigma$~\cite{LHCb:2024eyx}, providing support for the QCD sum rule predictions.

\item $\Omega_c(3000)$, $\Omega_c(3050)$, $\Omega_c(3065)$, $\Omega_c(3090)$, $\Omega_c(3120)$. These five narrow excited $\Omega_c$ states were observed in the invariant mass spectrum of $\Xi_c^{+} K^{-}$~\cite{LHCb:2017uwr}, whose spin-parity quantum numbers remain highly controversial. In HQET perspective~\cite{Yang:2021lce}, the authors suggested assigning $\Omega_c(3050)$, $\Omega_c(3065)$, $\Omega_c(3090)$, and $\Omega_c(3120)$ as $P$-wave $\Omega_c$ baryons with $J^P=\frac{1}{2}^{-}, \frac{3}{2}^{-}, \frac{3}{2}^{-}, \frac{5}{2}^{-}$, respectively. Moreover, the masses of the two $ P $-wave $ \Omega_c $ states with $J^P=\frac{1}{2}^{-}$ and $J^P=\frac{3}{2}^{-}$ are very close to $\Omega_c(3000)$, which can be further separated by the mass splitting $12 \pm 5 \,\mathrm{MeV}$. However, full QCD studies~\cite{Agaev:2017lip,Wang:2010it,Wang:2017zjw} proposed a different picture for these five excited $\Omega_c$ states. In Refs.~\cite{Wang:2010it,Wang:2017zjw}, the authors assigned $\Omega_c(3000)$, $\Omega_c(3050)$, $\Omega_c(3065)$, $\Omega_c(3090)$, and $\Omega_c(3120)$ to the $P$-wave baryon states with $J^P=\frac{1}{2}^{-}, \frac{1}{2}^{-}, \frac{3}{2}^{-}, \frac{3}{2}^{-}, \frac{5}{2}^{-}$, respectively. In contrast, $J^P=\frac{1}{2}^{-}, \frac{3}{2}^{-}, \frac{1}{2}^{+}, \frac{1}{2}^{+}, \frac{3}{2}^{+}$ were assigned to these states in Ref.~\cite{Agaev:2017lip}. The detailed results are summarized in Table~\ref{table:QCDSR-singly-charmbaryon-excited}. The mass results from HQET and full QCD are quite similar, but their predictions for decay constants exhibit significant differences.

\item $\Sigma_b(6097)$. Two excited $ \Sigma_b $ states $\Sigma_b(6097)^{\pm}$ were observed in the $\Lambda_b^0 \pi^{ \pm}$ invariant mass distribution by LHCb collaboration in 2018~\cite{LHCb:2018haf}, with the mass splitting to be $\Delta M=-2.2 \pm 2.4 \pm 0.3\, \mathrm{MeV}$. In Ref.~\cite{Aliev:2018vye}, the authors studied the mass spectrum of the $1P$ and $2S$ excited states of $\Sigma_b$ using QCD sum rules. Together with the analysis of their decay properties, the results support the interpretation of $\Sigma_b(6097)^{\pm}$ as the $\Sigma_b(1P)$ states with $J^P = \frac{3}{2}^{-}$. This assignment was later supported by studies within the framework of HQET~\cite{Cui:2019dzj,Yang:2020zrh}.

\item $\Xi_b(6227)$. The $\Xi_b(6227)^-$ was first observed in both $\Lambda_b^0 K^{-}$and $\Xi_b^0 \pi^{-}$invariant mass spectra by the LHCb collaboration in 2018~\cite{LHCb:2018vuc}, and its isospin partner $\Xi_b(6227)^0$ was later observed in $\Xi_b^{-} \pi^{+}$ final state~\cite{LHCb:2020xpu}. In Refs.~\cite{Aliev:2018lcs,Azizi:2020azq}, the authors studied the mass of $\Xi_b(6227)$ using QCD sum rules in full QCD, and found that it can be assigned as the flavor $\mathbf{6}_F\, \Xi_b^{\prime}(1 P)$ state with $J^P=\frac{3}{2}^{-}$. This assignment was supported by Refs.~\cite{Cui:2019dzj,Yang:2020zrh} within the framework of HQET. Additionally, the decay constant of $\Xi_b(6227)$ was also calculated, which shows considerable consistency.

\item $\Xi_b(6327)$, $\Xi_b(6333)$. These two narrow excited $ \Xi_b $ states were observed in the $\Lambda_b^0 K^-\pi^{+}$ invariant mass spectrum by the LHCb collaboration in 2021~\cite{LHCb:2021ssn}. In Ref.~\cite{Yu:2021zvl}, based on QCD sum rules in full QCD, the $\Xi_b(6327)$ and $\Xi_b(6333)$ were assigned as the $1D$ states with $J^P=\frac{3}{2}^{+}$ and $\frac{5}{2}^{+}$, respectively. Despite the proximity in mass between the two resonances, their decay constants exhibit significant differences~\cite{Yu:2021zvl}.

\item $\Omega_b(6316)$, $\Omega_b(6330)$, $\Omega_b(6340)$, $\Omega_b(6350)$. These four narrow excited $ \Omega_b $ states were observed in the $\Xi_b^0 K^-$ invariant mass spectrum by the LHCb collaboration in 2020~\cite{LHCb:2020tqd}. In the full QCD analysis~\cite{Wang:2020pri}, the authors assigned $\Omega_b(6316)$, $\Omega_b(6330)$, $\Omega_b(6340)$, and $\Omega_b(6350)$ as the $P$-wave $ \Omega_b $ states with $J^P=\frac{3}{2}^{-}$, $\frac{1}{2}^{-}$, $\frac{5}{2}^{-}$, and $\frac{3}{2}^{-}$, respectively. However, the conclusion from HQET does not support these assignments, and yields $J^P=\frac{1}{2}^{-}$, $\frac{1}{2}^{-}$, $\frac{3}{2}^{-}$ and $\frac{3}{2}^{-}$ for these states~\cite{Yang:2020zrh}. The mass results from HQET and full QCD are quite similar, but their predictions for decay constants exhibit significant differences~\cite{Wang:2020pri,Yang:2020zrh}.
\end{itemize}

\begin{table}[ht]
\centering
\caption{Same caption with Table~\ref{table:QCDSR-singly-heavybaryon}, but for excited singly bottom baryons.}
\renewcommand{\arraystretch}{1.3}
\resizebox{1\textwidth}{!}{
\tiny
\begin{tabular}{ccccc}
\hline\hline
State & Exp & HQET & Full QCD & Decay constant \\ \hline

\multirow{2}{*}{$\ket{\Lambda_b(5912),\frac{1}{2}^-} $} &\multirow{2}{*}{5.91}&5.87$\pm$0.12~\cite{Mao:2015gya}, &\multirow{2}{*}{5.91$\pm$0.13~\cite{Xin:2023usd}} &\multirow{2}{*}{0.116$\pm$0.025 GeV$^4$~\cite{Xin:2023usd}} \\
&&5.92$^{+0.13}_{-0.10}$~\cite{Tan:2023opd}&& \\\hline

\multirow{2}{*}{$\ket{\Lambda_b(5920),\frac{3}{2}^-} $} &\multirow{2}{*}{5.92}&5.88$\pm$0.11~\cite{Mao:2015gya}, &\multirow{2}{*}{5.92$\pm$0.15~\cite{Xin:2023usd}} &\multirow{2}{*}{0.101$\pm$0.025 GeV$^4$~\cite{Xin:2023usd}} \\
&&5.92$^{+0.13}_{-0.10}$~\cite{Tan:2023opd}&& \\\hline

\multirow{2}{*}{$\ket{\Lambda_b(6070),\frac{1}{2}^+} $} &\multirow{2}{*}{6.07}&\multirow{2}{*}{} &6.07$\pm$0.09~\cite{Azizi:2020ljx} &\multirow{2}{*}{0.051$\pm$0.007 GeV$^3$~\cite{Azizi:2020ljx}, 0.064$\pm$0.009 GeV$^3$~\cite{Azizi:2020ljx}} \\
&&&6.08$\pm$0.09~\cite{Wang:2020mxk}& \\\hline

\multirow{2}{*}{$\ket{\Lambda_b(6146),\frac{3}{2}^+} $} &\multirow{2}{*}{6.15}&\multirow{2}{*}{6.12$^{+0.10}_{-0.11}$~\cite{Mao:2020jln}} &6.14$\pm$0.07~\cite{Azizi:2020tgh} &0.114$^{+0.029}_{-0.026}$ GeV$^5$~\cite{Mao:2020jln}, \\
&&&6.13$^{+0.10}_{-0.09}$~\cite{Yu:2021zvl}&0.264$\pm$0.039 GeV$^5$~\cite{Azizi:2020tgh}, 0.145$^{+0.021}_{-0.022}$ GeV$^5$~\cite{Yu:2021zvl} \\\hline

\multirow{2}{*}{$\ket{\Lambda_b(6152),\frac{5}{2}^+} $} &\multirow{2}{*}{6.15}&\multirow{2}{*}{6.13$^{+0.10}_{-0.11}$~\cite{Mao:2020jln}} &\multirow{2}{*}{6.15$^{+0.13}_{-0.15}$~\cite{Yu:2021zvl}} &\multirow{2}{*}{0.048$^{+0.012}_{-0.011}$ GeV$^5$~\cite{Mao:2020jln}, 0.037$^{+0.005}_{-0.004}$ GeV$^5$~\cite{Yu:2021zvl}} \\
&&&& \\\hline

\multirow{2}{*}{$\ket{\Sigma_b(6097),?} $} &\multirow{2}{*}{6.10}&6.10$\pm$0.12~\cite{Cui:2019dzj}, &\multirow{2}{*}{6.09$^{+0.11}_{-0.12}$~\cite{Aliev:2018vye}} &0.102$\pm$0.022 GeV$^4$~\cite{Cui:2019dzj}, 0.102$\pm$0.028 GeV$^4$~\cite{Yang:2020zrh} \\
&&6.11$\pm$0.16~\cite{Yang:2020zrh}&&0.068$^{+0.010}_{-0.011}$ GeV$^3$~\cite{Aliev:2018vye} \\\hline

\multirow{2}{*}{$\ket{\Xi_b(6087),\frac{3}{2}^-} $} &\multirow{2}{*}{6.09}&\multirow{2}{*}{6.09$^{+0.13}_{-0.12}$~\cite{Tan:2023opd}} &\multirow{2}{*}{6.08$\pm$0.12~\cite{Xin:2023usd}} &\multirow{2}{*}{0.165$\pm$0.025 GeV$^4$~\cite{Xin:2023usd}} \\
&&&& \\\hline

$\ket{\Xi_b(6095),\frac{3}{2}^-} $ &\multirow{2}{*}{6.10}&\multirow{2}{*}{6.10$^{+0.13}_{-0.12}$~\cite{Tan:2023opd}} &\multirow{2}{*}{6.10$\pm$0.11~\cite{Xin:2023usd}} &\multirow{2}{*}{0.083$\pm$0.011 GeV$^4$~\cite{Xin:2023usd}} \\
$\ket{\Xi_b(6100),\frac{3}{2}^-} $ &&&& \\\hline

\multirow{2}{*}{$\ket{\Xi_b(6227),?} $} &\multirow{2}{*}{6.23}&6.27$\pm$0.12~\cite{Cui:2019dzj}, &\multirow{2}{*}{6.23$\pm$0.11~\cite{Azizi:2020azq}} &0.099$\pm$0.021 GeV$^4$~\cite{Cui:2019dzj}, 0.091$\pm$0.023 GeV$^4$~\cite{Yang:2020zrh} \\
&&6.23$\pm$0.15~\cite{Yang:2020zrh}&&0.081$\pm$0.010 GeV$^3$~\cite{Azizi:2020azq} \\\hline

\multirow{2}{*}{$\ket{\Xi_b(6327),?} $} &\multirow{2}{*}{6.33}&\multirow{2}{*}{} &\multirow{2}{*}{6.34$^{+0.12}_{-0.11}$~\cite{Yu:2021zvl}} &\multirow{2}{*}{0.298$^{+0.038}_{-0.032}$ GeV$^5$~\cite{Yu:2021zvl}} \\
&&&& \\\hline

\multirow{2}{*}{$\ket{\Xi_b(6333),?} $} &\multirow{2}{*}{6.33}&\multirow{2}{*}{} &\multirow{2}{*}{6.36$^{+0.11}_{-0.12}$~\cite{Yu:2021zvl}} &\multirow{2}{*}{0.067$^{+0.008}_{-0.007}$ GeV$^5$~\cite{Yu:2021zvl}} \\
&&&& \\\hline

\multirow{2}{*}{$\ket{\Omega_b(6316),?} $} &\multirow{2}{*}{6.32}&\multirow{2}{*}{6.32$\pm$0.11~\cite{Yang:2020zrh}} &\multirow{2}{*}{6.31$\pm$0.11~\cite{Wang:2020pri}} &\multirow{2}{*}{0.133$\pm$0.028 GeV$^4$~\cite{Yang:2020zrh}, 0.125$\pm$0.021 GeV$^4$~\cite{Wang:2020pri}} \\
&&&& \\\hline

\multirow{2}{*}{$\ket{\Omega_b(6330),?} $} &\multirow{2}{*}{6.33}&\multirow{2}{*}{6.34$\pm$0.10~\cite{Yang:2020zrh}} &\multirow{2}{*}{6.32$\pm$0.11~\cite{Wang:2020pri}} &\multirow{2}{*}{0.122$\pm$0.019 GeV$^4$~\cite{Yang:2020zrh}, 0.280$\pm$0.047 GeV$^4$~\cite{Wang:2020pri}} \\
&&&& \\\hline

\multirow{2}{*}{$\ket{\Omega_b(6340),?} $} &\multirow{2}{*}{6.34}&\multirow{2}{*}{6.34$\pm$0.09~\cite{Yang:2020zrh}} &\multirow{2}{*}{6.35$\pm$0.10~\cite{Wang:2020pri}} &\multirow{2}{*}{0.058$\pm$0.009 GeV$^4$~\cite{Yang:2020zrh}, 0.246$\pm$0.037 GeV$^4$~\cite{Wang:2020pri}} \\
&&&& \\\hline

\multirow{2}{*}{$\ket{\Omega_b(6350),?} $} &\multirow{2}{*}{6.35}&\multirow{2}{*}{6.35$\pm$0.13~\cite{Yang:2020zrh}} &\multirow{2}{*}{6.37$\pm$0.09~\cite{Wang:2020pri}} &\multirow{2}{*}{0.162$\pm$0.035 GeV$^4$~\cite{Yang:2020zrh}, 0.428$\pm$0.066 GeV$^4$~\cite{Wang:2020pri}} \\
&&&& \\\hline

\end{tabular}
}
\label{table:QCDSR-singly-bottombaryon-excited}
\end{table}

In Table~\ref{table:QCDSR-singly-charmbaryon-excited} and \ref{table:QCDSR-singly-bottombaryon-excited}, we note that the mass dimension of the decay constant differs across studies. According to Eq.~\eqref{2pt-phe}, the mass dimension of the decay constant depends on that of the correlation function, which in turn is determined by the structure of the interpolating current. Currents that involve partial derivatives or contain a larger number of quark fields generally lead to correlation functions with higher dimensions, and consequently, decay constants with higher mass dimensions. 

Very recently, the CMS experiment observed a near-threshold enhancement in top quark pair production~\cite{CMS:2025kzt}, providing the first experimental indication of a short-lived pseudoscalar $t\bar{t}$ bound state, referred to as toponium. Motivated by this observation, Ref.~\cite{Zhang:2025xxd} investigated the internal structure of ground-state singly topped baryons using QCD sum rules with heavy quark effective theory. Their results suggested the mass of ground-state singly topped baryons to be approximately $174\,\text{GeV}$, which is about $1\sim 1.5\,\text{GeV}$ higher than the pole mass of the top quark.

To conclude this subsection, we emphasize that although the structure of ground-state singly heavy baryons has been relatively well established, their decay properties are still not fully understood. In contrast, studies on the spectrum of excited singly heavy baryons are still at an early stage, with their internal structures remaining under debate. Even within the conventional three-quark picture, the classifications of excited states and their spin-parity assignments may differ across theoretical approaches. In particular, the decay properties play a key role in revealing their internal dynamics—a topic that will be discussed in detail in the next chapter. We anticipate that future experiments will provide more precise data on the heavy baryon spectrum, especially regarding decay constants, which are essential for deepening our understanding of the spectrum and the fundamental dynamics of baryons.

\subsubsection{Doubly heavy baryons}\

The doubly heavy baryons are composed of two heavy quarks and a light quark. In a naive picture of the doubly heavy baryons $ QQq $, the two heavy quarks form a color antitriplet, while the light quark acts as a spectator. Such systems are interesting since they combine the dynamics of charmonium or bottomonium states with those of $ D $ or $ B $ mesons, where the light quark remains relativistic and orbits a static tightly bound $ QQ $ pair~\cite{Bagan:1992za}. Therefore, the doubly heavy baryons provide an ideal platform for studying the heavy quark symmetry.

In 2002, the SELEX collaboration reported the first evidence of the doubly charmed baryon $ \Xi_{cc}^{+}(3519) $ in the decay process $ \Xi_{c c}^{+} \to \Lambda_c^{+} K^{-} \pi^{+} $~\cite{SELEX:2002wqn}. However, later experiments failed to confirm this state~\cite{Ratti:2003ez,BaBar:2006bab,Belle:2013htj,LHCb:2013hvt}. Until 2017, the LHCb collaboration observed the $ \Xi_{cc}^{++} $ baryon in a four-body decay channel $\Xi_{c c}^{++} \to \Lambda_c^{+} K^{-} \pi^{+} \pi^{+}$~\cite{LHCb:2017iph}, which was previously suggested by Yu et al.~\cite{Yu:2017zst}. Later the LHCb collaboration confirmed the existence of the $ \Xi_{cc}^{++} $ baryon in additional channels $ \Xi_{c c}^{++} \to \Xi_c^{+} \pi^{+}$~\cite{LHCb:2018pcs} and $\Xi_{c c}^{++} \to \Xi_c^{\prime+} \pi^{+}$~\cite{LHCb:2022rpd}. Its mass and lifetime in Particle Data Group~\cite{ParticleDataGroup:2024cfk} are
\begin{align} 
M_{\Xi_{cc}^{++}}&=3621.6\pm0.4\,\text{MeV}\,,\quad \tau_{\Xi_{cc}^{++}}=2.56\pm 0.27\,\text{s}\,,
\end{align}
where the mass value is significantly larger than the $ \Xi_{cc}^+ $ reported by SELEX~\cite{SELEX:2002wqn}. 

The first QCD sum rule calculation of doubly heavy baryon masses dates back to the 1990s by Bagan et al~\cite{Bagan:1992za}. The currents they used are:
\begin{align}
&j_{\Xi_{QQ}}=\varepsilon_{a b c}\big[\big(Q^{aT} C \gamma_5 u^b\big) Q^c+b\big(Q^{aT} C u^b\big) \gamma_5Q^c\big]\,,\nonumber\\[5pt]
&j^\mu_{\Xi_{QQ}^*}=\frac{1}{\sqrt{3}} \varepsilon_{a b c}\big[ 2\big(Q^{aT} C \gamma^\mu u^b\big) Q^c+\big(Q^{aT} C \gamma^\mu Q^b\big) u^c\big]\,,\nonumber\\[5pt]
&j^\mu_{\Xi_{QQ^\prime}}=\varepsilon_{a b c}\big[\big(u^{aT} C \gamma_5 Q^b\big) Q^{\prime c}+b\big(u^{aT} C Q^b\big) \gamma_5Q^{\prime c}\big]\,,
\label{Xi_QQ-Bagan}
\end{align}
which were later used to study the mass ratios of doubly heavy baryons through the analysis of double ratio QCD sum rules~\cite{Narison:2010py}. The authors of Ref.~\cite{Narison:2010py} found that the mass splitting between $\Xi_{Q Q}^*$ and $\Xi_{Q Q}$ is primarily driven by $ \alpha_s $ corrections and is largely independent of the baryon spin quantum numbers.

In Ref.~\cite{Zhang:2008rt}, the authors investigated the whole mass spectrum of doubly heavy baryons within a tentative structure of heavy-diquark-light-quark ($ QQ $-$q$) configuration. Their results are listed in Table~\ref{table:QCDSR-doublyheavybaryon}. The derived mass value for $ \Xi_{cc} $ was $ 4.26\pm 0.19\,\text{GeV} $, which is larger than other theoretical predictions and the later experimental value.

\begin{figure}[ht]
\centering
\includegraphics[width=0.4\textwidth]{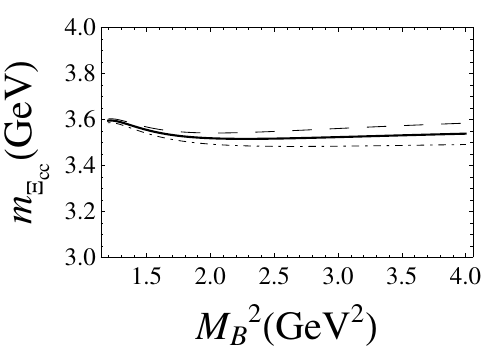}
\includegraphics[width=0.4\textwidth]{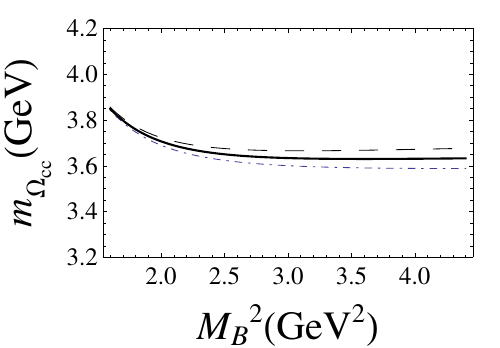}
\caption{The mass of $ \Xi_{cc} $ and $ \Omega_{cc} $ as a function of the Borel parameter $ M_B^2 $ in Ref.~\cite{Tang:2011fv}.}
\label{fig:Xi_cc_Omega_cc_mass}
\end{figure}

To test the $(QQ)$-$q$ configuration of the doubly heavy baryons, the authors adopted the currents based on the diquark model in Ref.~\cite{Tang:2011fv}:
\begin{align}
j(x) =\Phi_\mu^a(x) \Gamma_k^\mu q^a(x)\,, \quad j_\mu^*(x) =\Phi_\mu^a(x) \Gamma_k q^a(x)\,,\quad j^{\prime}(x) =\Phi^a(x) \Gamma_k q^a(x)\,,
\end{align}
where $\Phi^a_\mu(x)$ and $\Phi^a(x)$ denote axial vector and scalar diquarks, respectively. The interpolating current $J$ corresponds to $\Xi_{QQ^\prime}$ and $\Omega_{QQ^\prime}$, $J^*_\mu$ corresponds to $\Xi^*_{QQ^\prime}$ and $\Omega^*_{QQ^\prime}$ and $J^\prime$ corresponds to $\Xi^\prime_{QQ^\prime}$ and $\Omega^\prime_{QQ^\prime}$ respectively with $Q, Q^\prime=c,\;b$. The concrete definition of $\Gamma_k^{\mu}$ and $\Gamma_k$ are presented in Table~\ref{diquark-Gamma}. The Feynman diagrams are computed not only with the regular QCD Feynman rules, but also with the effective vertices for point-like diquarks. The masses of $ \Xi_{cc} $ and $ \Omega_{cc} $ as functions of the Borel parameter are shown in Fig.~\ref{fig:Xi_cc_Omega_cc_mass}. Their results are consistent with the predictions of other theoretical models, suggesting that the diquark model is reliable for describing the properties of doubly heavy baryons.

\begin{table}[ht]
\begin{center}
\caption{The choice of $\Gamma_k^{\mu}$ and $\Gamma_k$ in Ref.~\cite{Tang:2011fv}. $J_D^{P_D}$ means the spin-parity numbers of diquark. }\label{diquark-Gamma}
\vspace{0.5cm}
\begin{tabular}{cccccc}
\hline\hline Baryon & Constituent & $J^P$ & $J_D^{P_D}$ &
$\Gamma_k^{\mu}$ & $\Gamma_k$
\\
\hline $\Xi_{QQ^\prime}$ & $\Phi_{\{QQ^\prime\}}\;q$ & $\frac{1}{2}^+$
& $1^+$ & $\gamma^\mu\gamma_5$ &-
\\
$\Xi^*_{QQ^\prime}$ & $\Phi_{\{QQ^\prime\}}\;q$ & $\frac{3}{2}^+$
& $1^+$ & - & 1\\
$\Omega_{QQ^\prime}$ & $\Phi_{\{QQ^\prime\}}\;s$ & $\frac{1}{2}^+
$ & $1^+$ & $\gamma^\mu\gamma_5$ &- \\
$\Omega^*_{QQ^\prime}$ & $\Phi_{\{QQ^\prime\}}\;s$ & $\frac{3}{2}^+
$ & $1^+$ & - & 1\\
$\Xi^\prime_{QQ^\prime}$ & $\Phi_{[QQ^\prime]}\;q$ &
$\frac{1}{2}^+$ & $0^+$ & - & 1\\
$\Omega^\prime_{QQ^\prime}$ & $\Phi_{[QQ^\prime]}\;s$ &
$\frac{1}{2}^+$ & $0^+$ & -& 1
\\ \hline\hline
\end{tabular}
\end{center}
\end{table}

\begin{figure}[ht]
\centering
\includegraphics[width=0.32\textwidth]{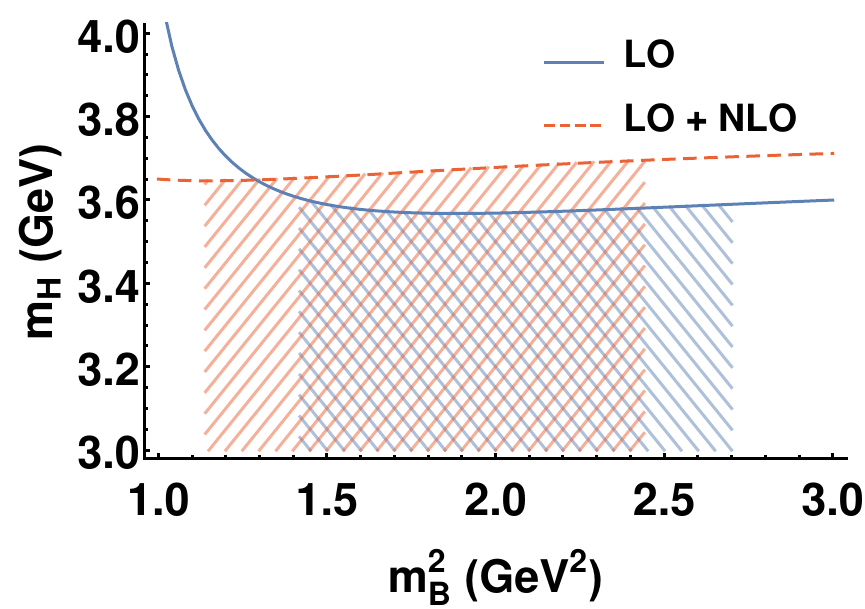}
\includegraphics[width=0.32\textwidth]{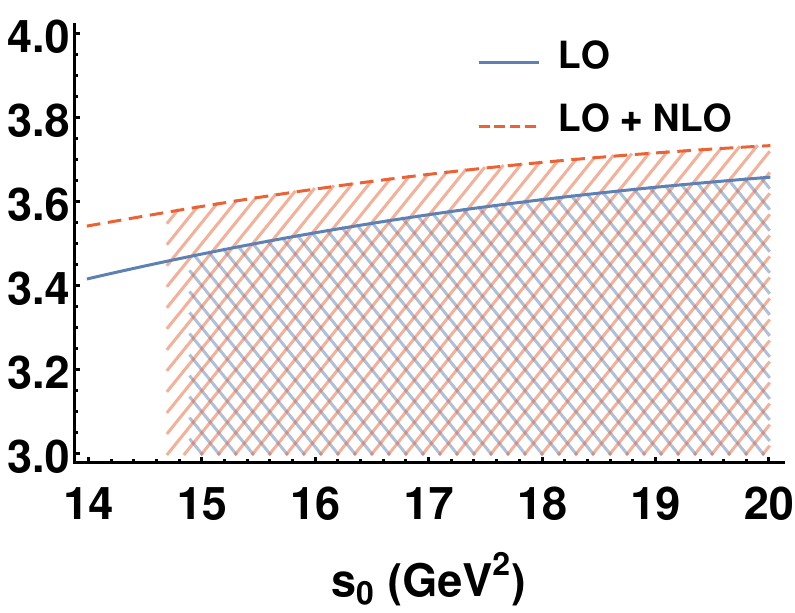}
\includegraphics[width=0.32\textwidth]{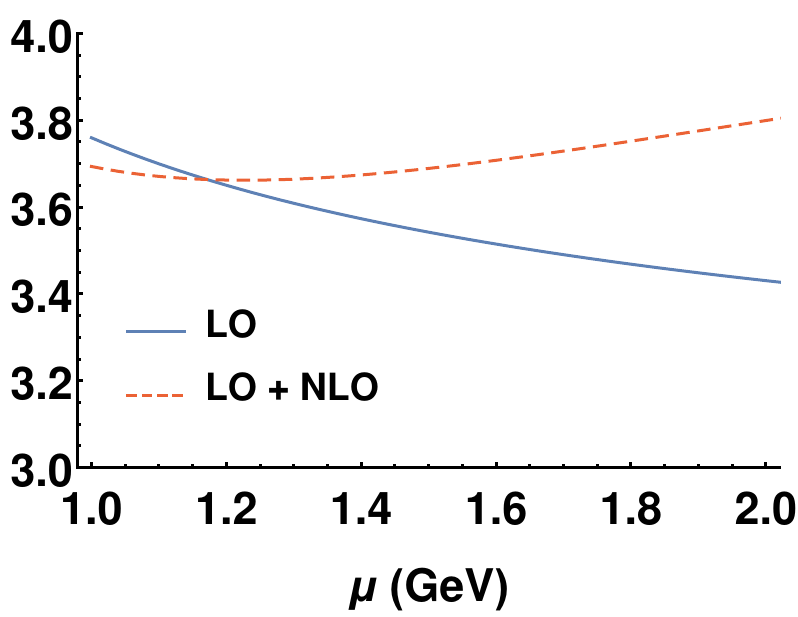}
\caption{The mass of $ \Xi_{cc}^{++} $ as a function of the Borel parameter $ m_B^2 $, the threshold parameter $ s_0 $, and renormalization scale $ \mu $ in Ref.~\cite{Wang:2017qvg}.}
\label{fig:Xi_cc_mass}
\end{figure}

QCD radiative corrections can also be introduced into QCD sum rules to reduce theoretical uncertainties. In Ref.~\cite{Wang:2017qvg}, the authors analytically obtained the NLO contribution to the perturbative part of the doubly charmed baryon $ \Xi_{cc}^{++} $, where the calculations for the loop integrals are carried out using the method of canonical differential equations~\cite{Kotikov:1990kg,Henn:2013pwa}. The mass value of $ \Xi_{cc}^{++} $ as functions of the Borel parameter $ m_B^2 $, the threshold parameter $ s_0 $, and the renormalization scale $ \mu $ are shown in Fig.~\ref{fig:Xi_cc_mass}.

\begin{table}[ht]
\centering
\caption{The predictions for the mass ($ \text{GeV} $) of doubly heavy baryons from QCD sum rules as well as other theoretical models. Results from different theoretical approaches are shown to illustrate the spread of predictions.}
\renewcommand{\arraystretch}{1.4}
\resizebox{\textwidth}{!}{
\tiny
\begin{tabular}{ccccc}
\hline\hline
State & QCDSR & RQM~\cite{Ebert:2002ig} & Bag model~\cite{He:2004px}&Mass formula~\cite{Lichtenberg:1995kg}\\\hline

\multirow{2}{*}{$\ket{\Xi_{cc},\frac{1}{2}^+} $}&3.48$\pm$0.06~\cite{Bagan:1992za}, 4.26$\pm$0.19~\cite{Zhang:2008rt}, 3.519~\cite{Tang:2011fv}, &\multirow{2}{*}{3.620}&\multirow{2}{*}{3.520}&\multirow{2}{*}{3.676}\\
&$3.66_{\scalebox{0.8}{$-0.10$}}^{\scalebox{0.8}{$+0.08$}}$~\cite{Wang:2017qvg}, 3.72$\pm$0.20~\cite{Aliev:2012ru}, $3.63_{\scalebox{0.8}{$-0.07$}}^{\scalebox{0.8}{$+0.08$}}$~\cite{Wang:2018lhz}&&&\\\hline

\multirow{2}{*}{$\ket{\Xi^*_{cc},\frac{3}{2}^+} $}&3.58$\pm$0.05~\cite{Bagan:1992za}, 3.90$\pm$0.10~\cite{Zhang:2008rt}, 3.62~\cite{Tang:2011fv},& \multirow{2}{*}{3.727}& \multirow{2}{*}{3.630}&\multirow{2}{*}{3.746}\\
& 3.69$\pm$0.16~\cite{Aliev:2012iv}, $3.75_{\scalebox{0.8}{$-0.07$}}^{\scalebox{0.8}{$+0.07$}}$~\cite{Wang:2018lhz}& &\\\hline

\multirow{2}{*}{$\ket{\Omega_{cc},\frac{1}{2}^+} $}&4.25$\pm$0.20~\cite{Zhang:2008rt}, 3.63~\cite{Tang:2011fv},&\multirow{2}{*}{3.778}&\multirow{2}{*}{3.619}&\multirow{2}{*}{3.787}\\
&3.73$\pm$0.20~\cite{Aliev:2012ru}, $3.75_{\scalebox{0.8}{$-0.09$}}^{\scalebox{0.8}{$+0.08$}}$~\cite{Wang:2018lhz}& &\\\hline

\multirow{2}{*}{$\ket{\Omega^*_{cc},\frac{3}{2}^+} $}&3.81$\pm$0.06~\cite{Zhang:2008rt}, 3.71~\cite{Tang:2011fv}, &\multirow{2}{*}{3.872}&\multirow{2}{*}{3.721}&\multirow{2}{*}{3.851}\\
& 3.78$\pm$0.16~\cite{Aliev:2012iv}, $3.85_{\scalebox{0.8}{$-0.08$}}^{\scalebox{0.8}{$+0.08$}}$~\cite{Wang:2018lhz}& &\\

\hline

\multirow{2}{*}{$\ket{\Xi_{bb},\frac{1}{2}^+} $}&9.94$\pm$0.91~\cite{Bagan:1992za}, 9.78$\pm$0.07~\cite{Zhang:2008rt}, 9.80~\cite{Tang:2011fv},&\multirow{2}{*}{10.202}&\multirow{2}{*}{10.272}&\multirow{2}{*}{}\\
& 9.96$\pm$0.90~\cite{Aliev:2012ru}, $10.22_{\scalebox{0.8}{$-0.07$}}^{\scalebox{0.8}{$+0.07$}}$~\cite{Wang:2018lhz}& &\\\hline

\multirow{2}{*}{$\ket{\Xi^*_{bb},\frac{3}{2}^+} $}&10.33$\pm$1.09~\cite{Bagan:1992za}, 10.35$\pm$0.08~\cite{Zhang:2008rt}, 9.84~\cite{Tang:2011fv}, &\multirow{2}{*}{10.237}&\multirow{2}{*}{10.337}&\multirow{2}{*}{10.398}\\
&10.40$\pm$0.10~\cite{Aliev:2012iv}, $10.27_{\scalebox{0.8}{$-0.07$}}^{\scalebox{0.8}{$+0.07$}}$~\cite{Wang:2018lhz}& &\\\hline

\multirow{2}{*}{$\ket{\Omega_{bb},\frac{1}{2}^+} $}&9.85$\pm$0.07~\cite{Zhang:2008rt}, 9.89~\cite{Tang:2011fv}, &\multirow{2}{*}{10.359}&\multirow{2}{*}{10.369}&\multirow{2}{*}{}\\
&9.85$\pm$0.07~\cite{Aliev:2012ru}, $10.33_{\scalebox{0.8}{$-0.08$}}^{\scalebox{0.8}{$+0.07$}}$~\cite{Wang:2018lhz}& &\\\hline

\multirow{2}{*}{$\ket{\Omega^*_{bb},\frac{3}{2}^+} $}&10.28$\pm$0.05~\cite{Zhang:2008rt}, 9.93~\cite{Tang:2011fv},&\multirow{2}{*}{10.389}&\multirow{2}{*}{10.429}&\multirow{2}{*}{10.483}\\
& 10.50$\pm$0.20~\cite{Aliev:2012iv}, $10.37_{\scalebox{0.8}{$-0.08$}}^{\scalebox{0.8}{$+0.07$}}$~\cite{Wang:2018lhz}& &\\

\hline

$\ket{\Xi_{cb},\frac{1}{2}^+} $ & 6.75$\pm$0.05~\cite{Zhang:2008rt}, 6.65~\cite{Tang:2011fv}, 6.72$\pm$0.20~\cite{Aliev:2012ru} & 6.933 & 6.838 & 7.053 \\\hline

$\ket{\Xi_{cb}^*,\frac{3}{2}^+} $ & 8.00$\pm$0.26~\cite{Zhang:2008rt}, 6.69~\cite{Tang:2011fv}, 7.25$\pm$0.20~\cite{Aliev:2012iv} & 6.980 & 6.986 & 7.083 \\\hline

$\ket{\Omega_{cb},\frac{1}{2}^+} $ & 7.02$\pm$0.08~\cite{Zhang:2008rt}, 6.75~\cite{Tang:2011fv}, 6.75$\pm$0.30~\cite{Aliev:2012ru} & 7.088 & 6.941 & 7.148 \\\hline

$\ket{\Omega_{cb}^*,\frac{3}{2}^+} $ & 7.54$\pm$0.08~\cite{Zhang:2008rt}, 6.77~\cite{Tang:2011fv}, 7.30$\pm$0.20~\cite{Aliev:2012iv} & 7.130 & 7.077 & 7.165 \\\hline

$\ket{\Xi_{cb}^\prime,\frac{1}{2}^+} $ & 6.95$\pm$0.08~\cite{Zhang:2008rt}, 6.61~\cite{Tang:2011fv}, 6.79$\pm$0.20~\cite{Aliev:2012ru} & 6.963 & 7.028 & 7.062 \\\hline

$\ket{\Omega_{cb}^\prime,\frac{1}{2}^+} $ & 7.02$\pm$0.08~\cite{Zhang:2008rt}, 6.69~\cite{Tang:2011fv}, 6.80$\pm$0.30~\cite{Aliev:2012ru} & 7.116 & 7.116 & 7.151 \\\hline

\end{tabular}
}
\label{table:QCDSR-doublyheavybaryon}
\end{table}

The authors pointed out that incorporating NLO perturbative corrections leads to a more stable plateau for the Borel parameter and the threshold parameter. Moreover, the dependence of the $ \Xi_{cc}^{++} $ mass on the heavy quark mass $ m_Q $ and the renormalization scale $ \mu $ is significantly reduced~\cite{Wang:2017qvg}.

Other QCD sum rules studies of doubly heavy baryons can be found in Refs.~\cite{Wang:2010hs,Wang:2010vn,Aliev:2012ru,Aliev:2012iv,Chen:2017sbg,Wang:2018lhz,Wang:2022ufh,ShekariTousi:2024mso}, where the values of decay constants and masses are provided. Since no other doubly heavy baryons have been observed experimentally, we summarize in Table~\ref{table:QCDSR-doublyheavybaryon} the mass predictions for them from QCD sum rules and other theoretical models, such as the relativistic quark model (RQM)~\cite{Ebert:2002ig}, the bag model~\cite{He:2004px}, and the mass formula~\cite{Lichtenberg:1995kg}.

The doubly heavy baryons with negative parities were studied in Ref.\cite{Wang:2010it}, where the authors employed a method similar to that of Jido et al.\cite{Jido:1996zw} to subtract contributions from positive-parity states. Additionally, the spectrum of excited doubly heavy baryons with spin-$\frac{1}{2}$ was investigated in Ref.\cite{Aliev:2019lvd}, including both the first radial ($2S$) and orbital ($1P$) excitations. The authors constructed symmetric and antisymmetric interpolating currents, assuming that each current couples simultaneously to the ground state and one of the excited states ($2S$ or $1P$). Under such an assumption, the masses of the $1P$ and $2S$ states cannot be determined separately, where the same mass value is obtained for both states of a given doubly heavy baryon~\cite{Aliev:2019lvd}.

Beyond QCD sum rules, other relevant studies on doubly heavy baryons can be found in Refs.~\cite{Ebert:1996ec,Kiselev:2001fw,Albertus:2006ya,Roberts:2007ni,Valcarce:2008dr,Giannuzzi:2009gh,Weng:2010rb,Eakins:2012jk,Karliner:2014gca,Yoshida:2015tia,Shah:2016vmd,Shah:2017liu,Weng:2018mmf,Mathur:2018rwu}. Despite significant progress, the overall properties of doubly heavy baryons remain to be fully explored. In the future, the interplay between high-statistics experiments and high-precision theoretical calculations will further unveil the rich physics of doubly heavy baryons, offering crucial insights into hadron structure as well as strong and weak interactions.

\subsubsection{Triply heavy baryons}\

The triply heavy baryons consist of three heavy $ b $ or $ c $ quarks, where the light quarks are absent. Free from light-quark effects, such baryons constitute an ideal system for investigating heavy-quark dynamics and provide a clean probe for exploring the interplay between perturbative and nonperturbative QCD interactions. However, unlike the well-established experimental cases of the triply light flavored $ \Delta^{++}(uuu) $ and strange $ \Omega^{-}(sss) $ baryons, the triply heavy baryons such as $ \Omega_{ccc}(ccc) $ currently lack experimental evidence. This absence of experimental signals is one reason why they have not garnered significant attention for a long time. 

Theoretical studies of triply heavy baryons can offer valuable insights into the three-body static potential. Due to the large masses of the heavy quarks, their relative motion within the baryon is non-relativistic, with a typical velocity $ v $ that is small in the rest frame of the baryon. As a result, the heavy quarks are tightly bound together~\cite{Chen:2011mb}. The production mechanisms of triply heavy baryons have been systematically studied in Refs.\cite{Baranov:2004er,GomshiNobary:2003sf,GomshiNobary:2004mq,GomshiNobary:2005ur,Chen:2011mb}, which include processes such as $ gg $ fusion\cite{Chen:2011mb}, $ e^+e^- $ annihilation~\cite{Baranov:2004er}, and heavy quark fragmentation~\cite{GomshiNobary:2003sf,GomshiNobary:2004mq,GomshiNobary:2005ur}.

\begin{table}[ht]
\centering
\caption{Same caption with Table~\ref{table:QCDSR-doublyheavybaryon}, but for triply heavy baryons. Only theoretical predictions are available; no triply heavy baryon has been experimentally observed.}
\renewcommand{\arraystretch}{1.5}
\resizebox{\textwidth}{!}{
\tiny
\begin{tabular}{ccccc}
\hline\hline
State & QCDSR & Lattice~\cite{Brown:2014ena} & Quark model~\cite{Roberts:2007ni} & Fadeev equation~\cite{Qin:2019hgk} \\\hline

\multirow{2}{*}{$\ket{\Omega_{ccc},\frac{3}{2}^+} $} 
& 4.67$\pm$0.15~\cite{Zhang:2009re}, 4.81$\pm$0.10~\cite{Wang:2020avt}, & \multirow{2}{*}{4.796} & \multirow{2}{*}{4.965} & \multirow{2}{*}{4.760} \\
& 4.72$\pm$0.12~\cite{Aliev:2014lxa}, $4.53_{\scalebox{0.8}{$-0.11$}}^{\scalebox{0.8}{$+0.26$}}$~\cite{Wu:2021tzo} & & & \\\hline

$\ket{\Omega_{ccc},\frac{3}{2}^-} $ 
&4.88$\pm$0.15~\cite{Wang:2011ae}, $5.17_{\scalebox{0.8}{$-0.13$}}^{\scalebox{0.8}{$+0.14$}}$~\cite{Najjar:2025dzl} & & 5.160 & 5.02 \\\hline

\multirow{2}{*}{$\ket{\Omega_{ccb},\frac{1}{2}^+} $} 
& 7.41$\pm$0.13~\cite{Zhang:2009re}, 8.02$\pm$0.08~\cite{Wang:2020avt}, & \multirow{2}{*}{8.007} & \multirow{2}{*}{8.245} & \multirow{2}{*}{7.867} \\
& $8.15_{\scalebox{0.8}{$-0.23$}}^{\scalebox{0.8}{$+0.27$}}$~\cite{Najjar:2024deh} & & & \\\hline

\multirow{2}{*}{$\ket{\Omega_{ccb},\frac{3}{2}^+} $} 
& 7.45$\pm$0.16~\cite{Zhang:2009re}, 8.03$\pm$0.08~\cite{Wang:2020avt},& \multirow{2}{*}{8.037} & \multirow{2}{*}{8.265} & \multirow{2}{*}{7.963} \\
& 8.07$\pm$0.10~\cite{Aliev:2014lxa}, $8.06_{\scalebox{0.8}{$-0.17$}}^{\scalebox{0.8}{$+0.16$}}$~\cite{Najjar:2025dzl}& & & \\\hline

$\ket{\Omega_{ccb},\frac{3}{2}^-} $ 
&7.73$\pm$0.13~\cite{Wang:2011ae}, $8.21_{\scalebox{0.8}{$-0.23$}}^{\scalebox{0.8}{$+0.20$}}$~\cite{Najjar:2025dzl} & & 8.420 & 8.27 \\\hline

\multirow{2}{*}{$\ket{\Omega_{bbb},\frac{3}{2}^+} $} 
& 13.28$\pm$0.10~\cite{Zhang:2009re}, 14.43$\pm$0.09~\cite{Wang:2020avt}, & \multirow{2}{*}{14.366} & \multirow{2}{*}{14.834} & \multirow{2}{*}{14.370} \\
&14.30$\pm$0.20~\cite{Aliev:2014lxa}, $14.27_{\scalebox{0.8}{$-0.32$}}^{\scalebox{0.8}{$+0.33$}}$~\cite{Wu:2021tzo}& & & \\\hline

$\ket{\Omega_{bbb},\frac{3}{2}^-} $ 
& 13.52$\pm$0.11~\cite{Wang:2011ae}, $14.10_{\scalebox{0.8}{$-0.13$}}^{\scalebox{0.8}{$+0.13$}}$~\cite{Najjar:2025dzl} & & 14.976 & 14.77 \\\hline

\multirow{2}{*}{$\ket{\Omega_{bbc},\frac{1}{2}^+} $} 
& 10.30$\pm$0.10~\cite{Zhang:2009re}, 11.22$\pm$0.09~\cite{Wang:2020avt},& \multirow{2}{*}{11.195} & \multirow{2}{*}{11.535} & \multirow{2}{*}{11.077} \\
& $11.13_{\scalebox{0.8}{$-0.19$}}^{\scalebox{0.8}{$+0.23$}}$~\cite{Najjar:2024deh} & & & \\\hline

\multirow{2}{*}{$\ket{\Omega_{bbc},\frac{3}{2}^+} $} 
& 10.54$\pm$0.11~\cite{Zhang:2009re}, 11.23$\pm$0.08~\cite{Wang:2020avt}, & \multirow{2}{*}{11.229} & \multirow{2}{*}{11.554} & \multirow{2}{*}{11.167} \\
&11.35$\pm$0.15~\cite{Aliev:2014lxa}, $11.02_{\scalebox{0.8}{$-0.12$}}^{\scalebox{0.8}{$+0.13$}}$~\cite{Najjar:2025dzl}& & & \\\hline

$\ket{\Omega_{bbc},\frac{3}{2}^-} $ 
& 10.59$\pm$0.11~\cite{Wang:2011ae}, $11.14_{\scalebox{0.8}{$-0.13$}}^{\scalebox{0.8}{$+0.13$}}$~\cite{Najjar:2025dzl} & & 11.711 & 11.52 \\\hline

\end{tabular}
}
\label{table:QCDSR-triplyheavybaryon}
\end{table}

Although experimentally producing and reconstructing a candidate for the triply heavy baryons is challenging, it is not unthinkable. In Ref.~\cite{Bjorken:1985ei}, Bjorken suggested that semileptonic decay processes, such as $\Omega_{ccc}^{++} \rightarrow \Omega_{sss}^{-} + 3 \mu^{+} + 3 \nu_\mu$ and $\Omega_{ccc}^{++} \rightarrow \Omega_{sss}^{-} + 3 \pi^{+}$, may provide signatures for a potential ``$ccc$" event. In addition, with the rapid development of baryon spectroscopy in recent years, more and heavier hadronic resonances have been observed~\cite{ParticleDataGroup:2024cfk}, renewing interest in the study of triply heavy baryons. Theoretical studies for the triply heavy baryons spectra have been conducted in different quark models~\cite{Migura:2006ep,Roberts:2007ni,Liu:2019vtx,Yang:2019lsg}, Lattice QCD~\cite{Meinel:2010pw,Meinel:2012qz,Padmanath:2013zfa,Brown:2014ena}, bag model~\cite{Hasenfratz:1980ka}, Fadeev equation~\cite{Gutierrez-Guerrero:2019uwa,Qin:2019hgk}, Regge trajectories~\cite{Wei:2015gsa,Wei:2016jyk} and QCD sum rules~\cite{Zhang:2009re,Wang:2011ae,Wang:2020avt,Aliev:2012tt,Aliev:2014lxa,Wu:2021tzo,Najjar:2024deh,Najjar:2025dzl}. Part of the mass predictions for the triply heavy baryons are summarized in Table~\ref{table:QCDSR-triplyheavybaryon}, while other detailed results can be found in the above literature.

We note that Ref.~\cite{Najjar:2024deh} provides a comparison of different mass renormalization schemes, including the pole mass and the $\overline{MS}$ mass. The results reveal a significant discrepancy between the two, with the masses obtained using the pole mass scheme being larger than those from the $\overline{MS}$ scheme. The authors pointed out that the $\overline{MS}$ results are more consistent with other frameworks. As discussed previously, this discrepancy can be reduced by incorporating higher-order corrections in the perturbative part, which have also been applied in the analysis of triply heavy baryons $ \Omega_{QQQ} $~\cite{Wu:2021tzo}.

The study of triply heavy baryons is still in its infancy, and further investigation is needed to explore the properties of these baryons. As experimental facilities such as the LHCb and future high-luminosity colliders continue to push the frontiers of hadron physics, the discovery of the triply heavy baryons may soon become a reality. The theoretical predictions discussed above will provide valuable guidance for future experimental searches.

\subsection{Hybrid baryons}\

The hybrid baryons are composed of three valence quarks and one valence gluon. In such systems, the gluon degree of freedom plays a crucial role in determining the baryon properties. However, defining the gluon degrees of freedom theoretically is highly challenging. Fortunately, one can circumvent this difficulty and instead work with a well-defined gluon field by employing QCD sum rules~\cite{Chen:2022asf}.

Another issue is the experimental distinction between hybrid states and conventional hadrons. In general, the quantum numbers $ J^{PC} $ can be used to differentiate between hybrid mesons and conventional mesons, as certain quantum numbers such as $1^{-+}$ are inaccessible to the meson system ($q\bar{q}$) but can be realized in hybrid mesons ($ q\bar{q}g $). However, this distinction does not apply to baryons and hybrid baryons, as both the baryons ($qqq$) and hybrid baryons ($qqqg$) can possess all $J^P$ quantum numbers. Therefore, the differentiation between baryons and hybrid baryons relies on other characteristics.

The decay constant may be a potential criterion for distinguishing baryons and hybrid baryons. In Ref.~\cite{Kisslinger:1995yw}, the authors calculated the mass and decay constants of the ground hybrid baryons using QCD sum rules with following current:
\begin{align}
j(x)=\varepsilon_{a b c}\big[u^{aT}(x) C \gamma^\mu u^b(x)\big] \gamma^\sigma G_{\mu \sigma}^A(x)\big[t^A d(x)\big]^c 
\label{eq:hybrid-baryon-current}
\end{align}
The numerical value of decay constant is $ 5.6\times 10^{-4}\,\text{GeV}^3 $, which is much smaller than for nucleons, which are of order $ 10^{-2} $. Additionally, its mass is predicted to be $ 1.8\,\text{GeV} $ with the nucleon contributions suppressed, making it significantly larger than the nucleon mass. 

In Ref.~\cite{Zhao:2023imq}, by replacing the light quarks with heavy quarks in current~\eqref{eq:hybrid-baryon-current}, the authors studied triply heavy hybrid baryon $ \Omega_{QQQg} $ using QCD sum rules. The mass predictions are
\begin{align}
M_{\Omega_{cccg}}=6.02_{-0.11}^{+0.11}\, \mathrm{GeV}\,,\quad M_{\Omega_{bbbg}}=14.68_{-0.06}^{+0.14}\, \mathrm{GeV}\,,
\end{align}
which are considerably larger than the predicted masses of the corresponding triply heavy baryons $ \Omega_{ccc} $ and $ \Omega_{bbb} $ listed in Table~\ref{table:QCDSR-triplyheavybaryon}. Thus, mass can also be used to distinguish heavy hybrid baryons from heavy baryons, since the spacing in the high-energy regime is generally larger, making experimental identification easier. Additionally, the $ P $-wave decay channels $\Xi_{c c}^{++} D^0$, $\Xi_{c c}^{+} D^{+}$, and $\Xi_{c c s}^{+} D_s^{+}$ were suggested to reconstruct the triply charmed hybrid baryon $ \Omega_{cccg} $, which are more accessible in experiments. Other analyses of hybrid baryons were performed in Refs.~\cite{Kisslinger:2003hk,Kisslinger:2009dr,Azizi:2017xyx,Wang:2024lnv,Yang:2025hzc}.

\section{Baryon semileptonic decays}\label{sec:decay}
Most hadrons are unstable and decay shortly after being produced. Their decays can proceed via different interactions, such as strong or weak decays. Strong decays, which do not involve quark flavor changes, are crucial for probing the hadronic structure of the parent particle. In contrast, weak decays involve quark flavor transitions and are essential for studying flavor physics, including measurements of CKM matrix elements and CP violation. In this review, we focus on weak decays, particularly semileptonic decays, while the information on strong decays of baryons can be found in Refs.~\cite{Cui:2019dzj,Yang:2020zjl,Yang:2020zrh,Yang:2021lce,Yang:2023fsc,Tan:2023opd,Aliev:2018vye,Agaev:2017lip,Aliev:2018lcs,Azizi:2020tgh}. These studies provide essential theoretical input for understanding the internal structure of baryons through their strong decay widths.

As for hadron weak decays, they can be classified into purely leptonic, semileptonic, and nonleptonic modes, depending on the final-state particles. Purely leptonic decays produce only leptons, semileptonic decays involve both hadrons and leptons, while nonleptonic decays produce only hadrons. Among these, semileptonic decays are particularly important. Compared to nonleptonic decays, strong interactions in semileptonic decays affect only the hadronic current, allowing their effects to be factorized and simplifying theoretical calculations. Unlike purely leptonic decays, the presence of a hadron in the final state enables studies of hadronic transitions and tests of the universality of weak interactions at the quark level. Experimentally, semileptonic decays are also advantageous due to their relatively large branching ratios, cleaner signatures, and manageable backgrounds~\cite{Richman:1995wm}.

The transition matrix element between the initial and final states is the key quantity in any decay process. Since semileptonic decays involve multiple energy scales, the transition matrix element can be factorized into a hadronic part and a leptonic part. The leptonic part is calculable using perturbation theory, while the hadronic part is parameterized by form factors: 
\begin{align}
\bra{B(q_2)} \bar{q} \gamma_\mu(1-\gamma_5) Q\ket{A(q_1)}&= \bar{u}_B(q_2)\big[f_1(q^2) \gamma_\mu+f_2(q^2) i\sigma_{\mu \nu}q^\nu/M_A+f_3(q^2) q_\mu/M_A\big] u_A(q_1) \nonumber\\[5pt]
& -\bar{u}_B(q_2)\big[g_1(q^2) \gamma_\mu+g_2(q^2) i\sigma_{\mu \nu}q^\nu/M_A+g_3(q^2) q_\mu/M_A\big] \gamma_5 u_A(q_1)\,.
\label{form-factors}
\end{align}
An alternate definition of the form factors in the literature is given by:
\begin{align}
\bra{B(q_2)} \bar{q} \gamma_\mu Q\ket{A(q_1)} &= \bar{u}_B(q_2)[f_0(q^2)(M_A-M_B) \frac{q^\mu}{q^2} +f_{+}(q^2) \frac{M_A+M_B}{s_{+}}(q_1^\mu+q_2^{\mu}-(M_A^2-M_B^2) \frac{q^\mu}{q^2}) \nonumber\\[5pt]
&+f_{\perp}(q^2)(\gamma^\mu-\frac{2 M_B}{s_{+}} q_1^\mu-\frac{2 M_A}{s_{+}} q_2^{\mu})] u_A(q_1)\,,\\[5pt]
\bra{B(q_2)} \bar{q} \gamma_\mu\gamma_5 Q\ket{A(q_1)} &= -\bar{u}_B(q_2)\gamma_5[g_0(q^2)(M_A+M_B) \frac{q^\mu}{q^2} +g_{+}(q^2) \frac{M_A-M_B}{s_{-}}(q_1^\mu+q_2^{\mu}-(M_A^2-M_B^2) \frac{q^\mu}{q^2}) \nonumber\\[5pt]
&+g_{\perp}(q^2)(\gamma^\mu+\frac{2 M_B}{s_{-}} q_1^\mu-\frac{2 M_A}{s_{-}} q_2^{\mu})] u_A(q_1)\,,
\end{align}
where $s_{\pm}=(M_A \pm M_B)^2-q^2$. The form factors depend on the momentum transfer and the specific decay process, and they can be extracted from experimental data or calculated using various theoretical approaches, such as Lattice QCD, QCD sum rules, chiral perturbation theory, or some quark models. The form factors encode all the kinematic information of a decay process, which are crucial for determining the decay rates and angular distributions of the final-state particles. Thus, the calculation of form factors is a central task for theorists.

However, since most theoretical models are only valid in limited kinematic regions, analytic functions, such as the dipole~\eqref{dipole} and the $z$-series~\eqref{BCL} parameterizations, are employed to extrapolate the form factors over the entire physical range. The decay observables can then be defined using helicity amplitudes. Certain asymmetry parameters such as the lepton forward-backward asymmetry and longitudinal polarization can probe the degree of parity violation. They are also sensitive to potential new physics, making them useful observables for indirect searches in experiments.

The technique of three-point QCD sum rules, originally developed by Ioffe et al. to investigate the electromagnetic form factors of light mesons~\cite{Ioffe:1982ia,Ioffe:1982qb,Nesterenko:1982gc}, has been extensively employed in the study of semileptonic decays. In this chapter, we will present the relevant decay observables and review the QCD sum rule studies on baryon semileptonic decays.

\subsection{Decay observables}\

In the decay process $A \to B+\ell+\nu_\ell$, the observables can be conveniently defined using helicity amplitudes, which offer a more intuitive physical picture and simplify the expressions for asymmetry parameters. The helicity amplitudes are related to the form factors as follows~\cite{Bialas:1992ny,Gutsche:2013pp}:
\begin{align}
& H_{\frac{1}{2}, 0}^V= \frac{\sqrt{Q_{-}}}{\sqrt{q^2}}\big(M_{+} f_1(q^2)-\frac{q^2}{M_{A}} f_2(q^2)\big)\,, \,\, H_{\frac{1}{2}, 0}^A= \frac{\sqrt{Q_{+}}}{\sqrt{q^2}}\big(M_{-} g_1(q^2)+\frac{q^2}{M_{A}} g_2(q^2)\big)\,, \nonumber\\
& H_{\frac{1}{2}, 1}^V= \sqrt{2 Q_{-}}\big(-f_1(q^2)+\frac{M_{+}}{M_{A}} f_2(q^2)\big)\,, \,\, H_{\frac{1}{2}, 1}^A= \sqrt{2 Q_{+}}\big(-g_1(q^2)-\frac{M_{-}}{M_{A}} g_2(q^2)\big)\,, \nonumber\\
& H_{\frac{1}{2}, t}^V= \frac{\sqrt{Q_{+}}}{\sqrt{q^2}}\big(M_{-} f_1(q^2)+\frac{q^2}{M_{A}} f_3(q^2)\big)\,,\,\, H_{\frac{1}{2}, t}^A= \frac{\sqrt{Q_{-}}}{\sqrt{q^2}}\big(M_{+} g_1(q^2)-\frac{q^2}{M_{A}} g_3(q^2)\big)\,.
\end{align}
Here, $H_{\lambda^\prime, \lambda_W}^{V(A)}$ denotes the helicity amplitudes for the weak transition induced by the vector ($V$) and axial-vector ($A$) currents, where $\lambda^\prime$ and $\lambda_W$ are the helicities of the final-state baryon and the virtual $W$ boson, respectively. The kinematic factors are defined as $Q_{\pm} = M_{\pm}^2 - q^2$, with $M_{\pm} = M_A \pm M_B$. The negative-helicity and total helicity amplitudes can be derived using the following relations:
\begin{align}
H_{-\lambda^\prime,-\lambda_W}^V&=H_{\lambda^\prime, \lambda_W}^V\,, \quad \quad H_{-\lambda^\prime,-\lambda_W}^A=-H_{\lambda^\prime, \lambda_W}^A\,,\\[5pt]
H_{\lambda^\prime, \lambda_W}&=H_{\lambda^\prime, \lambda_W}^V-H_{\lambda^\prime, \lambda_W}^A\,.
\end{align}

\subsubsection{Differential decay width}\

Based on the helicity amplitudes, the differential decay width of semileptonic decays can be expressed as:
\begin{align}
\label{differential-decay-width}
&\frac{d \Gamma\big(A\to B\ell\nu_\ell\big)}{d q^2}=\frac{G_F^2|V_{CKM}|^2 q^2\sqrt{Q_+Q_-}}{384 \,\pi^3 \,M_{A}^3}(1-\frac{m_\ell^2}{q^2})^2 H_{\text{tot}}\,,
\end{align}
where $G_F$ is the Fermi coupling constant, $V_{CKM}$ is the CKM matrix element, and $m_\ell$ is the mass of the lepton. $Q_{\pm}$ are defined as $Q_{\pm} = (M_A \pm M_B)^2 - q^2$, where $M_A$ and $M_B$ are the masses of the initial and final baryons, respectively. The total helicity amplitude $H_{\text{tot}}$ is defined as:
\begin{align}
&H_{\text{tot}} = \big(1+\frac{m_\ell^2}{2 q^2}\big)\big(H_{\frac{1}{2},1}^2+H_{-\frac{1}{2},-1}^2+H_{\frac{1}{2},0}^2+H_{-\frac{1}{2},0}^2\big)+\frac{3\,m_\ell^2}{2 q^2}\big(H_{\frac{1}{2},t}^2+H_{-\frac{1}{2},t}^2\big)\,.
\end{align}
The polarized decay widths can also be defined:
\begin{align}
\frac{\mathrm{d} \Gamma_L}{\mathrm{~d} q^2} & =\frac{G_F^2|V_{CKM}|^2 q^2\sqrt{Q_+Q_-}}{384 \,\pi^3 \,M_{A}^3}(1-\frac{m_\ell^2}{q^2})^2\Big((1+\frac{m_\ell^2}{2q^2})\big(H_{-\frac{1}{2},0}^2+H_{\frac{1}{2},0}^2\big)+\frac{3m_\ell^2}{2q^2}\big(H_{-\frac{1}{2},t}^2+H_{\frac{1}{2},t}^2\big)\Big)\\[5pt]
\frac{\mathrm{d} \Gamma_T}{\mathrm{~d} q^2}&=\frac{G_F^2|V_{CKM}|^2 q^2\sqrt{Q_+Q_-}}{384 \,\pi^3 \,M_{A}^3}(1-\frac{m_\ell^2}{q^2})^2(1+\frac{m_\ell^2}{2q^2})\Big(H_{\frac{1}{2},1}^2+H_{-\frac{1}{2},-1}^2\Big),
\end{align}

The total decay width can be obtained by integrating the differential decay width over the physical region of $q^2$:
\begin{align}
\label{total-decay-width}
\Gamma\big(A\to B\ell\nu_\ell\big)=\int_{m_\ell^2}^{(M_A-M_B)^2} \frac{d \Gamma\big(A\to B\ell\nu_\ell\big)}{d q^2} d q^2\,.
\end{align}
The corresponding branching fraction can also be derived:
\begin{align}
\label{branching-fraction}
Br\big(A\to B\ell\nu_\ell\big)=\frac{\Gamma\big(A\to B\ell\nu_\ell\big)}{\Gamma_{A}}\,,
\end{align}
where $\Gamma_{A}$ is the total decay width of the baryon $A$.

\subsubsection{Lepton forward-backward asymmetry}\

The lepton forward-backward asymmetry is defined as the difference between the forward and backward decay rates of the lepton, normalized by the total decay rate. It can be expressed as~\cite{Faustov:2016yza, Geng:2022fsr}:
\begin{align}
\label{AFB}
A_{F B}(q^2)&=\frac{\frac{d \Gamma}{d q^2}(\text{forward})-\frac{d \Gamma}{d q^2}(\text{backward})}{\frac{d \Gamma}{d q^2}}=\frac{\int_0^1 \frac{d^2 \Gamma}{d q^2 d \cos \theta_{\ell}} d \cos \theta_{\ell}-\int_{-1}^0 \frac{d^2 \Gamma}{d q^2 d \cos \theta_{\ell}} d \cos \theta_{\ell}}{\int_0^1 \frac{d^2 \Gamma}{d q^2 d \cos \theta_{\ell}} d \cos \theta_{\ell}+\int_{-1}^0 \frac{d^2 \Gamma}{d q^2 d \cos \theta_{\ell}} d \cos \theta_{\ell}}\nonumber\\
&=\frac{3}{4}\frac{H_{\frac{1}{2}, 1}^2-H_{-\frac{1}{2}, -1}^2-2\frac{m_\ell^2}{q^2}(H_{\frac{1}{2},0}H_{\frac{1}{2},t}+H_{-\frac{1}{2},0}H_{-\frac{1}{2},t})}{H_{\text{tot}}}\,, 
\end{align}
where $\theta_{\ell}$ is the angle between the lepton direction and the $\ell \nu_{\ell}$ system direction in the $A$ rest frame.

\subsubsection{Polarization asymmetry}\

The polarization asymmetry is defined as the difference between the decay rates of the baryon with positive and negative helicities, normalized by the total decay rate. It can be expressed as~\cite{Faustov:2016yza, Geng:2022fsr}:
\begin{align}
\label{PB}
\alpha(q^2)&=\frac{d \Gamma^{\lambda^\prime=\frac{1}{2}} / d q^2-d \Gamma^{\lambda^\prime=-\frac{1}{2}} / d q^2}{d \Gamma^{\lambda^\prime=\frac{1}{2}} / d q^2+d \Gamma^{\lambda^\prime=-\frac{1}{2}} / d q^2}\,,
\end{align}
where
\begin{align}
\frac{d \Gamma^{\lambda^\prime=\frac{1}{2}}}{d q^2}= &\, \frac{4 m_l^2}{3 q^2}\big(H_{\frac{1}{2}, 1}^2+H_{\frac{1}{2}, 0}^2+3 H_{\frac{1}{2}, t}^2\big)+\frac{8}{3}\big(H_{\frac{1}{2}, 1}^2+H_{\frac{1}{2}, 0}^2\big)\,,\\[5pt]
\frac{d \Gamma^{\lambda^\prime=-\frac{1}{2}}}{d q^2}= &\, \frac{4 m_l^2}{3 q^2}\big(H_{-\frac{1}{2}, -1}^2+H_{-\frac{1}{2}, 0}^2+3 H_{-\frac{1}{2}, t}^2\big)+\frac{8}{3}\big(H_{-\frac{1}{2}, -1}^2+H_{-\frac{1}{2}, 0}^2\big)\,.
\end{align}

Besides, there are also some other decay observables such as the convexity parameter, which can be found in Refs.~\cite{Gutsche:2015rrt,Faustov:2016yza}.

\subsection{Heavy baryon decays}\

With the successive discoveries of heavy quarks, heavy-flavor hadronic systems have emerged as a more advantageous platform than their light-flavor counterparts for probing the standard model. The heavy quarks in the standard model include the charm ($c$), bottom ($b$), and top ($t$) quarks. Among them, the top quark is too heavy to hadronize into stable bound states, rendering it inaccessible to hadron-level studies. In contrast, the charm and bottom sectors provide ideal energy ranges for testing the standard model and exploring possible new physics.

To facilitate such studies, numerous high-energy physics experiments have been established worldwide. For instance, $B$ factories such as BELLE in Japan and BABAR in the United States focus on the production and decay of bottom-flavored hadrons, while charm factories like BES in China and CLEO in the United States specialize in the study of charm-flavored hadrons. These facilities have yielded a wealth of data for precise measurements of decay branching ratios, $CP$ violation effects, and kinematic observables of heavy-flavor hadrons, thereby offering a robust experimental foundation for theoretical research.

Against this backdrop, significant progress has been made in the understanding of heavy-flavor mesons such as the $B$ and $D$ mesons. However, the study of heavy-flavor baryons, such as $\Lambda_b$ and $\Lambda_c$, remains experimentally challenging. Due to their relatively small production cross sections, the available data samples for heavy baryons are much smaller than those for mesons. This limits our understanding of their internal structure, decay dynamics, and nonperturbative QCD effects, leaving the physics of heavy-flavor baryons at a less developed stage. This experimental asymmetry makes heavy-flavor baryons a compelling and underexplored frontier in contemporary heavy-flavor physics.

In this section, we will review the QCD sum rule studies of semileptonic decays of heavy baryons. We will focus on the decays of heavy baryons containing $b$ and $c$ quarks.

\subsubsection{Bottom baryon decays}\

The semileptonic decays of bottom baryons are mainly induced by the bottom quark, such as the Cabibbo-suppressed process $b\to c$ and $b\to u$. The CKM matrix elements $V_{cb}$ and $V_{ub}$ can then be extracted from semileptonic decays of $b$-hadrons. Due to the relative rarity of $b$-baryon decays, current determinations of $V_{cb}$ and $V_{ub}$ mainly rely on $B$-meson decays, specifically, the $B \to D$ channel for $V_{cb}$ and the $B \to \pi$ channel for $V_{ub}$. However, the precision of these determinations remains limited~\cite{ParticleDataGroup:2024cfk}. As such, semileptonic decays of $b$-baryons remain an important area of study, offering valuable complementary information.

The decay processes of $b$-flavored hadrons can be conveniently described using heavy quark effective theory (HQET). For a $b \to u$ transition, such as $\Lambda_b \to p \ell \nu_\ell$, heavy quark symmetry imposes constraints on the general decomposition of the form factors in Eq.~\eqref{form-factors}. Under these constraints, Eq.~\eqref{form-factors} can be rewritten as:
\begin{align}
\label{heavy-light-form-factors}
\bra{p(q_2)}\bar{u}\,\Gamma\, b_v\ket{\Lambda_b(v)}=\bar{u}_{p}(q_2)\,\Big(F_1(z)+F_2(z) \slashed{v}\Big)\, \Gamma \,u_{\Lambda_b}(v)\,,
\end{align}
where $v$ denotes the four-velocities and $ z=q_2\cdot v $. The constraints $\slashed{v} u(v) = u(v)$ and $\slashed{q}_2 u(q_2) = M_p u(q_2)$ can be used to reduce the number of independent terms in the decomposition. The relations between $f_i$ and $F_i$ are given by:
\begin{align}
f_1=g_1=F_1+\frac{M_P}{M_{\Lambda_b}}F_2,\quad f_2=f_3=g_2=g_3=F_2\,.
\end{align}
For $b \to c$ decays, such as $\Lambda_b \to \Lambda_c \ell \nu_\ell$, the form factors are more strongly constrained by heavy quark symmetry, since it can be applied to both the initial and final heavy baryons. The form factors can be expressed in terms of the Isgur-Wise function $\zeta(w)$, which is a universal function that describes the transition between heavy baryons. The hadron matrix element can be expressed as:
\begin{align}
\label{heavy-heavy-form-factors}
\bra{\Lambda_c(v^\prime)} \bar{c}_{v^\prime} \Gamma b_v\ket{\Lambda_b(v)}=\zeta(w) \bar{u}_{\Lambda_c}(v^{\prime}) \Gamma u_{\Lambda_b}(v)\,,
\end{align}
where $w=v^\prime \cdot v$. Based on a single Isgur-Wise function $\zeta(w)$, all six form factors can be expressed:
\begin{align}
f_1=g_1=\zeta(w),\quad f_2=f_3=g_2=g_3=0\,.
\end{align}

The first QCD sum rule calculation for baryonic semileptonic decay was performed by Dai et al. in 1996~\cite{Dai:1996xv}. By combining HQET and QCD sum rules, they investigated the semileptonic decay process $\Lambda_b \to \Lambda_c \ell \nu_\ell$ and calculated the subleading baryonic Isgur-Wise function. They found that significant $1 / m_Q$ corrections in the decay process arise primarily from the weak current. The branching fraction was predicted to be $Br(\Lambda_b \to \Lambda_c \ell \nu_\ell)=9.8\%$ after taking $ V_{cb}=0.04 $.

To investigate the deviations between heavy quark symmetry and full QCD in $\Lambda_b \to \Lambda_c \ell \nu_\ell$ semileptonic decays, the authors of Ref.~\cite{Dosch:1997zx} employed the QCD sum rule method within the framework of full QCD, while also successfully reproducing the symmetry relations predicted by HQET. They have found that at the zero recoil point, heavy quark symmetry is violated by approximately 20\%. Although significant violations of heavy quark symmetry may occur for some form factors due to the large mass difference between the $b$ quark and the $\Lambda_b$ baryon, such large deviations for form factors protected by Luke’s theorem, such as $g_1(q^2_{max})$, remain quite unexpected. In their following paper~\cite{MarquesdeCarvalho:1999bqs}, the authors extended the analysis and changed the continuum model, which led to a more appropriate prediction of the branching fraction $Br(\Lambda_b \to \Lambda_c \ell \nu_\ell)=(7.38\pm 2.46)\%$.

In 2004, DELPHI collaboration measured the branching fraction of the decay $\Lambda_b \to \Lambda_c \ell \nu_\ell$ and found $Br(\Lambda_b \to \Lambda_c \ell \nu_\ell)=(5.0_{-0.8}^{+1.1}(stat)_{-1.2}^{+1.6}(sys)) \%$~\cite{DELPHI:2003qft}. The experimental measurement of this decay channel is generally consistent with previous theoretical predictions from QCD sum rules, although the QCD sum rule approach typically carries relatively large uncertainties.

Regarding the class of heavy-to-light semileptonic decays induced by $b\to u$, such as $\Lambda_b \to p \ell \nu_\ell$, Ball and Braun have pointed out that the application of full QCD sum rules without HQET may encounter significant limitations~\cite{Ball:1997rj}. In particular, the coefficient of the $\langle \bar{q}q \rangle^2$ condensate terms grows more rapidly with $m_b$ than that of the perturbative contribution, an effect absent in two-point QCD sum rules~\cite{Khodjamirian:1997lay}. This leads to a breakdown in the convergence of the operator product expansion, rendering the traditional QCD sum rules approach unreliable for heavy-to-light transitions. An effective way is to expand the correlation function on the light cone, leading to the light-cone sum rules method. Alternatively, QCD sum rules can still be applied within the framework of HQET to mitigate this issue.

\begin{figure}[ht]
\centering
\includegraphics[width=0.4\textwidth]{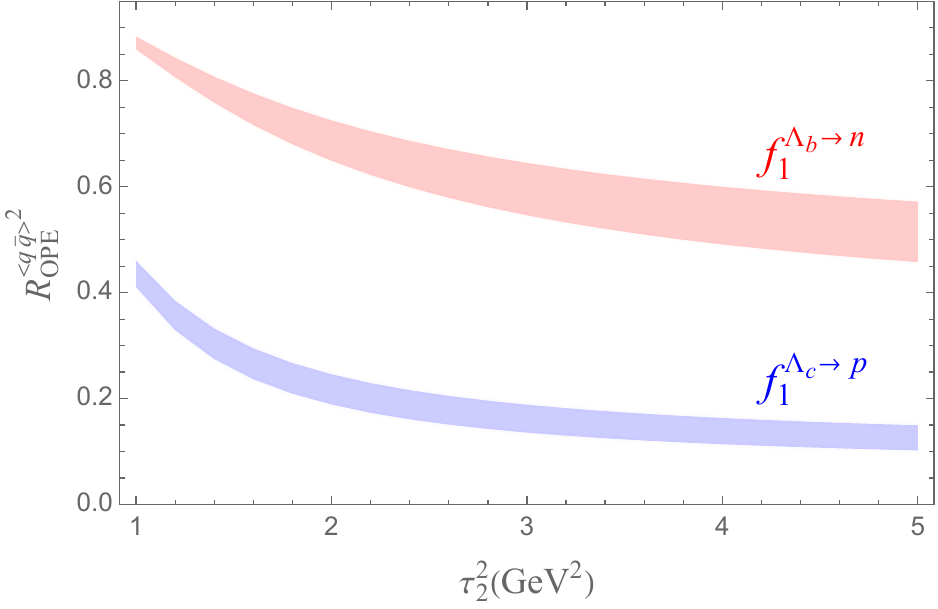}
\caption{The relative contributions of the highest-order condensate term $ \langle \bar{q}q\rangle^2 $ to the form factors in the processes $ \Lambda_b \to n $ (red) and $ \Lambda_c \to p $ (blue) taken from Ref.~\cite{Zhang:2024asb}.}
\label{fig:four-quark}
\end{figure}

In 1997, Huang, Qiao, and Yan performed the first analysis for the heavy-to-light transition $\Lambda_b \to p \ell \nu_\ell$ within the framework of HQET~\cite{Huang:1998rq}. They derived the analytic expressions of two form factors in Eq.~\eqref{heavy-light-form-factors}, here we show the results for some discussions:
\begin{align}
\label{F1}
-2 f_{\Lambda_b} f_p F_1 e^{-2 \bar{\Lambda} / M-m_p^2 / T}= & \int_0^{\nu_c} d \nu \int_{m_p^2}^{2 \nu z} d s \rho_{p e r t}^1 e^{-s / T-\nu / M}-\frac{1}{3}\langle\bar{q} q\rangle^2- \nonumber\\
& \frac{1}{3 \pi^4}\left\langle\alpha_s G G\right\rangle \int_0^{T / 4}(1-\frac{4 \beta}{T}) e^{-4 \beta(1-4 \beta / T) / M^2-8 \beta z /(T M)} d \beta, \\
\label{F2}
-2 f_{\Lambda_b} f_p m_p F_2 e^{-2 \bar{\Lambda} / M-m_p^2 / T}= & \int_0^{\nu_c} d \nu \int_{m_p^2}^{2 \nu z} d s \rho_{p e r t}^2 e^{-s / T-\nu / M}+ \nonumber\\
& \frac{1}{8 \pi^4}\left\langle\alpha_s G G\right\rangle \int_0^{T / 4}(1-\frac{4 \beta}{T}) \frac{\beta}{M} e^{-4 \beta(1-4 \beta / T) / M^2-8 \beta z /(T M)} d \beta\,,
\end{align}
where
\begin{align}
\begin{aligned}
& \rho_{\text {pert }}^1=\frac{1}{32 \pi^4 \sigma^3}\left[-2 z^3 \sigma^3-(-s+z(\nu+2 z))^3+3 z^2(-s+z(\nu+2 z)) \sigma^2\right], \\
& \rho_{\text {pert }}^2=\frac{-1}{64 \pi^4 \sigma^3}\left[s-2 z^2+z(-\nu+\sigma)\right]^2\left[\nu s+8 z^3-4 z^2(-2 \nu+\sigma)-2 z\left(-\nu^2+5 s+\nu \sigma\right)\right]\,.
\end{aligned}
\end{align}
According to Eqs.~\eqref{F1} and \eqref{F2}, the four-quark condensate $\langle \bar{q}q \rangle^2$ dominates the contribution to $F_1$, whereas $F_2$ receives no such nonperturbative contribution. The absolute magnitudes of both $F_1$ and $F_2$ depend on the decay constants $f_{\Lambda_b}$ and $f_p$, so the resulting uncertainty in the decay width is effectively doubled due to their combined influence. The total decay width was predicted to be:
\begin{align}
\Gamma=1.35 \times 10^{-11}\left|V_{u b}\right|^2 \mathrm{GeV}\,.
\label{Lambdab-p-HQET}
\end{align}
Since the final-state proton is not a heavy baryon, the full QCD calculation based on the three-point correlation function faces certain challenges, as discussed earlier. Nevertheless, it remains important to compare the result obtained within the partial HQET framework with that from the full QCD approach. In Ref.~\cite{MarquesdeCarvalho:1999bqs}, the authors also predicted the decay width of $\Lambda_b \to p \ell \nu_\ell$ with full QCD:
\begin{align}
\Gamma\left(\Lambda_b \rightarrow p+\ell^{-}+\bar{\nu}_{\ell}\right)=(1.7 \pm 0.7) \times 10^{-11}\left|V_{u b}\right|^2 \mathrm{GeV}\,,
\end{align}
which is in agreement with the HQET calculation~\eqref{Lambdab-p-HQET}. This result is intriguing, as the full QCD calculation is known to suffer from certain shortcomings, yet it manages to yield results consistent with those from HQET although relatively large uncertainties are involved. The authors of Ref.~\cite{MarquesdeCarvalho:1999bqs} did not show the contribution of four-quark condensate $\langle \bar{q}q \rangle^2$, while in Ref.~\cite{Zhang:2024asb} the authors compared the contributions from the highest-dimensional condensates to the form factors in the rare decays of charmed and bottom baryons, as shown in Fig.~\ref{fig:four-quark}. Within the allowed Borel window, the contribution of the four-quark condensate $\langle \bar{q}q\rangle^2$ in the $\Lambda_b \to n$ decay can reach as high as 50\%–90\%, which may compromise the validity of the operator product expansion. In contrast, its contribution in the $\Lambda_c \to p$ decay is significantly smaller.

\begin{figure}[ht]
\centering
\includegraphics[width=0.5\textwidth]{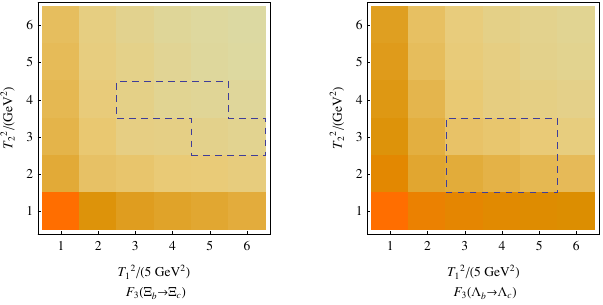}
\caption{$F_3(0)$ as functions of the free Borel parameters $T_1^2$ and $T_2^2$ taken from Ref.~\cite{Zhao:2020mod}.}
\label{fig:borel-plane}
\end{figure}

In Ref.~\cite{Zhao:2020mod}, the authors performed a QCD sum rule analysis of the semileptonic decay $\Lambda_b \to \Lambda_c \ell \nu_\ell$ and $\Xi_b\to \Xi_c\ell \nu_\ell$, where another form of the parametrization of the form factors were adopted:
\begin{align}
\bra{B(q_2)} \bar{q} \gamma_\mu(1-\gamma_5) Q\ket{A(q_1)}&= \bar{u}_B(q_2)\big[\frac{q_{1 \mu}}{M_A} F_1(q^2)+\frac{q_{2 \mu}}{M_B} F_2(q^2)+\gamma_\mu F_3(q^2)\big] u_A(q_1) \nonumber\\[5pt]
& -\bar{u}_B(q_2)\big[\frac{q_{1 \mu}}{M_A} G_1(q^2)+\frac{q_{2 \mu}}{M_B} G_2(q^2)+\gamma_\mu G_3(q^2)\big] \gamma_5 u_A(q_1)\,.
\label{form-factors-2}
\end{align}
The form factors $F_i$ and $G_i$ are related to the form factors in Eq.~\eqref{form-factors} as follows:
\begin{align}
& F_1=f_2+f_3\,, \quad F_2=\frac{M_B}{M_A}\left(f_2-f_3\right)\,, F_3=f_1-\frac{M_A+M_B}{M_A} f_2\,, \nonumber\\
& G_1=g_2+g_3\,, \quad G_2=\frac{M_B}{M_A}\left(g_2-g_3\right)\,, G_3=g_1+\frac{M_A-M_B}{M_A} g_2 \,.
\end{align}
They selected an appropriate region in the Borel parameter plane $T_1^2-T_2^2$, which is enclosed by dashed lines in Fig~\ref{fig:borel-plane}. Based on the form factors, the branching fractions were obtained to be $Br(\Lambda_b \to \Lambda_c \ell \nu_\ell)=(6.61 \pm 1.08)\%$ and $Br(\Xi_b\to \Xi_c\ell \nu_\ell)=(9.02 \pm 0.79)\%$. Results of other bottom baryon semileptonic decays~\cite{Neishabouri:2025abl,Neishabouri:2024gbc,Khajouei:2024frw} are summarized in Table~\ref{table:bottombaryon-decay}.

\begin{table}[ht]
\centering
\caption{The branching fractions of bottom baryon semileptonic decays obtained from QCD sum rules and other theoretical method, as well as the experimental data. The symbol $\ell$ in the first column specifically refers to either an electron or a muon.}
\renewcommand{\arraystretch}{1.2}
\resizebox{1\textwidth}{!}{
\begin{tabular}{llcccc}
\hline\hline
Channel & Method & $ Br $ \\ \hline

\multicolumn{1}{l}{\multirow{1}{*}{$\Lambda_b\rightarrow \Lambda_c \ell \bar{\nu}_\ell$}} & QCD sum rules &9.8$\%$~\cite{Dai:1996xv}, $(7.38\pm2.46)\%$~\cite{MarquesdeCarvalho:1999bqs}, $(6.61\pm 1.08)\%$~\cite{Zhao:2020mod}, $(6.04\pm 1.70)\%$~\cite{Azizi:2018axf} \\
& Lattice QCD &$(5.3\pm 0.8)\%$~\cite{Detmold:2015aaa} \\
& Relativistic quark model &$6.48\%$~\cite{Faustov:2016pal} \\
& Light-front quark model &$6.3\%$~\cite{Ke:2007tg}, 8.83\%~\cite{Zhao:2018zcb} \\
&Exp &$(5.0_{-0.8}^{+1.1}(stat)_{-1.2}^{+1.6}(sys))\%$~\cite{DELPHI:2003qft}\\
\hline

\multicolumn{1}{l}{\multirow{1}{*}{$\Lambda_b\rightarrow \Lambda_c \tau \bar{\nu}_\tau$}} & QCD sum rules &$(1.87 \pm 0.52) \%$~\cite{Azizi:2018axf} \\
& Lattice QCD &$(1.8\pm 0.2)\%$~\cite{Detmold:2015aaa} \\
& Relativistic quark model &$2.03\%$~\cite{Faustov:2016pal} \\
\hline

\multicolumn{1}{l}{\multirow{1}{*}{$\Lambda_b\rightarrow p \ell \bar{\nu}_\ell$}} & QCD sum rules &$(4.4^{+0.5}_{-0.4}) \times 10^{-4}$~\cite{Huang:1998rq}, $(5.5^{+3.1}_{-2.6}) \times 10^{-4}$~\cite{MarquesdeCarvalho:1999bqs} \\
& Lattice QCD &$(5.5_{-2.0}^{+2.3}) \times 10^{-4}$~\cite{Detmold:2015aaa} \\
&Light-front quark model&$3.14 \times 10^{-4}$~\cite{Zhao:2018zcb}\\
&Exp& $(4.1 \pm 1.0) \times 10^{-4}$~\cite{LHCb:2015eia}\\
\hline

\multicolumn{1}{l}{\multirow{1}{*}{$\Xi_b\rightarrow \Xi_c \ell \bar{\nu}_\ell$}} & QCD sum rules &$(9.02\pm 0.79)\%$~\cite{Zhao:2020mod}, $(8.18_{-3.34}^{+4.36})\%$~\cite{Neishabouri:2025abl} \\
& Relativistic quark model &$6.15\%$~\cite{Faustov:2018ahb}, $9.22\%$~\cite{Dutta:2018zqp} \\
&Light-front quark model&9.42\%~\cite{Zhao:2018zcb}\\
\hline

\multicolumn{1}{l}{\multirow{1}{*}{$\Xi_b\rightarrow \Xi_c \tau \bar{\nu}_\tau$}} & QCD sum rules &$(2.81_{-1.15}^{+1.50})\%$~\cite{Neishabouri:2025abl} \\
& Relativistic quark model &$2.00\%$~\cite{Faustov:2018ahb}, $2.35\%$~\cite{Dutta:2018zqp} \\
\hline

\multicolumn{1}{l}{\multirow{1}{*}{$\Omega_b\rightarrow \Omega_c \ell \bar{\nu}_\ell$}} & QCD sum rules &$(2.74_{-1.29}^{+1.64})\%$~\cite{Neishabouri:2024gbc} \\
& Large $N_c$ &2.82\%~\cite{Du:2011nj} \\
&Relativistic quark model & 3.06\%~\cite{Ivanov:1996fj}, 2.77\%~\cite{Ivanov:1999pz} \\
&Bethe-Salpeter&2.97\%~\cite{Ivanov:1998ya}\\
&Light-front quark model&2.72\%~\cite{Zhao:2018zcb}\\
\hline

\multicolumn{1}{l}{\multirow{1}{*}{$\Omega_b\rightarrow \Omega_c \tau \bar{\nu}_\tau$}} & QCD sum rules &$(0.78_{-0.36}^{+0.43})\%$~\cite{Neishabouri:2024gbc} \\
& Large $N_c$ &1.21\%~\cite{Han:2020sag} \\
\hline

\end{tabular}
}
\label{table:bottombaryon-decay}
\end{table}

At present, most of these predictions remain difficult to test due to the lack of corresponding experimental data. Moreover, for heavy-to-light semileptonic decay modes, there is currently no reliable method within the QCD sum rule framework to obtain full QCD results, apart from invoking HQET. It is therefore desirable to establish a complete treatment that does not rely on the heavy quark limit, allowing for a direct comparison with HQET.

\subsubsection{Charmed baryon decays}\

The semileptonic decays of charmed baryons are mainly induced by the charm quark, such as the Cabibbo-favored process $c\to s$ and the Cabibbo-suppressed process $c\to d$. Then the CKM matrix elements $V_{cs}$ and $ V_{cd} $ can be determined from such decay processes. Similar to the bottom sector, the measurement of $V_{cs}$ and $ V_{cd} $ mainly depends on charmed meson semileptonic decays, such as $D \rightarrow \pi \ell \nu_{\ell}$ and $D \rightarrow K \ell \nu_{\ell}$~\cite{ParticleDataGroup:2024cfk}. Hence, precise measurements and theoretical studies of charmed baryon semileptonic decays are highly desirable. Additionally, certain angular asymmetry observables can be involved to search for new physics effects.

Since charmed baryons are significantly lighter than bottom baryons, HQET is expected to be less effective in the charm sector compared to the bottom sector. For the same reason, the issue of the breakdown of QCD sum rules in heavy-to-light transitions, as pointed out by Ball and Braun, has a less significant impact on charmed hadron decays~\cite{Ball:1997rj}, which makes it possible to obtain full QCD results using QCD sum rules.

Among the charmed baryons, $ \Lambda_c $ is the lightest one, which can appear as the final state in decays of other heavier baryons. Its decays have therefore attracted considerable attention from both theorists and experimentalists. The Cabibbo-favored semileptonic decay mode $\Lambda_c \to \Lambda \ell \nu_\ell$ is the benchmark for those of other $ \Lambda_c $ semileptonic channels. In Ref.~\cite{MarquesdeCarvalho:1999bqs}, the authors employed QCD sum rules within the full QCD framework to determine the form factors for the decay process $\Lambda_c \to \Lambda \ell \nu_\ell$. The $q^2$ dependence of these form factors was subsequently modeled using a pole-type parametrization. Based on the form factors, the decay width was determined to be:
\begin{align}
\Gamma(\Lambda_c^{+} \rightarrow \Lambda+e^{+}+\nu_e)=(8.7 \pm 1.2) \times 10^{-14} \,\mathrm{GeV}\,.
\end{align}
Using the average lifetime of the $\Lambda_c$ baryon, $\tau_{\Lambda_c} = (201.5 \pm 2.7)\times 10^{-15}\,\text{s}$, the branching ratio can be estimated as $Br(\Lambda_c^{+} \rightarrow \Lambda+e^{+}+\nu_e)\approx (2.66 \pm 0.37)\%$. Other theoretical methods are also employed to investigate this decay channel, such as different quark models~\cite{Gutsche:2015rrt,Faustov:2016yza, Hussain:2017lir,Li:2021qod,Geng:2019bfz}, light-cone sum rules~\cite{Liu:2009sn}, Lattice QCD~\cite{Meinel:2016dqj}, and MIT bag model~\cite{Geng:2019bfz}.

Experimentally, compared to the rapid progress in charmed meson studies, our understanding of charmed baryons has advanced relatively slowly over the past five decades, largely due to their low production rates at colliders. This situation began to change in 2014, when the BESIII experiment set the center-of-mass energy to 4.6 GeV. At this energy, $\Lambda_c$ and $
\bar{\Lambda}_c$ baryons can be copiously produced in pairs without accompanying hadrons. This threshold setting provides a clean experimental environment, enabling systematic studies of $\Lambda_c$ production and decay~\cite{BESIII:2020nme}.

\begin{figure}[ht]
\centering
\includegraphics[width=0.6\textwidth]{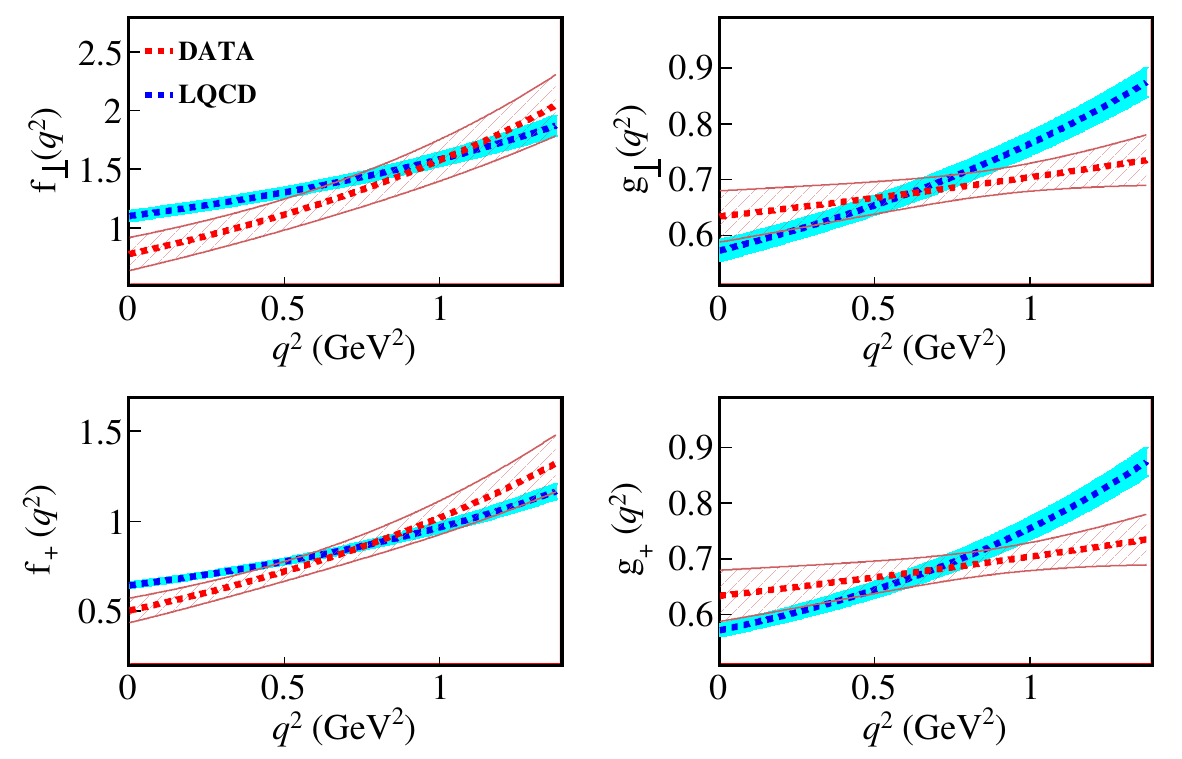}
\caption{Measured $q^2$ dependence of the form factors in $\Lambda_c \to \Lambda \ell \nu_\ell$ semileptonic decay mode taken from Ref.~\cite{BESIII:2023jxv}.}
\label{fig:lambdac-form-factor-distribution}
\end{figure}

\begin{figure}[ht]
\centering
\includegraphics[width=0.6\textwidth]{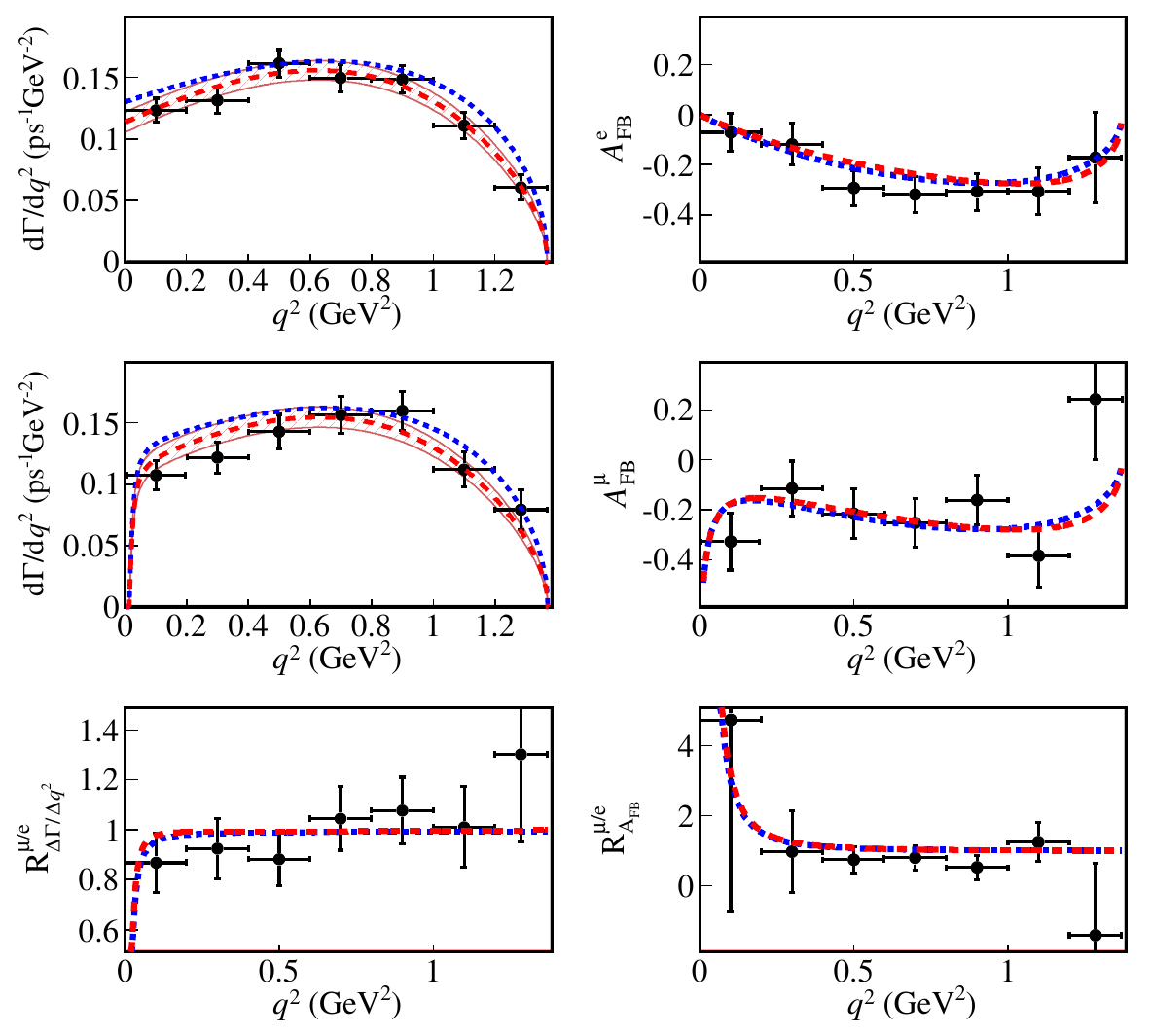}
\caption{Measured $q^2$ dependence of the observables in $\Lambda_c \to \Lambda \ell \nu_\ell$ semileptonic decay mode taken from Ref.~\cite{BESIII:2023jxv}.}
\label{fig:lambdac-gamma-distribution}
\end{figure}

In recent years, the BESIII experiment has leveraged its large $\Lambda_c$ sample to investigate the decay mode $\Lambda_c \to \Lambda \ell \nu_\ell$~\cite{BESIII:2015ysy, BESIII:2016ffj, BESIII:2022ysa, BESIII:2023jxv}, where the measured $q^2$ dependence of the form factors was presented in Refs.~\cite{BESIII:2022ysa,BESIII:2023jxv}, as shown in Fig~\ref{fig:lambdac-form-factor-distribution}. The $q^2$ dependence of the observables, such as the differential decay width and the forward-backward asymmetry, were also measured, as shown in Fig~\ref{fig:lambdac-gamma-distribution}. The most precise branching fractions currently reported are $Br\big(\Lambda_c \to \Lambda e^{+} \nu_e\big) = (3.56 \pm 0.11 \pm 0.07)\%$~\cite{BESIII:2022ysa} and $Br\big(\Lambda_c \to \Lambda \mu^{+} \nu_\mu\big) = (3.48 \pm 0.14 \pm 0.10)\%$~\cite{BESIII:2023jxv}. The average values of the angular asymmetry observables were determined to be $\langle A_{\mathrm{FB}}^e\rangle=-0.24 \pm 0.03_{\text {stat. }} \pm 0.01_{\text {syst }}$ and $\langle A_{\mathrm{FB}}^\mu\rangle=-0.22 \pm 0.04_{\text {stat. }} \pm 0.01_{\text {syst }}$, respectively. These experimental measurements are consistent with the results from QCD sum rules and Lattice QCD calculations, as summarized in Table~\ref{table:charmbaryon-decay}.

Comparing the electron-mode result with the inclusive measurement $Br\big(\Lambda_c \to X e^{+} \nu_e\big) = (3.95 \pm 0.34 \pm 0.09)\%$~\cite{BESIII:2018mug}, one infers the presence of additional semileptonic decay channels that remain unobserved. For instance, in 2022, the BESIII Collaboration reported for the first time two decay channels with excited $\Lambda$ states in the final state, namely $\Lambda_c \to \Lambda(1520) e^{+} \nu_e$ and $\Lambda_c \to \Lambda(1405) e^{+} \nu_e$~\cite{BESIII:2022qaf}. The branching fractions of these two modes are relatively small, measured to be $(1.02 \pm 0.52 \pm 0.11) \times 10^{-3}$ and $(0.42 \pm 0.19 \pm 0.04) \times 10^{-3}$, respectively. In addition, evidence for two five-body semileptonic decays, $\Lambda_c \to \Lambda \pi^+ \pi^- e^+ \nu_e$ and $\Lambda_c \to p K_s^0 \pi^- e^+ \nu_e$, has also been observed~\cite{BESIII:2023jem}, with upper limits on the branching fractions of $Br(\Lambda_c \to \Lambda \pi^+ \pi^- e^+ \nu_e) < 3.9 \times 10^{-4}$ and $Br(\Lambda_c \to p K_s^0 \pi^- e^+ \nu_e) < 3.3 \times 10^{-4}$. These decays, involving $\Lambda$ baryons or their excited states in the final state, correspond at the quark level to the weak transition $c \to s$. 

Moreover, as mentioned earlier, the standard model also allows for $c \to d$ transitions, corresponding at the hadronic level to the semileptonic decay $\Lambda_c \to n \ell \nu_\ell$, where a neutron appears in the final state. This decay mode had not been detected for a long time due to two main reasons: (1) it is driven by the Cabibbo-suppressed $c \to d \ell \nu_\ell$ transition, so its decay width is expected to be much smaller than that of the Cabibbo-favored process $c \to s \ell \nu_\ell$; and (2) the final state contains two neutral particles, making it experimentally challenging to distinguish the neutron signal from neutral backgrounds. Hopefully, with the improvement of detector performance and data analysis techniques, the BESIII Collaboration has made significant progress in measuring decay processes with neutrons in the final state~\cite{BESIII:2022onh, BESIII:2022xne, BESIII:2016yrc, BESIII:2022bkj}.

In Ref.~\cite{Zhang:2023nxl}, the authors performed a QCD sum rule analysis of the semileptonic decay $\Lambda_c \to n \ell \nu_\ell$. The value of the form factors at maximum recoil point $q^2=0$ are presented in Table~\ref{table:lambdac-f0}. Different theoretical models give generally consistent predictions for $f_1(0)$, $f_2(0)$, and $g_1(0)$, while significant discrepancies appear in the predictions for $g_2(0)$. In addition, the results obtained from QCD sum rules and light-cone sum rules show a high degree of consistency. 
\begin{table}[ht]
\centering
\caption{Theoretical predictions for the form factors of the semileptonic decay $ \Lambda_c\to n \ell \nu_\ell $ at the maximum recoil point $ q^2=0 $ with different approaches.}
\renewcommand{\arraystretch}{1.3}
\resizebox{\textwidth}{!}{
\begin{tabular}{lccccccc}
\hline\hline
Method & $ f_1(0) $ & & $ f_2(0) $ & & $ g_1(0) $ & & $ g_2(0) $ \\ \hline
QCD sum rules~\cite{Zhang:2023nxl} & $0.53 \pm 0.04$ & & $-0.25\pm 0.03$ & & $0.53 \pm 0.04$ & & $-0.25\pm 0.03$ \\
Light-cone sum rules~\cite{Khodjamirian:2011jp} & $0.59^{+0.15}_{-0.16}$ & & $-0.43^{+0.13}_{-0.12}$ & & $0.55^{+0.14}_{-0.15}$ & & $-0.16^{+0.08}_{-0.05}$ \\
Light front quark model~\cite{Zhao:2018zcb} & 0.513 & & $-0.266$ & & 0.443 & & $-0.034$ \\
Covariant confined quark model~\cite{Gutsche:2014zna} & 0.47 & & $-0.246$ & & 0.414 & & 0.073 \\
Relativistic quark model~\cite{Faustov:2016yza} & 0.627 & & $-0.259$ & & 0.433 & & 0.118 \\
Bag model~\cite{Geng:2020fng} & 0.40 & & $-0.22$ & & 0.43 & & 0.07 \\
Lattice QCD~\cite{Meinel:2017ggx} & $0.672\pm 0.039$ & & $-0.321\pm 0.038$ & & $0.602\pm 0.031$ & & $0.003 \pm 0.052$ \\
\hline
\end{tabular}
}
\label{table:lambdac-f0}
\end{table}

\begin{figure}[ht]
\centering
\includegraphics[width=5.8cm]{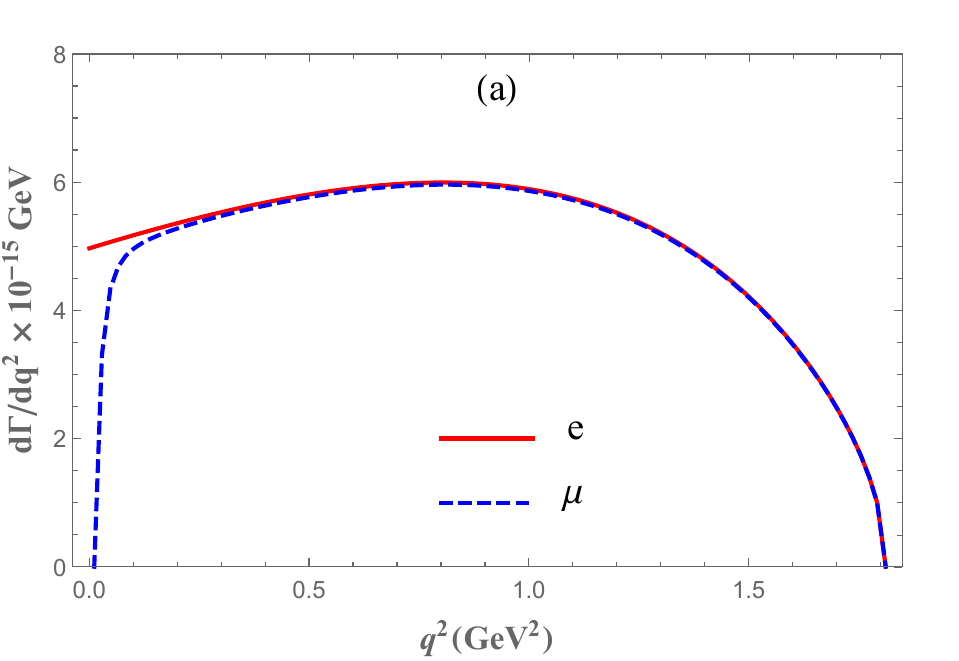}
\includegraphics[width=6cm]{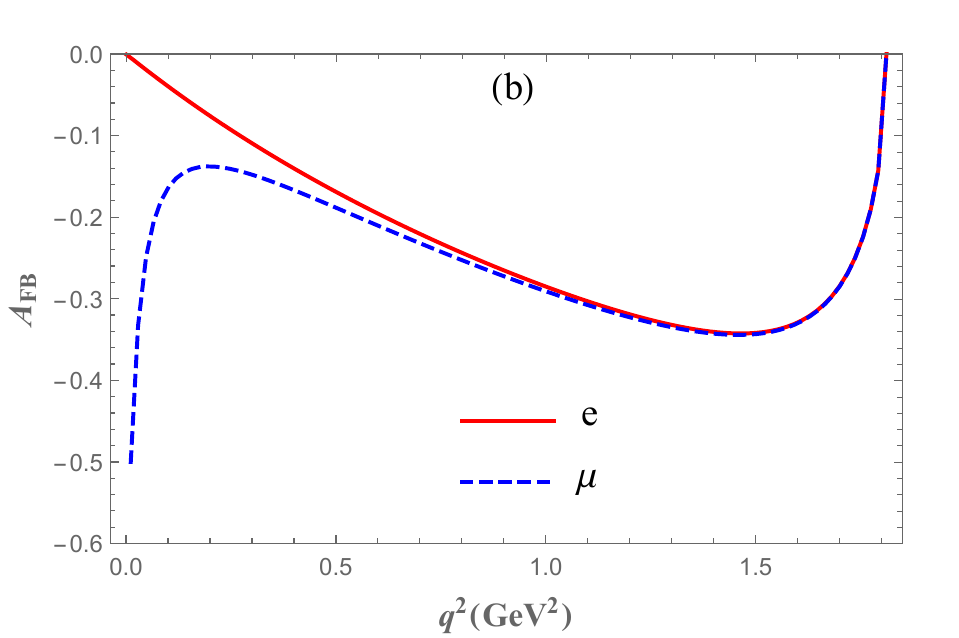}
\includegraphics[width=6cm]{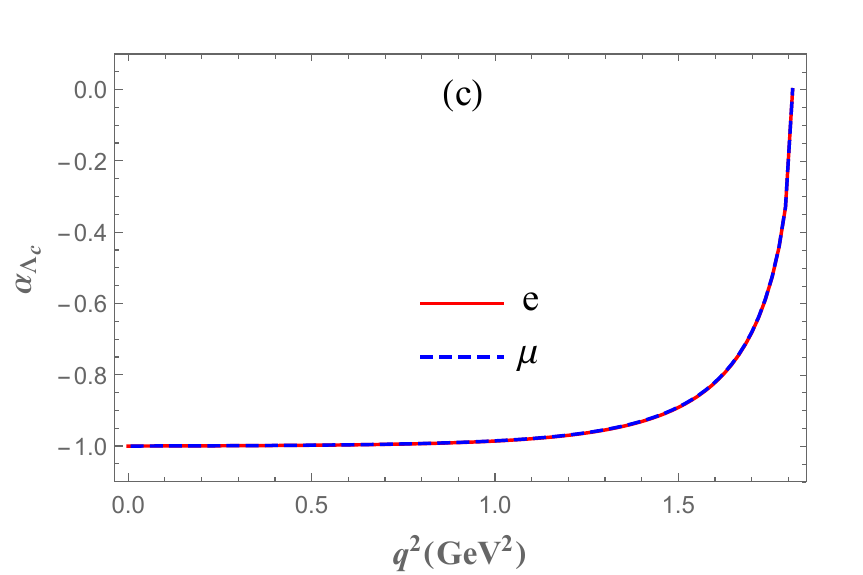}
\caption{The $ q^2 $ dependence of the differential decay width and the relevant decay observables for $ \Lambda_c \to n \ell \nu_\ell$ semileptonic decay taken from Ref.~\cite{Zhang:2023nxl}. The red solid line denotes $ \ell=e^+ $, while the blue dashed line denotes $ \ell=\mu^+ $.}
\label{fig:lambdac-width}
\end{figure}

The $q^2$ dependence of the differential decay width, the forward-backward asymmetry, and the polarization asymmetry are shown in Fig.~\ref{fig:lambdac-width}.
As shown in the figure, the differential decay widths and asymmetry parameters for the electron and muon final states exhibit similar behavior near the zero recoil point, $q^2 = (M_{\Lambda_c} - M_n)^2$. However, in the low-$q^2$ region, the lepton forward-backward asymmetry shows significant differences between the two modes. Specifically, for the electron final state, the asymmetry approaches zero as $q^2 \to m_e^2$, while for the muon final state, it tends toward $-0.5$ as $q^2 \to m_\mu^2$. Regarding the polarization parameter, $\alpha_{\Lambda_c}$ increases from $-1$ to $0$ as $q^2$ varies from zero to the maximum recoil point $q^2_{\text{max}}$. Nevertheless, across the entire physical region, it is nearly impossible to distinguish between the electron and muon final states based on this observable.

The functional behavior of the decay observables can be understood from the asymptotic properties of the kinematic variable $q^2$. As $q^2 \to (M_{\Lambda_c} - M_n)^2$, the contributions of lepton mass terms in the expressions for the differential decay width~\eqref{differential-decay-width}, the lepton forward-backward asymmetry~\eqref{AFB}, and the polarization parameter ~\eqref{PB} are suppressed by a factor of $m_\ell^2/q^2$. Consequently, as $q^2$ increases, the influence of the lepton mass on the observables becomes increasingly negligible, leading to the convergence of the behaviors of the electron and muon modes. Moreover, in this limit, the kinematic factor $Q_{-} = (M_{\Lambda_c} - M_n)^2 - q^2$ approaches zero, and thus:
\begin{align}
\frac{d\Gamma}{dq^2}&\sim \sqrt{Q_{-}}=0\,,\nonumber\\[5pt]
A_{FB}&\sim H_{\frac{1}{2}, 1}^2-H_{-\frac{1}{2}, -1}^2=-4H_{\frac{1}{2}, 1}^VH_{\frac{1}{2}, 1}^A\sim \sqrt{Q_{-}}=0\,,\nonumber\\[5pt]
\alpha_{\Lambda_c}&\sim H_{\frac{1}{2}, 1}^2-H_{-\frac{1}{2}, -1}^2+H_{\frac{1}{2}, 0}^2-H_{-\frac{1}{2}, 0}^2\nonumber\\[5pt]
&=-4\big(H_{\frac{1}{2}, 1}^VH_{\frac{1}{2}, 1}^A+H_{\frac{1}{2}, 0}^VH_{\frac{1}{2}, 0}^A\big)\sim \sqrt{Q_{-}}=0\,.
\end{align}
Therefore, all three observables vanish at $q^2 = (M_{\Lambda_c} - M_n)^2$. In the opposite limit, as $q^2 \to m_\ell^2$, it follows directly from the definition of the differential decay width \eqref{differential-decay-width} that $d\Gamma/dq^2 \to 0$, while explicit calculations yield $A_{FB} \to -0.5$ and $\alpha_{\Lambda_c} \to -1$. For the electron final state, the electron mass squared is negligible compared to the typical scale of $q^2$ near $q^2 \approx m_e^2$, resulting in a sharp but practically indistinguishable drop in the differential decay width and the forward-backward asymmetry. The branching fractions and the average of the asymmetry observables in $ \Lambda_c\rightarrow n \ell \nu_\ell $ semileptonic decays are summarized in Table~\ref{table:charmbaryon-decay}, where results from other theoretical models are also shown for comparison. In 2024, the BESIII collaboration applied a graph neural network approach to analyze the decay process $\Lambda_c \to n e^+ \nu_e$, obtaining an absolute branching fraction of $0.357 \pm 0.034 \pm 0.14$~\cite{BESIII:2024mgg}. This measurement is consistent with the prediction based on the QCD sum rule approach within one standard deviation, thereby validating the reliability of QCD sum rules in the study of hadronic decays.
\begin{table}[ht]
\centering
\caption{Theoretical predictions of branching fractions, the forward-backward asymmetry, and the asymmetry parameter for $ \Lambda_c\rightarrow n \ell \nu_\ell $ and $ \Lambda_c\rightarrow \Lambda \ell \nu_\ell $ semileptonic decay with different methods.}
\renewcommand{\arraystretch}{1.3}
\resizebox{\textwidth}{!}{
\begin{tabular}{llcccc}
\hline\hline
Channel & Method & & $ Br(\%) $ & $\langle A_{FB} \rangle$ & $\langle \alpha \rangle$ \\ \hline
\multicolumn{1}{l}{\multirow{1}{*}{$\Lambda_c\rightarrow \Lambda e^+ \nu_e$}} & QCD sum rules~\cite{Zhang:2023nxl} & & $ 3.49\pm 0.65 $ & $ -0.20\pm 0.01 $ & $ -0.90\pm 0.03 $ \\
%\multicolumn{1}{c}{} & Light-cone sum rules~\cite{Liu:2009sn} & & $ 3.0\pm 0.3 $ \\
\multicolumn{1}{c}{} & Relativistic quark model~\cite{Faustov:2016yza} & & $ 3.25 $& $ -0.209 $ & $ -0.86 $ \\
\multicolumn{1}{c}{} & Light-front quark model~\cite{Li:2021qod} & & $ 4.04\pm 0.75 $& $ -0.20\pm 0.05 $ & $ -0.87\pm 0.09 $ \\
\multicolumn{1}{c}{} & Lattice QCD~\cite{Meinel:2016dqj} & & $ 3.80\pm 0.22 $ & $ -0.20\pm 0.06 $ & $ -0.87\pm 0.10 $ \\
\multicolumn{1}{c}{} & Exp~\cite{BESIII:2022ysa, BESIII:2023jxv} & & $ 3.56\pm 0.11\pm 0.07 $ & $ -0.24\pm 0.03 $ & $ -0.86\pm 0.04 $ \\
\hline

\multicolumn{1}{l}{\multirow{1}{*}{$\Lambda_c\rightarrow \Lambda \mu^+ \nu_\mu$}} & QCD sum rules~\cite{Zhang:2023nxl} & & $ 3.37\pm 0.54 $ & $ -0.24\pm 0.01 $ & $ -0.90\pm 0.02 $ \\
\multicolumn{1}{c}{} & Relativistic quark model~\cite{Faustov:2016yza} & & $ 3.14 $& $ -0.242 $ & $ -0.86 $ \\
\multicolumn{1}{c}{} & Light-front quark model~\cite{Li:2021qod} & & $ 3.90\pm 0.73 $& $ -0.16\pm 0.04 $ & $ -0.87\pm 0.09 $ \\
\multicolumn{1}{c}{}& Lattice QCD~\cite{Meinel:2016dqj} & & $ 3.69\pm 0.22 $ & $ -0.17\pm 0.07 $ & $ -0.87\pm 0.10 $ \\
\multicolumn{1}{c}{} &Exp~\cite{BESIII:2022ysa, BESIII:2023jxv}& & $ 3.48\pm 0.17 $ & $ -0.22\pm 0.04 $ & $ -0.94\pm 0.08 $ \\
\hline

\multicolumn{1}{l}{\multirow{1}{*}{$\Lambda_c\rightarrow n e^+ \nu_e$}} & QCD sum rules~\cite{Zhang:2023nxl} & & $ 0.281\pm 0.056 $ & $ -0.23\pm 0.01 $ & $ -0.93\pm 0.03 $ \\
& Covariant confined quark model~\cite{Gutsche:2014zna} & & $ 0.207 $ & $-0.236$ & \\
\multicolumn{1}{c}{} & Relativistic quark model~\cite{Faustov:2016yza} & & $ 0.268 $ & $-0.251$ & $ -0.91 $ \\
\multicolumn{1}{c}{} & $ SU(3) $~\cite{Geng:2019bfz} & & $ 0.51\pm 0.04 $ & & $ -0.89\pm 0.04 $ \\
\multicolumn{1}{c}{} & MIT bag model~\cite{Geng:2020fng} & & $ 0.279 $ & & $ -0.87 $ \\
\multicolumn{1}{c}{} & Light front constituent quark model~\cite{Geng:2020fng} & & $ 0.36\pm 0.15 $ & & $ -0.96\pm 0.04 $ \\
\multicolumn{1}{c}{} & Lattice QCD~\cite{Meinel:2017ggx} & & $ 0.410\pm 0.026 $ & & \\
\multicolumn{1}{c}{} & Exp~\cite{BESIII:2024mgg} & & $ 0.357\pm 0.034\pm 0.014 $ \\
\hline

\multicolumn{1}{l}{\multirow{1}{*}{$\Lambda_c\rightarrow n \mu^+ \nu_\mu$}} & QCD sum rules~\cite{Zhang:2023nxl} & & $ 0.275\pm 0.055 $ & $ -0.25\pm 0.02 $ & $ -0.93\pm 0.03 $ \\
& Covariant confined quark model~\cite{Gutsche:2014zna} & & $ 0.202$ & $-0.260$ & \\
\multicolumn{1}{c}{} & Relativistic quark model~\cite{Faustov:2016yza} & & $ 0.262$& $-0.276$ & $ -0.90 $ \\
\multicolumn{1}{c}{} & Lattice QCD~\cite{Meinel:2017ggx} & & $ 0.400\pm 0.026 $ & & \\
\hline

\multicolumn{1}{l}{\multirow{1}{*}{$\Xi_c^0 \rightarrow \Xi^{-} e^{+} \nu_e$}} & QCD sum rules~\cite{Zhao:2021sje} &&$1.83 \pm 0.45$ \\
\multicolumn{1}{c}{} & Lattice QCD~\cite{Zhang:2021oja} & & $2.38 \pm 0.30 \pm 0.33$ \\
\multicolumn{1}{c}{} & Exp~\cite{Belle:2021crz} & & $1.31 \pm 0.04 \pm 0.07 \pm 0.38$ \\
\hline

\multicolumn{1}{l}{\multirow{1}{*}{$\Xi_c^0 \rightarrow \Xi^{-} \mu^{+} \nu_\mu$}} & QCD sum rules~\cite{Zhao:2021sje} &&$1.77 \pm 0.43$ \\
\multicolumn{1}{c}{} & Lattice QCD~\cite{Zhang:2021oja} & & $2.29 \pm 0.29 \pm 0.31$ \\
\multicolumn{1}{c}{} & Exp~\cite{Belle:2021crz} & & $1.27 \pm 0.06 \pm 0.10 \pm 0.37$ \\
\hline

\end{tabular}
}
\label{table:charmbaryon-decay}
\end{table}

\begin{figure}[ht]
\centering
\includegraphics[width=0.6\textwidth]{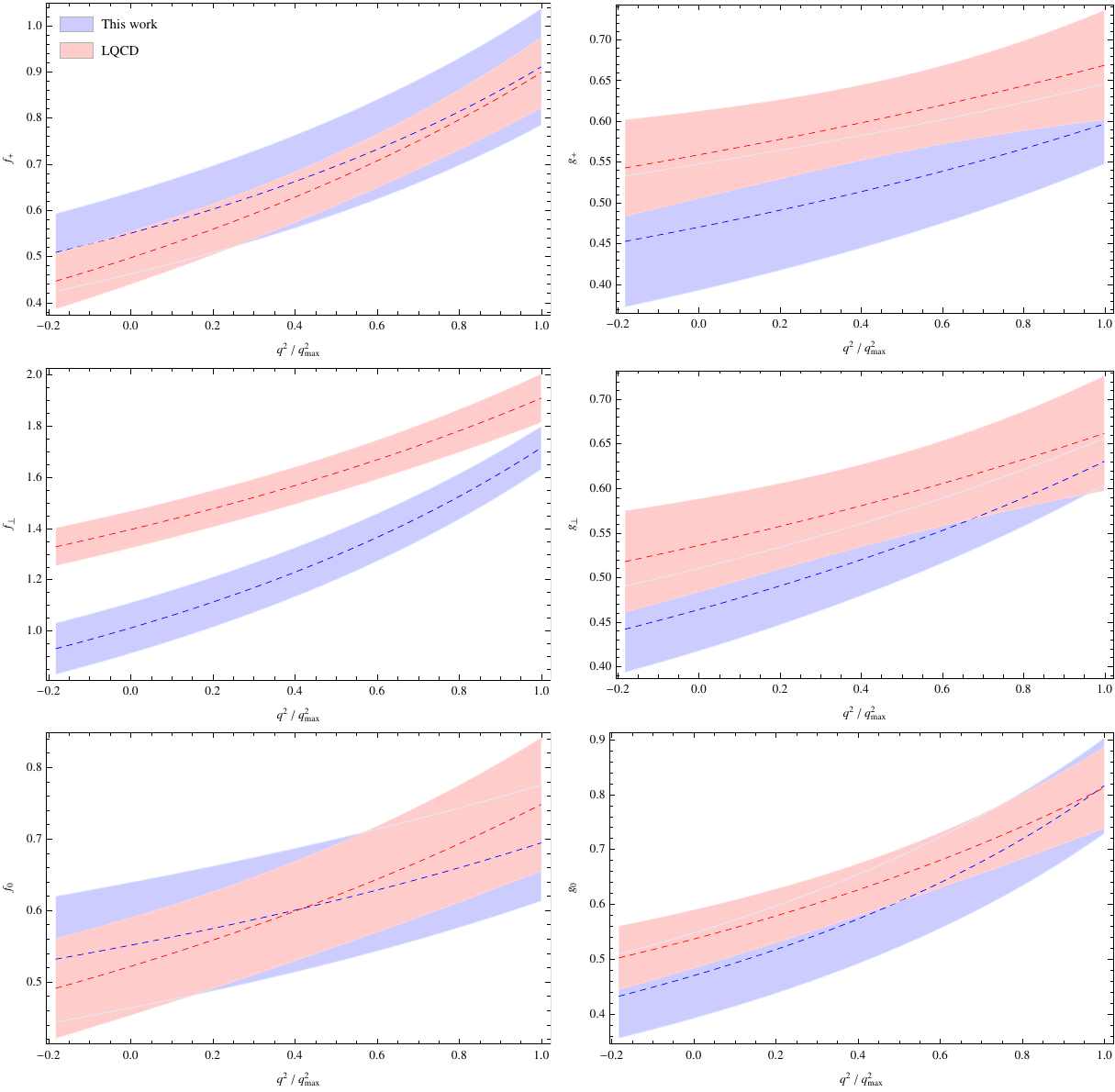}
\caption{Helicity form factors obtained by QCD sum rules~\cite{Zhao:2021sje} and Lattice QCD~\cite{Zhang:2021oja} for the semileptonic decay $\Xi_c^0 \rightarrow \Xi^{-} \ell^{+} \nu_\ell$. The purple bands denote the results from QCD sum rules, while the red bands denote those from Lattice QCD.}
\label{fig:form-factors-comparison}
\end{figure}

Motivated by the observed discrepancy between the experimental measurement and the Lattice QCD prediction of the branching fraction for the semileptonic decay $\Xi_c^0 \rightarrow \Xi^{-} \ell^{+} \nu_\ell$, Zhao et al. conducted a QCD sum rule analysis of the semileptonic decays of another class of charmed baryons, $\Xi_c$~\cite{Zhao:2021sje}. The $q^2$ dependence of the helicity form factors obtained by QCD sum rules is consistent with those of Lattice QCD within error except for $f_{\perp}$, as shown in Fig~\ref{fig:form-factors-comparison}. The results of branching fractions are summarized in Table~\ref{table:charmbaryon-decay}.

From Table~\ref{table:charmbaryon-decay}, one can see that the theoretical predictions and experimental measurements of the semileptonic branching fractions are generally of the same order of magnitude. In fact, branching fractions alone are insufficient to rigorously test the consistency between theory and experiment. Form factors, which encode more detailed information about the decay dynamics, offer a more sensitive probe. Fig~\ref{fig:lambdac-form-factor-distribution} illustrates the differences in the $\Lambda_c \to \Lambda \ell \nu_\ell$ form factors between Lattice QCD calculations and BESIII measurements. Moreover, discrepancies in the helicity form factor $f_{\perp}$ have been identified as one of the sources of deviation between QCD sum rule and Lattice QCD predictions for the branching fraction of $\Xi_c^0 \to \Xi^{-} \ell^{+} \nu_\ell$~\cite{Zhao:2021sje}. Thus, precise measurements of form factors in future experiments are crucial for robust tests of theoretical models.

QCD sum rule analyses of semileptonic decays have also been extended to doubly heavy baryons~\cite{Shi:2019hbf,Xing:2021enr,Tousi:2024usi,ShekariTousi:2025fjf,Yu:2026tbk} and triply heavy baryons~\cite{Najjar:2024ngm}, providing valuable theoretical predictions for their decay form factors and branching fractions. These studies offer insight into the internal dynamics of multiquark systems and test the applicability of nonperturbative methods in the heavy quark sector. However, due to the extremely low production rates of such baryons at current colliders, there are no experimental measurements available to date. The lack of data poses a significant challenge for validating theoretical approaches, and highlights the need for future high-luminosity experiments capable of exploring this largely uncharted territory.

\subsection{Light baryon decays}\

The semileptonic decays of light baryons are mainly driven by the strange and down quarks, such as the Cabibbo-suppressed process $s\to u$ and $d\to u$. These decays are crucial for understanding the weak interactions of light quarks, as well as the determination of CKM matrix elements $V_{us}$ and $V_{ud}$. Unlike the heavy sector where the bottom and charm quarks are much heavier than the light quarks, the flavor $SU(3)$ symmetry is well preserved in the light hadron sector. In 1963, Cabibbo made important theoretical predictions for the form factors in hyperon semileptonic decays based on flavor $SU(3)$ symmetry~\cite{Cabibbo:1963yz}. These predictions make the measurement of form factors in hyperon decays a key probe for testing flavor symmetry and its breaking mechanisms. Based on the Cabibbo model, the form factors governing the semileptonic decays of spin-$\frac{1}{2}$ light baryon octet members acquire definite values under exact flavor $SU(3)$ symmetry, as shown in Table~\ref{table:SU3}. Details of hyperon semileptonic decays can be found in two well-known review articles~\cite{Gaillard:1984ny, Cabibbo:2003cu}.

\begin{table}[ht]
\centering
\caption{Form factors of hyperon semileptonic decays predicted by the Cabibbo theory. $ \mu_p $ and $ \mu_n $ stand for the anomalous magnetic moments of the proton and neutron, respectively. F and D denote the coupling constants, resulting from Cabibbo’s fundamental assumption that the weak hadronic currents belong to a single self-conjugate representation of $SU(3)$.}
\renewcommand{\arraystretch}{1.4}
\begin{tabular}{lcccc}
\hline\hline
& $f_1(0)$ & $g_1(0)$ & $g_1(0)/f_1(0)$ & $f_2(0)/f_1(0)$ \\
\hline
$n\to p$ & $1$& $F+D$ &$F+D$& $\frac{M_{n}}{M_p}\frac{(\mu_p-\mu_n)}2$ \\
$\Sigma^{ \pm} \to \Lambda$& $0$& $\sqrt{\frac{2}{3}}D$ &$\sqrt{\frac{2}{3}}D$& $-\frac{M_{\Sigma^{ \pm}}}{M_p} \sqrt{\frac{3}{2}} \frac{(\mu_n)}{2}$ \\
$\Sigma^{-} \to \Sigma^0$& $\sqrt{2}$& $\sqrt{2}F$ &$F$& $\frac{M_{\Sigma^{-}}}{M_p} \frac{(2 \mu_p+\mu_n)}{4}$ \\
$\Sigma^{0} \to \Sigma^+$& $\sqrt{2}$& $-\sqrt{2}F$ &$-F$& $\frac{M_{\Sigma^{0}}}{M_p} \frac{(2 \mu_p+\mu_n)}{4}$ \\
$\Xi^{-} \to \Xi^0$& $-1$& $D-F$ &$F-D$&$\frac{M_{\Xi^{-}}}{M_p} \frac{(\mu_p+2 \mu_n)}{2}$ \\

$\Lambda\to p$ & $-\sqrt{\frac{3}{2}}$& $-\sqrt{\frac{3}{2}}(F+\frac{D}{3})$ &$F+\frac{D}{3}$& $\frac{M_\Lambda}{M_p}\frac{\mu_p}2$\\
$\Sigma^{0}\to p$ & $-\sqrt{\frac{1}{2}}$& $-\sqrt{\frac{1}{2}}(F-D)$ &$F-D$& $\frac{M_{\Sigma^0}}{M_p}\frac{(\mu_p+2\mu_n)}2$ \\
$\Sigma^{-}\to n$ & $-1$& $-(F-D)$ &$F-D$& $\frac{M_{\Sigma^-}}{M_p}\frac{(\mu_p+2\mu_n)}2$ \\
$\Xi^{-}\to \Lambda$ & $\sqrt{\frac{3}{2}}$& $\sqrt{\frac{3}{2}}(F-\frac{D}{3})$ &$F-\frac{D}{3}$& $-\frac{M_{\Xi^-}}{M_p}\frac{(\mu_p+\mu_n)}2$ \\
$\Xi^{-}\to \Sigma^{0}$ & $\sqrt{\frac{1}{2}}$& $\sqrt{\frac{1}{2}}(F+D)$ &$F+D$& $\frac{M_{\Xi^-}}{M_p}\frac{(\mu_p-\mu_n)}2$ \\
$\Xi^{0}\to \Sigma^{+}$ & $1$& $F+D$ &$F+D$& $\frac{M_{\Xi^0}}{M_p}\frac{(\mu_p-\mu_n)}2$ \\

\hline\hline
\end{tabular}
\label{table:SU3}
\end{table}

Flavor $SU(3)$ symmetry is only an approximate symmetry under the strong interaction. In the exact $SU(3)$ symmetry limit, mesons or baryons belonging to the same flavor multiplet are expected to have identical masses. However, experimental data reveal significant mass splittings among different isospin multiplets within the same hadron multiplet~\cite{ParticleDataGroup:2024cfk}, and the constituent quark masses of the fundamental representation, $u$, $d$, and $s$, also differ substantially. These observations indicate that flavor $SU(3)$ symmetry is not exact, but rather broken to a certain extent.

The mass splittings within the spin-$\frac{1}{2}$ light baryon octet can be described by the Gell-Mann-Okubo formula~\cite{Okubo:1961jc, Gell-Mann:1968hlm}, which originates from the breaking of flavor $SU(3)$ symmetry. In semileptonic decay processes, the effects of symmetry breaking are manifested in the behavior of the form factors. For example, in the decays $n \to p e \bar{\nu}_e$ and $\Xi^0 \to \Sigma^+ e \bar{\nu}_e$, the ratio of the axial-vector to vector form factors, $g_1(0)/f_1(0)$, is measured to be $1.2754 \pm 0.0013$ and $1.22 \pm 0.05$, respectively~\cite{ParticleDataGroup:2022pth, KTeV:2001djr, NA48I:2006yat}. According to Table~\ref{table:SU3}, these two decay modes are expected to yield identical values for this ratio under exact flavor $SU(3)$ symmetry. The observed discrepancy in the experimental values thus may indicate a degree of symmetry breaking. Of course, gaining deeper insight into the origin of this $SU(3)$ breaking requires more precise experimental measurements.

Current determinations of $V_{us}$ and $V_{ud}$ heavily depends on pion and kaon decays~\cite{ParticleDataGroup:2024cfk}. In 2023, the BESIII Collaboration reported the latest measurement of the decay $ \Lambda \to p \mu^- \bar{\nu}_{\mu} $, with the absolute branching fraction determined to be $Br(\Lambda \to p \mu^- \bar{\nu}_\mu) = (1.48 \pm 0.21 \pm 0.08) \times 10^{-4}$~\cite{BESIII:2021ynj}. This result, together with earlier studies, opens the possibility for a more precise extraction of $|V_{us}|$ and provides a valuable basis for testing the quark flavor mixing mechanism within the hyperon sector.

From a theoretical perspective, a model-independent evaluation of $SU(3)$ breaking effects plays a crucial role in precision tests of the unitarity of the CKM matrix~\cite{Gaillard:1984ny, Cabibbo:2003cu}. For such a test, chiral perturbation theory ($\chi \mathrm{PT}$)~\cite{Kaiser:2001yc, Villadoro:2006nj,Geng:2009ik,Ledwig:2014rfa} and Lattice QCD~\cite{Guadagnoli:2006gj,Sasaki:2008ha,Sasaki:2012ne,Sasaki:2017jue,Cooke:2012xv,Cooke:2013qqa} have historically played an important role in advancing the understanding of $SU(3)$-breaking effects. In Refs.~\cite{Geng:2009ik,Ledwig:2014rfa}, Geng et al. applied covariant baryon chiral perturbation theory to calculate the hyperon vector coupling $f_1(0)$ and the ratios $g_1 / f_1$ in hyperon semileptonic decays, providing a systematic framework to quantify the $SU(3)$ breaking effects. The first quenched Lattice QCD investigation of the form factors in $\Sigma^- \to n\ell\nu_\ell$ semileptonic decays was carried out in Ref.~\cite{Guadagnoli:2006gj}. Building on this approach, Sasaki and Yamazaki later also performed a quenched Lattice calculation for the $\Xi^0 \to \Sigma^+\ell\nu_\ell$ decay, extracting the relevant form factors~\cite{Sasaki:2008ha}. Full Lattice QCD studies with (2+1)-flavors of dynamical domain-wall fermions for $\Xi^0 \to \Sigma^+\ell\nu_\ell$ and $\Sigma^- \to n \ell \nu_\ell$ were also conducted in Refs.~\cite{Sasaki:2012ne,Sasaki:2017jue}. In contrast, no Lattice QCD results are currently available for the $\Lambda \to p\ell\nu_\ell$ and $\Xi^- \to \Lambda \ell \nu_\ell$ decays. Other theoretical approaches on the analysis of $SU(3)$ breaking effects in hyperon semileptonic decays include the quark model~\cite{Donoghue:1986th, Schlumpf:1994fb, Faessler:2008ix, Migura:2006en}, the soliton model~\cite{Yang:2015era, Ledwig:2008ku}, $1/N_c$ and large $N_c$ expansion~\cite{Flores-Mendieta:1998tfv, Flores-Mendieta:2004cyh} and $SU(3)$ Skyrme model~\cite{Park:1989nz, Kondo:1991fc}.

The QCD sum rule approach discussed in this review can also be directly applied to determine the form factors of hyperon semileptonic decays and to analyze the effects of flavor $SU(3)$ symmetry breaking. As emphasized in Ref.~\cite{Ball:1991bs}, the underlying physical assumptions of QCD sum rules are more closely aligned with the framework of quantum field theory, which may resonate with the argument of ``model-independent'' methods as discussed in Ref.~\cite{Cabibbo:2003cu}.

In Ref.~\cite{Zhang:2024ick}, the authors calculated the form factors for $|\Delta S| = 1$ hyperon semileptonic decays using QCD sum rules. In total, there are six relevant decay channels: $\Lambda \to p$, $\Sigma^0 \to p$, $\Xi^- \to \Lambda$, $\Xi^- \to \Sigma^0$, $\Xi^0 \to \Sigma^+$, and $\Sigma^- \to n$. Among them, the first four processes were explicitly analyzed, while the remaining two can be inferred via isospin symmetry relations~\cite{Yang:2015era, Ledwig:2008ku}. The results of the form factors are presented in Table~\ref{table:hyperon-ratios}.

\begin{table}[ht]
\centering
\caption{Form factors in hyperon semileptonic decays, taken from Ref.~\cite{Zhang:2024ick}.}
\renewcommand{\arraystretch}{1.3}
\resizebox{0.9\textwidth}{!}{
\begin{tabular}{lcccc}
\hline\hline
& $\Lambda\rightarrow N$ & $\Sigma\rightarrow N$ & $\Xi\rightarrow \Lambda$ & $\Xi\rightarrow \Sigma$ \\
\hline
$ f_1(0)/f_1^{SU(3)} $ \\
QCD sum rules~\cite{Zhang:2024ick} & $0.963\pm 0.061$ & $0.993\pm 0.059$ & $ 0.993\pm 0.061$ & $0.956\pm 0.049$\\
Quark model~\cite{Donoghue:1986th}& $0.987$ & $0.987$ & $0.987$ &$0.987$ \\
Quark model~\cite{Schlumpf:1994fb} & $0.976$ & $0.975$ & $0.976$ &$0.976$ \\
$ \chi $PT~\cite{Villadoro:2006nj} & $1.027$ & $1.041$ & $1.043$ & $ 1.009 $\\
$ \chi $PT~\cite{Geng:2009ik} & $1.001^{+0.013}_{-0.010}$ & $1.087^{+0.042}_{-0.031}$ & $1.040^{+0.028}_{-0.021}$ &$1.017^{+0.022}_{-0.016}$\\
$ 1/N_c $ expansion~\cite{Flores-Mendieta:1998tfv}& $1.02\pm 0.02$ & $1.04\pm 0.02$ & $1.10\pm 0.04$ & $1.12\pm 0.05$\\
Lattice QCD~\cite{Sasaki:2012ne,Bacchio:2025auj} & $0.9674\pm 0.0047$ & $0.957\pm 0.01$ & &$0.976\pm 0.005$\\
\hline
$ g_1(0)/f_1(0) $ \\
QCD sum rules~\cite{Zhang:2024ick} & $0.708\pm 0.047$ & $-0.327\pm 0.046$ & $ 0.271\pm 0.045$ & $1.293\pm 0.100$\\
Cabibbo model~\cite{Cabibbo:2003cu} & $0.731$ & $-0.341$ & $ 0.195$ & $1.267$\\
Quark model~\cite{Faessler:2008ix} & $0.724$ & $-0.260$ & $ 0.265$ & $1.20$\\
Soliton model~\cite{Yang:2015era} & $0.718\pm 0.003$ & $-0.340\pm 0.003$ & $ 0.250\pm 0.002$ & $1.210\pm 0.005$\\
Soliton model~\cite{Ledwig:2008ku} & $0.68$ & $-0.27$ & $ 0.21 $ & $1.16$\\
$ 1/N_c $ expansion~\cite{Flores-Mendieta:1998tfv} & $0.73$ & $-0.34$ & $ 0.22$ & $1.03$\\
Lattice QCD~\cite{Guadagnoli:2006gj, Sasaki:2008ha,Bacchio:2025auj} & $0.6902\pm 0.0044$ & $-0.287\pm 0.052$ & & $1.248\pm 0.029$\\
Exp~\cite{ParticleDataGroup:2022pth}& $0.718\pm 0.015$ & $-0.340\pm 0.017$ & $ 0.25\pm 0.05$ & $1.22\pm 0.05$\\
\hline
$ f_2(0)/f_1(0) $ \\
QCD sum rules~\cite{Zhang:2024ick} & $0.752\pm 0.074$ & $-1.042\pm 0.090$ & $ 0.118\pm 0.032$ & $1.957\pm 0.255$\\
Cabibbo model~\cite{Cabibbo:2003cu} & $1.066$ & $-1.292$ & $ 0.085$ & $2.609$\\
Quark model~\cite{Faessler:2008ix} & $1$ & $-0.962$ & $ 0.129$ & $2.402$\\
Soliton model~\cite{Yang:2015era} & $0.637\pm 0.041$ & $-0.709\pm 0.036$ & $ -0.069\pm 0.027$ & $1.143\pm 0.061$\\
Soliton model~\cite{Ledwig:2008ku} & $0.71$ & $-0.96$ & $ -0.02$ & $2.02$\\
$ 1/N_c $ expansion~\cite{Flores-Mendieta:1998tfv} & $0.90$ & $-1.02$ & $ -0.06$ & $1.85$\\
Lattice QCD~\cite{Sasaki:2008ha,Bacchio:2025auj} & $0.693\pm 0.017$ & $-1.52\pm 0.81$ & & \\
Exp~\cite{ParticleDataGroup:2022pth}& $ 1.32\pm 0.81 $~\cite{Bristol-Geneva-Heidelberg-Orsay-Rutherford-Strasbourg:1983jzt} & $-0.97\pm 0.14$ & $ -0.24\pm 0.25 $~\cite{Bristol-Geneva-Heidelberg-Orsay-Rutherford-Strasbourg:1983jzt} & $2.0\pm 0.9$\\
\hline\hline
\end{tabular}
}
\label{table:hyperon-ratios}
\end{table}

For the calculation of $f_1(0)$, early studies based on quark models~\cite{Donoghue:1986th, Schlumpf:1994fb} found that the ratio $f_1(0)/f_1^{SU(3)}$ exhibits notable similarity across the four categories of hyperon semileptonic decay modes, with $SU(3)$ symmetry-breaking corrections to $f_1$ consistently appearing with a negative sign. This observation has been supported by recent results from Lattice QCD~\cite{Sasaki:2012ne,Bacchio:2025auj} and QCD sum rule calculations (central values)~\cite{Zhang:2024ick}. In contrast, studies based on chiral perturbation theory~\cite{Villadoro:2006nj, Geng:2009ik} and the $1/N_c$ expansion approach~\cite{Flores-Mendieta:1998tfv} generally predicted a positive correction to $f_1(0)$. Resolving this discrepancy remains important for a deeper understanding of symmetry-breaking effects in hyperon semileptonic decays.

According to the Cabibbo model, the ratio $g_1(0)/f_1(0)$ is determined by the coupling constants $F$ and $D$ (see Table~\ref{table:SU3}). Based on experimental data, the fitted values of the couplings are approximately $F \approx 0.463$ and $D \approx 0.804$~\cite{Cabibbo:2003cu}. As shown in Table~\ref{table:hyperon-ratios}, the values of $g_1(0)/f_1(0)$ obtained from different theoretical approaches and experimental measurements are generally consistent within uncertainties. Notably, the $\Xi \to \Sigma$ decay mode occupies a particularly unique position. In the exact $SU(3)$ flavor symmetry limit, exchanging the $d$ and $s$ quarks makes this process fully equivalent to neutron $\beta$ decay. Under such symmetry, the only difference between the decay amplitudes lies in the distinct CKM matrix elements~\cite{Ledwig:2008ku}. Therefore, comparing $\Xi \to \Sigma$ decays with neutron $\beta$ decay provides a sensitive probe of $SU(3)$ symmetry breaking effects~\cite{Sasaki:2008ha}. In Table~\ref{table:hyperon-ratios}, the experimental result for $\Xi \to \Sigma$ is taken from the world average reported in the Particle Data Group~\cite{ParticleDataGroup:2024cfk}. The QCD sum rule result shows better agreement with the measurement reported by the KTeV Collaboration, which gives $g_1(0)/f_1(0) = 1.32^{+0.21}_{-0.17} \pm 0.05$~\cite{KTeV:2001djr}. For comparison, the current best experimental determination of this ratio in neutron $\beta$ decay is $g_1(0)/f_1(0) = 1.2754 \pm 0.0013$~\cite{ParticleDataGroup:2022pth}, which shows slight deviations from both theoretical predictions and measurements for the $\Xi^0 \to \Sigma^+$ process. Furthermore, the isospin-partner process $\Xi^- \to \Sigma^0$ yields $g_1(0)/f_1(0) = 1.25^{+0.14}_{-0.16}$~\cite{Bristol-Geneva-Heidelberg-Orsay-Rutherford-Strasbourg:1983jzt}. A more precise measurement of the form factors in this decay mode would provide valuable insights into isospin symmetry breaking effects.

As for the ratio $f_2(0)/f_1(0)$, the relatively large experimental uncertainties, as shown in Table~\ref{table:hyperon-ratios}, pose challenges for direct comparison with theoretical predictions. It is also worth noting that different theoretical approaches yield significantly different results for this quantity. Therefore, further investigations aimed at determining the weak magnetism form factor $f_2$ with higher precision are essential.

A recent Lattice QCD study~\cite{Bacchio:2025auj} has provided the first determination of $\Lambda \to N$ form factors using gauge ensembles with physical quark masses, where the results are also presented in Table~\ref{table:hyperon-ratios}. This analysis achieves unprecedented precision and reveals a striking consistency with prior QCD sum rule results~\cite{Zhang:2024ick}.

\begin{figure}[ht]
\centering
\includegraphics[width=7.0cm]{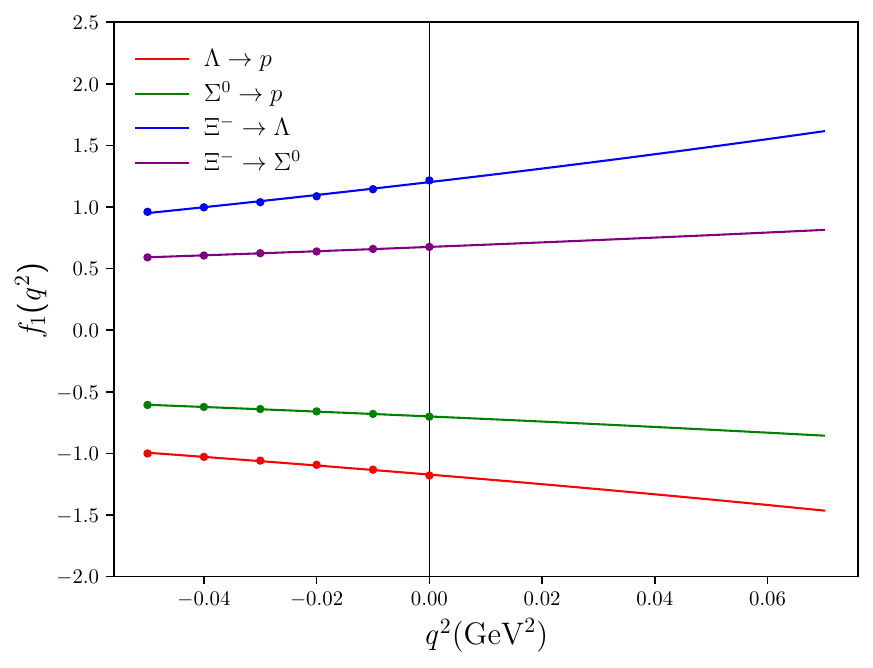}
\includegraphics[width=7.0cm]{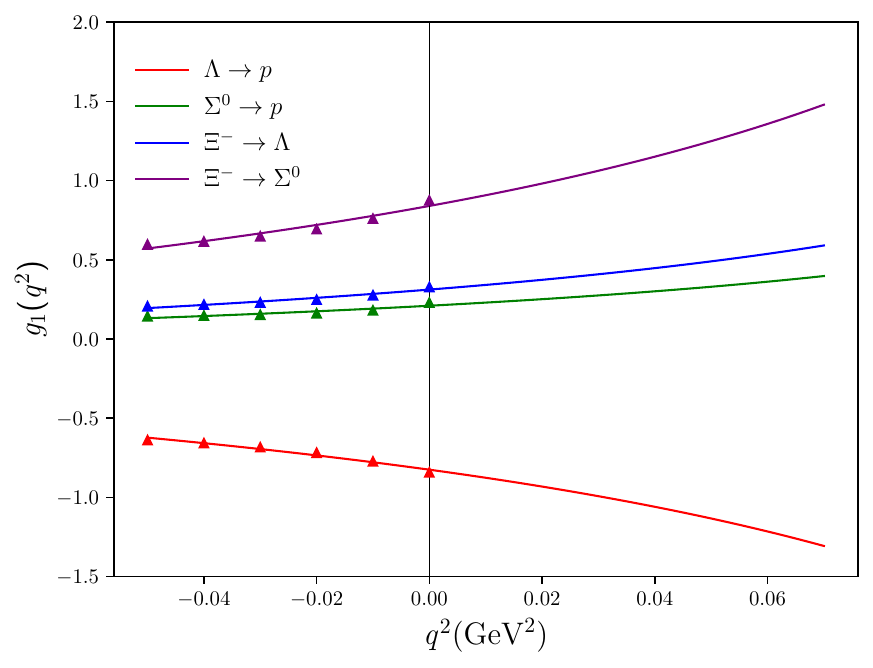}
\caption{The $ q^2 $ dependence of form factors in hyperon semileptonic decays taken from Ref.~\cite{Zhang:2024ick}. The symbol points and the triangles represent the fitted points for each of the form factors.}
\label{fig:hyperon-ft}
\end{figure}

Based on the $q^2$ dependence of the form factors, as shown in Fig~\ref{fig:hyperon-ft}, the QCD sum rule analysis in Ref.~\cite{Zhang:2024ick} also provides results of the branching fractions for hyperon semileptonic decays. The results are summarized in Table~\ref{table:hyperon-br}. The direct branching ratios are listed in the second column of Table~\ref{table:hyperon-br}, while the results in the third column are obtained by neglecting the $q^2$ dependence of the form factors. The $q^2$ dependence has a significant impact on the branching ratios and improves the agreement with experimental data. Specifically, it contributes approximately 10\% to 20\% to the branching ratios. Moreover, compared to the electron final states, the $q^2$ dependence of the form factors typically has a more pronounced effect in the case of muon final states.

\begin{table}[ht]
\centering
\caption{Branching fractions for hyperon semileptonic decays. The superscript 0 in the third column represents the results without $ q^2 $ dependence of the form factors.}
\renewcommand{\arraystretch}{1.4}
\resizebox{0.9\linewidth}{!}{
\begin{tabular}{lccccccc}
\hline\hline
&& & QCD sum rules & & QCD sum rules$ ^0 $ & & Exp~\cite{ParticleDataGroup:2024cfk}\\
\hline
$\mathcal{B}(\Lambda\to p e^- \bar{\nu}_e)\times 10^{-4}$& & & $7.72\pm 0.64$&&$7.12\pm 0.70$ & &$8.34\pm 0.14$ \\
$\mathcal{B}(\Lambda\to p\mu^-\bar{\nu}_\mu)\times 10^{-4}$& & & $1.35\pm 0.11$& & $1.15\pm 0.11$ & &$1.51\pm 0.19$ \\
$\mathcal{B}(\Sigma^{-}\to n e^- \bar{\nu}_e)\times 10^{-3}$& & & $1.00\pm 0.18$& & $0.87\pm 0.14$ & &$1.017\pm 0.034$ \\
$\mathcal{B}(\Sigma^{-}\to n\mu^-\bar{\nu}_\mu)\times 10^{-4}$& & & $4.73\pm 0.88$& & $3.74\pm 0.62$ & &$4.5\pm 0.4$ \\
$\mathcal{B}(\Sigma^{0}\to p e^- \bar{\nu}_e)\times 10^{-13}$& & &$2.50\pm 0.70$ && $2.18\pm 0.58$ && $2.46\pm 0.32$~\cite{Wang:2019alu}\\
$\mathcal{B}(\Sigma^{0}\to p\mu^-\bar{\nu}_\mu)\times 10^{-13}$& & &$1.18\pm 0.34$ && $0.94\pm 0.25$ &&$1.13\pm 0.15$~\cite{Wang:2019alu} \\
$\mathcal{B}(\Xi^{-}\to \Lambda e^- \bar{\nu}_e)\times 10^{-4}$& & & $5.17\pm 0.73$ & & $4.67\pm 0.66$ & & $5.63\pm 0.31$ \\
$\mathcal{B}(\Xi^{-}\to \Lambda\mu^-\bar{\nu}_\mu)\times 10^{-4}$& & & $1.54\pm 0.23$ & & $1.24\pm 0.18$ & & $3.5^{+3.5}_{-2.2}$ \\
$\mathcal{B}(\Xi^{0}\to \Sigma^{+} e^- \bar{\nu}_e)\times 10^{-4}$& & &$2.41\pm 0.28$ & & $2.39\pm 0.32$ & & $2.52\pm 0.08$\\
$\mathcal{B}(\Xi^{0}\to \Sigma^{+}\mu^-\bar{\nu}_\mu)\times 10^{-6}$& & &$2.07\pm 0.27$ && $1.91\pm 0.25$& & $2.33\pm 0.35$\\
$\mathcal{B}(\Xi^{-}\to \Sigma^{0} e^- \bar{\nu}_e)\times 10^{-5}$& & & $7.80\pm 0.77$ & & $7.72\pm 0.86$ & & $8.7\pm 1.7$\\
$\mathcal{B}(\Xi^{-}\to \Sigma^{0}\mu^-\bar{\nu}_\mu)\times 10^{-6}$& & & $1.06\pm 0.12$ & & $0.97\pm 0.11$ & & $ <800$\\
\hline\hline
\end{tabular}
}
\label{table:hyperon-br}
\end{table}

Due to the absence of experimental data for the decay process $\Sigma^0 \to p$, the result listed in the fourth column of Table~\ref{table:hyperon-br} is derived based on $SU(3)$ flavor symmetry~\cite{Wang:2019alu}. It is important to emphasize that most of the experimental data extracted from the particle data group are relatively dated and subject to sizable uncertainties. Many of these values are not direct measurements but are instead inferred from corresponding two-body hadronic decays~\cite{ParticleDataGroup:2024cfk}. Moreover, events with muon final states are extremely rare, especially for the transitions $\Xi^{-} \to \Lambda \mu^- \bar{\nu}_\mu$ and $\Xi^{-} \to \Sigma^0 \mu^- \bar{\nu}_\mu$, where the former exhibits considerable experimental uncertainty, and the latter has only an upper limit reported for its branching ratio.

Due to the small branching fractions of hyperon semileptonic decays~\cite{ParticleDataGroup:2024cfk} and the significant challenge of extracting semileptonic signals from dominant hadronic two-body decay backgrounds~\cite{Cabibbo:2003cu}, experimental studies of these processes remain highly demanding. Therefore, further investigations into the mechanisms of flavor $SU(3)$ symmetry breaking, both experimentally and theoretically, continue to be of great importance. Encouragingly, the proposed Super Tau-Charm Factory (STCF) may offer a breakthrough in hyperon physics. STCF is expected to produce large-statistics samples of hyperon-antihyperon pairs and enable high-precision measurements of time-like form factors. These capabilities would allow for a detailed examination of symmetry-breaking effects~\cite{Achasov:2023gey}. Such experimental advancements will not only sharpen tests of the standard model but may also provide crucial input for the determination of CKM matrix elements and the search for possible new physics contributions.

\subsection{Rare decays}\

Another kind of special semileptonic decay is induced by the neutral current interactions, which are called flavor-changing neutral current (FCNC) processes. The Cabibbo model describes only charged current processes and does not include FCNC processes. This leads to a contradiction with experimental observations, such as the significantly suppressed decay mode $K_L \to \mu^+ \mu^-$, which cannot be explained within the Cabibbo framework. To resolve this issue, Glashow, Iliopoulos, and Maiani proposed the GIM mechanism in 1970~\cite{Glashow:1970gm}, predicting the existence of the charm quark. By introducing the charm quark as a weak isospin partner of the strange quark in the second quark doublet, the GIM mechanism naturally incorporates neutral current couplings and successfully addresses the shortcomings of the Cabibbo model. 

The introduction of the GIM mechanism extended weak interactions to include transitions between different generations of quarks, and demonstrated that FCNC processes are strictly forbidden at tree level. Such processes can only occur via loop diagrams, which results in strong suppression. This naturally explains why FCNC processes observed experimentally generally exhibit very small branching fractions. Consequently, FCNC processes are often referred to as rare decays. Additionally, rare decays may interfere with contributions from non-standard model particles, providing a unique window into the possible existence of yet-unobserved particles such as supersymmetric partners or dark matter candidates~\cite{Hewett:1996ct, Buchalla:2000sk, Bird:2004ts, Davidson:1993qk}.

\begin{figure}[ht]
\centering 
\includegraphics[width=0.4\textwidth]{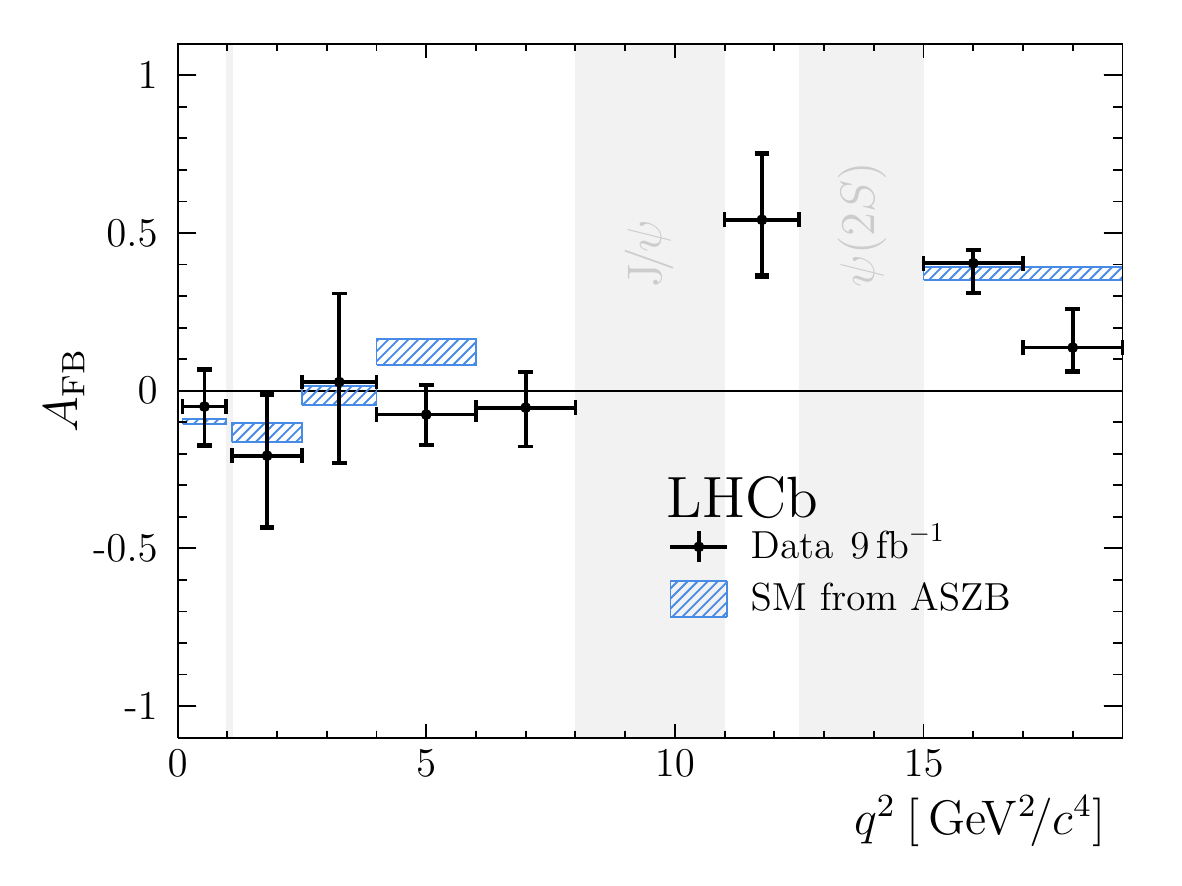}
\includegraphics[width=0.4\textwidth]{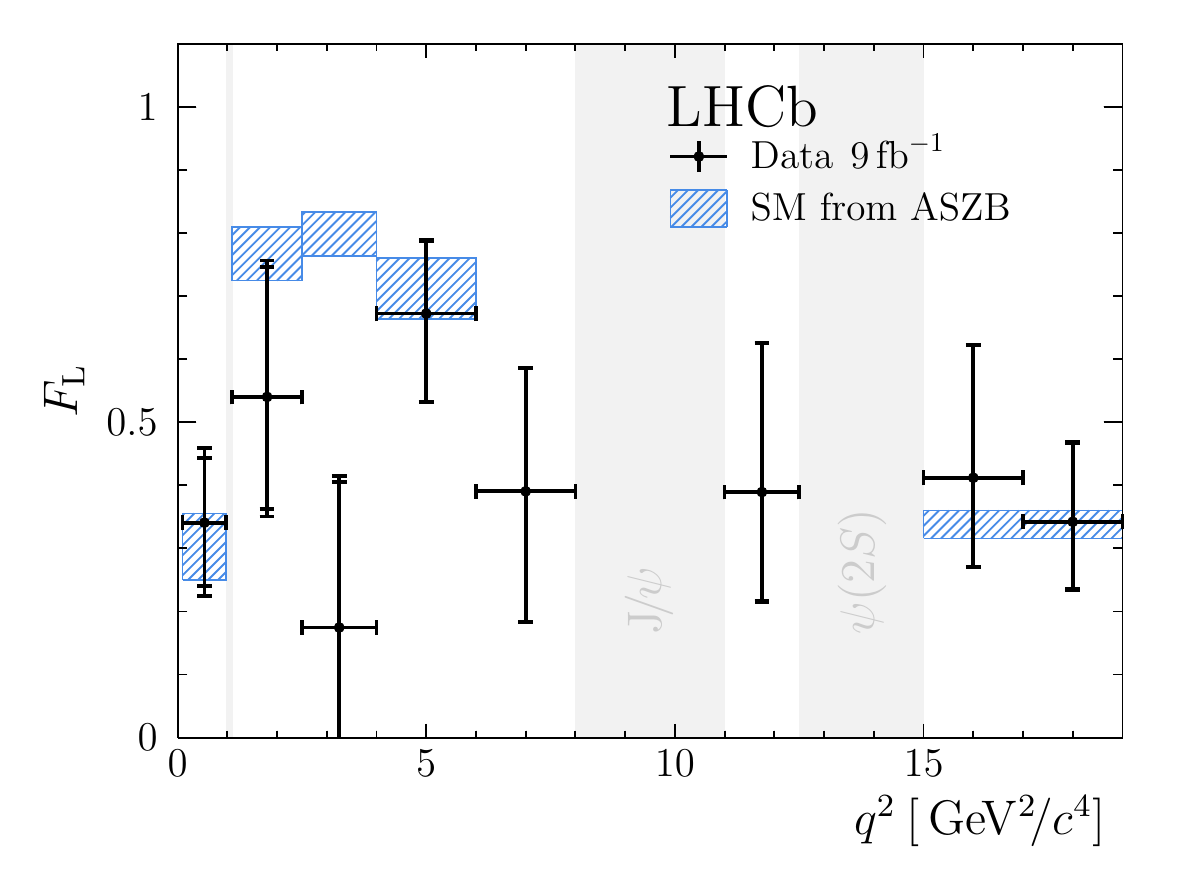}
\caption{The angular observables in the decay process $B^{+} \rightarrow K^{*+} \mu^{+} \mu^{-}$, taken from Ref.~\cite{LHCb:2020gog}.}
\label{fig:rare-decay-deviations}
\end{figure}

At the quark level, rare decays typically proceed via transitions such as $b \to s \ell \ell$, $s \to d \ell \ell$, and $c \to u \ell \ell$. Due to the strong suppression imposed by the GIM mechanism within the standard model, these processes serve as ideal probes for exploring physics beyond the standard model~\cite{Glashow:1970gm}. For instance, certain angular observables in rare decays are highly sensitive to potential new physics effects. Precision measurements of these angular distributions can reveal subtle deviations from standard model predictions and thus serve as indirect signals of new physics. In 2020, the LHCb collaboration performed an angular analysis of the decay process $B^{+} \rightarrow K^{*+} \mu^{+} \mu^{-}$~\cite{LHCb:2020gog}, where local deviations from the standard model predictions were observed, as shown in Fig~\ref{fig:rare-decay-deviations}. Prior to ascribing the observed anomalies to new physics, a meticulous re-examination of both experimental data and standard model predictions is imperative. It is necessary to ascertain whether the discrepancies originate from intrinsic standard model phenomena, such as higher-order perturbative corrections or soft-gluon contributions~\cite{Wang:2008sm}. Furthermore, exploring the potential manifestation of similar deviations in rare baryon decays is warranted.

Owing to the complexity of baryon structure and their relatively low production rates, experimental progress in studying rare baryon decays has been comparatively slow. Nevertheless, such decays offer several unique advantages. First, they allow for a direct extraction of the helicity structure in the effective Hamiltonian, which is often obscured in mesonic systems~\cite{Mannel:1997xy}. Second, baryon decays involve a diquark spectator system, in contrast to the single-quark spectator in meson decays, thus exhibiting different hadronic dynamics. Moreover, baryonic modes may receive contributions from $W$-boson exchange involving both two-quark and three-quark transitions~\cite{Wang:2021uzi}. As a result, rare decays in baryon systems serve as an important complement to those in mesons.

The rare decay mode of the $b$-baryon, $\Lambda_b \to \Lambda \ell^+ \ell^-$, has long attracted considerable attention. Significant progress has been made in studying this process using various theoretical approaches, including QCD sum rules~\cite{Huang:1998ek,Chen:2001ki,Chen:2001zc,Chen:2001sj}
, light-cone sum rules~\cite{Wang:2008sm,Aliev:2010uy,Gan:2012tt,Wang:2015ndk}, quark model~\cite{Mott:2011cx,Gutsche:2013pp,Faustov:2017wbh}, Lattice QCD~\cite{Detmold:2012vy,Detmold:2016pkz}, and Bethe-Salpeter equation~\cite{Liu:2019igt,Liu:2022xud}. Part of the results for the branching fractions and the mean values of the forward-backward asymmetry $A_{FB}$ and the polarization asymmetry $\alpha$ are summarized in Table~\ref{table:rare-bottom-decay}.
\begin{table}[ht]
\centering
\caption{Theoretical predictions for the branching fractions and the mean values of the forward-backward asymmetry $A_{FB}$ and the polarization asymmetry $\alpha$ in the rare decay process $\Lambda_b \to \Lambda \ell^+ \ell^-$, where $\ell = e, \mu, \tau$. The script ``No'' and ``Res'' represent the results without and with long-distance resonance contributions, respectively.}
\renewcommand{\arraystretch}{1.5}
\resizebox{\linewidth}{!}{
\begin{tabular}{lcccccc}
\hline\hline
& \multicolumn{2}{c}{$Br (\times 10^{-6})$} & \multicolumn{2}{c}{$ \langle A_{FB} \rangle $} & \multicolumn{2}{c}{$ \langle \alpha\rangle $} \\ 
\hline
&No&Res &No &Res &No&Res \\
\hline
$ \Lambda_b\to \Lambda e^+e^- $& & & &\\
QCD sum rules~\cite{Chen:2001ki} &$ 2.3 $&$ 53 $& && & \\ 
Light-cone sum rules~\cite{Aliev:2010uy} &$ 4.6\pm 1.6 $&& && & \\ 
Quark model~\cite{Faustov:2017wbh} &$ 1.07 $&&$-0.288$ &$-0.294$& & \\ 
Bethe-Salpeter equation~\cite{Liu:2019igt,Liu:2022xud} &$ 0.252\sim 0.392 $&&$-0.1371$ && & \\
\hline

$ \Lambda_b\to \Lambda \mu^+\mu^- $& & & &\\
QCD sum rules~\cite{Chen:2001ki,Chen:2001zc} &$ 2.1 $&$ 53 $&$-0.1338$ &&$-0.583$ & \\ 
Light-cone sum rules~\cite{Wang:2008sm} &$ 6.1_{-1.7}^{+5.8} $&$ 39_{-11}^{+23} $&$(-1.22_{-0.73}^{+1.42})\%$ &$(-0.99_{-0.68}^{+1.32})\%$&$-0.36_{-0.02}^{+0.05}$ &$-0.50_{-0.01}^{+0.04}$ \\
Light-cone sum rules~\cite{Aliev:2010uy} &$ 4.0\pm 1.2 $&& && & \\
Quark model~\cite{Mott:2011cx} &$ 0.6 $&$ 21 $&$-0.1255$ &$-0.1128$& & \\ 
Quark model~\cite{Faustov:2017wbh} &$ 1.05 $&&$-0.286$ &$-0.266$& & \\ 
BS equation~\cite{Liu:2019igt,Liu:2022xud} &$ 1.051\sim 1.098 $&&$-0.1376$ && & \\ 
Exp~\cite{CDF:2011buy} &$ 1.73 \pm 0.42 \pm 0.55 $ \\
\hline

$ \Lambda_b\to \Lambda \tau^+\tau^- $& & & &\\
QCD sum rules~\cite{Chen:2001ki,Chen:2001zc} &$ 0.18 $&$ 11 $&$-0.04$ &&$-0.1084$ & \\ 
Light-cone sum rules~\cite{Wang:2008sm} &$ 2.1_{-0.6}^{+2.3} $&$ 4.0_{-1.1}^{+3.7} $&$(-0.67_{-0.21}^{+0.23})\%$ &$(-0.62_{-0.21}^{+0.22})\%$&$-0.28_{-0.03}^{+0.03}$ &$-0.27_{-0.03}^{+0.03}$ \\ 
Light-cone sum rules~\cite{Aliev:2010uy} &$ 0.8\pm 0.3 $&& && & \\
Quark model~\cite{Mott:2011cx} &$ 0.22 $&$ 0.59 $&$-0.0342$ &$-0.0319$& & \\ 
Quark model~\cite{Faustov:2017wbh} &$ 0.26 $&&$-0.161$ &$-0.127$& & \\ 
BS equation~\cite{Liu:2019igt,Liu:2022xud} &$ 0.286\sim 0.489 $&&$-0.1053$ && & \\ 
\hline\hline
\end{tabular}
}
\label{table:rare-bottom-decay}
\end{table}

Theoretical descriptions of rare decay processes rely on the framework of the effective Hamiltonian. The short- and long-distance contributions are separated using the operator product expansion. The short distance effects are encoded in the Wilson coefficients, while the long distance contributions arise from the hadronic matrix elements of the relevant weak currents between baryon states, which are typically parameterized in terms of invariant form factors. The explicit form of effective Hamiltonian for the inclusive decay $b \to s \ell^+ \ell^-$ is given by~\cite{Buchalla:1995vs}:
\begin{align}
H_{eff}=-\frac{4 G_F}{\sqrt{2}} V_{t s}^* V_{t b} \sum_{i=1}^{10} C_i(\mu)\, O_i(\mu),
\end{align}
where $C_i$ are the Wilson coefficients and $O_i$ denote the standard model operators. Then the matrix element can be written as:
\begin{align}
M(\Lambda_b\to \Lambda\ell\ell) = \frac{G_F\,\alpha_{em}}{2\sqrt{2}\pi}&\bigg\{C_9^{eff}(\mu)\,\big[\bar{s} \gamma^\mu\big(1-\gamma_5\big) b\big]\big[\bar{\ell} \gamma_\mu \ell\big]+C_{10}\big[\bar{s} \gamma^\mu\big(1-\gamma_5\big) b\big]\big[\bar{\ell} \gamma_\mu \gamma_5 \ell\big]\nonumber\\[5pt]
&-C_7^{eff}(\mu)\frac{2 m_b}{q^2}\big[\bar{s}i\sigma_{\mu\nu}q^\nu(1+\gamma_5)b\big]\big[\bar{\ell} \gamma_\mu \ell\big] \bigg\}\,,
\end{align}
where the manifest expressions and values for $C_7^{eff}$ and $C_9^{eff}$ can be found in Refs.~\cite{Misiak:1992bc,Ciuchini:1993vr,Buras:1993xp,Buras:1994dj,Ligeti:1997aq,Ali:1999mm,Altmannshofer:2008dz,Chen:2001zc,Chen:2001ki,Wang:2008sm,Faustov:2017wbh,Mott:2011cx}. Since the operator $O_{10}$ cannot be generated through the insertion of four-quark operators due to the absence of the $Z$ boson in the effective theory, the Wilson coefficient $C_{10}$ remains unaffected by QCD corrections and is therefore independent of the energy scale $\mu$~\cite{Wang:2008sm}. Besides the $V-A$ interaction, the effective Hamiltonian also includes the contributions from the tensor-type current, which can be parameterized as:
\begin{align}
\bra{p(q_2)}\bar{u}i\sigma_{\mu\nu}q^\nu\big(1+\gamma_5\big)c\ket{\Lambda_c(q_1)}&=\bar{u}_{2}(q_2) \bigg[\frac{f_1^{T}(q^2)}{M_{\Lambda_c}}\big(\gamma^\mu q^2-q^\mu\slashed{q}\big)-f_2^{T}(q^2)i\sigma_{\mu\nu}q^\nu \bigg] u_1(q_1)\nonumber\\[5pt]
&+\bar{u}_{2}(q_2) \bigg[\frac{g_1^{T}(q^2)}{M_{\Lambda_c}}\big(\gamma^\mu q^2-q^\mu\slashed{q}\big)-g_2^{T}(q^2)i\sigma_{\mu\nu}q^\nu \bigg]\gamma_5 u_1(q_1)\,,
\label{rare-matrix}
\end{align}

Because of the strong suppression imposed by the GIM mechanism, rare decays receive relatively sizable long distance contributions, which can be absorbed into the effective Wilson coefficients $C_9^{eff}$:
\begin{align}
C_9^{eff}(\mu)=C_9(\mu)+Y_{\mathrm{SD}}(q^2)+Y_{\mathrm{LD}}(q^2)\,.
\end{align}
The long distance contributions $Y_{\mathrm{LD}}(q^2)$ are assumed to originate from intermediate vector resonances such as the $c\bar{c}$ structures $J/\psi$, $\psi(2S)$..., etc., which are usually modeled as a Breit-Wigner formula~\cite{Ali:1999mm,Chen:2001ki,Chen:2001zc,Faustov:2017wbh,Mott:2011cx,Gutsche:2013pp,Melikhov:1998ws}. $Y_{\mathrm{SD}}(q^2)$ describes the short distance contributions from four-quark operators far away from the $c\bar{c}$ resonances, which can be calculated reliably using perturbative QCD methods. From Table~\ref{table:rare-bottom-decay}, one can infer that the long distance contributions can enhance the branching fractions by $1\sim 2$ orders of magnitude, while their impact on the asymmetry parameters remains relatively small and tends to be negative (especially for $\langle A_{FB} \rangle$).

Furthermore, non-factorizable effects~\cite{Melikhov:1998ws, Soares:1995ri} arising from the charm-quark loop may lead to additional corrections to the radiative $b \to s \gamma$ transition. These corrections can be effectively absorbed into the redefinition of the Wilson coefficient $C_7^{eff}$~\cite{Chen:2001zc}:
\begin{align}
C_7^{eff}(\mu)=C_7(\mu)+C_{b \to s \gamma}^{\prime}(\mu)\,,
\end{align}
where $C_{b \to s \gamma}^{\prime}(\mu)$ is the absorptive part of the $b \to s \gamma$ after neglecting the small contribution from CKM sector $V_{u b} V_{u s}^*$~\cite{Soares:1991te,Asatrian:1996as}. In Ref.~\cite{Chen:2001zc}, the authors found that the non-factorizable effects have a minimal impact on the branching fractions with approximately $1\%\sim 5\%$ corrections, while the asymmetry parameters which are directly related to non-factorizable effects such as $CP$ asymmetry are significantly affected, with corrections of about $30\%\sim 60\%$.

Based on the form factors obtained by QCD sum rules~\cite{Huang:1998ek}, the authors in Refs.~\cite{Chen:2001ki,Chen:2001zc,Chen:2001sj} systematically studied the phenomenology of rare decay process $\Lambda_b \to \Lambda \ell^+ \ell^-$. By considering cases with Wilson coefficients that differ from those in the standard model, they investigated the new physics effects beyond the standard model. Their results show that the $CP$-odd transverse polarization of $\Lambda$ is zero within the standard model, while it is expected to be sizable in new physics~\cite{Chen:2001ki}. Moreover, significant $CP$ violating effects may appear after considering an explicit supersymmetric extension of the standard model~\cite{Chen:2001zc}.

In 2011, the decay $\Lambda_b \to \Lambda \mu^+ \mu^-$ was first observed by the CDF collaboration~\cite{CDF:2011buy}, and was subsequently confirmed by the LHCb collaboration~\cite{LHCb:2013uqx}. In 2015, LHCb further reported measurements of the binned differential branching fractions and angular observables for this decay~\cite{LHCb:2015tgy}.

Studies on another rare decay mode of the $\Lambda_b$, $\Lambda_b \to n \ell^+ \ell^-$, can be found in Refs.~\cite{Azizi:2010zzb,Faustov:2017ous,Aliev:2018hyy,Liu:2019rpm}. A similar analysis can be carried out for the rare decay process $\Lambda_c \to p \ell^+ \ell^-$, which proceeds via the $c \to u \ell^+ \ell^-$ transition. The corresponding Wilson coefficients related to charm scale are given by~\cite{deBoer:2015boa,deBoer:2016dcg,deBoer:2017que}.

\begin{table}[ht]
\centering
\caption{Theoretical predictions of the branching fractions, and the mean value of $ F_L $ for the rare decay $ \Lambda_c\to p \ell \ell $ and $ \Xi_c\to (\Sigma,\;\Lambda) \ell \ell $ with and without resonances contributions. The branching fractions of $ \Xi_c\to (\Sigma,\;\Lambda) \ell \ell $ are taken from Ref.~\cite{Zhang:2024asb}.} 
\renewcommand{\arraystretch}{1.3}
\resizebox{\linewidth}{!}{
\begin{tabular}{lcccccc}
\hline\hline
& & \multicolumn{2}{c}{$Br$} & & \multicolumn{2}{c}{$ \langle F_L\rangle $} \\
\hline
& &No&Res & &No &Res \\
\hline
$ \Lambda_c\to p e^+e^- $& & & &\\
QCD sum rules~\cite{Zhang:2024asb} &&$ (3.2\pm 2.3)\times 10^{-13} $&$ (4.9\pm 1.4)\times 10^{-7} $ &&$ 0.54\pm 0.01 $& $ 0.55\pm 0.01 $ \\
Lattice QCD~\cite{Azizi:2010zzb} & &$(2.46\pm 1.25)\times 10^{-14}$& & && \\ 
Light-cone sum rules~\cite{Azizi:2010zzb} & &$(4.19\pm 2.35)\times 10^{-14}$& & && \\ 
Light-cone sum rules~\cite{Sirvanli:2016wnr}& & $(4.5\pm 2.4)\times 10^{-14}$&$ (4.2\pm 0.7)\times 10^{-6} $ && & \\ 
Relativistic quark model~\cite{Faustov:2018dkn} && $ (3.8\pm 0.5)\times 10^{-12} $&$ (3.7\pm 0.8)\times 10^{-7}$ &&& \\ 
\hline
$ \Lambda_c\to p \mu^+\mu^- $& & & &\\
QCD sum rules~\cite{Zhang:2024asb} &&$ (2.4\pm 1.8)\times 10^{-13} $&$ (4.9\pm 1.4)\times 10^{-7} $ & &$ 0.47\pm 0.01 $& $ 0.48\pm 0.01 $ \\
Lattice QCD~\cite{Meinel:2017ggx}&&$(4.1\pm 0.4^{+6.1}_{-1.9})\times 10^{-11}$&$(3.7\pm 1.3)\times 10^{-7}$ & && \\ 
Lattice QCD~\cite{Azizi:2010zzb} & &$(1.94\pm 1.05)\times 10^{-14}$& & && \\ 
Light-cone sum rules~\cite{Azizi:2010zzb} & &$(3.87\pm 2.26)\times 10^{-14}$& & && \\ 
Light-cone sum rules~\cite{Sirvanli:2016wnr} & &$(3.77\pm 2.28)\times 10^{-14}$&$ (3.2\pm 0.7)\times 10^{-6} $ & && \\ 
Relativistic quark model~\cite{Faustov:2018dkn} && $ (2.8\pm 0.4)\times 10^{-12} $&$(3.7\pm 0.8)\times 10^{-7}$ &&& $ 0.52\pm 0.02 $ \\ 
\hline 
$ \Xi_c^+\to \Sigma^+ e^+e^- $& & $(6.9\pm 4.9)\times 10^{-13}$&$(3.0\pm 1.5)\times 10^{-6} $ & &$ 0.57\pm 0.02 $&$ 0.57\pm 0.01 $\\
$ \Xi_c^+\to \Sigma^+ \mu^+\mu^- $& & $(5.2\pm 3.7)\times 10^{-13}$&$(3.0\pm 1.5)\times 10^{-6} $ & &$ 0.49\pm 0.02 $&$ 0.49\pm 0.01 $\\
$ \Xi_c^0\to \Sigma^0 e^+e^- $& & $(1.1\pm 0.8)\times 10^{-13}$&$ (4.9\pm 2.4)\times 10^{-7} $ & & $ 0.57\pm 0.02 $&$ 0.57\pm 0.01 $\\
$ \Xi_c^0\to \Sigma^0 \mu^+\mu^- $& &$(8.6\pm 6.2)\times 10^{-14}$&$ (4.9\pm 2.4)\times 10^{-7} $ & & $ 0.49\pm 0.02 $&$ 0.49\pm 0.01 $ \\
$ \Xi_c^0\to \Lambda e^+e^- $& & $(4.5\pm 2.8)\times 10^{-14}$&$ (2.0\pm 1.0)\times 10^{-7} $ & &$ 0.56\pm 0.02 $&$ 0.56\pm 0.01 $\\
$ \Xi_c^0\to \Lambda \mu^+\mu^- $& & $(3.4\pm 2.2)\times 10^{-14}$&$ (1.9\pm 1.0)\times 10^{-7} $ & &$ 0.48\pm 0.01 $&$ 0.49\pm 0.01 $\\
\hline\hline
\end{tabular}
}
\label{table:rare-charm-decay}
\end{table}

\begin{figure}[ht]
\centering
\includegraphics[width=0.4\textwidth]{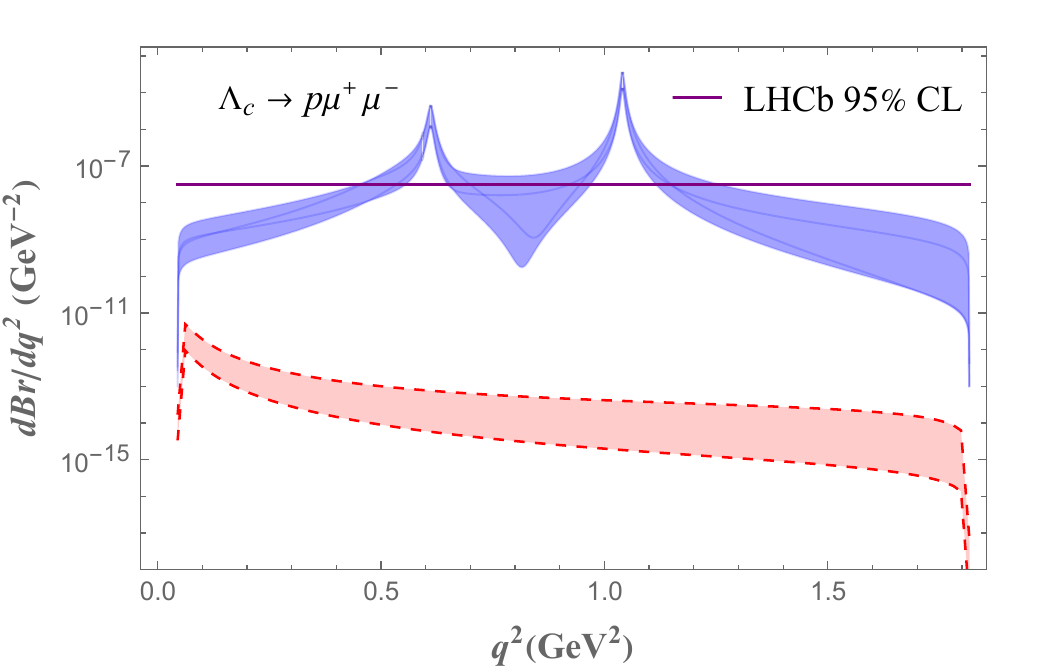}
\includegraphics[width=0.35\textwidth]{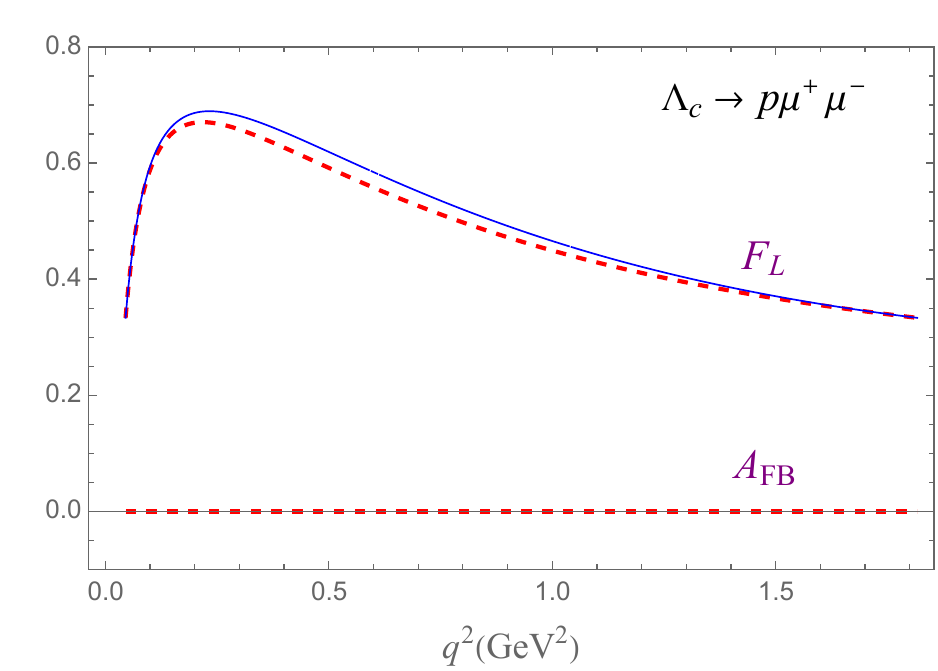}
\caption{The $ q^2 $ dependence of the differential branching fractions, forward-backward asymmetry and longitudinally polarized dileptons in $ \Lambda_c \to p \ell \ell $ decays, taken from Ref.~\cite{Zhang:2024asb}. The red line represents the contribution excluding resonances, while the blue line includes the contributions from resonances. The shaded area represents the uncertainty of the theoretical predictions.}
\label{fig:observable-rare}
\end{figure}

A QCD sum rules analysis for $\Lambda_c \to p \ell^+ \ell^-$ was performed in Ref.~\cite{Zhang:2024asb}. The $q^2$ dependence of the differential branching fractions, forward-backward asymmetry and longitudinally polarized dileptons is shown in Fig~\ref{fig:observable-rare}. The peaks in the differential branching fractions represent the contributions from the intermediate vector resonances $\rho$, $\omega$ and $\phi$. Theoretical predictions for the branching fractions of $\Lambda_c \to p \ell^+ \ell^-$ are summarized in Table~\ref{table:rare-charm-decay}. The long-distance resonance contributions can enhance the branching fraction by as much as 4 to 8 orders of magnitude, which is significantly larger than the corresponding effect in the $\Lambda_b$ case. This is because the $c \to u \ell^+ \ell^-$ transition is subject to a much stronger GIM suppression, making the resonance contributions more pronounced. 

\begin{figure}[ht]
\centering
\includegraphics[width=0.4\textwidth]{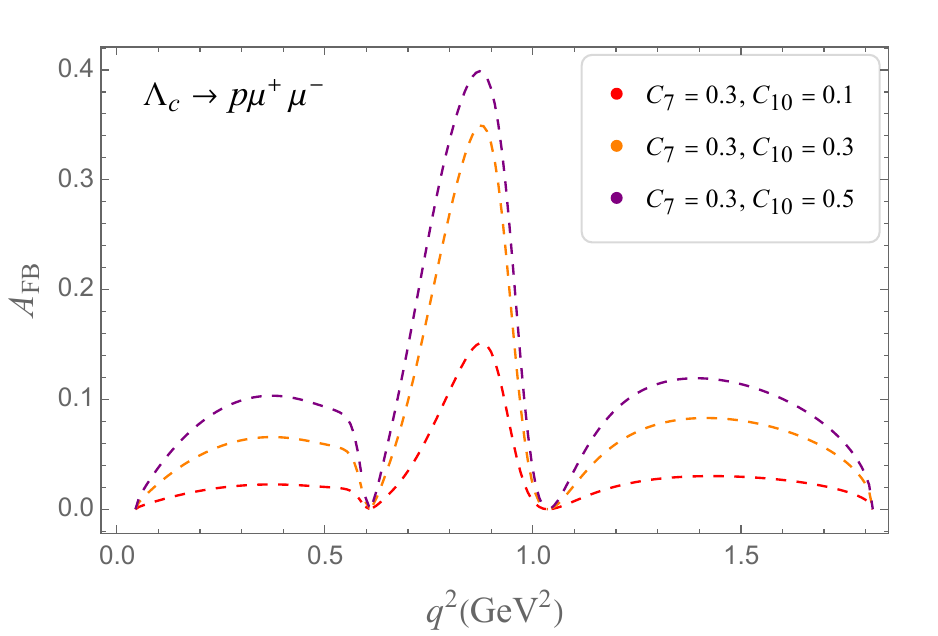}
\includegraphics[width=0.4\textwidth]{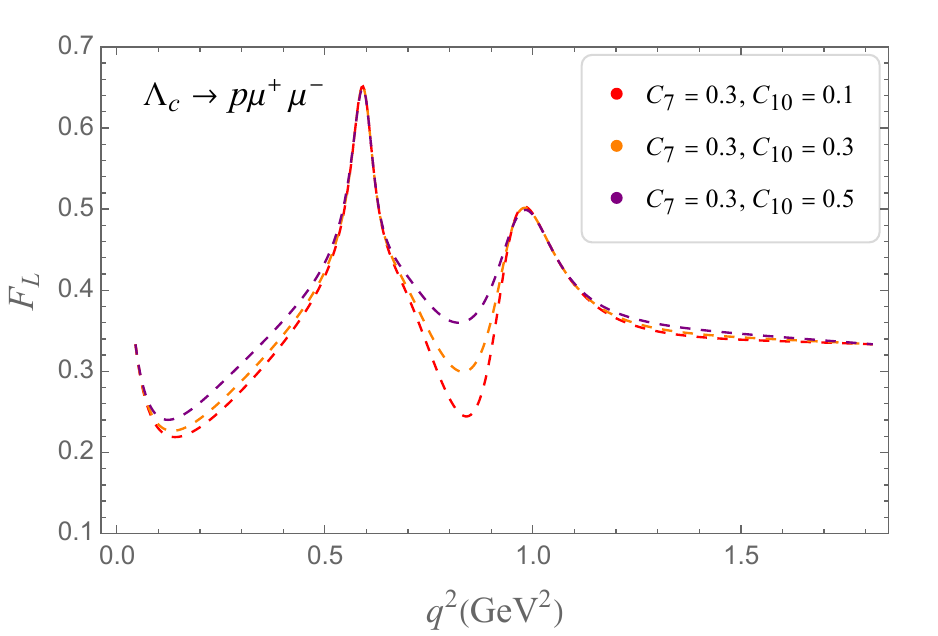}
\caption{New physics effects for the forward-backward asymmetry, and the fraction of longitudinally polarized dileptons with different values of the Wilson coefficients, taken from Ref.~\cite{Zhang:2024asb}}
\label{fig:AFB-rare}
\end{figure}

Two angular observables are found to be sensitive to new physics effects as shown in Fig~\ref{fig:AFB-rare}: 1) The lepton forward-backward asymmetry $A_{FB}$ is predicted to vanish across the entire kinematic range within the standard model. Therefore, any observed nonzero value would serve as a clear indication of possible new physics; 2) The longitudinal polarization fraction of the lepton $F_L$ shows no visible contribution from meson resonances under standard model assumptions. However, in the presence of new physics, significant resonance contributions can emerge~\cite{Zhang:2024asb}. Consequently, measurements of the $F_L$ distribution can be used as an indirect probe for new physics signatures.

\section{Baryoniums}\label{sec:baryonium}
With the advent of high-luminosity experiments and precise detection techniques, the past two decades have marked a ``golden age'' for exotic hadron spectroscopy. Key milestones include the identification of the $Z_c(3900)$ tetraquark by BESIII~\cite{BESIII:2013ris} and the $P_c(4450)$ pentaquark by LHCb~\cite{LHCb:2015yax}. These findings have spurred extensive theoretical efforts, particularly using non-perturbative tools like QCD sum rules and Lattice QCD, to decipher the internal structure of these states. In light of these discoveries, the exploration of hexaquark states emerges as the next natural and crucial frontier in hadron physics.

A prime example is the well-known deuteron. Its binding energy is approximately 2.225 MeV, indicating a loosely bound molecular configuration. Although the binding is weak, the residual strong interaction is sufficient to confine the proton and neutron, thereby laying the foundation for cosmic nucleosynthesis from hydrogen to heavier elements. The strong interaction not only ensures the stability of the deuteron but also allows for the possible existence of other stable deuteron-like dibaryon states. The $H$-dibaryon, originally proposed by Jaffe in 1976 based on bag model~\cite{Jaffe:1976yi}, was historically viewed as a classic example of a deeply bound state. Nevertheless, recent Lattice QCD investigations~\cite{Inoue:2010es,NPLQCD:2010ocs}, most notably the 2019 results obtained at the physical pion mass~\cite{HALQCD:2019wsz}, challenge this view. These findings indicate that the $H$-dibaryon is likely unbound, favoring an interpretation as a resonance above the mass threshold. Consequently, one is led to ask: does nature allow for the formation of hexaquark structures with significantly tighter binding?

Within the framework of the one-boson-exchange (OBE) potential model, the interactions in dibaryon ($BB$) and baryon-antibaryon ($B\bar{B}$) systems are related via $G$-parity~\cite{Dover:1979zj,Shapiro:1978wi,Cote:1982gr}. For meson exchanges with specific quantum numbers, the $G$-parity transformation induces a sign reversal of the potential. Consequently, interactions that manifest as short-range repulsion in the $BB$ system may transform into strong attraction in the $B\bar{B}$ sector. This suggests that $B\bar{B}$ systems are more likely to form deeply bound states compared to their dibaryon counterparts. Such bound states or resonances, consisting of baryons and antibaryons, are collectively termed baryoniums. In principle, their composition is not strictly limited to pairs of color-singlet hadrons but may also involve color multiplets, such as triquark clusters. In the latter case, the clusters are bound not by the weak residual interaction, but by the direct strong interaction through gluon exchange. This mechanism results in a significantly more tightly bound structure, known as compact multiquark state.

The study of baryoniums can be traced back to the 1940s, when Fermi and Yang first proposed that the pion might be a composite particle formed by a nucleon-antinucleon pair~\cite{Fermi:1949voc}, an idea that was later superseded by the quark model. In the early 21st century, with the discovery of numerous exotic hadrons, the baryonium picture was revisited and employed to explain the properties of some of these states. In Refs.~\cite{Datta:2003iy,Zou:2003zn,Zhu:2005ns,Ding:2005gh}, the authors proposed a $p$-$\bar{p}$ baryonium picture for the $X(1860)$~\cite{BES:2003aic} and $ X(1835) $~\cite{BES:2005ega}. In Refs~\cite{Qiao:2005av,Qiao:2007ce}, Qiao proposed a hidden charm baryonium picture $ \Lambda_c $-$ \bar{\Lambda}_c $ for the $Y(4260)$, which was observed in the initial-state radiation process $e^{+} e^{-} \to \gamma \pi^{+} \pi^{-} J / \psi$ by the BaBar collaboration~\cite{BaBar:2005hhc}. The study of baryonium states has since gained momentum, with various theoretical approaches being employed to investigate their properties and interactions~\cite{Gao:2003ka,Rosner:2003bm,Kerbikov:2004gs,Yan:2004xs,Liu:2004er,Sibirtsev:2004id,Chang:2004us,Ding:2005ew,Wang:2006sna,Chen:2008ee,Ma:2008hc,Dedonder:2009bk,Wang:2010vz,Chen:2010an,Niu:2024cfn,Zhao:2013ffn,Deng:2013aca,Dong:2017rmg,Liu:2007tj,Cotugno:2009ys,Chen:2011cta,Chen:2013sba,Lee:2011rka,Karliner:2015ina,Chen:2017vai,Lu:2017dvm,Liu:2021gva,Song:2022svi,Salnikov:2023qnn,Qi:2024dqz,Cao:2024hnn,Wang:2024riu,Meguro:2011nr}. These predicted structures may provide theoretical interpretations for a number of experimentally observed resonances. We also refer the readers to earlier review articles~\cite{Shapiro:1978wi,Montanet:1980te,Richard:1999qh,Klempt:2002ap} for a comprehensive overview of the theoretical and experimental aspects in this field.

In the following sections, we will separately discuss two distinct color configurations of baryonium. Furthermore, the QCD sum rules approach represents a well-established method for investigating these states. Consequently, this chapter provides a review of QCD sum rules alongside other theoretical frameworks applied to the study of both heavy and light baryonium systems. We also address potential experimental signatures and the broader implications for our understanding of multiquark dynamics.

\subsection{Baryon-antibaryon}

\subsubsection{Light sector}\label{sec:light_baryonium}\

The study of light baryon-antibaryon bound states traces back to early investigations of the nucleon-antinucleon system. Fermi and Yang were among the first to propose that the pion could be understood as a bound state of a nucleon ($N$) and an antinucleon ($\bar{N}$)~\cite{Fermi:1949voc}. Although the original interpretation of the pion as a baryon-antibaryon bound state was later ruled out, the underlying idea of baryon-antibaryon, i.e., the baryonium, survived and evolved in various theoretical contexts.

In the 1970s, theoretical studies began exploring the possibility of loosely bound $N\bar{N}$ states, motivated by the presence of a strongly attractive potential between the nucleon and antinucleon~\cite{Dalkarov:1970qb}. Various potential models subsequently predicted a spectrum of deeply bound isoscalar baryoniums with quantum numbers $J^{PC} = 2^{++}, 1^{--}$, and $0^{++}$, among which the $2^{++}$ state was often considered the most strongly bound~\cite{Buck:1977rt,Dover:1979zj}. Experimentally, several candidate states were reported in the seventies at CERN, BNL, and KEK, some of which exhibited statistically significant signals~\cite{ParticleDataGroup:2024cfk} (for details see Refs.~\cite{Montanet:1980jy,Amsler:1987qqd}). Among these, the $f_2(1565)$ resonance, mostly observed in $p\bar{p}$ annihilation~\cite{ParticleDataGroup:2024cfk}, has been considered a particularly interesting candidate for the $2^{++}$ $p\bar{p}$ bound-state~\cite{Dover:1990kn}.

The $N$-$\bar{N}$ system is also of considerable interest in nuclear physics, particularly because it may offer complementary insights into the nucleon-nucleon ($N$-$N$) interaction. Although the nuclear forces in the two systems are not identical, they both originate from similar light boson exchange mechanisms. As a result, their interactions can be related to some extent through the $G$-parity transformation~\cite{Shapiro:1978wi,Richard:1999qh}. At present, conclusions about the $N$-$N$ nuclear force derived from $NN$ scattering phase shifts and deuteron structure data remain somewhat ambiguous. Even under the guidance of specific theoretical models, it is difficult to reliably determine the precise form of the $N$-$N$ interaction based solely on existing experimental data, as these forces are inherently weak in two-body systems~\cite{Shapiro:1978wi}. In contrast, the situation for the $N$-$\bar{N}$ system is markedly different, as the attraction between the nucleon and antinucleon is expected to be significantly stronger than that in the $N$-$N$ case. This stronger interaction may give rise to more pronounced binding effects and enhance the likelihood of forming deeply bound states. For this reason, a strongly attractive $N\bar{N}$ bound state is expected to emerge near threshold in the nucleon-antinucleon system, much like the deuteron in the $NN$ system. The existence of such states would offer valuable insight into the underlying mechanisms of the strong force in the baryon-antibaryon sector. 

\begin{figure}[ht]
\centering
\includegraphics[width=0.4\textwidth]{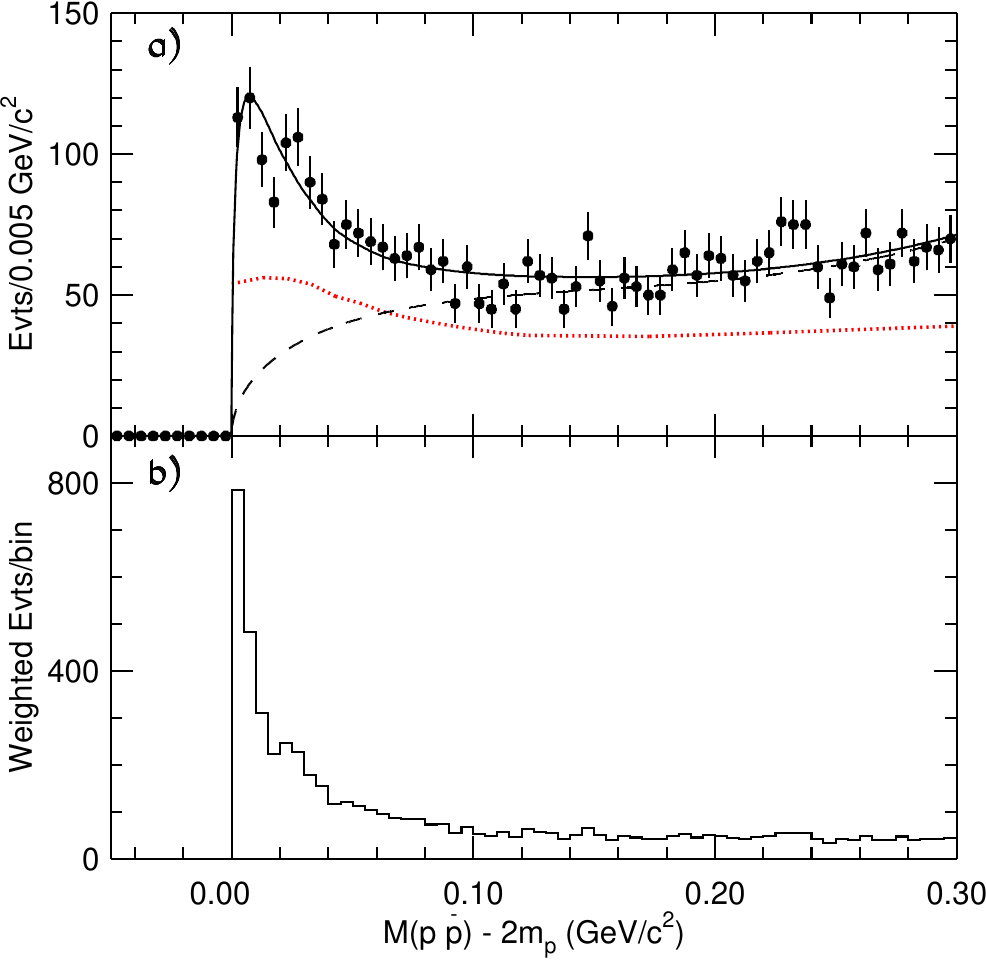}
\hspace{1cm}
\includegraphics[width=0.3\textwidth]{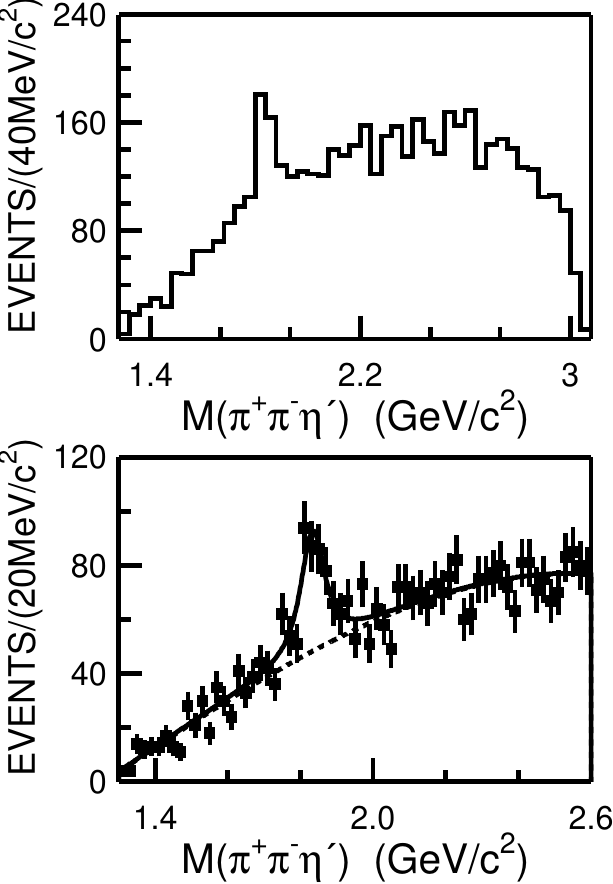}
\caption{The enhancement observed in $p\bar{p}$ invariant mass spectrum from Ref.~\cite{BES:2003aic} (left) and the $X(1835)$ state observed in $\eta^\prime\pi^+\pi^-$ invariant mass spectrum from Ref.~\cite{BES:2005ega} (right).}
\label{fig:1860&1855}
\end{figure}

In the early 2000s, strong enhancements near the $\bar{p}p$ threshold were observed in the $J / \psi \to \gamma p \bar{p}$ radiative decays~\cite{BES:2003aic}, as shown in Fig~\ref{fig:1860&1855} (left), while no similar structure was found in $\pi^0 p \bar{p}$ and $\omega p \bar{p}$ channels~\cite{Klempt:2007cp}. This enhancement can be fit with an $S$-wave Breit Wigner resonance function, where the peak mass is below the $p\bar{p}$ threshold at $M=1859_{-10}^{+3} (stat)_{-25}^{+5} (sys)$ MeV. The total decay width was measured to be less than 30 MeV at the 90\% confidence level. We hereafter denote the observed enhancement in the $p \bar{p}$ invariant mass spectrum as $X(p\bar{p})$.

Later in 2005, the BES collaboration reported a narrow resonance called $X(1835)$ in $J / \psi \to \gamma \pi^{+} \pi^{-} \eta^{\prime}$ channel with a statistical significance of 7.7$\sigma$~\cite{BES:2005ega}, as shown in Fig~\ref{fig:1860&1855} (right). The mass, total decay width, and the product branching fraction of $ X(1835) $ were measured to be: 
\begin{align}
&M=1833.7\pm 6.1 (stat) \pm 2.7 (sys) \,\mathrm{MeV}\,,\quad \Gamma=67.7\pm 20.3 (stat) \pm 7.7 (sys) \,\mathrm{MeV}\,,\nonumber\\[5pt]
&B(J / \psi \to \gamma X) \,B(X \to \pi^{+} \pi^{-} \eta^{\prime})=\big( 2.2 \pm 0.4(stat) \pm 0.4(syst) \big)\times 10^{-4}\,.
\end{align}
This state was further confirmed with high statistical significance by the BESIII experiment~\cite{BESIII:2010gmv}.

Based on a larger dataset, BES collaboration reanalyzed the $ X(p\bar{p}) $ in the radiative decay process of charmonium states $J/\psi$ and $\psi^\prime $~\cite{BESIII:2010vwa,BESIII:2011aa}. In Ref.~\cite{BESIII:2010vwa}, the mass of $ X(p\bar{p}) $ in $\psi^{\prime} \to \pi^{+} \pi^{-} J / \psi(J / \psi \to \gamma p \bar{p})$ channel was fitted to be $M=1861_{-13}^{+6}(stat)_{-26}^{+7}(syst)$ MeV, which was consistent with the former result~\cite{BES:2003aic}. In Ref.~\cite{BESIII:2011aa}, a partial wave analysis of $J / \psi \to \gamma p \bar{p}$ and $\psi^{\prime} \to \gamma p \bar{p}$ decays were performed. The spin-parity number of the enhancement $X(p\bar{p})$ was determined to be $0^{-+}$. The mass, decay width, and the product branching fraction were measured to be within the final state interactions effects, which were:
\begin{align}
&M=1832_{-5}^{+19}(stat)_{-17}^{+18}(syst) \pm 19(model) \,\mathrm{MeV}\,,\nonumber\\[5pt]
&\Gamma=13 \pm 39 (stat)_{-13}^{+10}(syst) \pm 4 (model) \,\mathrm{MeV}\,,\nonumber\\[5pt]
&B(J / \psi \to \gamma X) \,B(X \to p \bar{p})=\big( 9.0_{-1.1}^{+0.4}(stat)_{-5.0}^{+1.5} (syst) \pm 2.3(model) \big)\times 10^{-5}\,.
\end{align}
Other experimental findings for the enhancement near $p\bar{p}$ threshold can be found in Refs.~\cite{Belle:2002bro,Belle:2002fay,Belle:2005mke,Belle:2007oni,BaBar:2005pon,BaBar:2013ves,CMD-3:2015fvi,BESIII:2016fbr}.

Neither the $X(p\bar{p})$ observed in the $p\bar{p}$ invariant mass spectrum nor the $X(1835)$ seen in the $\eta^\prime\pi^+\pi^-$ channel can be easily identified with any known hadronic states based on their masses and decay properties. Among the various theoretical interpretations, one particularly intriguing possibility is that these states are bound systems of a proton and an antiproton. In Ref.\cite{Datta:2003iy}, Datta and O'Donnell proposed that $X(p\bar{p})$ could be a deuteron-like singlet $^1S_0$ state with quantum numbers $J^{PC} = 0^{-+}$, and also predicted a $0^{-+}$ $\Lambda$-$\bar{\Lambda}$ bound state with a mass around 2.2 GeV and a binding energy of 31 MeV. In Ref.\cite{Zou:2003zn}, Zou and Chiang emphasized the crucial role of final-state interactions in generating the $p\bar{p}$ near-threshold enhancement. Zhu and Gao argued that the $X(1835)$ may be a $p$-$\bar{p}$ baryonium state, based on its strong coupling to the $p\bar{p}$ channel, suppressed coupling to mesonic final states, and the absence of significant decays into strange three-body channels~\cite{Zhu:2005ns}.

Several other studies further support the baryonium interpretation of $X(p\bar{p})$ and $X(1835)$~\cite{Gao:2003ka,Rosner:2003bm,Kerbikov:2004gs,Yan:2004xs,Liu:2004er,Sibirtsev:2004id,Chang:2004us,Ding:2005ew,Wang:2006sna,Chen:2008ee,Ma:2008hc,Dedonder:2009bk,Wang:2010vz,Chen:2010an,Niu:2024cfn}. For instance, Ref.~\cite{Yan:2004xs} employed a phenomenological Skyrme-type potential to analyze the $p\bar{p}$ enhancement in $J/\psi \to \gamma p \bar{p}$ and proposed a weakly bound $p$-$\bar{p}$ state with a small binding energy and narrow decay width. They also found that the dominant decay mechanism arises from nucleon-antinucleon annihilation, a characteristic feature of baryonium decays. The mesonic decays of the $p$-$\bar{p}$ system due to the nucleon-antinucleon annihilation were discussed in Ref.~\cite{Ding:2005ew}, with the authors suggesting that the decay into $\eta \,4\pi$ final state could be the most favorable channel. Wang performed a QCD sum rules calculations for the mass of $ X(1835) $ based on the $ 0^{-+} $ $p$-$\bar{p}$ baryonium picture~\cite{Wang:2006sna}, where the interpolating current was constructed as:
\begin{align}
&j^{0^{-+}}_{p\text{-}\bar{p}}(x)=i \bar{\eta}_{p}(x) \gamma_5 \eta_{p}(x)\,,\\[5pt]
&\eta_p(x)=\varepsilon_{a b c}\big(u^{a T}(x) \mathcal{C} \gamma^\mu u^b(x)\big) \gamma_5 \gamma_\mu d^c(x)\,.
\end{align}
The mass was estimated to be $ 1.9\pm 0.1 $ GeV, indicating a qualitative agreement with experimental data.

Both the $X(p\bar{p})$ and $X(1835)$ states were observed in radiative decays of the $J/\psi$, which are known to be rich in gluonic content. Hence, glueball explanations for these states have also been proposed in the literature. However, one major challenge to this scenario is the discrepancy between the observed mass of these states and the theoretical predictions for glueball masses. Lattice QCD and QCD sum rules calculations suggested that the mass of a pure pseudoscalar two-gluon glueball lies in the range of 2.2$\sim$2.6 GeV~\cite{Morningstar:1999rf,Chen:2005mg,Meyer:2004gx,Gregory:2012hu,Athenodorou:2020ani,Chen:2021bck}, which is significantly higher than the observed mass of $X(p\bar{p})$ and $X(1835)$. Thus, to interpret the $X(p\bar{p})$ and $X(1835)$ as glueballs or gluon-rich hadrons, additional hadronic mixing mechanisms are needed to significantly lower the mass of a pure glueball to match the observed values. This possibility has been discussed in detail in Refs.~\cite{Kochelev:2005tu,Li:2005vd,Kochelev:2005vd,He:2005nm,Hao:2005hu}. In particular, in Ref.~\cite{Hao:2005hu}, the authors proposed a mixing scheme between the $X(p\bar{p})$ and a $0^{-+}$ glueball state based on QCD sum rules results, where they provided several arguments in support of this idea: 1) the observed $p\bar{p}$ enhancement lies within the mass range predicted for the $0^{-+}$ glueball in their calculation; 2) the large branching ratio to $\gamma p\bar{p}$ and three-meson final states, compared to two-meson channels, suggested a strong coupling to six-quark configurations, which can be naturally explained if the state has significant three-gluon glueball content; 3) theoretically, $J/\psi$ radiative decays into two- and three-gluon final states occur at comparable rates, after applying appropriate cuts to regulate infrared divergences; 4) the absence of a $\Lambda\bar{\Lambda}$ threshold enhancement suggested significant $SU(3)$ flavor breaking, favoring a $p\bar{p}$ baryonium-glueball mixed state, as pure baryonium faces $SU(3)$ breaking issues and pure glueball scenarios predicted a mass that is too high.

\begin{figure}[ht]
\begin{center}
\includegraphics[width=0.4\linewidth]{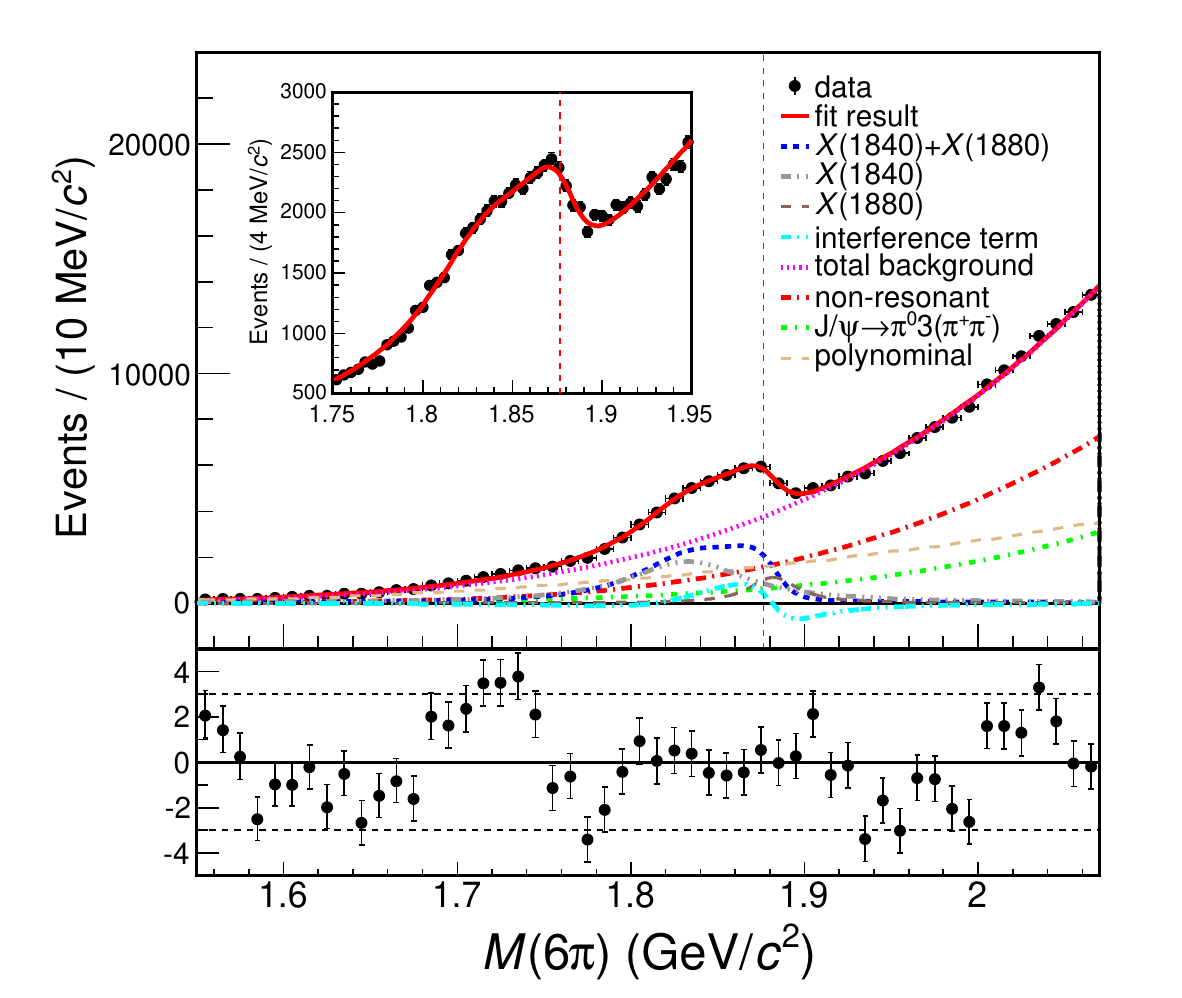}
\caption{The fitting curves for $X(1840)$ and $X(1880)$ in $3(\pi^+\pi^-)$ invariant mass spectrum in Ref.~\cite{BESIII:2023vvr}.} 
\label{fig:1840&1880}
\end{center}
\end{figure}

Another intriguing question is whether $X(p\bar{p})$ and $X(1835)$ represent the same state. Current experimental data show that the decay width of $X(p\bar{p})$ is much smaller than that of $X(1835)$, suggesting they may be distinct states, or that a coupling mechanism could exist between them. However, after considering the zero isospin and final-state-interactions of Ref.~\cite{Sibirtsev:2004id}, the $X(p\bar{p})$ state may yield a broad structure ($\Gamma\sim 100$ MeV)~\cite{BES:2005ega,BESIII:2010vwa}. Measuring the relative production ratios of $X(1835)$ in $J/\psi$ and $\psi^{\prime}$ radiative decays may offer further insights into whether $X(p\bar{p})$ and $X(1835)$ are indeed the same state, based on differences in their production ratios~\cite{BESIII:2011aa}. In 2013 and 2023, the BESIII collaboration observed two new resonance structures, $X(1840)$~\cite{BESIII:2013sbm} and $X(1880)$~\cite{BESIII:2023vvr}, in the $J/\psi \to \gamma 3(\pi^+ \pi^-)$ decay channel, as shown in Fig~\ref{fig:1840&1880}. Their mass and decay width were measured to be:
\begin{align}
&X(1840):\quad M = 1832.5 \pm 3.1\pm 2.5 \, \text{MeV}\,, \quad \Gamma = 80.7 \pm 5.2 \pm 7.7 \, \text{MeV}\,, \nonumber\\
&X(1880):\quad M = 1882.1 \pm 1.7\pm 0.7 \, \text{MeV}\,, \quad \Gamma = 30.7 \pm 5.5 \pm 2.4 \, \text{MeV}\,, 
\end{align}
The observation of such two states provides tentative support for the proton-antiproton baryonium interpretation, though alternative explanations such as threshold effects or final-state interactions cannot be excluded at present, where the details can be found in a recent review~\cite{Ma:2024gsw}. Additionally, the anomalous line shapes observed in $3(\pi^+\pi^-)$ and $\eta^\prime\pi^{+}\pi^{-}$ spectrum near the $p\bar{p}$ threshold suggested complex resonant structures. To clarify the relationships among the resonances in the $[1.8, 1.9]~\mathrm{GeV}$ region and uncover their nature, more data as well as other measurements such as the determination of spin-parity and the coupled-channel analyses are highly needed~\cite{BESIII:2023vvr}. 

In addition to the $p$-$\bar{p}$ configuration, light baryonium states may also be composed of other baryon-antibaryon pairs from the baryon octet and decuplet, such as $\Lambda$-$\bar{\Lambda}$, $\Sigma$-$\bar{\Sigma}$, and $\Xi$-$\bar{\Xi}$. In Ref.~\cite{Zhao:2013ffn}, the authors used a one-boson-exchange potential model to explore whether the $\phi(2170)$ and $\eta(2225)$ could be interpreted as $\Lambda \bar{\Lambda}$ bound states in the $^3S_1$ and $^1S_0$ channels, respectively. $\phi(2170)$, also noted as $Y(2175)$, was observed in the initial-state-radiation process $e^{+} e^{-} \to \gamma_{I S R}\, \phi(1020)\, f_0(980)$ by BABAR collaboration~\cite{BaBar:2006gsq}, while $\eta(2225)$ was observed in the radiative decay process $J / \psi \to \gamma \phi \phi$ by the MARK-III collaboration ~\cite{MARK-III:1990bpj}. The properties of such two hidden strange states listed in Particle Data Group~\cite{ParticleDataGroup:2024cfk} are:
\begin{align}
&\phi(2170):\quad J^{PC}=1^{--}\,,\quad M = 2164 \pm 6 \, \text{MeV}\,, \quad \Gamma = 106^{+24}_{-18} \, \text{MeV}\,, \nonumber\\
&\eta(2225):\quad J^{PC}=0^{-+}\,,\quad M = 2221^{+13}_{-10} \, \text{MeV}\,, \quad \Gamma = 185^{+40}_{-20} \, \text{MeV}\,, 
\end{align}
The strong decays of these two states as $\Lambda$-$\bar{\Lambda}$ bound states were discussed in Ref.~\cite{Dong:2017rmg}, where $\eta(2225) \to K^* K$ and $\phi(2170) \to K K$ were identified as their dominant decay channels.

\begin{figure}[ht]
\begin{center}
\includegraphics[width=0.4\linewidth]{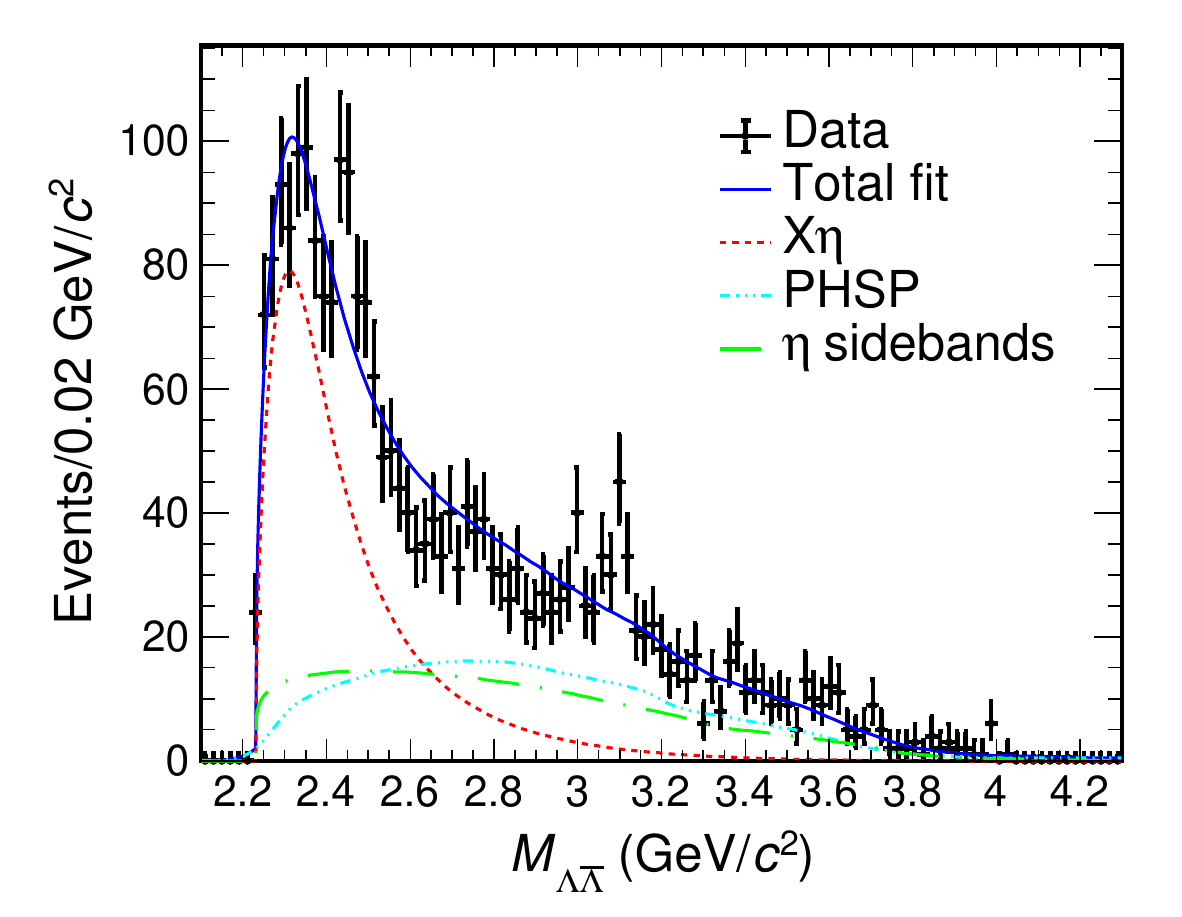}
\caption{The $\Lambda\bar{\Lambda}$ enhancement observed by BESIII in Ref.~\cite{BESIII:2022tvj}.} 
\label{fig:lambda-lambda}
\end{center}
\end{figure}

A systematic study of light baryonium states using QCD sum rules was presented in Ref.~\cite{Wan:2021vny}. The authors calculated the masses of various baryonium states, including $N$-$\bar{N}$, $\Lambda$-$\bar{\Lambda}$, $\Sigma$-$\bar{\Sigma}$, and $\Xi$-$\bar{\Xi}$, with different quantum numbers. The interpolating currents for these light baryonium states were constructed as:
\begin{align}
j^{0^{-+}}(x)&= i\,\bar{\eta}_{\mathcal{B}}(x) \gamma_5 \eta_{\mathcal{B}}(x) \,,\label{lightbaryonium-Ja0-+}\\
j^{1^{--}}_\mu(x)&=\bar{\eta}_{\mathcal{B}}(x) \gamma_\mu \eta_{\mathcal{B}}(x)\,, \label{lightbaryonium-Ja1--}\\
j^{0^{++}}(x)&= \bar{\eta}_{\mathcal{B}}(x) \eta_{\mathcal{B}}(x)\,,\label{lightbaryonium-Ja0++}\\
j^{1^{++}}_\mu(x)&= i\,\bar{\eta}_{\mathcal{B}}(x) \gamma_\mu\gamma_5 \eta_{\mathcal{B}}(x) \,, \label{lightbaryonium-Ja1++}\\
\eta_{\mathcal{B}}(x)&=i \varepsilon_{a b c}\big[ q_a^{i,T}(x) C \gamma_5 q_b^j(x) \big]q^k_c(x)\,,
\end{align}
where the superscripts $i,j,k$ denote the flavor indices of the quarks, with $(i,j,k) = (u,d,u)$, $(u,d,s)$, $(u,s,d)$, and $(s,u,s)$ corresponding to the $p$, $\Lambda$, $\Sigma$, and $\Xi$ states, respectively. The numerical results for the masses of light baryonium states are summarized in Table~\ref{table:mass_baryonium}. The predicted mass of the $0^{-+}$ $N$-$\bar{N}$ baryonium agrees well with the observed $X(p\bar{p})$ and $X(1835)$. The $0^{-+}$ $\Lambda$-$\bar{\Lambda}$ state lies close to the mass of $\eta(2225)$, while the $1^{--}$ $\Lambda$-$\bar{\Lambda}$ state is slightly heavier than $\phi(2170)$, suggesting that the latter is unlikely to be a pure $\Lambda$-$\bar{\Lambda}$ baryonium. Notably, the predicted mass of the $1^{--}$ $\Lambda$-$\bar{\Lambda}$ baryonium better matches the enhancement seen by BESIII in $e^{+}e^{-} \to \Lambda \bar{\Lambda} \eta$ with a measured mass of $2356 \pm 7 \pm 17$ MeV~\cite{BESIII:2022tvj}, which are shown in Fig~\ref{fig:lambda-lambda}. 

The $\Omega$-$\bar{\Omega}$ spectrum was analyzed in Ref.~\cite{Wan:2025zau}, where the $0^{-+}$ and $1^{--}$ configurations were found to be sub-threshold, indicative of bound states. On the other hand, the $0^{++}$ and $1^{++}$ configurations exhibit masses exceeding the respective baryon-antibaryon thresholds, which is characteristic of resonance states.

\begin{table}[ht]
\centering
\caption{The predicted masses of light baryonium states in Refs.~\cite{Wan:2021vny,Wan:2025zau, Zhang:2024ulk}.}
\renewcommand{\arraystretch}{1.3}
\begin{tabular}{ccccc}\hline\hline
States&$J^{PC}$&$\sqrt{s_0}\,(\text{GeV})$&$M_B^2\,(\text{GeV}^2)$&$M^X_{J^{PC}}\,(\text{GeV})$ \\ \hline
$N$-$\bar{N}$&$0^{-+}$ &$2.5\pm0.1$&$1.6-2.2$&$1.81\pm0.09$ \\
&$1^{--}$ &$2.5\pm0.1$&$1.6-2.2$ &$1.82\pm0.10$ \\ \hline
$\Lambda$-$\bar{\Lambda}$ &$0^{-+}$ &$2.9\pm0.1$ &$1.9-2.5$&$2.27\pm0.13$ \\
&$1^{--}$ &$2.9\pm0.1$&$1.9-2.5$ &$2.34\pm0.12$ \\ \hline
$\Sigma$-$\bar{\Sigma}$ &$0^{-+}$ &$3.0\pm0.1$ &$2.0-2.6$&$2.39\pm0.13$ \\
&$1^{--}$ &$3.1\pm0.1$&$2.0-2.6$ &$2.48\pm0.12$ \\
\hline
$\Xi$-$\bar{\Xi}$ &$0^{-+}$ &$3.2\pm0.1$ &$2.1-2.8$ &$2.67\pm0.12$ \\
&$1^{--}$ &$3.3\pm0.1$ &$2.1-2.9$ &$2.79\pm0.11$ \\
\hline
$\Omega$-$\bar{\Omega}$ &$0^{-+}$&$3.5\pm0.1$ &$1.8-2.6$ &$3.22\pm0.07$ \\
&$0^{++}$ &$3.8\pm0.1$ &$1.7-2.6$ &$3.46\pm0.09$ \\
&$1^{--}$ &$3.6\pm0.1$ &$1.8-2.6$ &$3.28\pm0.08$ \\
&$1^{++}$ &$3.9\pm0.1$ &$2.2-2.7$ &$3.54\pm0.11$ \\
\hline
$p$-$\bar{\Lambda}$ &$0^{+}$ &$2.9\pm0.1$ &$2.4-2.9$ &$1.96\pm0.03$ \\
&$0^{-}$ &$2.8\pm0.1$ &$1.6-2.2$ &$2.00\pm0.20$ \\
&$1^{-}$ &$2.8\pm0.1$ &$1.6-2.1$ &$2.05\pm0.18$ \\
\hline
$p$-$\bar{\Sigma}$ &$0^{+}$ &$2.9\pm0.1$ &$2.4-2.9$ &$1.98\pm0.04$ \\
&$0^{-}$ &$2.8\pm0.1$ &$1.6-2.1$ &$1.99\pm0.18$ \\
&$1^{-}$ &$2.8\pm0.1$ &$1.7-2.2$ &$2.06\pm0.18$ \\
\hline
\hline
\end{tabular}
\label{table:mass_baryonium}
\end{table}

A distinct class of baryonium consists of non-conjugate baryon-antibaryon pairs, such as the proton-antihyperon ($p$-$\bar{\Lambda}$) system, which expands the landscape of potential exotic states. The existence of such configurations is substantiated by the observation of the $X(2075)$ and $X(2085)$ resonances in the $p\bar{\Lambda}$ final state~\cite{BES:2004fgd, BESIII:2023kgz}. Possible baryonium configurations have been discussed in Ref.~\cite{Zhang:2024ulk}. The authors predicted the possible existence of six $ p \bar{\Lambda}$ and $p \bar{\Sigma}$ molecular states with quantum numbers $J^P=0^{-}, 0^{+}, 1^{-}$, where the values of these states are summarized in Table~\ref{table:mass_baryonium}. Moreover, the authors found the $X(2075)$ lies in the vicinity of $1^{-} p \bar{\Lambda}$ and $p \bar{\Sigma}$ molecular states, which suggested $X(2075)$ may yield large components of $p \bar{\Lambda}$ and $p \bar{\Sigma}$.

Despite extensive theoretical investigations, the existence of light baryonium states remains inconclusive to date. Light baryonium states are generally expected to lie near the 2 GeV energy region, where strong mixing between conventional and exotic hadrons complicates their identification. Due to their relatively low masses and typically narrow decay widths, such states are difficult to observe directly in experiments. Nevertheless, with ongoing advances in experimental techniques and the accumulation of high-precision data, future discoveries and studies of light baryonium states remain promising.

\subsubsection{Heavy sector}\label{sec:heavy_baryonium}\

\begin{figure}[ht]
\centering
\includegraphics[width=0.35\textwidth]{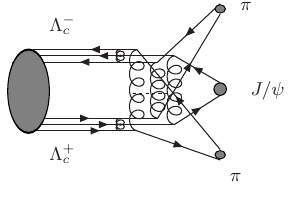}
\caption{Schematic diagram of the decay of the $\Lambda_c$-$\bar{\Lambda}_c$ baryonium state into the $J/\psi\,\pi^{+}\pi^{-}$ final state in Ref.~\cite{Qiao:2005av}.}
\label{fig:Lc-Lc-decay}
\end{figure}

Since the $\Lambda_c$ is the lightest charmed baryon, it is natural to infer that the lowest-lying heavy baryonium state is most likely formed by the $\Lambda_c$-$\bar{\Lambda}_c$ system. In a naive picture, this baryonium state is expected to decay readily into the $J/\psi\,\pi^{+}\pi^{-}$ final state, as illustrated schematically in Fig.~\ref{fig:Lc-Lc-decay}. Similarly, the $J/\psi$ in the final state may be replaced by its radial excitation $\psi^\prime$, albeit with a smaller branching fractions. Open-charm processes may also contribute to decay channels such as $D^{(*)}\bar{D}^{(*)}\pi$, which could offer further insight into the internal structure and decay dynamics of baryonium states.

In contrast, a decay mode involving only the $D^{(*)}\bar{D}^{(*)}$ final state is almost impossible. A possible mechanism for generating such a final state is through the annihilation of a light quark-antiquark pair into a photon, followed by electromagnetic production of the $D^{(*)}\bar{D}^{(*)}\gamma$ final state. As this process proceeds via electromagnetic interaction, its rate is expected to be two orders of magnitude lower than that of strong decay modes. Furthermore, due to the absence of strange quarks in the initial $\Lambda_c$-$\bar{\Lambda}_c$ system, decays into final states containing strangeness, such as $D_s^{(*)}\bar{D}_s^{(*)}$ or $J/\psi \,K^{+}K^{-}$, should also be suppressed.

Based on the properties of the $\Lambda_c$-$\bar{\Lambda}_c$ baryonium system, Qiao proposed this hadronic picture~\cite{Qiao:2005av}, further extended to include configurations involving $\Sigma_c$-$\bar{\Sigma}_c$~\cite{Qiao:2007ce}, to account for several experimentally observed neutral charmonium-like states with quantum numbers $J^{PC}=1^{--}$, such as $Y(4260)$~\cite{BaBar:2005hhc}, $Y(4360)$~\cite{BaBar:2006ait,Belle:2007umv}, and $Y(4660)$~\cite{Belle:2007umv}. These $Y$ states appear to exceed the expected number of conventional $c\bar{c}$ charmonium states, presenting a challenge to traditional quark model classifications~\cite{Nielsen:2009uh,Albuquerque:2018jkn}. The baryonium picture offers a plausible alternative framework, though this interpretation remains one of several competing scenarios, in which the predicted allowed and suppressed decay modes are broadly consistent with current experimental observations. For instance, the $Y$ states do not correspond to enhancements in the $e^{+}e^{-} \to D^{(*)\pm}D^{(*)\mp}$ cross sections measured by the Belle~\cite{Belle:2006hvs} and BaBar~\cite{BaBar:2009elc} collaborations, decay channels that are also inaccessible or strongly suppressed within the $\Lambda_c$-$\bar{\Lambda}_c$ baryonium scenario.

Later in Ref.~\cite{Liu:2007tj}, the authors predicted the mass of the lowest heavy baryonium states within the framework of the large $N_c$ QCD combining with the HQET. The mass of $1^-$ $\Lambda_c$-$\bar{\Lambda}_c$ baryonium was estimated to be $4.78\pm0.15$ GeV, which is much larger than the mass of the $Y(4260)$ and $Y(4360)$ states. Furthermore, an intriguing theoretical insight from their analysis is that a baryonium state in the large $N_c$ limit can be identified with a glueball composed of $N_c$ valence gluons, suggesting a deep connection between baryon-antibaryon bound states and purely gluonic excitations.

In Refs.~\cite{Chen:2011cta, Chen:2013sba}, the authors studied the interactions between heavy baryon-antibaryon pairs within the framework of heavy baryon chiral perturbation theory (HBCPT), and demonstrated that bound states such as $\Lambda_c$-$\bar{\Lambda}_c$, $\Sigma_c$-$\bar{\Sigma}_c$, and $\Lambda_b$-$\bar{\Lambda}_b$ may exist under suitable choices of coupling constants and cutoff parameters. They further predicted the masses of these baryonium states with spin-0 and 1, and found them to be highly sensitive to the model inputs. Within this framework, their results suggested that the $Y(4260)$ and $Y(4360)$ resonances could be interpreted as spin-triplet $\Lambda_c$-$\bar{\Lambda}_c$ baryonium states. In contrast, the $Y(4660)$ could not be easily accommodated as either a spin-singlet or spin-triplet $\Lambda_c$-$\bar{\Lambda}_c$ bound state within this framework.

In Ref.~\cite{Lee:2011rka}, the one boson exchange model was extended to study possible molecular states composed of a pair of heavy baryons. The authors considered the interactions between $\Lambda_c$-$\bar{\Lambda}_c$, $\Sigma_c$-$\bar{\Sigma}_c$, and $\Xi_c$-$\bar{\Xi}_c$ systems, incorporating contributions from $\pi$, $\eta$, $\rho$, $\omega$, $\phi$, and $\sigma$ meson exchanges. Their analysis revealed that the $\Lambda_c$-$\bar{\Lambda}_c$ system could form a bound state with a binding energy of a few to several tens of MeV, depending on the choice of cutoff parameters. Similarly, the $\Sigma_c$-$\bar{\Sigma}_c$ and $\Xi_c$-$\bar{\Xi}_c$ systems were found to exhibit attractive interactions, suggesting the possibility of forming molecular states under suitable conditions. In addition, the authors emphasized that the short-range interactions require careful treatment, although a weakly bound heavy baryonium state can still exist even in the absence of these short-range contributions.

In Ref.~\cite{Deng:2013aca}, the color flux-tube model was employed to investigate baryonium states composed of strange, charm, and bottom quarks. The analysis indicates that many low-spin baryonium configurations exhibit sizable binding energies, making them stable against dissociation into free baryon-antibaryon pairs. Instead, these states are expected to decay via flux-tube breaking and recombination processes, leading to final states consisting of three mesons. The authors further proposed that the $Y(4260)$ and $Y_b(10890)$ resonances could be interpreted as $\Lambda_c$-$\bar{\Lambda}_c$ and $\Lambda_b$-$\bar{\Lambda}_b$ bound states.

In Ref.~\cite{Lu:2017dvm}, the authors employed effective field theory to investigate heavy baryon-antibaryon systems. It was found that pion exchange plays a significant role in isoscalar $\Sigma_Q^{(*)}$-$\bar{\Sigma}_Q^{(*)}$ systems, whereas it may be absent in $\Xi_Q$-$\bar{\Xi}_Q$ and $\Lambda_Q$-$\bar{\Lambda}_Q$ molecular configurations. Another key conclusion of this work is the identification of the quantum number channels in which the formation of heavy baryonium states is most likely. The most favorable candidates for forming bound states were theoretically suggested to be the isoscalar $\Lambda_Q$-$ \bar{\Lambda}_Q$, $\Sigma_Q$-$\bar{\Sigma}_Q$, $\Sigma_Q^*$-$\bar{\Sigma}_Q$, and $\Sigma_Q^*$-$ \bar{\Sigma}_Q^*$ systems, as well as the isovector $\Lambda_Q$-$\bar{\Sigma}_Q$ and $\Lambda_Q$-$\bar{\Sigma}_Q^*$ configurations. 

In Ref.~\cite{Liu:2021gva}, the mass spectrum of hidden-charm hexaquark states was investigated within the framework of the color-magnetic interaction (CMI) model. Mass predictions were provided for a variety of baryonium configurations. The analysis indicated that hexaquark states with lower isospin tend to be more compact. In particular, hidden-charm hexaquark states with exotic quantum numbers, such as $J^{PC} = 0^{--}, 1^{-+},$ and $3^{-+}$, were identified as plausible candidates deserving further investigation. Furthermore, the authors discussed the possible strong two-body decay modes of these baryonium states and their relevance to future experimental searches.

In Ref.~\cite{Salnikov:2023qnn}, by taking into account the BESIII experimental data on the cross section of $e^+ e^-\to\Lambda_c \bar{\Lambda}_c$ process, the authors carried out the analysis of various contributions to the $\Lambda_c \bar{\Lambda}_c$ interaction potential. The authors predicted the existence of a narrow sub-threshold resonance in the $\Lambda_c \bar{\Lambda}_c$ system with a mass of 38 MeV below the threshold. More recently, a comprehensive coupled-channel analysis was carried out in Ref.~\cite{Nakamura:2023exd}, in which a global fit to the $e^+e^-\to c\bar{c}$ cross sections in $\sqrt{s}=3.75-4.7$\,GeV was performed using BESIII data and a $\Lambda_c\bar{\Lambda}_c$ bound state was extracted. Due to coupled-channel effects, signatures of this bound state also manifest in other processes, such as a sharp dip at 4.57\,GeV in $e^+e^-\to\pi D^*\bar{D}^*$. Moreover, the $\Lambda_c\bar{\Lambda}_c$ bound state was found to decay significantly into $D_1\bar{D}^*$, providing an interesting insight from the analysis of the BESIII data.

Besides, the QCD sum rule approach provides a powerful framework for studying baryonium states. By constructing suitable interpolating currents with well-defined $J^{PC}$ quantum numbers, one can systematically analyze the mass spectra and decay properties of these states within the QCD sum rule formalism.

In Ref.~\cite{Chen:2016ymy}, Chen et al.~\cite{Chen:2016ymy} investigated hidden-charm baryonium states and identified the $J^{PC}=3^-$ channel as a prime candidate for experimental search based on QCD sum rules results. Unlike other configurations, the $3^-$ state is inaccessible via $S$-wave $[\bar{c} c+\pi \pi]$ and $[\bar{c} c \bar{q} q+\pi]$ structures, meaning its signal would not be obscured by common scattering backgrounds. This decoupling implies that observing a $3^-$ resonance would provide strong motivation for a compact exotic interpretation, since this channel is inaccessible via common $S$-wave scattering backgrounds. Therefore, the study advocated for a dedicated search in the $D$-wave $J/\psi \pi \pi$ and $P$-wave $J/\psi \rho(\omega)$ modes.

Later, an improved calculation incorporating higher-order nonperturbative terms $\langle\bar{q} q\rangle^2\langle G^2\rangle$ was performed by Wan et al~\cite{Wan:2019ake}. The authors constructed appropriate interpolating currents for $\Lambda_c$-$\bar{\Lambda}_c$ and $\Lambda_b$-$\bar{\Lambda}_b$ baryonium states:
\begin{align}
j^{0^{-+}}(x)&= i\,\bar{\eta}_{\mathcal{B}}(x) \gamma_5 \eta_{\mathcal{B}}(x) \,,\label{baryonium-Ja0-+}\\
j^{1^{--}}_\mu(x)&=\bar{\eta}_{\mathcal{B}}(x) \gamma_\mu \eta_{\mathcal{B}}(x)\,, \label{baryonium-Ja1--}\\
j^{0^{++}}(x)&= \bar{\eta}_{\mathcal{B}}(x) \eta_{\mathcal{B}}(x)\,,\label{baryonium-Ja0++}\\
j^{1^{++}}_\mu(x)&= i\,\bar{\eta}_{\mathcal{B}}(x) \gamma_\mu\gamma_5 \eta_{\mathcal{B}}(x) \,, \label{baryonium-Ja1++}
\end{align}
where $\eta_{\mathcal{B}}$ denotes the lowest-lying heavy baryon interpolating currents, identical to that defined in Eq.~\eqref{current-Lambda_Q}. The spin-parity quantum numbers of the resulting baryonium states are determined by the specific Lorentz structures inserted into the current. By evaluating the two-point correlation functions, the authors identified the possible existence of two $\Lambda_b$-$\bar{\Lambda}_b$ baryonium states with quantum numbers $J^{PC} = 0^{++}$ and $1^{--}$, along with two hidden-charm counterparts. They further pointed out that the $ Y(4660) $ state observed in $\psi(2 S)\pi^{+} \pi^{-} $ invariant mass spectrum~\cite{Belle:2007umv} is close in magnitude to $\Lambda_c$-$\bar{\Lambda}_c$ systems with $J^{PC}=1^{--}$. They also suggested searching for signals of the $\Lambda_b$-$\bar{\Lambda}_b$ baryonium in the $\Upsilon\, \pi^{+} \pi^{-}$ final state.

Similar analyses were also performed by Wang et al.~\cite{Wang:2021qmn,Wang:2021pua}, with the $\Sigma_c$-$\bar{\Sigma}_c$ baryonium states involved. The detailed results are summarized in Table~\ref{table:QCDSR-heavy-baryonium}. In addition, the exotic hidden-charm and -bottom baryonium states with quantum numbers $0^{--}$ and $0^{+-}$ were explored in Ref.~\cite{Wan:2023epq}. The masses of the $0^{--}$ and $0^{+-}$ hidden-charm baryonium states were predicted to lie in the range of 5.2$\sim$5.5 GeV and 4.8$\sim$5.7 GeV, respectively. The corresponding hidden-bottom partners were found to span 11.68$\sim$12.28 GeV and 11.38$\sim$12.23 GeV, respectively. 

The non-conjugate $\Lambda_c$-$\bar{\Sigma}_c$ hexaquark configuration has also been investigated. Zhang and Qiao~\cite{Zhang:2025jqx} performed a QCD sum rule analysis of the $\Lambda_c\bar{\Sigma}_c$ and $\Lambda_b\bar{\Sigma}_b$ systems, finding ground-state masses around $5.7$--$5.8$ GeV and $11.9$ GeV, respectively, which do not support a bound-state interpretation, consistent with the BESIII findings~\cite{BESIII:2025zgc}. The $\bar{\Lambda}_c\Sigma_c$ system was also investigated via QCD sum rules by Wang, Wang, and Yu~\cite{Wang:2026lta}, who identified possible molecular bound states with $J^P=0^-$ and $1^-$ and resonance states for other configurations.

\begin{table}[ht]
\centering
\caption{The predictions for the mass ($ \text{GeV} $) of heavy baryoniums from QCD sum rules as well as other theoretical models.}
\renewcommand{\arraystretch}{1.3}
\resizebox{\textwidth}{!}{
\begin{tabular}{ccccccc}
\hline\hline
State&$J^{PC}$ &QCDSR& Large $ N_c $~\cite{Liu:2007tj} & HBCPT~\cite{Chen:2013sba} & CMI model~\cite{Liu:2021gva} \\ \hline

$\Lambda_c$-$\bar{\Lambda}_c$&$0^{+}$ &5.00~\cite{Chen:2016ymy}, 5.19$\pm$0.24~\cite{Wan:2019ake}, 5.11$^{+0.15}_{-0.12}$~\cite{Wang:2021qmn}&&\multirow{2}{*}{4.41$\sim$4.57}& \\
&$0^{-}$ &4.66$^{+0.10}_{-0.06}$~\cite{Wang:2021qmn}&&&4.65$\sim$4.77 \\ 
&$1^{+}$ &4.89~\cite{Chen:2016ymy}, 4.99$^{+0.10}_{-0.09}$~\cite{Wang:2021qmn}&&\multirow{2}{*}{4.29$\sim$4.56}& \\ 
&$1^{-}$ &4.78$\pm$0.23~\cite{Wan:2019ake}, 4.68$^{+0.08}_{-0.08}$~\cite{Wang:2021qmn} &4.78$\pm$0.15&&4.58$\sim$4.94 \\
&$2^{+}$ &5.15~\cite{Chen:2016ymy} \\
&$2^{-}$ &4.83~\cite{Chen:2016ymy} \\ 
&$3^{+}$ &5.68~\cite{Chen:2016ymy} \\
&$3^{-}$ &5.04~\cite{Chen:2016ymy} \\ \hline 

$\Sigma_c$-$\bar{\Sigma}_c$&$0^{+}$ &$5.23_{-0.07}^{+0.07}$~\cite{Wang:2021pua}&&\multirow{2}{*}{4.88$\sim$4.90}& \\ 
&$0^{-}$ &$4.88_{-0.08}^{+0.08}$~\cite{Wang:2021pua}&&& \\ 
&$1^{+}$ &$5.31_{-0.07}^{+0.07}$~\cite{Wang:2021pua}&&\multirow{2}{*}{4.87$\sim$4.90}& \\ 
&$1^{-}$ &$4.88_{-0.08}^{+0.09}$~\cite{Wang:2021pua}\\\hline

$\Lambda_b$-$\bar{\Lambda}_b$&$0^{+}$ &11.84$\pm$0.22~\cite{Wan:2019ake}&&\multirow{2}{*}{11.21$\sim$11.23}& \\
&$0^{-}$ & \\ 
&$1^{+}$ &&&\multirow{2}{*}{11.20$\sim$11.23}& \\ 
&$1^{-}$ & 11.72$\pm$0.26~\cite{Wan:2019ake} \\ \hline

$\Lambda_c$-$\bar{\Sigma}_c$&$0^{-}$ &$5.72\pm0.08$~\cite{Zhang:2025jqx}$$ &&& \\[5pt]
&$0^{+}$ &$5.77\pm0.06$~\cite{Zhang:2025jqx}$$ &&& \\[5pt]
&$1^{-}$ &$5.79\pm0.09$~\cite{Zhang:2025jqx}$$, $4.72\pm0.09$~\cite{Wang:2026lta} &&& \\[5pt]
&$1^{+}$ &$5.82\pm0.09$~\cite{Zhang:2025jqx}$$ &&& \\ \hline
$\Lambda_b$-$\bar{\Sigma}_b$&$0^{-}$ &$11.86\pm0.10$~\cite{Zhang:2025jqx}$$ &&& \\[5pt]
&$0^{+}$ &$11.89\pm0.07$~\cite{Zhang:2025jqx}$$ &&& \\[5pt]
&$1^{-}$ &$11.92\pm0.11$~\cite{Zhang:2025jqx}$$ &&& \\[5pt]
&$1^{+}$ &$11.95\pm0.07$~\cite{Zhang:2025jqx}$$ &&& \\ \hline

\hline\hline
\end{tabular}
}
\label{table:QCDSR-heavy-baryonium}
\end{table}

Furthermore, heavy baryonium states may also be formed from doubly heavy baryon-antibaryon pairs. In Refs.~\cite{Wang:2021wjd,Wan:2022uie}, motivated by the $X(7200)$ enhancement observed in the $J/\psi$-pair spectrum and its proximity to the $\bar{\Xi}_{cc} \Xi_{cc}$ threshold, the authors proposed a $\Xi_{cc}$-$\bar{\Xi}_{cc}$ baryonium interpretation for this state. They constructed suitable interpolating currents and performed a QCD sum rule analysis, which showed that the predicted mass of a $\Xi_{cc}$-$\bar{\Xi}_{cc}$ baryonium with $J^{PC}=0^{-+}$ is consistent with the observed $X(7200)$ mass. Possible decay channels were also discussed, including transitions into $J/\psi$-pair and $D^{(*)}\bar{D}^{(*)}$ final states.

In summary, heavy baryonium represents a compelling class of multiquark configurations that extend our understanding of hadronic matter beyond conventional mesons and baryons. Although theoretical investigations have predicted a variety of possible heavy baryon-antibaryon bound states with distinctive quantum numbers and decay patterns, firm experimental evidence remains limited. Further progress requires not only refined theoretical treatments, such as more accurate handling of short-range dynamics and nonperturbative effects, but also dedicated experimental efforts to explore relevant final states near heavy baryon-antibaryon thresholds. As future facilities continue to provide high-precision data, especially in the heavy-flavor sector, the heavy baryonium picture may play an increasingly important role in deciphering the rich structure of the hadron spectrum.

\subsection{Triquark-antitriquark}\

As mentioned earlier, baryonium states do not necessarily consist of two color-singlet baryons, they may also arise from two color non-singlet clusters, i.e., within a triquark-antitriquark picture. In the diquark picture, a triquark ($q_1 q_2 \bar{q}_3$) is viewed as a bound state composed of one diquark ($q_1 q_2$) and an antiquark ($\bar{q}_3$). The diquark is composed of two quarks, where the color decomposition is:
\begin{align}
3_c \otimes 3_c=\bar{3}_c \oplus 6_c\,.
\end{align}
Note that, due to the stronger QCD attraction in the antisymmetric configuration $\bar{3}_c$, the physically more relevant diquark structure is typically taken to be in the color antitriplet. In this review, however, we consider both the $\bar{3}_c$ and $6_c$ configurations. Accordingly, the color structure of the triquark is given by:
\begin{align}
q_1 q_2 \bar{q}_3=(q_1q_2)+\bar{q}_3=(3_c \otimes 3_c) \otimes \bar{3}_c=(\bar{3}_c \oplus 6_c)\otimes \bar{3}_c=3_c \oplus \bar{6}_c\oplus 3_c\oplus 15_c\,.
\end{align}
Then the general form of the interpolating current for the triquark can be written as:
\begin{align}
\label{triquark-1}
j^{triquark}_{\bar{3}_c\otimes \bar{3}_c}&\simeq \varepsilon^{d e c} \varepsilon^{e a b}\big[q_{1a}^T \Gamma_1 q_{2b}\big] \Gamma_2 \bar{q}_{3c}\,, \\[5pt]
\label{triquark-2}
j^{triquark}_{6_c\otimes \bar{3}_c}&\simeq\big[q_{1a}^T \Gamma_1 q_{2}^b\big] \Gamma_2 \bar{q}_{3b}\,.
\end{align}
Triquark correlations were calculated in QCD sum rules~\cite{Lee:2004dp}, where the authors adopted the following interpolating currents:
\begin{align}
j^1_{\bar{3}_c\otimes \bar{3}_c}&=\frac{1}{4} \varepsilon^{a b c} \varepsilon^{b d e}\big[u_{d}^T \Gamma d_{e}\big] C \bar{s}_{c}^T\,, \quad j^2_{\bar{3}_c\otimes \bar{3}_c}=\frac{1}{4} \varepsilon^{a b c} \varepsilon^{b d e}\big[u_{d}^T \Gamma_1 d_{e}\big] \gamma_5 C \bar{s}_{c}^T\,,\nonumber\\[5pt]
j_{6_c\otimes \bar{3}_c}&=\frac{1}{4 \sqrt{3}}\left[u_{a}^T C \gamma_\mu d^{b}+u^{bT} C \gamma_\mu d_{a}\right] \gamma_5 \gamma^\mu C \bar{s}_{b}^T\,.
\end{align} 

On the other hand, a triquark can also be composed of a quark-antiquark pair coupled to an additional quark. The color decomposition of quark-antiquark pair is:
\begin{align}
3_c \otimes \bar{3}_c=1_c \oplus 8_c\,.
\end{align}
By further coupling to another quark in a color triplet, one obtains:
\begin{align}
q_1 q_2 \bar{q}_3=(q_1\bar{q}_2)+q_3=3_c \otimes \bar{3}_c \otimes 3_c=(1_c \oplus 8_c)\otimes 3_c=3_c \oplus 3_c \oplus \bar{6}_c\oplus 15_c\,,
\end{align}
where the corresponding interpolating currents can be expressed as:
\begin{align}
\label{triquark-3}
j^{triquark}_{1_c\otimes 3_c}&\simeq \big[q_{1a} \Gamma_1 \bar{q}_{2}^a\big] \Gamma_2 \bar{q}_{3b}\,, \\[5pt]
\label{triquark-4}
j^{triquark}_{8_c\otimes 3_c}&\simeq\big[q_{1a} \Gamma_1 (t^A_{ab}) \bar{q}_{2}^b\big] \Gamma_2 q_{3b}\,.
\end{align}
Here, $t^A=\lambda^A/2$ are the Gell-Mann matrices.

\begin{table}[ht]
\centering
\caption{Comparison of $p$-$\bar{\Lambda}$ states between baryon-antibaryon and triquark-antitriquark configurations from QCD sum rules~\cite{Zhang:2024ulk, Zhang:2025vqg}. All values are in units of GeV.}
\renewcommand{\arraystretch}{1.4}
\begin{tabular}{cccc}\hline\hline
States&$J^{PC}$&Baryon-antibaryon & Triquark-antitriquark \\ \hline
$p$-$\bar{\Lambda}$ &$0^{+}$ &$1.96\pm0.03$ & $1.99\pm0.02$ \\
&$0^{-}$ &$2.00\pm0.20$ & $1.84\pm0.21$ \\
&$1^{-}$ &$2.05\pm0.18$&$2.01\pm0.20$ \\
&$1^{+}$ &&$1.97\pm0.03$ \\
\hline
$p$-$\bar{\Sigma}$ &$0^{+}$ &$1.98\pm0.04$ &$1.99\pm0.02$ \\
&$0^{-}$ &$1.99\pm0.18$ & $1.84\pm0.21$ \\
&$1^{-}$ &$2.06\pm0.18$&$2.02\pm0.20$ \\
&$1^{+}$ &&$1.97\pm0.02$ \\
\hline
\hline
\end{tabular}
\label{table:mass_pLambda}
\end{table}

The interpolating currents of the triquark~\eqref{triquark-1}, \eqref{triquark-2}, \eqref{triquark-3}, and \eqref{triquark-4} can be used to construct the corresponding triquark-antitriquark interpolating currents. Discussion of such configurations is necessary, as the use of QCD sum rules in calculating molecular states has been subject to skepticism. The first attempt to study the compact hexaquark states using QCD sum rules was made by Zhang et al.~\cite{Zhang:2025vqg}, where the authors employed the current~\eqref{triquark-2} to construct triquark-antitriquark states. As a result, a total of six independent and non-degenerate hexaquark candidates were predicted as potential. While the two $J^P=1^{-}$ states exhibit masses compatible with the $X(2075)$, the two $J^P=1^{+}$ states differ significantly from the observed $X(2075)$ and $X(2085)$ signals. The remaining $J^P=0^{+}$ and $0^{-}$ states are proposed as potential compact hexaquark candidates. 

Furthermore, the detailed mass predictions are summarized in Table~\ref{table:mass_pLambda}, alongside the corresponding baryon-antibaryon results from Ref.~\cite{Zhang:2024ulk} for comparison. The $1^+$ $p$-$\bar{\Lambda}$ and $p$-$\bar{\Sigma}$ systems offer a unique opportunity to discriminate between the molecular and compact hexaquark descriptions. According to QCD sum rules, the baryon-antibaryon configuration precludes the formation of these states, whereas the triquark-antitriquark framework accommodates them. Consequently, the observation of such states would provide supporting evidence for a compact hexaquark configuration, though definitive structural determination requires complementary information from decay patterns and production mechanisms, serving as a valuable guide for future experiments.

Fundamentally, the molecular and compact hexaquark interpretations differ in their underlying binding mechanisms. Although the two pictures may yield similar mass predictions, in the molecular picture, loosely bound $B\bar{B}$ states are held together by residual strong interactions (e.g., one-boson exchange), with binding energies typically on the order of tens of MeV. In contrast, compact hexaquark states are bound directly by gluon exchange between color-non-singlet clusters, resulting in larger binding energies and distinct decay patterns---molecular states tend to dissociate into their constituent hadron pairs, while compact states may decay via quark rearrangement into mesons. For a given experimentally observed resonance, distinguishing between these scenarios requires not only mass matching but also analysis of binding energies, decay channels, and production mechanisms. Furthermore, near-threshold resonant structures may also arise from threshold cusp effects or final-state interactions without involving a genuine bound state, further complicating the structural identification. Spin-parity determinations and coupled-channel analyses are therefore essential for disentangling these possibilities.

Consequently, a comprehensive investigation into compact hexaquark states remains a crucial frontier in hadron physics. Systematically exploring the remaining triquark-antitriquark configurations may help clarify the nature of the $X(2075)$ and $X(2085)$ and provide a basis for distinguishing compact multiquark structures from hadronic molecules. Such studies will ultimately deepen our understanding of non-perturbative QCD dynamics and color confinement, offering vital guidance for identifying new exotic states in future experimental searches.

\section{Summary and outlook}

In this review, we have discussed the progress made over the past decades in the study of the baryon spectrum and semileptonic decays, as well as the baryonium states within the framework of QCD sum rules. As one of the most widely applied theoretical approaches in hadron physics, QCD sum rules have achieved remarkable success in the study of hadron spectroscopy and decays. In general, in the study of hadron spectroscopy, QCD sum rules are an effective tool when a newly observed resonance requires confirmation of its existence. Moreover, the method allows for theoretical construction and analysis of hypothetical resonances that have not yet been observed experimentally, providing insight into their possible existence. In the study of semileptonic decays, QCD sum rules enable the calculation of form factors, thus facilitating phenomenological analysis such as the calculation of branching fractions and the exploration of potential signals of new physics. A brief summary and discussion of this review are given as follows:
\begin{itemize}[leftmargin=*, align=left]
\item The basic methodology of QCD sum rules is outlined in chapter~\ref{sec:method}, where phenomenological parameters such as hadron masses, decay constants, and form factors can be extracted.

\item A QCD sum rule analysis of baryon spectra is reviewed in chapter~\ref{sec:spectra}. Our understanding of conventional ground-state baryons is relatively well established, while some newly observed baryonic states remain subject to considerable debate. On the one hand, these states may be interpreted as excited baryons. On the other hand, some resonances with masses lying above the thresholds of two-hadron systems are proposed to be pentaquark molecular states, where the baryon candidates for hadronic molecules can be found in Ref~\cite{Guo:2017jvc}. QCD sum rules provide mass predictions and decay analysis of different hadron states, offering valuable insights into the internal structure of these resonances. In addition, QCD sum rules have played a significant role in motivating the discovery of the doubly charmed baryon $\Xi_{cc}$, and have also predicted the possible existence of triply heavy baryons and hybrid baryons.

\item A QCD sum rule analysis of baryon semileptonic decays is reviewed in chapter~\ref{sec:decay}. The study of baryon semileptonic decays had long progressed slowly, but QCD sum rules have provided significant theoretical support to this field. Within this framework, the form factors, which are key quantities closely related to the decay kinematics, can be directly calculated. These form factors then can be used to construct decay observables, which play a central role in flavor physics analyses, such as the extraction of CKM matrix elements, the investigation of flavor symmetry breaking effects, and the search for potential signals of new physics.

\item A QCD sum rule analysis of baryonium states is reviewed in chapter~\ref{sec:baryonium}. Baryonium states have long attracted attention from physicists, particularly the hypothetical bound state of a proton and antiproton, often referred to as protonium. Experimentally observed structures such as $X(1835)$, $X(1840)$, and $X(1880)$ have been suggested as potential candidates for such states. The concept of baryonium is relatively broad, allowing for constituents that are not strictly color singlets, which is different from hadronic molecular states. Most existing theoretical studies focus on baryoniums composed of two color-singlet hadrons, while investigations of baryonium with compact multiquark configurations remain limited. Within the framework of QCD sum rules, one can construct appropriate interpolating currents for baryoniums and compute their masses, enabling comparison with predictions from other theoretical approaches. Currently, no conclusive experimental evidence confirms the existence of baryonium as a specific internal structure. While the observation of several near-threshold structures in baryon-antibaryon channels is consistent with baryonium interpretations, such structures may also arise from threshold effects, final-state interactions, or conventional hadronic resonances. Distinguishing these possibilities requires determination of spin-parity quantum numbers and coupled-channel analyses. In the light-quark sector, strong mixing between conventional and exotic resonances makes it difficult to disentangle their nature. In contrast, the heavy-quark sector may offer better prospects for observing such exotic states due to relatively larger energy level spacings.

\end{itemize}

We have tried our best to include the key theoretical developments and experimental findings that have shaped our current understanding of these topics, however, many valuable contributions could not be included due to space limitations. Nonetheless, several challenges and open questions remain, which also point toward promising directions for future development:
\begin{itemize}[leftmargin=*, align=left]

\item Theoretical uncertainties: As pointed out in many works, the major problem of QCD sum rules is the theoretical uncertainties. The numerical results are often sensitive to the values of input parameters, such as quark masses, condensate values, the Borel window, and the continuum threshold, leading to sizable theoretical uncertainties. Thus, the predictive power of QCD sum rules for high-precision experiments has been subject to skepticism by some researchers. An effective approach to improve the precision of QCD sum rule calculations is to include next-to-leading order corrections to the perturbative part. NLO corrections have been preliminarily applied in the mass spectrum calculations of baryons and certain tetraquark states~\cite{Wang:2017qvg,Wu:2021tzo,Wu:2022qwd,Wu:2023ntn,Fanomezana:2014qya,Albuquerque:2020ugi,Albuquerque:2021erv}, yet their application to a broader range of systems, including more baryon species, multiquark states, and semileptonic decay form factors, remains an important priority. Additionally, a kind of inverse problem method is expected to show the possibility of handling the uncertainty issue~\cite{Li:2020xrz,Li:2020ejs,Xiong:2022uwj,Li:2021gsx,Zhao:2024drr}.

\item Continuum threshold uncertainties: The choice of $s_0$ currently relies on heuristic criteria such as $s_0\approx(M_H+1\,\text{GeV})^2$ combined with $\tau^2$ stability. A more quantitative approach is self-consistency: requiring the extracted $M_H$ to satisfy $s_0\approx(M_H+\Delta E)^2$, where $\Delta E$ can be constrained by experimental or lattice QCD input for the first excited state. The $\tau^2$-dependent threshold method of Refs.~\cite{Wu:2021tzo,Wu:2022qwd,Wu:2023ntn} and Bayesian parameter-inference frameworks offer further avenues for reducing this key source of systematic uncertainty.

\item Higher-order OPE with lattice-grounded condensates: Most baryonic QCD sum rule studies treat higher-dimensional condensates through the vacuum saturation approximation. Improved lattice QCD determinations of four-quark and mixed condensates could directly reduce uncertainties in the QCD representation, especially for exotic states where higher-dimensional operators are not negligible.

\item Molecular limitations: When QCD sum rules are used to study the structure of molecular states, the interaction between the two components of the molecular state can be described by gluon condensates, as shown in Fig~\ref{fig:molecular} (a). According to the one-meson-exchange model, the interaction between the two components in a molecular state can be described by single meson exchange, as illustrated in Fig~\ref{fig:molecular} (b). In QCD sum rule calculations for molecular states, the meson exchange contributions to the binding dynamics are not fully considered, which may pose certain limitations when dealing with molecular states.

\begin{figure}[ht]
\centering
\raisebox{0.8cm}{\includegraphics[width=0.25\textwidth]{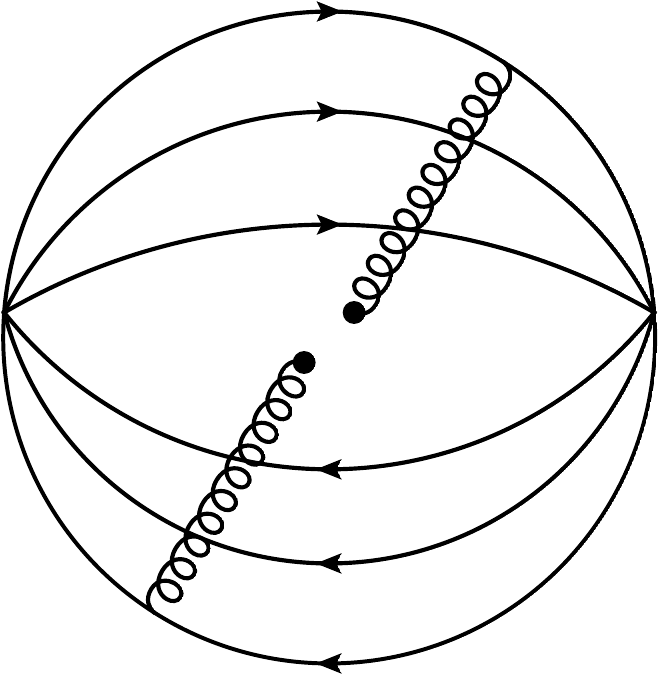}}\hspace{2cm}\includegraphics[width=0.4\textwidth]{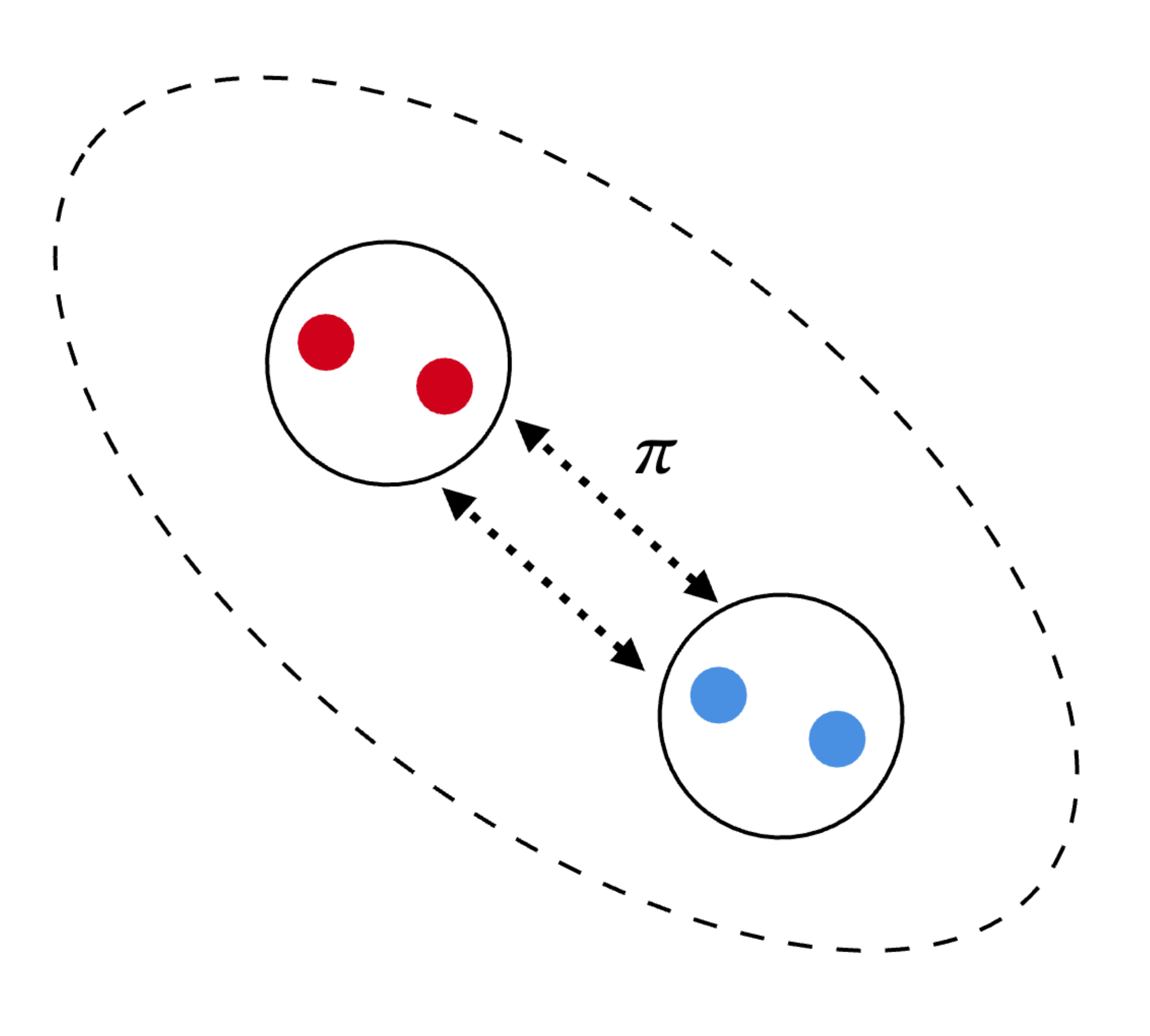}

(a) \hspace{7.5cm} (b)
\caption{(a) The gluon condensate and (b) the one-meson-exchange picture of the molecular state.}
\label{fig:molecular}
\end{figure}

\item Physical range of form factors: Due to kinematic constraints of the decay process, QCD sum rules calculation for form factors only works in the low-$q^2$ region. Therefore, certain analytic functions are required to extrapolate the results to the full physical kinematic range. However, various functional forms can be chosen for the analytic extrapolation, which introduces additional model-dependent uncertainties. Combining results from other theoretical approaches, such as lattice QCD which is reliable in the high-$q^2$ region, offers a promising way to address this issue. 

\item Experimental prospects for baryonium: Identifying and prioritizing experimental channels most suitable for testing baryonium assignments is essential. Promising channels include: (i) $J/\psi \to \gamma p\bar{p}$ and $J/\psi \to \gamma \eta'\pi^+\pi^-$ for light baryoniums; (ii) $e^+e^- \to J/\psi\pi^+\pi^-$ and $e^+e^- \to \Lambda_c\bar{\Lambda}_c$ for hidden-charm baryoniums; (iii) $\Upsilon\pi^+\pi^-$ for hidden-bottom baryoniums. Dedicated coupled-channel analyses and spin-parity determinations near $B\bar{B}$ thresholds will be crucial for disentangling baryonium signals from threshold effects and conventional resonances.

\item Applicability issues in $b$-hadron decays: Ball and Braun pointed out that the conventional three-point QCD sum rules lose their effectiveness in the study of $b$-baryon decays~\cite{Ball:1997rj}. Specifically, as the $b$-quark mass increases, the coefficient of the $\langle \bar{q}q\rangle^2$ condensate grows faster than that of the perturbative term~\cite{Khodjamirian:1997lay}, leading to a breakdown of the convergence of the operator product expansion. Although alternative approaches such as light-cone sum rules and heavy quark effective theory have been applied to overcome these difficulties, achieving a full QCD description of the $b$-hadron decays remains a fundamental objective.

\end{itemize}

Concerning these open questions and limitations, QCD sum rules alone are not sufficient to provide a complete understanding of the hadron physics. Therefore, it is essential to combine QCD sum rules with other non-perturbative approaches, such as lattice QCD, relativistic quark model, and light-cone sum rules, to achieve a more comprehensive understanding of these topics. Systematic benchmarking of QCD sum rule predictions against lattice QCD results for baryon masses, decay constants, and form factors, particularly in the heavy-flavor sector where lattice calculations are reaching higher precision, will be increasingly valuable in this regard.

Additionally, the introduction of other new technologies such as artificial intelligence (AI) may offer new opportunities for research in hadron physics. For instance, in Ref.~\cite{Ghalenovi:2024scv}, a deep neural network was employed to solve the Schrödinger equation, from which the authors investigated the mass spectra and semileptonic decays of doubly heavy baryons within the hypercentral quark model. Conditional generative adversarial networks have also been developed to predict the masses and decay widths of fully-heavy tetraquarks~\cite{Rostami:2025sff,Malekhosseini:2025hyx}. Applications of AI to other areas of physics can be found in Refs.~\cite{Ghosh:2022zdz,Chan:2023ume,Zhou:2023pti,Jiao:2024dst,Fang:2025fmv,Wu:2025wvv}. We believe that with the introduction of new technologies and encouraging improvements in experiments, many of the outstanding problems in hadron physics will eventually be resolved.

\vspace{0.5cm}
%%%%%%%%%%%%%%%%%%%%%%%%%%%%%%%%%%%%%%%%%%%%%%%%%%%%%%%%%%
{\Large\bf Declaration}

{\bf Acknowledgments}

We would like to express our sincere appreciation to Zi-Qiang Chen, Qi-Ming Feng, Liang Tang, Bing-Dong Wan, and Hao Yang for their valuable discussions and contributions.

{\bf Authors’ contributions}

Congfeng Qiao conceived the idea and supervised the project. Shengqi Zhang and Congfeng Qiao contributed to the writing of the manuscript. All authors read and approved the final manuscript.

{\bf Funding}

This work is supported by the National Key Research and Development Program of China under Contract No. 2025YFA1613900, by the National Natural Science Foundation of China (NSFC) under the Grants 12475087, 12235008, and 12547114.

{\bf Data availability}

Data are available from the corresponding author upon reasonable request.

{\bf Competing interests}

The authors have no competing interests.

\end{document}